\documentclass[12pt,letterpaper]{article}
\pdfoutput=1
\usepackage{graphicx,array}
\usepackage{color}
\usepackage{latexsym}
\usepackage{amsthm}
\usepackage{amsmath}
\usepackage{enumitem}
\usepackage{amssymb}
\usepackage{hyperref}
\usepackage[hang,flushmargin]{footmisc} 
\usepackage{youngtab}
\usepackage{tikz}
\def\simgt{\mathrel{\lower2.5pt\vbox{\lineskip=0pt\baselineskip=0pt
           \hbox{$>$}\hbox{$\sim$}}}}
\def\simlt{\mathrel{\lower2.5pt\vbox{\lineskip=0pt\baselineskip=0pt
           \hbox{$<$}\hbox{$\sim$}}}}

\newcommand{\beq}{\begin{equation}}
\newcommand{\eeq}{\end{equation}}
\newcommand{\bea}{\begin{eqnarray}}
\newcommand{\eea}{\end{eqnarray}}
\newcommand{\be}{\begin{eqnarray}}
\newcommand{\ee}{\end{eqnarray}}

\newcommand{\Z}{\mathbb{Z}}

\def\tilde{\widetilde}
\def\hat{\widehat}

\def\CO{{\mathcal O}}

\def\CS{{\mathcal S}}

\def\cf{\mathcal{F}}

\def\ts{{\mathtt s}}
\def\tv{{\mathtt v}}
\def\tt{{\mathtt t}}

\newcommand{\polp}{\epsilon^\parallel}
\newcommand{\polo}{\epsilon^\perp}

\newcommand{\tyng}{\tiny\yng}

\newcommand{\syng}[1]{\scalebox{0.2}{\yng(#1)}}

\definecolor{nicered}{rgb}{0.7,0.1,0.1}
\definecolor{nicegreen}{rgb}{0.1,0.5,0.1}

\def\denom{{\mathtt D}}
\definecolor{mGreen}{rgb}{0,0.6,0}
\definecolor{mgray}{rgb}{0.6,0.6,0.6}
\definecolor{mpurple}{rgb}{0.58,0,0.82}
\definecolor{backgroundColour}{rgb}{0.95,0.95,0.92}

\definecolor{mred}{rgb}{0.5,0.0,0.0}
\definecolor{mgreen}{rgb}{0.0,0.4,0.0}
\definecolor{mblue}{rgb}{0.0,0.0,0.6}
\definecolor{myellow}{rgb}{0.4,0.4,0.0}
\definecolor{mpink}{rgb}{0.4,0.0,0.4}
\definecolor{mcyan}{rgb}{0.0,0.4,0.4}
\definecolor{mblack}{rgb}{0.0,0.0,0.0}

\def\e{{\epsilon}}

\def\g{{\gamma}}

\def\a{{\alpha}}
\def\b{{\beta}}

\def\d{{\delta}}

\def\CO{{\mathcal O}}

\def\CS{{\mathcal S}}

\newcommand{\bd}[1]{\begin{fmffile}{#1}\begin{fmfgraph*}}
		\newcommand{\ed}{\end{fmfgraph*}\end{fmffile}}

\def\0{{(0)}}
\def\1{{(1)}}
\def\2{{(2)}}
\def\3{{(3)}}
\def\4{{(4)}}

\def\+{{(+)}}
\def\-{{(-)}}

\newcommand{\ba}{\begin{align}}
\newcommand{\ea}{\end{align}}
\def\be{\begin{equation}}
\def\ee{\end{equation}}
\def\beq{\be\begin{array}{c}}
	\def\eeq{\end{array}\ee}

\numberwithin{equation}{section}

\begin{document}
\thispagestyle{empty}
\hfill
TIFR/TH/19-37
\vspace{2cm}
\begin{center}
{\LARGE\bf
Classifying and constraining local four photon 
and four graviton S-matrices}\\
\bigskip\vspace{1cm}{
{\large Subham Dutta Chowdhury${}^{1,a}$, Abhijit Gadde${}^{2,a}$, Tushar Gopalka${}^{3,a,c}$, Indranil Halder${}^{4,a}$, Lavneet Janagal${}^{5,a,b}$, Shiraz Minwalla${}^{6,a}$}\let\thefootnote\relax\footnotetext{email:\\ ${}^{1}$subham@theory.tifr.res.in,\,${}^{2}$abhijit@theory.tifr.res.in,\,${}^{3}$tushar.gopalka@students.iiserpune.ac.in,\\${}^{4}$indranil.halder@tifr.res.in,\,${}^{5}$lavneet@kias.re.kr,\,${}^{6}$minwalla@theory.tifr.res.in} 
} \\[7mm]
 {\it${}^{a}$ Department of Theoretical Physics, \\ 
 Tata Institute for Fundamental Research, Mumbai 400005}\\
[4mm]
{\it ${}^{b}$School of Physics, Korea Institute for Advanced Studies,\\
\it	Seoul 02455, Korea.}\\
[4mm]
{\it ${}^c$ Indian Institute of Science Education and Research, Pune - 411008}
 \end{center}
\bigskip
\centerline{\large\bf Abstract}

\begin{quote} \small
We study the space of all kinematically allowed four photon and four graviton S-matrices, polynomial in scattering momenta. We demonstrate that this space is the permutation invariant sector of a module over the ring of 
polynomials of the Mandelstam invariants $s$, $t$ and $u$. 
We construct these modules for 
every value of the spacetime dimension $D$, and so explicitly count and parameterize the most general four photon and four graviton S-matrix at any given derivative order. We also  explicitly list the local Lagrangians that give rise to these S-matrices. We then conjecture that the Regge growth of S-matrices in all physically acceptable classical theories is bounded by $s^2$ at fixed $t$. A four parameter subset of the polynomial photon S-matrices constructed above satisfies this Regge criterion. For gravitons, on the other hand, no polynomial addition to the Einstein S-matrix obeys this bound for $D \leq 6$. For $D \geq 7$  there is a single six derivative polynomial Lagrangian consistent with our conjectured Regge growth bound. Our conjecture thus implies that the Einstein four graviton S-matrix does not admit any physically acceptable polynomial modifications for $D\leq 6$. A preliminary analysis also suggests that every finite sum of pole exchange contributions to four graviton scattering also such violates our conjectured Regge growth bound, at least when $D\leq 6$, even when the exchanged particles have low spin. 
\end{quote}

\newpage

\tableofcontents

\newpage

\section{Introduction}

\subsection{Motivation} 

Consider a compactification of Type II string theory on 
$R^p \times M_{10-p}$\footnote{As usual $M_{10-p}$ must be a manifold whose worldsheet sigma model is a $(1,1)$ superconformal field theory with ${\hat c}=10-p$. For example, when $p=4$ $M_{6}$ could be a  Calabi-Yau manifold.}. The string spectrum on this background includes four dimensional gravitons. Graviton scattering amplitudes at loop level are sensitive probes of the detailed structure of the manifold $M_{10-p}$. At genus zero, however, graviton scattering amplitudes depend on $M_{10-p}$ only through an overall multiplicative factor.	
When expressed in terms of $G_p$, the effective $p$ dimensional Newton constant, these tree amplitudes are completely independent of $M_{10-p}$. These amplitudes are also the same in IIA theory, IIB theory  and Type I theory (see Appendix \ref{uct} for a discussion).

The universality of tree level graviton scattering amplitudes is a special case of a broader phenomenon. Consider the set of all worldsheet vertex operators that are identity in the 
$M_{10-p}$ sector and are invariant separately under $(-1)^{F_L}$, $(-1)^{F_R}$ and $\Omega$ in the $R^p$ sector\footnote{$F_L$ and  $F_R$ 
	are the left and right moving worldsheet Fermion number operators while $\Omega$ is the worldsheet orientation reversal operator.}. Let the collection of spacetime particles corresponding to the BRST cohomology classes of all such vertex operators be denoted by $C^{{\rm II}}_p$~\footnote{$C^{{\rm II}}_p$ is, of course,  an infinite collection of particles of and includes particles with arbitrary spins.}. Consider the collection of all tree level S-matrices\footnote{Expressed in terms of $G_p$.} with {\it every} external particle in $
C^{{ \rm II}}_p$. The general result is that these scattering amplitudes are  {\it all} universal\footnote{In the sense that they are independent of $M_{10-p}$, and are also the same in IIA theory, IIB theory  and Type I theories.}; moreover {\it every} pole in {\it each} of these S-matrices results from the exchange of a particle in $C^{{\rm II}}_p$. There are no poles from the exchange of particles outside the sector 
$C^{{ \rm II}}_p$. These facts - which follow immediately from the general structure of string worldsheet perturbation theory (see Appendix \ref{uct})- have a striking target space interpretation.  They imply that the classical target space dynamics of the sub-sector $C^{{\rm II}}$ -which we schematically denote by  $S(C^{{ \rm II}}_p)$-  is a {\it universal consistent truncation} of classical type II (or type I) string theory on $R^p \times M_{10-p}$.

The discussion of the previous paragraphs has an immediate generalization to Heterotic compactifications. Classical graviton scattering amplitudes for the Heterotic string on $R^p \times M_{10-p}$ are universal (independent of $M_{10-p}$)
once they are expressed in terms of $G_p$. Once again gravitons in such Heterotic compactifications belong to a  
collection of particles $C^{{\rm H}}_p$
Heterotic theories admits a consistent truncation to the universal dynamical 
system $S(C^{{\rm H}}_p)$. $S(C^{{\rm H}}_p)$ is a universal consistent truncation of the classical dynamics of all Heterotic string theories on $R^p \times M_{10-p}$.

In the limit $g_s \to 0$ \footnote{$g_s$ is the string coupling.} graviton scattering amplitudes for type II/ Heterotic theory on $R^p \times M_{10-p}$ reduce to the tree amplitudes computed using $S(C^{{ \rm II}}_p)$ or $S(C^{{ \rm H }}_p)$. Similarly, in the low energy limit string scattering amplitudes in a wide class of compactifications 
(not necessarily at small $g_s$) reduce to tree amplitudes 
computed using the Einstein action
$$S^{\rm Einstein}= \int \sqrt{-g} R. $$
We are not aware of any other classical theory which accurately captures string scattering amplitudes in any parametric limit of string theory.\footnote{Classical theories that admit a consistent truncation to one of the three theories - say e.g. 
	$S(C^{{ \rm II}}_p)$ - yield the same result for gravitational S-matrices as $S(C^{{ \rm II}}_p)$ itself. 
	For our purposes we thus regard all such theories as equivalent to $S(C^{{ \rm II}}_p)$.}.

The richness of the now known examples of consistent string compactifications has dampened early hopes that constraints imposed by consistency alone would permit a simple classification of {\it quantum} theories of gravity. The discussion above suggests that the situation is much more hopeful in the 
{\it classical} limit. As discussed above, the zoo of known quantum theories of gravity displays a surprising universality in the classical limit. Indeed, to the best of our knowledge, the data from all known string compactifications is consistent with the following bold conjecture\footnote{Conjectures 1-3 were outlined by one of us in a talk at String 2018 \cite{ShirazTalk}. Suggestions very similar to these have also been made over a period of several years by Nima Arkani Hamed (private communication); similar considerations may also have partly motivated the analysis of \cite{Camanho:2014apa}}
\begin{itemize} 
	\item
	{\bf Conjecture 1:} There exist exactly three classical gravitational S-matrices\footnote{i.e. S-matrices that are analytic functions of momenta apart from poles.}
	that are consistent with a set of physically motivated `low energy' constraints (including stability of the vacuum,  factorization on poles, causality and positivity of energy). These are the Einstein S-matrix generated by $S^{\rm Einstein}$, the type II S-matrix generated by $S(C^{{ \rm II}}_p)$ and the Heterotic S-matrix generated by $S(C^{{ \rm H}}_p)$. 
	\newline\\
	By  the phrase `classical gravitational S matrices' in the statement above we mean the collection of all S matrices with external particles taken to be any members of minimal classical truncation that includes gravity in the theory under study and not just the S matrices of gravitons themselves.
	
	Conjecture 1 is very striking but it is also extremely bold, and the evidence in its favour is, as yet, rather limited (see Appendix \ref{cono} for a brief discussion). We will not directly study this conjecture - which may well turn out to be incorrect as stated - 
	in this paper. We have nonetheless included a discussion of Conjecture 1 because it implies (but is not implied by) the considerably weaker Conjecture 2 below, which is directly relevant to the analysis of this paper:
	\item
	{\bf Conjecture 2:} The only consistent classical gravitational S-matrix whose exchange poles are 
	bounded in spin is the Einstein S-matrix.
	\newline\\
	Conjecture 2 in turn implies (but is not implied by) the 
	still  weaker conjecture:
	\item
	{\bf Conjecture 3:} The only consistent classical gravitational S-matrix with only graviton exchange poles 
	is the Einstein S-matrix.
\end{itemize} 

We have used observations about string compactifications to motivate this `Russian doll set' of three successively weaker conjectures. Once motivated, however, these conjectures can be studied on their own terms without any reference to string theory. The direct study, refinement  and possible eventual proof of these conjectures 
appears to us to be a very interesting research program. In this paper we will initiate (or more accurately continue) a study of these conjectures. We will make some progress towards establishing conjectures 3 and 2. We will not directly study conjecture 1 in this paper; however our technical results may have bearing on related studies in the future. 

\subsection{Three graviton Scattering} 

The three conjectures described in the previous subsection 
apply to $n$ graviton scattering for all $n \geq 3$. The case $n=3$ is particularly simple. It follows 
from kinematical considerations that the most general 
3 graviton S-matrix is a linear combination of the 
two derivative structure ( see \eqref{R1sm}), the four derivative 
structure (see \eqref{R2sm}) and the six derivative structure 
(see \eqref{R3sm}). In other words the most general 3 graviton S-matrix in any theory of gravity - classical or quantum - is specified by three real numbers. 

In a classic paper whose results have partly motivated the  current work \footnote{Conversations with N Arkani-Hamed, over a period of several years, also form part of the motivation for the current work.}, Camanho, Edelstein, Maldacena and Zhibeodov (CEMZ) \cite{Camanho:2014apa} demonstrated that a classical theory with 3 gravitational S-matrices that include a non-zero admixture of the four derivative and six derivative three graviton structures {\it necessarily} violates causality unless its four graviton scattering amplitude include contributions from the exchange of poles of arbitrarily high spin. The constraints follow from a particular sign of the Shapiro-time delay which in turn corresponds to the sign of the phase shift in flat space. In AdS similar argument has been made in \cite{Kulaxizi:2017ixa}. 

Constraints of causality have also been used in the past to constrain sign of certain higher derivative terms in the low energy Lagrangian \cite{Adams:2006sv, Bellazzini:2015cra}. 

It thus follows from the results of \cite{Camanho:2014apa} that the three graviton scattering amplitude is necessarily two derivative - i.e. that of the Einstein theory  - in any causal classical theory of gravity whose four graviton S-matrices have exchange contributions that are bounded in spin. In other words CEMZ have already established Conjecture 2 of the previous subsection for the special case of 3 graviton 
scattering\footnote{The authors of \cite{Camanho:2014apa} also rule out exchanges of particles of spin $>2$ when spectrum of such particles is bounded in spin. They do this using the classical Regge growth conjecture stated at the beginning of section \ref{crg-conj}.}. The uniqueness of graviton three point function  has also been demonstrated using causality in conformal field theory \cite{Afkhami-Jeddi:2016ntf, Afkhami-Jeddi:2017rmx}. In an alternative approach, used in \cite{Kologlu:2019bco}, the authors show that if the scattering amplitude obeys a bound in the Regge limit then the effect of shockwaves on a probe commutes which in turn implies the uniqueness of the Einstein gravity three point function.

The CEMZ result already  makes a case for the validity of 
Conjecture 2. We should, however, be careful 
not to overstate the strength of this evidence. 
As we have reviewed above, three graviton S-matrix is specified by a finite number of parameters because 
it is kinematically special. On the other hand four and higher point scattering amplitudes are specified by a finite number of functions of kinematical invariants ($s$ and $t$ in the case of four graviton scattering) and so an infinite number of real 
parameters. While the Einstein three graviton scattering amplitude describes a surface of co-dimension 2 in the space of all kinematically allowed 3 graviton scattering amplitudes, the four point (or higher) scattering amplitude describes 
a surface of infinite codimension in the space 
of all kinematically allowed classical four
(or higher) graviton scattering amplitudes. 

It follows that CEMZ type result for four graviton scattering would qualitatively strengthen the evidence for Conjectures 2 or 3. In the 
rest of this paper we will focus on the study 
of four graviton scattering.  We will first 
present an exhaustive kinematical classification and parameterization of local classical four graviton S-matrices and then attempt to cut down the space of allowed S-matrices by proposing a physical criterion 
that acceptable S-matrices must obey. 

\subsection{Classification and parameterization of polynomial four graviton 
	scattering amplitudes} \label{cpafca}

Consider any classical theory of gravity
interacting with other fields. We assume 
that the equations of motion of our theory 
are local, i.e. the number of derivatives 
is finite. When expressed as functions of 
the polarizations $\epsilon_i$ and momenta 
$p_i$ the four graviton S-matrices of any such theory may have poles corresponding to the exchange of particles. Let us focus on the pole corresponding the exchange of the particle $P$. The residue of this pole is completely fixed by the on-shell 
three particle $g g P$ scattering amplitude. 

Now all three particle scattering amplitudes 
are kinematically fixed to be a linear combination of a finite number - in this case lets say $d_{g}(P)$ - structures. It thus follows that the residue of the four graviton scattering amplitude is a quadratic form 
in $d_{g}(P)$ undetermined constants (the coefficients behind the $d_{g}(P)$ structures in the three point function). The precise structure of this quadratic form is completely fixed by kinematical considerations. The most general exchange S-matrix is a sum over such a structure for every exchanged particle.
Once all pole contributions have been subtracted out, the rest of the amplitude is a polynomial
in polarizations and momenta. Below we often 
refer to this as the `analytic part' or the `polynomial part' of the scattering amplitude.

As there is no definite bound on the degree of 
the polynomial part of the S-matrix, the number of parameters needed to specify the 
most general polynomial S-matrix is not finite. However polynomial S-matrices can 
be graded by their dimension (i.e. number of 
powers of $p_i$). The number of parameters, 
$n(m)$, needed to specify the most general 
dimension $m$ S-matrix is finite. It is 
convenient to encapsulate the information of 
$n(m)$ for all $m$ in the S-matrix partition 
function 
\begin{equation}\label{Smpfintro}
Z_{ {\text {S-matrix}}}(x) = \sum_{m=0}^\infty n(m)x^m.
\end{equation}
In the classification part of this paper, among other things, we present explicit results for the partition
function  \eqref{Smpfintro} separately 
for parity odd and parity even structures, 
and separately in every dimension. We now 
briefly explain how we obtain these results, 
and at the same time  accomplish an explicit construction of the space of polynomial S-matrices. 

Not every polynomial in the polarizations and 
momenta is an acceptable S-matrix. S-matrices
have to satisfy three additional constraints. 
First, all acceptable S-matrices must be 
Lorentz invariant (all indices must contract 
in pairs or with a Levi-Civita tensor). 
Second, the S-matrices must obey 
the constraints of gauge invariance 
( see \eqref{tranfeph}). Finally, the S-matrices of four identical Bosonic particles must enjoy invariance under the permutation group $S_4$ that permutes the data $(\epsilon_i, p_i)$, 
$i=1 \ldots 4$ of the four scattering particles.

For a reason that will soon become clear, it 
turns out to be useful to impose the constraint of $S_4$ invariance in two steps. $S_4$ has a $\Z_2\times \Z_2$ normal subgroup that leaves the kinematical variables $s$, $t$, and $u$ (see \eqref{stu}) invariant. The coset space obtained by modding $S_4$ out 
by $\Z_2\times \Z_2$ is $S_3$ (see \eqref{permgp}). It follows that we can 
impose $S_4$ invariance by first imposing the  
constraint of $\Z_2\times \Z_2$ invariance 
and later imposing the constraint of 
$S_3$ invariance on the resultant structure. We refer to the set of Lorentz, gauge 
and $\Z_2\times \Z_2$ invariant polynomials 
of $\epsilon_i$ and $p_i$ as `{\it quasi invariant}' S-matrices. As $s$, $t$ and $u$ 
are all individually Lorentz, gauge and 
$\Z_2 \times \Z_2$ invariant, it follows 
that the product of a quasi invariant S-matrix 
and any polynomial of $s$, $t$ and $u$  is 
itself also a quasi invariant S-matrix. 
In mathematical parlance the set of gauge and 
Lorentz invariant polynomials of $p_i$ and 
$\epsilon_i$ is a {\it module} - which we 
call the local module (see subsection \ref{lsm})  -  over the ring of polynomials  $s$, $t$ and $u$. It turns out 
that the most useful way to think of the 
space of polynomial S-matrices is to think of it as the  $S_3$ invariant part of the local 
module. 

The local module is finitely generated and so  is completely characterized by its generators, which in turn are labeled by their $S_3$ 
transformations properties and their dimension (i.e. derivative order). In mathematics, the simplest modules are freely generated. It turns out that the local module of quasi invariant S-matrices is freely generated when $D\geq 5$. In this case the 
partition function \eqref{Smpfintro} is 
completely determined by the spectrum of 
generators of the module by
\begin{equation} 
Z_{ \text{S-matrix}}(x) = \sum_{J} x^{\Delta_J} 
Z_{{\bf R_J}}(x)
\end{equation} 
 where the index $J$ labels the generators of the local module (note, of course, that $J$ has nothing to do with angular momentum), $\Delta_J$ is the `dimension' (more accurately derivative order) of the $J^{th}$ generator, 
 ${\bf R_J}$ is the representation of $S_3$ in which 
 the $J^{th}$ generator transforms and the  functions $Z_{{\bf R}}(x)$ are listed in \eqref{partfnsss}.  

The local module is not freely generated 
when a linear combination of the descendants 
of different generators vanishes. We refer 
to all such linear combinations as `relations'
or sometimes as `null states'. The set of 
relations themselves form a module. The 
relation module is completely characterized 
by {\it its} generators.
In the context under study in this paper it 
turns out that the relation modules are 
themselves always freely generated (i.e. there
are no relations for relations). When the local module is not 
freely generated - this turns out to be the case when $D\leq 4$ -   it follows that

\begin{equation} 
Z_{ \text{S-matrix}}(x) = \sum_{J} x^{\Delta_J} 
Z_{{\bf R_J}}(x) - \sum_{I} x^{\Delta_I} 
Z_{{\bf R_I}}(x)
\end{equation} 
where the sum over $J$ runs over all module generators, while the sum over $I$ runs over all relation generators;
$\Delta_I$ and ${\bf R_I}$ are the dimension and $S_3$ 
label of the $I^{th}$ relation generator.

The reason that the local module is not freely generated in 
$D=4$ (and $D=3$) presumably has to do with the fact that 
scattering in these dimensions is very special. In particular scattering in $D=4$ is often more conveniently 
expressed in the spinor helicity formalism than in terms 
of polarizations as in this paper. Using this formalism a great deal is known about photon and graviton S matrices 
in $D=4$ (see e.g. \cite{Arkani-Hamed:2017jhn} and references therein). It would 
be interesting to re derive the results of this paper - specialized to $D=4$ - using this formalism. We leave this 
to future work.

It follows that a complete identification of the generators of the local module (and of the 
relation module when it exists) immediately 
determines the partition function over 
S-matrices \eqref{Smpfintro}. In sections 
\ref{mgls}, \ref{gravsm} and related Appendices
we have explicitly presented all generators 
of the local module, and also all generators 
of the relations modules (when they exist). 
We have also presented an explicit parameterization of all the $S_3$ invariant
descendants of these generators, and so 
an explicit parameterization of all polynomial 
S-matrices both for the case of 4 photon as 
well as the case of 4 graviton scattering. The 
results of these sections permit an immediate 
computation of the partition function over 
S-matrices. The final results for the case of 
photons and  gravitons are given in Table \ref{introresp}  and \ref{introresg}. The 
results in this table are presented in terms 
of the quantity $\denom$. \footnote{$\denom=\frac{1}{(1-x^4)(1-x^6)}$ is not to be confused with the number of Space-time dimensions $D$.}  
\begin{table}
	\begin{center}
		\begin{tabular}{|l|l|l|}
			\hline
			dimension & even partition function & odd partition function \\
			\hline
			$D\geq  10$ & $x^4(2+3x^2+2x^4)\denom $ & $0$\\
			\hline
			$D=9$ & $x^4(2+3x^2+2x^4)\denom $ & $0$\\
			\hline
			$D=8$ & $x^4(2+3x^2+2x^4)\denom $ & $0$ \\
			\hline
			$D=7$ & $x^4(2+3x^2+2x^4)\denom $ & $x^3\denom $ \\
			\hline
			$D=6$ & $x^4(2+3x^2+2x^4)\denom $ & $x^{12}\denom $  \\
			\hline
			$D=5$ & $x^4(2+3x^2+2x^4)\denom$ & 0\\
			\hline
			$D=4$ & $x^4(2+3x^2+2x^4-(x^4+x^6))\denom$ & $x^4(1+2x^2+x^4-(x^4+x^6))\denom$  \\
			\hline
			$D=3$ & $x^4(1+x^2+x^4-(x^4+x^6))\denom $  & $x^9\denom$    \\
			\hline
		\end{tabular}
	\end{center}
	\caption{Final Result for the partition function over 4 photon S-matrices. Here $\denom=\frac{1}{(1-x^4)(1-x^6)}$ }
	\label{introresp}
\end{table}

\begin{table}
	\begin{center}
		\begin{tabular}{|l|l|l|}
			\hline
			dimension & Even partition function & Odd partition function  \\
			\hline
			$D\geq 10$ & $x^{8}(x^{-2}+6+9x^2+10x^4+3x^6)\denom$ & 0 \\
			\hline
			$D=9$ & $x^{8}(x^{-2}+6+9x^2+10x^4+3x^6)\denom$ & 0 \\
			\hline
			$D=8$ & $x^{8}(x^{-2}+6+9x^2+10x^4+3x^6)\denom$ & 0 \\
			\hline
			$D=7$ & $x^{8}(x^{-2}+6+9x^2+10x^4+3x^6)\denom$ & $x^8(2x^{-1}+3 x+2x^3)\denom$\\
			\hline
			$D=6$ & $x^{8}(6+9x^2+10x^4+3x^6)\denom$ & $3x^{10}(x^2+x^4+x^6)\denom$  \\
			\hline
			$D=5$ & $x^{8}(4+7x^2+8x^4+3x^6)\denom$ & $x^{11}(x^2+x^4+x^6)\denom$  \\
			\hline
			$D=4$ & $x^{8}(2+2x^2+3x^4-x^6-x^8)\denom$ & $x^8(1+x^2+2x^4-x^6-x^8)\denom$ \\
			\hline
		\end{tabular}
	\end{center}
	\caption{Final Result for the partition function over 4 graviton S-matrices}
	\label{introresg}
\end{table}

In this paper we have also obtained the 
results of Tables \ref{introresp} and
\ref{introresg} in a second independent way
which we now describe. Every polynomial 
gauge invariant S-matrix can be obtained 
from the four graviton (or four photon) part of  
a finite derivative  gauge invariant contact term in  Lagrangian. As far as 
four photon / graviton terms are concerned, 
{\it almost} \footnote{We describe the exceptions below.}  
all gauge invariant Lagrangians 
are obtained by taking the products of derivatives of four field strengths / Riemann 
tensors. Lagrangians that differ off-shell but agree on-shell produce the same S-matrix (see section \ref{cll} for an explanation of a more careful version of this statement). For 
this reason the partition function over all 
polynomial S-matrices can be evaluated (up to a small error that is not difficult to separately account for, see below) 
 by computing the partition function 
over degree four polynomials of derivatives 
of the field strength / the Riemann tensor, 
modulo equations of motion and modulo total 
derivatives. We have explicitly evaluated these `Plethystic' partition functions in subsection \ref{pleth-spin} using matrix model techniques. The results of this evaluation are presented in Tables \ref{photon-plethystic} and \ref{graviton-plethystic} respectively. Similar techniques were used in \cite{Henning:2015daa} to compute the partition function on ``operator bases" in effective field theories and were generalized to compute partition function on S-matrices in \cite{Henning:2017fpj}. \footnote{We thank 
T. Hartman for drawing our attention to the papers  \cite{Henning:2015daa} and \cite{Henning:2017fpj} while commenting on a preliminary version of this paper.}

We have explained that the plethystic partition function evaluates the S-matrix partition function \eqref{Smpfintro} only up to a small error. The error has to do with the fact that it is sometimes possible to find gauge invariant Lagrangians  - e.g. Chern Simons Lagrangians - that cannot be constructed out of the product of (derivatives of) four field 
strengths or four Riemann tensors. Another source of error is that some local Lagrangians generate trivial S-matrices whenever they are total 
derivatives - even if they are total derivatives of objects (like Chern Simons terms) which themselves cannot be written as products of field strength operators. In sections 
\ref{mgls} and \ref{gravsm} we have carefully 
corrected for each of these errors. Once the corrections are taken 
into account Tables \ref{photon-plethystic} and \ref{graviton-plethystic} reduce to Tables 
\ref{introresp} and \ref{introresg} presented 
above, providing a highly non-trivial check of our construction of the most general  
four graviton and four photon S-matrices. 

In sections \ref{mgls} and \ref{gravsm} and 
relevant appendices we have also presented 
an explicit listing of the non linearly gauge 
invariant Lagrangian whose four photon / 
four graviton contact term produces each 
of the polynomial S-matrices that we have 
explicitly classified and listed in the same 
sections. 

\subsection{A conjectured bound on Regge scattering} \label{crg-conj}

The work reviewed above yields a complete classification 
and parameterization of all local four graviton S-matrices, analogous to (but much more complicated than) the 3 parameter parameterization of 3 graviton S-matrices. 
In their classic analysis \cite{Camanho:2014apa} were able to  the 
use physical criteria (the requirement of causality) to constrain the parameters that
appeared in the most general three point function. 
We will now attempt to do the same for the 
the four point function. More specifically, we will constrain 
classical theories using a conjecture on a bound of the Regge growth 
of classical scattering amplitudes that we now state
\begin{itemize} 
	\item {\bf Classical Regge Growth Conjecture:} The S-matrix of a consistent classical theory never grows faster than $s^2$ at fixed $t$ - at all physical values of momenta and for every possible choice of the normalized polarization vector $\epsilon_i$.
\end{itemize} 
The conditions for `consistency' of a classical theory in this conjecture are the same as those spelt out in Conjecture 1 earlier in this introduction. In the rest of this subsection we will summarize the evidence for the Classical Regge Growth (CRG) Conjecture. In the 
next subsection we will explore its consequences. 

The first piece of evidence in favor of the CRG conjecture 
is that it is always obeyed by two derivative theories involving particles of spin no greater than two - theories that we independently expect to be consistent. The two derivative nature of interactions ensures that both contact contributions as well as $s$ and $u$ channel exchange contributions  
grow no faster than $s$ in the Regge limit. $t$ channel exchange graphs on the other hand grow no faster than $s^J$ where 
$J$ is the spin of the exchanged particle. Since we have 
assumed $J \leq 2$, all contributions obey the CRG conjecture. 

The next piece of evidence in support of the conjecture described in this subsection is that it is obeyed by all classical string scattering amplitudes. Recall that, for instance the Type II string scattering amplitudes grow in the Regge limit like 
\begin{equation}\label{brgu}
s^{2 + \frac12 \alpha' t} 
\end{equation}
 As $t$ is negative at 
physical values of momenta it follows that this behavior 
obeys our conjecture. Note also that \eqref{brgu} reduces 
to $s^2$ in the limit $\alpha' \to 0$, matching with the 
fact that gravitational amplitudes, which grow like 
$\frac{s^2}{t}$ in the Regge limit, saturate the CRG bound.

The strongest evidence for the CRG conjecture follows from 
the observation that the CRG conjecture is tightly related to the chaos bound\cite{Maldacena:2015waa}. We pause to review how this works. Working in a large $N$ unitary CFT in $D \geq 2$ consider a four point $\langle OOOO\rangle $ where $O$ is a real scalar operator and the insertion points are taken to be 
$$\left(\pm\sinh\left(\frac{\tau}{2} \right), \pm\cosh\left(\frac{\tau}{2} \right) \right).$$
 All insertions are denoted by the doublet $(t,x)$; insertions all lie 
 completely in the $(t,x)$ plane.  The 
 authors of \cite{Maldacena:2015waa} used the unitarity of the CFT to demonstrate 
 that the growth of the connected correlator $\langle OOOO\rangle $ with boost times $\tau$ cannot be faster than $e^{\tau}$. This result holds in the large $N$ limit for boost times $\tau$ large compared to unity but small compared to $\ln N$.
 
 The chaos bound has also been used by Simon Caron-Huot to show that the OPE  coefficients are analytic functions of spin and to derive a powerful  ``inversion formula" for the same \cite{Caron-Huot:2017vep}. Physical aspects of the inversion formula have been clarified in \cite{Simmons-Duffin:2017nub}.
The application of the chaos bound to correlators of spinning operators was considered in \cite{Kravchuk:2018htv} \footnote{See \cite{Halder:2019ric, Mezei:2019dfv, Poojary:2018esz, Jahnke:2019gxr} for generalizations.}. 
 
 Let us now study a situation in which the CFT under study has a bulk dual. Let the
 bulk field dual to the operator $O$ be denoted by $\phi$. Let us suppose that 
 the fields $\phi$ in the bulk has a standard quadratic term and also has a have a local four point self interaction of the schematic form $ \phi \partial ...\partial \phi  \partial ...\partial \phi \partial...\partial \phi$. One can use the usual 
 rules of AdS/CFT to directly compute the correlator $\langle OOOO\rangle $ of the previous  paragraph, and so evaluate its growth with $\tau$. The authors of \cite{Cornalba:2007fs, Heemskerk:2009pn} were able 
 to carry through this computation for the most general bulk 
 contact term using the classical bulk theory. They discovered the following interesting fact.  Any bulk vertex which, in flat space, would give rise to an S-matrix that grows like  $s^{m+1}$ in the Regge limit,  turns out to give a contribution to $\langle OOOO\rangle $ that grows with boost time like $e^{m \tau}$.  It follows that at least in AdS space, any bulk interaction associated with a flat-space S-matrix that grows faster than $s^2$ at fixed $t$ leads to a boundary correlator that violates a field theory theorem, and so must classically inconsistent. A similar connection between CFT correlators and bulk exchange diagrams has been analyzed in \cite{Shenker:2014cwa}.
 
 Although the results described above have been carefully verified only for scalar operator insertions, we feel it is likely that they will continue to hold for insertions of all spins. The tight connection between the CRG and the Chaos bound, is, in our opinion, striking evidence in favor of the
 CRG conjecture. 

Note that the CRG conjecture immediately implies the non existence of a consistent interacting theory of higher spin particles (of bounded spin) propagating in flat space. Let the highest spin in the theory be $J>2$. As the spin $J$ particle is assumed to be interacting, there exists an S-matrix that receives 
contributions from spin $J$ exchange. In the $t$-channel this exchange contribution scales like $s^{J}$, violating 
the CRG conjecture. This argument has been used in \cite{Camanho:2014apa} to rule out the possibility of spectrum of particles of spin $>2$ that is bounded in spin.

\subsection{Consequences of the CRG conjecture} 

It is now possible to scan through the explicit results
of the classification of S-matrices reviewed in subsection 
\ref{cpafca} and determine the subclass of S-matrices that 
obeys the CRG conjecture. We first present the complete result of this exercise for polynomial S-matrices and then present 
some preliminary partial results for exchange amplitudes. 

\subsubsection{Polynomial amplitudes} 

\subsubsection*{The scattering of four identical scalars}

For $D \geq 3$ the most general local contact scalar interaction term 
that obeys the CRG bound is given by 
\begin{equation} \label{mjsit} 
S= \int d^Dx \left( a_1 \left( \phi^4 \right) + a_2\left( \phi^2 \partial_\mu \partial_\nu  \phi~ \partial_\mu \partial_\nu \phi  \right) + a_3\left( \phi^2  \partial_\mu \partial_\nu \partial_\rho \phi
~\partial_\mu \partial_\nu \partial_\rho \phi \right)  \right).
\end{equation} 
This implication of the CRG has been used effectively to compute the four point function of certain scalar operators in a theory with slightly broken higher spin symmetry \cite{Turiaci:2018dht}.

\subsubsection*{The scattering of four identical photons}

For $D \geq 4$ the most general parity even local contact four photon interaction term that obeys the CRG bound is given by 
\begin{equation} \label{mjsitp} 
S= \int d^Dx \left( a_1{\rm Tr}\left( (F^2) \right)^2  + a_2 {\rm Tr} \left( F^4\right)  + 
a_3{\rm Tr}\left( \partial_\mu F F \partial^\mu F F \right) + a_4 \left(F_{ab} {\rm Tr}\left( \partial_\a F \partial_b F  F \right) \right) \right).
\end{equation} 
(where matrix multiplication is implied and traces are over Lorentz indices of $F_{\mu\nu}$, and $a_1$, $a_2$, $a_3$ and $a_4$ are constants).
In $D =3$ the most general parity even local four photon interaction term that obeys the CRG bound is given by 
\begin{equation} \label{mjsitpth} 
S= \int d^3x \left( a_2 {\rm Tr} \left( F^4\right)  + 
a_3{\rm Tr}\left( \partial_\mu F F \partial^\mu F F \right) \right).
\end{equation} 

Additionally, in special dimensions there are some parity odd local contact Lagrangians that obey the CRG bound. These are 
the action
\begin{equation}\label{acthh} 
S= \int d^7x \left(A\wedge F \wedge F \wedge F\right)
\end{equation} 
in $D=7$ 
and the action  
\begin{equation} \label{dufact}
S= \int d^4x \left( b_1 {\rm Tr} \left(F \wedge F \right)  
{\rm Tr} \left( F^2 \right) + 
b_2 \epsilon_{mnab} F_{ab} Tr \left(\partial_m F \partial_n F F
\right) \right).
\end{equation} 
in $D=4$, 
where $b_1$ and $b_2$ are constants.

\subsubsection*{The scattering of four identical gravitons}

In $D \leq 6$ there are no contact four graviton Lagrangians 
consistent with the CRG conjecture. For $D \geq 7$ the unique 
such Lagrangian is the second Lovelock Lagrangian
\begin{equation} \label{sllintro}
\chi_6 = \int \sqrt{-g} \left(  \frac{1}{8}\,\, \delta_{[a}^{g}\delta_{b}^{h}\delta_{c}^{i}\delta_{d}^{j}\delta_{e}^{k}\delta_{f]}^{l}\,\,R_{ab}^{\phantom {ab}gh} R_{cd}^{\phantom{cd}ij}R_{ef}^{\phantom{ef}kl} \right). 
\end{equation} 
The S-matrix that follows from this Lagrangian is proportional 
to 
\begin{equation} \label{Smatrix} 
\left( \epsilon_1 \wedge \epsilon_2 \wedge \epsilon_3  \wedge 
\epsilon_4 \wedge p_1 \wedge p_2 \wedge p_3 \right)^2.
\end{equation} 
If we assume the validity of the CRG conjecture, the results just presented
amount to a proof of Conjecture 3 for the special case of 4 graviton 
scattering for $D \leq 6$.

\subsubsection{Exchange contributions} 

We have not yet completed a thorough analysis of the Regge
behavior of all possible exchange amplitudes; we leave this 
exercise for future work. In this subsection we summarize a 
few preliminary results. 

Exchange diagrams in the $t$ channel generically scale like 
$s^J$ where $J$ is the largest number of symmetrized indices 
in the representation that labels the exchanged particle. 
This $s^J$ scaling holds independent of the nature of the external particles, and so applies equally to scalar, photon and graviton scattering. It follows that the exchange of particles with $J \geq 3$ generically violates the CRG conjecture. Exchange of particles with $J \leq 2$ never 
violates the CRG bound in the $t$ channel but may violate 
this conjecture in the $s$ and $u$ channels. As the contribution in the $s$ channel and $u$ channel is analytic in $t$, the violations in these channels mean violations at zero impact parameter. This is in contrast with the t-channel where the CRG bound for $J\geq 2$ is violated at finite impact parameter\footnote{We thank Douglas Stanford for highlighting this issue to us.}.

The behavior of scattering in the $s$ and $u$ channels is 
sensitive to the nature of the external scattering particles. In the case of four external scalars or four external photons it is easy to find examples of exchange contributions that do not violate the CRG bound. We have, for example, explicitly computed the contribution to four photon scattering from the exchange of a massive particle
of arbitrary mass and demonstrated that its Regge growth 
is slower than $s^2$. In the case of external gravitons, on the other hand, we have shown by explicit computation that the exchange of massive scalars or massive spin two particles always leads to an S matrix that violates the CRG bound - and moreover violates it in a manner that cannot be canceled by a compensating local contribution\footnote{The AdS version of this statement has previously been argued in  \cite{Meltzer:2017rtf, Afkhami-Jeddi:2018own}. They have shown that in a large  $N$ CFT with a spectrum that has a large gap $\Delta_{\rm gap}$ to higher spin (spin $3$ or higher) single trace primaries, the three point function coefficient $\langle T_{\mu\nu} T_{\rho \sigma} O\rangle$ where $O$ is either a scalar, a spin two operator,  is shown to be suppressed by inverse powers of $\Delta_{\rm gap}$. In \cite{Meltzer:2017rtf}, the authors use  conformal Regge theory and in \cite{Afkhami-Jeddi:2018own}, the authors use  average null energy condition to establish the bounds on the three point function coefficients. In the present context,  the result should mean that coupling between two gravitons and $\phi$, where $\phi$ is either a scalar,  a spin two particle, must vanish when there are no exchange poles corresponding to higher spin particles. }. The same is also true of exchange of a massless spin two graviton whenever the three graviton scattering amplitude deviates from the Einstein form. More generally we have demonstrated in section \ref{ec} - under some hopefully reasonable assumptions - that {\it every} exchange contribution to four graviton scattering in  $D \leq 6$ - other than graviton exchange with the Einstein three point scattering - violates the CRG bound in a way that cannot be compensated for by local counter term contributions. 

Tightened up versions of the arguments described above  would amount to a proof of Conjecture 2 for the special case of four graviton scattering for $D \leq 6$. Of course the proof would rely on the validity 
of the CRG conjecture.

\subsection{Organization of this paper}

In section \ref{gsms} below we present a discussion of the 
general structural features of S-matrices for four identical 
bosons, focusing especially on the module structure of polynomial S-matrices. In section \ref{index-structure} we present an explicit 
construction of the so called bare module, a freely generated 
module of `index structures'. The importance of this construction lies in the fact that the physical local module is a submodule of the bare module. In section 
\ref{cll} we present a discussion of local Lagrangians 
that give rise to local S-matrices, and present 
the results of the plethystic counting of local Lagrangians described above. In sections \ref{mgls} and \ref{gravsm} we present a detailed explicit 
construction of the local modules of quasi invariant 
S-matrices for four photon and four graviton scattering. We also present an explicit parameterization of the most general polynomial S-matrices and the contact Lagrangians that give rise to 
these S-matrices. In section \ref{ec} we present a
preliminary discussion of the contribution of exchange diagrams to four scalar, four photon and four graviton scattering. We also argue that all such contributions
to four graviton scattering appear to violate the CRG bound at least for $D \leq 6$. In section \ref{conc} 
we end with a discussion of open questions and future directions. The details of several calculations and 
expressions are relegated to appendices. 

\section{Generalities of $2 \rightarrow 2$ S-matrix}
\label{gsms}

In this section we review\footnote{In particular  \ref{scat} \ref{Permsfst}, \ref{irrep} contain only well known results, and have been included in this
paper only to establish notation and to jog the reader's memory.} and discuss the general structural features  of $2 \rightarrow 2$ S-matrices of four identical bosonic scalars, photons or gravitons in an arbitrary number of spacetime dimensions. 

\subsection{Scattering data} \label{scat}

\subsubsection{Momenta} \label{mom}

Consider the scattering of four massless particles in $D$-dimensional Minkowski space. Let $p_i^\mu$ be momentum of the $i^{\rm th}$ particle. The masslessness of the scattering particles and momentum conservation means
\begin{equation}\label{osmc}
p_i^2=0,\qquad \sum_{i=1}^{4}p_i^\mu=0.
\end{equation}
 
We use the mostly positive convention and define Mandelstam variables,
\begin{equation}\label{stu} \begin{split}
s&:=-(p_1+p_2)^2=-(p_3+p_4)^2=-2 p_1.p_2 =-2 p_3.p_4\\
t&:=-(p_1+p_3)^2=-(p_2+p_4)^2=-2 p_1.p_3=-2 p_2.p_4\\
u&:=-(p_1+p_4)^2=-(p_2+p_3)^2=-2 p_1.p_4=-2 p_2.p_3.
\end{split}
\end{equation}
The equalities in \eqref{stu} follow from \eqref{osmc}. Thanks to momentum conservation $s+t+u=0$. In the rest of the paper when we need to make a specific choice of independent Mandelstam variables we will usually take these to $s$ and $t$\footnote{In the special case $D=2$ the kinematics of four particle scattering degenerates and  variables  $u$ and $t$ can be solved for in terms of $s$.  Through this paper we will, however, always assume that $D\geq3$, and so $s$ and $t$ are always independent variables. }.

\subsubsection{Polarizations} \label{pol}

The $2 \rightarrow 2$ S-matrix is a Lorentz invariant complex valued function of the momenta $p_i$, together with the data  that specifies the internal or spin degree of freedom of each scattering particle. 

\subsubsection*{Scalars}

Scalar particles have no internal degrees of freedom, so $2 \rightarrow 2$ scalar S-matrices are functions only of momenta. For $D \geq 4$ Lorentz invariance ensures that S-matrices are, in fact, functions only of 
$s$ and $t$. In the special case $D=3$, four scalar S-matrices can be either parity even or parity odd. Parity even S-matrices are simply a function of $s$ and $t$ as in higher dimensions. Parity odd S-matrices are given by 
$\epsilon_{\mu \nu \rho} p_1^\mu p_2^\nu p_3^\rho$ 
times a second function of $s$ and $t$.

\subsubsection*{Photons}

The internal degree of freedom of a photon may be taken to be its polarization vector $\epsilon_i^\mu$. In the Lorentz gauge (which we use throughout this paper in order to preserve manifest Lorentz invariance),  
\begin{equation}\label{mli}
\epsilon_i\cdot p_i=0
\end{equation}
The S-matrix must also be  invariant under residual gauge transformations i.e. under the transformations, 
\begin{equation}\label{tranfep}
\epsilon_i^\mu \rightarrow   \epsilon_i^\mu + \zeta(p_i) p_i^\mu
\end{equation} 
We will sometimes use the notation 
$$\zeta(p_i)= \zeta_i.$$
As $\zeta(p_i)$ is a completely arbitrary function of $p_i$ \footnote{And as two of the particles participating in a four particle scattering event never have identical $D$-momenta.} the four numbers $\zeta_i$ 
can be varied independently of each other. 
It follows that the requirement of gauge invariance is simply the condition that the S-matrix is separately invariant under each of the transformations 
\begin{equation}\label{tranfeph}
\epsilon_i^\mu \rightarrow   \epsilon_i^\mu +
\zeta_i p_i^\mu
\end{equation} 
separately for each $i$.

To summarize, a $4$-photon S-matrix is a Lorentz invariant complex valued function $\CS(p_i,\epsilon_i)$, subject to the condition \eqref{mli}. It depends linearly on each of the four polarizations $\epsilon_i^\mu$ and is separately invariant under each of the four shifts \eqref{tranfeph}.

\subsubsection*{Gravitons}

The internal degrees of freedom of a graviton can be parameterized by its traceless symmetric polarization tensor  $h_i^{\mu\nu}$. In 
Lorentz gauge, 
\begin{equation}\label{gpi}
h_i^{\mu\nu} p^\nu_i=0
\end{equation}
As before, the S-matrix enjoys invariance under residual gauge transformations,  
\begin{equation}\label{gth}
h^{\mu\nu}_i \rightarrow   h_i^{\mu\nu} + \zeta_i^\mu p^\nu_i +\zeta_i^\nu p^\mu_i,  \qquad {\rm where}\quad \zeta_i\cdot p_i=0.
\end{equation}
Through most of this paper we will find it convenient to specialize to the special choice of polarization 
\begin{equation} \label{spcpol}
h^i_{\mu\nu}=
\epsilon^i_\mu \epsilon^i_\nu \qquad {\rm where}\quad k_i\cdot \epsilon_i=0, ~~\epsilon_i\cdot \epsilon_i=0
\end{equation} 
The gauge transformation parameter $\zeta_i^\mu=\zeta_i \epsilon^i_\mu$, preserves the choice of the polarization \eqref{spcpol} and induces the gauge 
transformations  
\begin{equation} \label{spcgaugeeff}
\epsilon_i^\mu \rightarrow \epsilon_i^\mu + \zeta_i p_i^\mu.
\end{equation}
These transformations are same as the ones in \eqref{tranfep} \footnote{The special choice \eqref{spcpol} does not result in loss of generality. Let $S(\epsilon)$ denote the S-matrix with a single special choice of polarization,  $h_1^{\mu\nu}=\epsilon^\mu \epsilon^\nu$. Then the linear combination $$S(u+v) -S(u) -S(v),$$ where $u$ and $v$ are orthogonal polarization vectors, yields the S-matrix for the choice of polarization $h_1^{\mu\nu}=u^\mu v^\nu+v^\mu u^\nu$ and this sort of polarizations form  a basis for general symmetric traceless tensors $h_1^{\mu\nu}$. As the S-matrix is linear in  $h_1^{\mu\nu}$, the S-matrix with the choice \eqref{spcpol} carries the same information as the most general $4$-graviton S-matrix.}.

In conclusion, with the choice \ref{spcpol}, a $4$-graviton S-matrix $\CS(p_i,\epsilon_i)$, like the 
photon, is a Lorentz invariant complex valued function $\CS(p_i,\epsilon_i)$ but this time one that is a bilinear function of each of the $\epsilon_i$'s, subjected to the tracelessness condition $\epsilon_i\cdot \epsilon_i=0$.

\subsubsection{Unconstrained polarizations}\label{unconst}

In the previous subsubsection we have expressed the S-matrix 
as shift invariant functions of the  polarization vectors $\epsilon_i$. It is possible to simultaneously `solve' for the constraints on $\epsilon_i$ (tracelessness) and the 
constraints on the S-matrix (shift or gauge invariance) and 
re express the S-matrix as a function of independent unconstrained variables as follows. 

The momenta $p_i$ span a three dimensional subspace of $D$-dimensional Minkowski space. We refer to this subspace as the scattering 3-plane. The polarization vectors $\epsilon_i$ can be decomposed into part transverse to the scattering plane $\polo_i$ and part parallel to the scattering plane $\polp_i$
\begin{equation}\label{epsilondecomp} 
\epsilon_i= \polo_i + \polp_i
\end{equation} 
The condition $\epsilon_i .p_i= \polp_i.p_i=0$ forces
$\polp_i$ to lie in a two dimensional subspace of the scattering plane. Moreover the constraint that S-matrices are invariant under the shifts 
$\polp_i \rightarrow \polp_i + p_i$ 
tells us that the S-matrix is a function only of one of the 
two free components of $\polp_i$. It follows that - for the 
purpose of evaluating gauge invariant S-matrices - the set of 
inequivalent vectors $\polp_i$ may be parameterized by a single complex number $\alpha_i$. We choose the following (arbitrary) parameterization that obeys $\polp_i.p_i=0$.
\begin{equation}\label{perpp}
\begin{split}
&\polp_1 = \alpha_1\sqrt{\frac{st}{u}} \left( \frac{p_2}{s} - 
\frac{p_3}{t} \right) + a_1 p_1 \\
&\polp_2 = \alpha_2\sqrt{\frac{st}{u}} \left( \frac{p_1}{s} - 
\frac{p_4}{t} \right) + a_2 p_2\\
&\polp_3 = \alpha_3\sqrt{\frac{st}{u}} \left( \frac{p_4}{s} - 
\frac{p_1}{t} \right) + a_3 p_3\\
&\polp_4 = \alpha_4\sqrt{\frac{st}{u}} \left( \frac{p_3}{s} - 
\frac{p_2}{t} \right) + a_4 p_4.\\
\end{split}
\end{equation}
The numbers $a_i$ represent the freedom to shift $\epsilon_i$ 
by gauge transformations; $a_i$ are redundancies of description 
and will not show up in any gauge invariant physical result. 
On the other hand the parameters $\alpha_i$ are physical. In particular 
\begin{equation} \label{normalphai}
\polp_i. (\polp_i)^*=|\alpha_i|^2
\end{equation} 
(see also \eqref{manifest} below). 

With these definitions in place we can write 
\begin{equation}\label{exppol}
\epsilon_i= \polo_i + \polp_i.
\end{equation} 
Equations \eqref{exppol} and \eqref{perpp} express the 
$D$ component vector $\epsilon_i$ in terms of the $D-3$ 
component vector $\polo_i$ and the single parameter $\alpha_i$ and the redundant variables $a_i$.

Unlike $\epsilon_i$, the pair $(\polo_i, \alpha_i)$ are unconstrained data in the case of photons. 
In the case of gravitons the data still has 
to obey the single constraint\footnote{It is sometimes also useful to view the polarizations
$\epsilon_i$ as normalized according to the condition
\be
\epsilon_i\cdot\epsilon_i^*=1\qquad \Rightarrow \qquad |\polo_i|^2+|\alpha_i|^2=1.
\ee
Notice that $\epsilon_i$ and $\epsilon_i + p_i$ have the 
same norm, so this condition is gauge invariant.
We will not need to impose this normalization condition in this paper.} 
- which is 
a consequence of the tracelessness of $\epsilon_i$
\begin{equation}\label{constrem}
\polo_i. \polo_i + \alpha_i^2=0
\end{equation}   
This constraint can be used to solve for 
$\polo_i.\polo_i$ in terms of $\alpha_i^2$. In 
enumerating contraction structures we simply 
omit all terms containing factors of 
$\polo_i.\polo_i$. For counting purposes, 
therefore, $\polo_i$ can effectively be treated as null.

The expressions \eqref{perpp} and \eqref{exppol} allow us to convert any Lorentz and gauge invariant expression parity even expression for a photon/graviton S-matrix, initially presented as a function of $\epsilon_i$ and $p_i$, into an function of $(\polo_i, \alpha_i)$
and $(s, t)$. This function is separately linear/bilinear in $(\polo_i, \alpha_i)$. Note that the individual momenta $p_i$ enter 
into this reduced form of the S-matrix only through\footnote{This follows from the fact $\polo_i. p_j =0$. While $\polp_i.p_j \neq 0$ the result of this dot product 
is given by $\alpha_i$ times an easily computed function of $(s, t)$.} 
$(s, t)$. 

The converse of the assertion of the previous paragraph also holds. The translation formulae\footnote{ In  \eqref{perpinv} we have presented one of many inequivalent looking - but actually equal - gauge and Lorentz invariant expressions for 
	$\alpha_i$ and $\polo_i$. For instance the second of \eqref{perpinv} follows from the observation that the expression 
	$-2\frac{p_2.F^1}{s}$ equals $\polo_1$ plus a vector in the scattering three plane. $\polo_1$ is then isolated by removing 
	the unwanted vector via the subtractions in the last three terms on the RHS of the second line of \eqref{perpinv}.  
	It is also true, however,  that $-2\frac{p_3.F^1}{t}$ is proportional to $\polo_1$ plus a vector in the scattering 3 plane. This observation leads to a different looking - but 
	actually equal - expression of the schematic form 
	$\polo_1=-2\frac{p_3.F^1}{t} +{\rm subtractions}$.}

\be\label{manifest}\
\alpha_1= 2\frac{p_2. F^1.  p_3}{\sqrt{stu}},\qquad \alpha_2= 2\frac{p_1.F^2. p_4}{\sqrt{stu}}, \qquad \alpha_3= 2\frac{p_4.F^3. p_1}{\sqrt{stu}},\qquad \alpha_4= 2\frac{p_3.F^4. p_2}{\sqrt{stu}}.
\ee
\begin{eqnarray}\label{perpinv}
\polo_1 &=& -2\frac{p_2.F^1}{s}+2\frac{p_2.F^1.p_3}{tu}p_{3}-2\frac{p_2.F^1.p_3}{ts}p_{1}-2\frac{p_2.F^1.p_3}{su}p_{2}\nonumber\\
\polo_2 &=& -2\frac{p_1.F^2}{s}+2\frac{p_1.F^2.p_3}{tu}p_{3}-2\frac{p_1F^2.p_3}{us}p_{2}-2\frac{p_1.F^2.p_3}{ts}p_{1}\nonumber\\
\polo_3 &=& -2\frac{p_2.F^3}{u}+2\frac{p_2.F^3.p_1}{ts}p_{1}-2\frac{p_2.F^3.p_1}{tu}p_{3}-2\frac{p_2.F^3.p_1}{su}p_{2}\nonumber\\
\polo_4 &=& -2\frac{p_2.F^4}{t}+2\frac{p_2.F^4.p_3}{su}p_{3}-2\frac{p_2.F^4.p_3}{ts}p_{4}-2\frac{p_2.F^4.p_3}{tu}p_{2}.
\end{eqnarray}
where 
\begin{equation}\label{defef} F_{\mu\nu}^i=p^i_\mu\epsilon^i_\nu-p^i_\nu\epsilon^i_\mu
\end{equation}  is the manifestly gauge invariant Field strength and $a.F.b\equiv a^\mu F_{\mu\nu}b^{\nu}$, allow us to re-express any  $SO(D-3)$ parity invariant function of  $(\polo_i, \alpha_i)$ and $(s, t)$ with the correct homogeneity
properties as a manifestly Lorentz and gauge invariant S-matrix.

There is a slight subtlety in the discussion of parity odd 
S-matrices, i.e. S-matrices constructed out of a single factor 
of the $D$-dimensional Levi-Civita tensor $\varepsilon$. The reason these 
structures are subtle is simply that $\varepsilon$ tensors 
in different numbers of dimensions have different numbers of 
indices and so do not simply map to one another. In order to 
resolve this subtlety it will prove convenient to formally 
regard the $\varepsilon$ tensor as one of the arguments of 
parity odd $S$-matrices. From this viewpoint, a parity odd 
S-matrix is a Lorentz and gauge invariant function of $p_i$,  $\epsilon_i$ and $\varepsilon$ that has the property that
is linear in $\varepsilon$ (it also has the usual homogeneities in $\epsilon_i$). We now define a $D-3$ dimensional ${\tilde \varepsilon}^{D-3}$
tensor by the equation, 
\begin{equation}\label{dmt}
{\tilde \varepsilon}^{D-3}
=\varepsilon_{\mu_1 \ldots \mu_{D-3} \mu_{D-2} \mu_{D-1}\mu_{D} }
	 p_1^{\mu_{D-2}} p_2^{\mu_{D-1}} p_3^{\mu_{D}}.
\end{equation}
Note that ${\tilde \varepsilon}^{D-3}$ is totally anti-symmetric under the $S_4$ permutation of particles. It has momentum degree three.  Note that ${\tilde \varepsilon}^{D-3}$ is proportional to $(stu)^{\frac12}\varepsilon^{D-3}$, where $\varepsilon^{D-3}$ is a $D-3$ dimensional Levi-Civita tensor. For this reason it is sometimes useful to work with 
the `normalized' tensor 
\begin{equation}\label{dmtn}
N\left( {\tilde \varepsilon}^{D-3} \right)=\frac{{\tilde \varepsilon}_{\mu_1 \ldots \mu_{D-3}}
	}{\sqrt{stu}}
\end{equation} 
$N({\tilde \varepsilon}^{D-3})$ is proportional to the Levi-Civita 
tensor in $D-3$ dimensions up to a sign\footnote{Note that the LHS of \eqref{dmtn} is precisely defined 
(unlike the $D-3$ Levi-Civita tensor which is precisely defined only once we specify an orientation in the $D-3$ 
plane orthogonal to the scattering plane).}. The action 
of the permutation group on ${\tilde \varepsilon}^{D-3}$ and $N({\tilde \varepsilon}^{D-3})$ is given by 
\begin{equation} \label{transfundperm} 
P ({\tilde \varepsilon}^{D-3})= (-1)^{{\rm sgn} (P)}
{\tilde \varepsilon}^{D-3}, ~~~P \left(N({\tilde \varepsilon}^{D-3}) \right)= (-1)^{{\rm sgn} (P)}
N({\tilde \varepsilon}^{D-3})
\end{equation} 
where $P$ is an arbitrary permutation in $S_4$ of the momenta $p_1 \ldots p_4$. In other words both ${\tilde \varepsilon}^{D-3}$ and $N({\tilde \varepsilon}^{D-3})$ pick up a sign under 
every odd permutation (e.g. under single exchange permutations).
 
When $D$ is odd, we will choose to regard every parity odd S-matrix a function of ${\tilde \varepsilon}^{D-3}$, 
$\polo_i, \alpha_i$ and $(s,t)$; the 
function in question is linear in ${\tilde \varepsilon}^{D-3}$ (it also has the usual homogeneities in $(\polo_i, \alpha_i)$).   When $D$ is even, on the other hand, we will 
choose to regard every parity  odd S-matrix a function of $N({\tilde \varepsilon}^{D-3})$, $\polo_i, \alpha_i$ and $(s,t)$; once again the function is linear in $N({\tilde \varepsilon}^{D-3})$ and has the usual homogeneities in $(\polo_i, \alpha_i)$\footnote{The reason we make this distinction between odd and even $D$ will become clearer later in this paper.  Roughly speaking the reason goes as follows. We will see below that S-matrices can be expanded in a sort of Taylor Series in momenta. In every dimension the basis functions of this expansion for parity even S-matrices all have even powers of momenta. As far as parity odd S-matrices go, however, the basis functions are even in momenta when $D$
is even; the fact that $N({\tilde \varepsilon}^{D-3})$ is also even 
in powers of momenta makes \eqref{dmtn} a natural building 
block of such S-matrices. On the other hand the building 
blocks for parity odd S-matrices in odd $D$ are odd in momenta; the fact that  \eqref{dmt} is cubic in momenta
makes it a natural building block for S-matrices in this case.}.

For parity odd structures, the formulae \eqref{perpinv}, \eqref{manifest}, supplemented with \eqref{dmt} and \eqref{dmtn}, 
map any $SO(D-3)$ invariant function of  $(\polo_i, \alpha_i)$, $(s, t)$ and ${\tilde \varepsilon}^{D-3}$ (or $N({\tilde \varepsilon}^{D-3})$) that is linear in ${\tilde \varepsilon}^{D-3}$ - and with the appropriate homogeneity properties in $(\polo_i, \alpha_i)$ - to a manifestly Lorentz and gauge invariant S-matrix that is linear in $\varepsilon$ and 
therefore is parity odd.

\subsection{$S_4$ permutation symmetry}
\label{epp}

Apart from the requirements of gauge and Lorentz invariance, the $2\to 2$ S-matrix for identical bosonic particles also enjoys invariance under under $S_4$, the group that  permutes the four scattering particles. We now turn to a brief analysis of the action of this Bose symmetry
and its consequences.

\subsubsection{Action of permutations on unconstrained data}
\label{apud}

In this subsection we discuss the action of the permutation group $S_4$ act on the scattering data $(p_i, \polo_i, \alpha_i)$. It turns out that this action is not as trivial as one might first 
guess.

 Let $B=(p_1, p_2, p_3, p_4)$,  $C=(\alpha_1, \alpha_2, \alpha_3, \alpha_4)$ and $D=(\polo_1, \polo_2, \polo_3, \polo_4)$
each represent column vectors of length 4. Let $P$ be any 
permutation in $S_4$. Let $M(P)$ denote the representation of $S_4$ in its `defining' (reducible) 4 dimensional representation. 
Clearly the column vectors $B$ and $D$ transform under permutations as 
\begin{equation}\label{rob}
B \rightarrow M(P) B, ~~~~~~D \rightarrow M(P) D.
\end{equation} 
\eqref{rob} asserts that $P$ acts on $p_i$ and 
$\polo_i$ by permuting the $i$ indices; this follows from 
definitions.

On the other hand permutations act on $C$ in a more complicated 
way. This is because the quantities $\alpha_i$ are not completely 
physical; the equations \eqref{manifest} that define $\alpha_i$ 
in terms of gauge invariant data express $\alpha_i$ as 
$F^i_{\mu\nu}$ with indices contracted with two of the other 
(non $i$) scattering momenta. Which momenta we choose to contract
$F^i_{\mu\nu}$ changes the answer only up to a sign. As there 
is no permutation invariant way of making a choice of contracted 
momenta, the action of permutations on $\alpha_i$ permutes 
$\alpha_i$ but only up to a sign\footnote{We can also see this 
in another way. It follows from \eqref{normalphai} that 
$|\alpha_i|^2$ is completely physical, so permutations 
act on $|\alpha_i|^2$ in the usual way; once again we see that 
the ambiguity is in the sign.}. It follows that on general grounds 
the action of permutations on $C$ must take the form 
$$ C \rightarrow D(P) M(P) C$$
where $D(P)$ is a diagonal matrix with diagonal entries 
$\pm 1$. In other words $C$ transforms in a projective representation of $S_4$ with the phase ambiguities given by 
$\pm 1$. A simple explicit computation using \eqref{manifest} 
demonstrates that $D(P)=(-1)^{{\rm sgn(P)}}$ so that the 
action of permutations on $C$ is given by 
\begin{equation}\label{robb}
C \rightarrow (-1)^{{\rm sgn(P)}} M(P) C.
\end{equation}

Note that the `anomaly' in the transformation properties 
of $\alpha_i$ is extremely simple. With our choice of basis 
it is a phase that acts in the same way on all four $\alpha_i$. 
It follows immediately that the action of permutations on 
even polynomials of $\alpha_i$ is `non-anomalous' whereas 
the action of permutations on odd polynomials of $\alpha_i$ 
is the naive action times $(-1)^{{\rm sgn(P)}}$.

Parity even S-matrices are even functions of $\alpha_i$ in 
every dimension\footnote{The overall S-matrix is 
of combined degree 4 in $\alpha, \polo$ variables (in the 
case of photon scattering) and of combined degree 8 in 
$\alpha$ and $\polo$  (for the case of graviton scattering). 
Because the $\polo_i$ always contract in pairs in even 
scattering amplitudes, it follows that every term in such 
a scattering amplitude is an even function of $\alpha_i$. }.
On the other hand  parity odd S-matrices are even 
function of the $\polo_i$ (and hence of $\alpha_i$) when 
$D$ is odd, but an odd function of both $\polo_i$ and 
$\alpha_i$ when $D$ is even\footnote{This follows from the observation that in parity odd S-matrices exactly $D-3$ of indices of the $\varepsilon$ tensor have to be contracted with the $\polo_i$; 
other $\polo_j$s then contract in pairs. Thus the number 
of $\polo_i$ - hence $\alpha_i$ - that appear in any given 
term in a parity odd S-matrix is odd if $D-3$ is odd but 
even if $D-3$ is even.}.

We have already explained (see around \eqref{transfundperm}
) that it is convenient to regard parity odd S-matrices 
as functions of ${\tilde \varepsilon}^{D-3}$ (see \eqref{dmt}) when 
$D$ is odd, but of $N({\tilde \varepsilon}^{D-3})$ (see \eqref{dmtn})
when $D$ is even. We have also seen that ${\tilde \varepsilon}^{D-3}$ and $N({\tilde \varepsilon}^{D-3})$ both transform under permutations 
as \eqref{transfundperm}. As far as S-matrices go, it is  thus possible to absorb the anomalous phase in the transformation of $\alpha_i$ (see \eqref{robb}) into a
renormalized permutation transformation rule for 
$N({\tilde \varepsilon}^{D-3})$ as we summarize in detail in the next
subsubsection.

\subsubsection{Effective transformations under permutations}
\label{etp}

It follows from the analysis of the previous subsubsection 
that, as far as S-matrices go the correct permutation  transformation laws are captured the following effective rules. $p_i$,  $\alpha_i$ and $\polo_i$ all 
transform under permutations in the simple `non-anomalous' manner
\begin{equation}\label{robeff}
B \rightarrow M(P) B, ~~~~~~D \rightarrow M(P) D, ~~~~~~ C \rightarrow M(P) C
\end{equation}
Parity even S-matrices are separately of homogeneity one (two) 
in $\alpha_i, \polo_i$ for photon (graviton) S-matrices. 
Parity odd S-matrices also have the same homogeneity in $\alpha_i, \polo_i$, but in addition are linear in ${\tilde \varepsilon}^{D-3}$  (see \eqref{dmt}) when $D$ is odd and in $N({\tilde \varepsilon}^{D-3})$ 
when $D$ is even.  ${\tilde \varepsilon}^{D-3}$
transforms under permutations as 
\begin{equation} \label{tmo} 
P ({\tilde \varepsilon}^{D-3})= (-1)^{{\rm sgn} (P)}
{\tilde \varepsilon}^{D-3},
\end{equation}
while $N({\tilde \varepsilon}^{D-3})$  are assigned the transformation properties
\begin{equation} \label{tmoo} 
P \left(N({\tilde \varepsilon}^{D-3}) \right)= 
N({\tilde \varepsilon}^{D-3})
\end{equation}
i.e. is invariant under permutations. We will use the rules 
summarized in this subsubsection to compute the transformations 
of S-matrices under permutations in the rest of this paper.

\subsection{Permutations: $\Z_2\times \Z_2$ and $S_3$}
\label{Permsfst}

The  permutation group $S_4$ has a special abelian subgroup $\Z_2\times \Z_2$  generated by $(2143)$, $(3412)$ i.e. the subgroup of double transpositions\footnote{We label an element of $S_4$ by the image of $(1234)$ under that element. The group $\Z_2 \times \Z_2$ consists of the two listed generators together with the identity permutation $(1234)$ and a fourth element $(4321)$.}.
The importance of this subgroup is that it  leaves all the Mandelstam variables $s,t$ and $u$ invariant. 

Another feature of this subgroup is that it is normal\footnote{Recall that a subgroup $H\subset G$ is normal if it obeys the property that for any $h\in H$, $ghg^{-1}\in H$ for all $g\in G$. In other words, the normal subgroup is fixed by the adjoint action of the group.}. As a result, the coset\footnote{Here the coset is either by left action or by right action, both cosets are equivalent because subgroup is normal.} $S_4/(\Z_2\times \Z_2)$ inherits the group structure of $S_4$. The coset  group is easily identified. Every $S_4$ group element is $\Z_2\times \Z_2$ equivalent to a unique element of the form $(abc4)$. It follows that the $\Z_2\times \Z_2$ `gauge invariance' can be fixed by adopting the `gauge fixing condition' that  particle $4$ is not permuted. This choice of gauge fixing clearly reveals the coset to be simply the $S_3$ that permutes particles $1$, $2$ and $3$. Thus we conclude that 
\begin{equation} \label{permgp}
\frac{S_4}{\left( \Z_2 \times \Z_2 \right)}=S_3.
\end{equation}

It follows that the condition of $S_4$ invariance on the S-matrices can be imposed in two steps. In the first step, we impose only $\Z_2\times \Z_2$ symmetry on the gauge invariant functions of $\epsilon_i^\mu$ and $p_i^\mu$ (with the necessary homogeneity properties in $\epsilon_i^\mu$). We call the S-matrices thus obtained, \emph{quasi-invariant} S-matrices. The coset group $S_3$ acts linearly on the space of quasi-invariant S-matrices. In order to obtain fully $S_4$ invariant S-matrices we must further project the space of quasi-invariant S-matrices down to its $S_3$ invariant subspace. 
The importance of the notion of quasi-invariant S-matrices lies in the fact that the multiplication of a quasi-invariant S-matrix by a function of $s, t$ leaves it quasi-invariant because $(s,t)$ are themselves individually invariant under $\Z_2\times \Z_2$. It follows that the  space of not-necessarily-polynomial quasi-invariant S-matrices forms a vector space over functions of $(s,t)$. It is intuitively clear - and we will see in great detail below - that this vector space is finite dimensional.

\subsection{Local S-matrices and a module structure} \label{lsm}

This paper is principally devoted to a study of a special class of S-matrices which we call local S-matrices. We define these objects as  
S-matrices that are polynomial functions of $\epsilon_i$ and $p_i$. In this subsection we turn our attention to such S-matrices. We focus, first on quasi-invariant S-matrices, postponing the task of enforcing $S_3$ invariance to later.

\subsubsection{The local module} \label{lmod} 

It is clear that the set of local S-matrices is closed under multiplication by any polynomial $p(s,t)$ and addition. This structure is reminiscent of the vector space except for one important difference. Polynomials of $(s,t)$ do not form a field but rather only a ring \emph{i.e.} they do not have multiplicative inverse\footnote{This is a very important difference. In a genuine vector space if a vector $a$ is a multiple of a vector $b$ then it is also true that the vector $b$ is a multiple of the vector $a$. In a module, on the other hand, if $a$ equals a ring element times $b$ then it is usually not true that $b$ equals a ring element times $a$. In other words the notion of proportionality is inherently hierarchical in a ring. We elaborate on this below. }.
 Consequently, the set of local quasi-invariant S-matrices forms a \emph{module}, over the ring of polynomials of $(s,t)$ and not a vector space. The identification of space of local S-matrices with a module over a ring of polynomials of Mandelstam variables has also been made in \cite{Henning:2017fpj}.

Viewed as a vector space over the field of complex numbers, the space of local quasi-invariant S-matrices is, of course, infinite dimensional. Viewed as a module, however, this space is `finitely generated' as we now explain. We pause to introduce  some (standard) mathematical terminology.

The  elements of the form $r\cdot m$, where $m$ is a given element of the module and $r$ is any element of the ring, are said to form the span of $m$. We call the elements in  the span of $m$, the descendants of $m$\footnote{This is non-standard mathematical terminology but being physicists we connect well to the word ``descendant". Note that the set of basis vectors of a  conformal multiplet can be thought of as a module generated by the primary operator over the ring of polynomials in $P_\mu$. From this point of view, conformal descendants are descendants in our sense. Similarly, the Verma module can be thought of as the module generated by the primary operator over the ring of Virasoro creation operators.}. Sometimes we denote the descendant of $m$ in a more physical notation $r|m\rangle$.
A subset $G=\{g_i\}$ of the module $M$ is said to generate $M$ if the smallest submodule which contains $G$ is $M$ itself. In other words, the union of spans of all descendants of $g_i$ is $M$ itself.
A module $M$ is said to be finitely generated if it has a finite generator (i.e. a generator with a finite number of elements). A generator set $G$ is said to generate $M$ freely if the following condition holds,
\be
\sum_{i} r_i\cdot g_i=0 \qquad\qquad {\rm iff}\qquad \qquad {\rm all}\,\,\,r_i=0.
\ee
In other words, every element of $m$ is a unique linear combination of $g_i$ over the ring. A module $M$ is a free module if there exists a $G$ that generates it freely. In this case the generator set $G$ is called the basis of $M$. A free module is the next best thing after a vector space. Understanding its structure is equivalent to understanding its basis elements. When the module is not free, one has to characterize the module by giving its generators and their relations\footnote{If the relations do not form a free module, then one has to characterize the relation module in the same way and so on. This is called the free resolution of a module.}.

We can find the generators of the module of local quasi-invariant S-matrices in following way. These S-matrices are obtained from local Lagrangians. We first  look for a basis over complex numbers of local quasi-invariant S-matrices of the lowest degree. Next we again look for a basis over complex numbers of local quasi-invariant S-matrices of lowest degree that are not in the span of the previously chosen elements, and so on.  This process terminates at a finite degree - intuitively because the gauge invariant field strengths built out of $\epsilon_i$ have 
a finite number of indices\footnote{Once all these indices are contracted with momenta, the remaining momentum indices have 
to contract with each other yielding  powers of $s, t$  and hence belonging to the span of lower degree structures.} - more on this later. It follows that the module of local quasi-invariant S-matrices 
is finitely generated. We will call this module, the local module for short and will label its generators as $E_J(p_i,\epsilon_i)$ and the generator set as $L$.
 
As the local S-matrices are constrained by transversality and shift invariance of polarizations $\epsilon_i$, and as these constraints involve the momenta $p_i$ that in turn define $s$, $t$ and $u$, it is  much less clear that this module is freely generated, and, indeed we will find  below that this is not always the case.

\subsubsection{The bare module} \label{fbm}

 As we have already 
 explained in subsubsection \ref{unconst}, the equations 
\eqref{perpp} and \eqref{exppol} allow us to re-express any 
photon/graviton quasi-invariant S-matrix as a
polynomial of $(\polo_i, \alpha_i)$ that is simultaneously 
of degree one/two in each of these pairs of variables; the 
coefficients of this polynomial expressions are functions of 
$(s, t)$.

We now turn to a crucial point. If we start with a local quasi-invariant S-matrix, it is possible to show that the resulting expression, written as a polynomial of $(\polo_i, \alpha_i)$, using \eqref{perpp} and \eqref{exppol}, has coefficients that are polynomials (rather than generic functions) of $(s,t)$\footnote{This statement is 
not obvious as the RHS of \eqref{perpp} involves expressions that are not polynomial in $s$, $t$ and $u$. And indeed individual 
Lorentz invariant building blocks of S-matrices - like 
$p_2 . F^1. p_3$ (see \eqref{manifest}) are not individually 
polynomials in $p$ and $\epsilon$. When we put these building 
blocks together, however, we always recover polynomials. 
For instance the product of four terms - $p_2 . F^1. p_3$ and a 
a similar term for particle 2, 3 and 4 - is proportional to 
$(stu)^2 \alpha_1 \alpha_2 \alpha_3 \alpha_4$ (see \eqref{manifest}). We have checked by explicit computation in many many different contexts, that this is the case. See, e.g.,\eqref{explicit-embeddingo}, \eqref{oot}, \eqref{henum} for explicit examples worked out in detail.}. We will discuss this observation momentarily in the next subsubsection. 

This motivates us to define a new module. We define 
the parity even part of the module of \emph{bare} quasi-invariant photon/graviton S-matrices, or \emph{bare} module for short, over the ring of polynomials of $(s,t)$, to be the set of parity even (i.e. ${\tilde \varepsilon}^{D-3}$ independent ) rotationally invariant and $\Z_2\times \Z_2$ invariant polynomials of $(\polo_i, \alpha_i)$ and $(s,t)$ that are simultaneously of degree one/two in each of the 
pair of variables $(\polo_i, \alpha_i)$\footnote{In the case of 
the gravitational S-matrix, the variables $(\polo_i, \alpha_i)$  
are also constrained to obey \eqref{constrem}.}. Any basis of the vector space - over the field of complex numbers - of rotationally invariant polynomials of $(\polo_i, \alpha_i)$ (subjected to the requirement of $\Z_2 \times \Z_2$ invariance and appropriate homogeneity requirements) forms a generating set for this module\footnote{At the end of subsection \eqref{Permsfst} we had explained that the set of not necessarily local quasi-invariant S-matrices constitute a vector space over the field of functions of $s,t,u$. The generators of the bare module clearly also define a basis for this vector space.}. Let us denote this generator set as $B$ and its elements as $e_I(\alpha_i,\polo_i)$.
We sometimes call $e_I$ ``index structures".
Notice that our generators are all independent of
$s$, $t$ and so, in particular, are of zero homogeneity 
in derivatives.  As the variables 
$\polo_i$ and $\alpha_i$ are completely unconstrained (in the case 
of photons) or obey only the momentum independent constraint 
\eqref{constrained} (in the case of gravitons), it is clear that this choice generates our module finitely and freely. This makes $B$ the basis of the bare module.
The key point - made at the beginning of this subsubsection - is that the local module  is a submodule of the free bare module. 
 
The parity odd part of the bare module is defined in odd/even $D$ in a similar manner, to be set of rotationally invariant and $\Z_2\times \Z_2$ invariant polynomials of $(\polo_i, \alpha_i)$ , $(s,t)$ and ${\tilde \varepsilon}^{D-3}$ (resp. $N({\tilde \varepsilon}^{D-3})$) that are linear in ${\tilde \varepsilon}^{D-3}$ (resp. $N({\tilde \varepsilon}^{D-3})$) and are also simultaneously of degree one or two (corresponding to photons and gravitons)  in each of the pair of variables $(\polo_i, \alpha_i)$. 
Basis elements are functions of ${\tilde \varepsilon}^{D-3}$, (resp. $N({\tilde \varepsilon}^{D-3})$)  $\alpha_i, \polo_i$ only; there is no further dependence on $s$ and $t$. 
Note that these basis elements are of dimension zero in even 
$D$ but of dimension 3 in odd $D$. 

Like its even component, the  parity odd part of the bare module is also free. 
Once again the module of parity odd local S-matrices is a submodule of the parity odd part of the bare module. 
 
These observations allows us to carry out the program of 
characterizing the local module 
in two steps. In the first step - carried out in detail 
in Section \ref{index-structure}, we first completely characterize the bare module by identifying $e_I$'s. This step is relatively simple. In the second step (which we briefly discuss in the next subsubsection and then 
implement in detail in the subsequent sections of this paper) 
we understand the embedding of the local module into the free bare module, 
and thereby understand its structure and enumerate its elements. 

\subsubsection{Embedding the local module into the bare module} \label{elmbm}
Remarkably we find that the generators of the local module $E_J(p_i,\epsilon_i)\in L$ are related to the basis elements $e_I(\alpha_i,\polo_i)\in B$ of the bare module as,
\begin{equation} \label{relationship}
E_J(p_i,\epsilon_i)=\sum_{e_I\in B} p_{IJ}(s,t) e_I(\alpha_i,\polo_i).
\end{equation}
where $p_{IJ}(s,t)$ are polynomials\footnote{The remarkable fact is that $p_{IJ}(s,t)$ is simply a polynomial without any negative or fractional powers.  Unfortunately we do not have a simple abstract proof of this statement, and it would be nice to find one. Nonetheless we have explicitly verified it in every dimension and for both photon and graviton scattering. The verification has proceeded on Mathematica; in each dimension we have used  Mathematica to explicitly re express the local generators in terms of the bare generators and verified that the coefficients of this expansion are polynomials in $s, t, u$. }
of $s, t$. 

In our study of photon/graviton S-matrices later in this paper we encounter two cases. In the first case, (this holds for both photon and graviton scattering when $D\geq 5$), $|L|=|B|$\footnote{The symbol $|A|$ denotes the number 
of elements in the set $A$.}. 
In the second case, which turns out to apply to both photon 
and graviton scattering in $D=4$ and also photon scattering in $D=3$, $|L|>|B|$. In this 
subsection we briefly discuss these two cases in turn. 

Let us first consider the case $|L|=|B|$. In this case the local module is freely generated 
if and only if the equation
\begin{equation}\label{freecond}
\sum_{E_J\in L}r^J(s,t) E_J(p_i,\epsilon_i) =0,
\end{equation}
has no non-trivial solutions for polynomials $r^J(s,t)$.  
Plugging the expansion \eqref{relationship} into \eqref{freecond}
and equating coefficients of $e_I$ we find that 
\eqref{freecond} turns into 
\begin{equation}\label{freecondp}
\sum_{J}p_{IJ}(s,t)r^J(s,t) =0.
\end{equation}
For each value of $s$ and $t$, \eqref{freecondp} is a set of 
$|B|$ linear equations for $|B|$ variables. This set of 
equations has non-trivial solutions if and only if 
\begin{equation}\label{freeconds}
{\rm Det} \left[ p_{IJ}(s,t) \right]=0.
\end{equation}
Equation \eqref{freecondp} has no solutions unless \eqref{freeconds} 
holds for every value of $s$ and $t$. Equation \eqref{freeconds} is, 
of course, an extremely onerous condition, and we find that it is not met for the $p_{IJ}$ matrix that arises in the study of $S$ matrices with $D\geq 5$. 
It follows, as a consequence, that the local module is also  freely generated 
for $D\geq 5$. 

In the case that $|L| > |B|$ (which we encounter for $D= 4$ and also $D=3$ for photons),
on the other hand, it is very easy to see that $|L|$ cannot 
be freely generated\footnote{The argument for this is the 
	following. If $L$ were freely generated then the number 
	of local quasi-invariant S-matrices of degree $d$ would 
	grow like $|L|d$ at large $d$. This is larger than 
	the number of bare quasi-invariant S-matrices, which grows 
	like $|B|d$ at large $d$, contradicting the fact that 
	the local module is a submodule of the bare module.}.
Given that the local module is not freely generated in $D=4$, it is important to 
discover the relations in this module. We will return 
to this question later in this paper. 

\subsection{Irreducible representations of $S_3$ and fusion rules } \label{irrep}

As we have explained above, the space of physical (hence $S_4$ invariant) S-matrices is the projection of the local module of S-matrices onto $S_3$ singlets. In this subsection we discuss
the nature of this projection. As preparation for our discussion 
we first review elementary facts about $S_3$ representation 
theory.

The permutation group of three elements, $S_3$, has three irreducible representations the one dimensional totally symmetric representation which we call ${\bf 1_S}$, the one dimensional totally anti-symmetric representation which we call 
${\bf 1_A}$ and a two dimensional representation with mixed symmetry which we call ${\bf 2_M}$. The subscript for ${\bf 2_M}$ emphasizes the mixed symmetry.  It is standard to associate the following Young diagrams with these representations,
\begin{equation} \label{yts}
{\bf 1_S}={\tyng(3)}\qquad\qquad {\bf 1_A}={\tyng(1,1,1)}\qquad \qquad {\bf 2_M}={\tyng(2,1)}.
\end{equation}
It is easy to decompose an representation of $S_3$, into the subspaces that transform, respectively, in the ${\bf 1_S}$, the ${\bf 1_A}$ and ${\bf 2_M}$ representations. 
Complete symmetrization project onto the  ${\bf 1_S}$ subspace, complete anti symmetrization projects onto the ${\bf 1_A}$ subspace and whatever is left over, \emph{i.e.} the part that is annihilated by both complete symmetrization and complete anti symmetrization, transforms in the ${\bf 2_M}$ representation.

In order to get some familiarity with these representations, 
let us first consider a 3 dimensional column vector whose elements are $q_1, q_2, q_3$ respectively. The permutation 
group has a natural action on this column vector; any given 
element $\sigma$ of $S_3$ maps this vector to the column with 
entries $q_{\sigma(1)}, q_{\sigma(2)}, q_{\sigma(3)}$\footnote{For instance 
the $\Z_2$ element that flips one and two gives $(q_2, q_1, q_3)$.}. This linear map is generated by a (unique) $3 \times 3$ matrix $M(\sigma)$ acting on acting on the column $(q_1, q_2, q_3)$.
The collection of matrices $M(\sigma)$ yields a representation of 
$S_3$. We use the symbol ${\bf 3}$ to denote this `defining' representation of $S_3$. This representation is {\it not}
irreducible but can be decomposed as\footnote{To see why this is the case, note that the  
complete symmetrization the column $x_1, x_2, x_3$ yields a column whose elements are all equate to $2(x_1+x_2+x_3)$. This column is permutation invariant and so transforms in the one dimensional completely symmetric representation of $S_3$. 
Removing this column one is left with the action of $S_3$ 
on a column $(y_1, y_2, y_3)$ whose elements are subject to the constraint $y_1+y_2+y_3=0$, which generates the ${\bf 2_M}$ representation.}
\begin{equation}\label{decomp3}
{\bf 3}={\bf 2_M} + {\bf 1_S}.
\end{equation}
As a second exercise let us study the 6 dimensional representation, ${\bf 6}_{\rm left}$ generated by the left action of $S_3$ onto itself.  It is not difficult to demonstrate (see appendix \ref{s3permutations}) that 
\begin{equation}\label{sdr}
{\bf 6}_{\rm left}={\bf 1_S} + 2  \cdot {\bf 2_M}   + {\bf 1_A}.
\end{equation} 
Of course the same decomposition also applies for the ${\bf 6}_{\rm right}$ representation 
generated by the right action of $S_3$ on itself. 

As our next example consider the adjoint action of $S_3$ on itself $\sigma \rightarrow  g^{-1} \sigma g,$ which  also yields a 6 dimensional representation ${\bf 6}_{\rm adj}$. The adjoint representation can be decomposed into the ${\bf 1_S}$ (which acts on the identity element which is invariant under adjoint action) a ${\bf 3}$ (which acts on the 3 exchange permutations $(213), (321), (132)$), and  a ${\bf 2_M}$ (which acts on the two cyclical permutations $(231)$ and $(312)$). In equations 
\begin{equation} \label{adjact}
{\bf 6}_{\rm adj}= {\bf 1_S}+ {\bf 2_M} + {\bf 3}.
\end{equation} 
Of course the ${\bf 3}$ can itself be further decomposed 
using \eqref{decomp3}.

The action of the adjoint representation on the permutation group is particularly important, because the elements that transform in the same representation in \eqref{adjact} have equal eigenvalues when acting on any representation.  The identity element (symmetric representation in \eqref{adjact}), of course, acts as unity on every representation. The cyclic elements (which transform in the ${\bf 2_M}$ in \eqref{adjact}) obey the identity $g^3=1$. It follows that the eigenvalues of the cyclic elements are cube roots of unity in any representation of the symmetric group. It is not difficult to verify that each of these elements act as unity on 
the ${\bf 1_S}$ and ${\bf 1_A}$ representations but have eigenvalues ($e^{\frac{2 \pi i}{3}}, e^{-\frac{2 \pi i}{3}})$ in the ${\bf 2_M}$ representation. Finally the exchange elements (which transform in the ${\bf 3}$ in \eqref{adjact}) obey $g^2=1$ and so always have eigenvalue $\pm 1$. 
These elements are represented by unity in the ${\bf 1_S}$, by $-1$ in the ${\bf 1_A}$, and have eigenvalues 
$(+1, -1)$ in the ${\bf 2_M}$ dimensional representation. 
Note also that the ${\bf 2_M}$ (cyclical elements) together with 
the identity make up the abelian group $\Z_3$ - the maximal abelian subgroup of $S_3$.

Note that it is possible to choose a basis of the ${\bf 3}$ representation
such that the three basis elements are respectively invariant 
under the three distinct exchange permutations. A converse of this statement is also true. Consider a representation vector that is invariant under (say) $(213)$. Construct a set of three vectors given by this vector and its images under the two cyclic permutations\footnote{In more detail, the first basis element is invariant under $(213)$ the second element under $(321)$ and the third element under $(132)$.}. If the resultant set of 3 vectors are linearly independent then they transform in the ${\bf 3}$ representation of $S_3$.

Recall that every $S_3$ element, so in particular every two 
particle exchange in $S_3$, is represented by unity in the 
${\bf 1_S}$ representation. On the other hand every two particle 
exchange in $S_3$ is represented by $-1$ in the ${\bf 1_A}$ representation. Finally the ${\bf 2_M}$ representation somewhere
right in between the ${\bf 1_S}$ and ${\bf 1_A}$ in the following 
sense: every two particle exchange element is represented by 
a $ 2\times 2$ matrix whose eigenvalues are $\pm 1$.

Second, given a collection of objects, 
$n_{\bf 1_S}$ of which transform in the completely symmetric representation, $n_{\bf 1_A}$ of which transform in the completely antisymmetric representation and $n_{\bf 2_M}$ of which transform in the mixed two dimensional representation. It is clear from the discussions of this subsection that our $n_{\bf 1_S}+n_{\bf 1_A} +2n_{\bf 2_M}$ dimensional vector space of objects can be decomposed into 
a $n_{\bf 1_S}+n_{\bf 2_M}$ dimensional subspace in which 
any particular exchange operator,  say $(213)$, has eigenvalue plus one, and the complementary $n_{\bf 1_A}+n_{\bf 2_M}$ dimensional subspace in which the same 
exchange operator has eigenvalue $-1$. It follows that 
the $\Z_2$ invariant subspace of our collection of objects is $n_{\bf 1_S}+n_{\bf 2_M}$ dimensional.

We end this subsection with a presentation of the fusion rules of $S_3$, i.e. the rules for the decomposition of the direct products of every pair of irreducible representations  of $S_3$  into the sum of irreducible representations. It is clear that the direct product of any representation $\bf R$ with the symmetric representation is $\bf R$. The remaining direct products of irreducible representations are easily  verified to be (see appendix \ref{s3permutations}), 
\be
{\bf 1_A}\otimes {\bf 2_M}= {\bf 2_M},\qquad {\bf 1_A}\otimes {\bf 1_A}={\bf 1_S},\qquad {\bf 2_M}\otimes {\bf 2_M}={\bf 2_M}\oplus{\bf 1_S}\oplus {\bf 1_A}.
\ee

\subsection{Projecting onto $S_3$ singlets}

We now return to a discussion of the local module and its 
$S_3$ projection.
It follows from definitions that every bare and local quasi-invariant S-matrix, denoted as $m(p_i,\epsilon_i)$ and $M(p_i,\epsilon_i)$ respectively, can be expressed as 
\begin{equation} \label{bq} \begin{split} 
&m(p_i,\epsilon_i)=\sum_{e_I\in B} p_I(s,t)e_I(\alpha_i,\polo_i).\\
&M(p_i,\epsilon_i)=\sum_{E_J\in L} P_J(s,t)E_J(p_i,\epsilon_i).\\
\end{split}
\end{equation}
where $p_I(s,t)$ and $P_J(s,t)$ are polynomials of $(s,t)$. The $S_4$ invariant local S-matrices are obtained by simply projecting the elements of the local module onto the trivial representation of $S_3$.
\begin{equation} \label{cgi} 
\boxed{	
\CS(p_i,\epsilon_i)=\sum_{\sigma\in S_3} M^\sigma(p_i,\epsilon_i)=\sum_{E_J\in L} \sum_{\sigma\in S_3} P_J^\sigma(s,t)E_J^\sigma(p_i,\epsilon_i).
}
\end{equation} 
The superscript $\sigma$ denotes the action of $\sigma$ permutation.
As the local module admits the action of $S_3$, its generators  $E_J$'s can be decomposed 
into irreducible representations of $S_3$. 
Moreover the space of functions of $(s, t)$ can also be decomposed into irreducible representations of $S_3$. It follows from  \eqref{cgi} that if a subset of $E_J$'s  transforms in any given irreducible representation ${\bf R}$ of $S_3$ then the functions $P_J(s,t)$  must also transform in the same representation ${\bf R}$\footnote{This conclusion follows from the fusion rules of $S_3$ listed in the previous subsection.}.

In order to understand the detailed structure of the projection of the local module onto $S_3$ invariants we thus need to 
understand the decomposition of the space of polynomials of 
$(s,t)$ into representations of $S_3$. We turn to this question now.

\subsection{Action of $S_3$ on polynomials of $(s,t)$}
The action of $S_3$ on polynomials in $(s,t)$ is best realized as permutations of $(s,t,u)$, that belong to the defining representation ${\bf 3}$ of $S_3$, subjected to the condition $s+t+u=0$. 
To gain some familiarity, let us work with some examples. To start with we ignore the constraint $s+t+u=0$ but impose it later. 

At degree $0$, we only have $1$ which clearly transforms in the symmetric representation. At degree $1$, the space of polynomials is three dimensional, it decomposes into a symmetric and a mixed representation. The explicit form of the symmetric is $s+t+u$ and the two dimensional mixed representation is formed by $2s-t-u$ and $2t-s-u$. In order to verify the mixed symmetry, note that complete symmetrization or anti symmetrization of these polynomials vanish. At degree $2$, the space of polynomials is $6$ dimensional. It decomposes as two symmetric $s^2+t^2+u^2$ and $st+tu+us$ and two mixed representations. The mixed representations are spanned by $(2s^2-t^2-u^2, 2t^2-s^2-u^2)$ and $(2st-tu-us,2tu-st-us)$ respectively.  The space of polynomials at degree $3$ is $10$ dimensional. It is decomposed as $3$ symmetric,  $3$ mixed symmetric and $1$ anti-symmetric. Here we will give explicit expression of the anti-symmetric only, $s^2t+t^2u+u^2s-st^2-tu^2-us^2$.

 Some of the representations discussed above vanish after imposing momentum conservation $s+t+u=0$. At degree 1, one mixed representation and at degree $2$, one symmetric and one mixed and at degree $3$, one symmetric, one mixed and one anti-symmetric survive.

It is useful to count these polynomials by their representations into a partition function. Again, we first ignore the constraint $s+t+u=0$. The single variable partition function is 
\be
z(x):={\rm Tr}\, x^{2\Delta}=\frac{1}{1-x^2}.
\ee
Here $\Delta$ is the degree of momentum homogeneity. The partition functions over polynomials of three variables with given transformation property are
\begin{eqnarray} \label{defz} 
\tilde Z(x)&:=&Z_\text{no-sym}=z(x)^3,\qquad \tilde Z_{{\bf 1_S},{\bf 1_A}}(x)\,\,=\,\,\frac16 z(x)^3\pm\frac12 z(x)z(x^2)+\frac13z(x^3),\nonumber\\
\tilde Z_{{\bf 2_M}}(x)&=& \frac{\tilde Z(x)-\tilde Z_{{\bf 1_S}}(x)-\tilde Z_{{\bf 1_A}}(x)}{2} .\label{tildez}
\end{eqnarray}
The quantities ${\tilde Z}_{{\bf R}}$ that appear in 
\eqref{defz} are defined by 
\begin{equation} \label{deftz}
{\tilde Z}_{{\bf R}}= \sum_m n_{\bf R}(m) x^{2m}
\end{equation} 
where $n_{\bf R}(m)$ is the number of $S_3$ representations of type ${\bf R}$ that appear in the 
decomposition of  polynomials of $s, t, u$ into representations of $S_3$ at degree $m$\footnote{The factor of 
	$1/2$ in th last of \eqref{defz} follows from the 
	fact that the ${\bf 2_M}$ representation is two dimensional; the partition function that counts 
	representations is thus half the partition function that counts polynomials in this representation. }.

The space of polynomials in representation $\bf R$ without the constraint is generated by the space of polynomials in representation $\bf R$ with the constraint by multiplying it with polynomials of a single variable $s+t+u$. For partition functions it means,
\be\label{constrained}
Z_{\bf R}(x)=(1-x^2)\tilde{Z}_{\bf R}(x).
\ee
Here $Z_{\bf R}$ denotes the partition function over polynomials in the representation ${\bf R}$ with the constraint $s+t+u=0$. Using equations \eqref{tildez} and \eqref{constrained}, we compute all the relevant partition functions. We find 
\begin{equation} \label{partfnsss}
\begin{split} 
&Z_{{\bf 1_S}}(x)=\denom, \qquad Z_{{\bf 1_A}}(x)=x^{6}\denom
,\qquad Z_{{\bf 2_M}}(x)=(x^2+x^4)\denom, \\
& Z_{{\bf 3}}(x)=
Z_{{\bf 1_S}} + Z_{{\bf 2_M}}(x)=\left( 1+ x^2 +x^4 \right) \denom \\
& Z_{{\bf 3_A}}(x)=
Z_{{\bf 1_A}} + Z_{{\bf 2_M}}(x)=\left( x^6+ x^2 +x^4 \right) \denom \\
&Z_{{\bf 6}}(x)=
Z_{{\bf 1_S}}+ Z_{{\bf 1_A}}  + 2Z_{{\bf 2_M}}(x)=\left( 1+ 2x^2 +2x^4 +x^6 \right) \denom \\
&{\rm where}\quad \denom=\frac{1}{(1-x^4)(1-x^6)} \\
\end{split}
\end{equation} 
(the representation ${\bf 3_A}$ is defined around 
\eqref{tof2} below).

The results \eqref{partfnsss} are easy to understand. The fully symmetric polynomials of $s, t$ and $u$ are arbitrary polynomials of the dimension 4 letter $s^2+t^2+u^2$ and 
the dimension 6 letter $stu$. The lowest dimensional anti-symmetric polynomial occurs at dimension 6 and is given by $s^2t-t^2s-s^2u+s u^2-u^2t+t^2u$. The most general antisymmetric 
function is obtained by multiplying this element by an 
arbitrary polynomial of $s^2+t^2+u^2$ and $stu$. Finally the 
two lowest doublets of mixed polynomials occur at dimension 2 
($s+t, s-t$) and dimension 4 ($(s^2+t^2 -2 u^2 ), (t^2+u^2 -2s^2)$). The most general mixed polynomial is given by multiplying these doublets by an arbitrary polynomial of 
$s^2+t^2+u^2$ and $stu$. These comments give a complete explanation of the formulae \eqref{partfnsss}. See subsection \ref{rgss} for a related discussion.

\subsection{Standard Bases for representations of $S_3$} \label{cbr}

Any function of $s$, $t$ and $u$ may be viewed as a function 
of $t$, $u$ only. The action of the permutation group on such 
functions generically produces several new functions. The set of $f(t,u)$ and all its permutations clearly transforms in 
a representation of the permutation group. The representations
thus obtained differ depending on the symmetry properties of the function $f(t,u)$ that generate the orbit. We consider 
various cases in turn. 
\begin{itemize} 
\item If $f(t,u)=f(u,t)=f(-t-u, u)$ then the the action of 
the permutation group leaves $f$ invariant. $f$ transforms in 
the ${\bf 1_S}$ representation of the permutation group.
\item If $f(t,u)=-f(u,t)=-f(-t-u, u)$ then the permutation 
group acts on $f$ as $f \rightarrow {\rm sgn} (P)f$. In this case $f$ transforms in ${\bf 1_A}$ representation of the permutation 
group. 
\item  If $f(t,u)=f(u,t)$ and 
\begin{equation}\label{sumcond}
f(t,u)+ f(-t-u, t)+f(u,-t-u)=0 
\end{equation}
Then the triplet  
\begin{equation} \label{tof1}
\left(N^{{\bf 2_M}, (1)}, N^{{\bf 2_M}, (2)}, N^{{\bf 2_M}, (3)} \right)
\equiv (f(t,u), f(u, s), f(s,t))
\end{equation}
transforms in the ${\bf 2_M}$ representation. Note that 
$$ \sum_i N^{{\bf 2_M}, (i)}=0.$$
consistent with the fact that the ${\bf 2_M}$ representation 
is two dimensional. In the rest of this paper we will refer 
to \eqref{tof1} as the `symmetric' basis  for  ${\bf 2_M}$ representation. Note the 
following property of the symmetric basis: the permutation $3 \leftrightarrow 4$ (equivalently $t \leftrightarrow u)$ leaves
the $(1)$ basis element invariant. In a similar way the  permutations $2 \leftrightarrow 4$ (resp.$2 \leftrightarrow 3$) leaves the $(2)$ (resp.$(3)$) basis element invariant. 
\item  If $f(t,u)=-f(u,t)$ and 
\begin{equation}\label{asm}
f(t,u)+ f(-t-u, t)+f(u,-t-u)=0 
\end{equation}
then the triplet  
\begin{equation} \label{tof3}
\left({\tilde N}^{{\bf 2_M}, (1)}, {\tilde N}^{{\bf 2_M}, (2)}, {\tilde N}^{{\bf 2_M}, (3)} \right)
\equiv (f(t,u), f(u, s), f(s,t))
\end{equation}
also transforms in the ${\bf 2_M}$ representation. Note that 
once again 
$$ \sum_i {\tilde N}^{{\bf 2_M}, (i)}=0.$$
We will refer to \eqref{tof3} as the `antisymmetric basis' for  ${\bf 2_M}$ representation. In the antisymmetric basis the permutation $3 \leftrightarrow 4$ (equivalently $t \leftrightarrow u)$ takes 
the $(1)$ basis element to minus itself. In a similar way the  permutations $2 \leftrightarrow 4$ (resp.$2 \leftrightarrow 3$) take the $(2)$ (resp.$(3)$) basis elements to minus themselves. 
\item  If $f(t,u)=f(u,t)$ but \eqref{sumcond} is not obeyed 
then the triplet \eqref{tof1}, which we now denote by 
\begin{equation} \label{tof2}
\left(N^{{\bf 3}, (1)}, N^{{\bf 3}, (2)}, N^{{\bf 3}, (3)} \right)
\equiv (f(t,u), f(u, s), f(s,t)),
\end{equation}
 transforms in the ${\bf 3}$
representation. Of course the ${\bf 3}$ is no more than 
the sum of a ${\bf 1_S}$ - namely the quantity on the LHS of 
\eqref{sumcond} - and a ${\bf 2_M}$ listed in the symmetric basis. \eqref{tof1} is the only basis we will use 
for the ${\bf 3}$ representation. 
\item  Similarly, if $f(t,u)=-f(u,t)$ but \eqref{asm} is not obeyed then the triplet \eqref{tof3}, which we now denote by 
\begin{equation} \label{tof4}
\left({\tilde N}^{{\bf {3_A}}, (1)}, {\tilde N}^{{\bf 
		{3_A}}, (2)}, {\tilde N}^{{\bf {3_A}}, (3)} \right)
\equiv (f(t,u), f(u, s), f(s,t))
\end{equation}
 transforms in the ${\bf 3}_{\bf A}$ representation. The ${\bf 3}_{\bf A}$ is no more than 
the sum of a ${\bf 1_A}$ - namely the quantity on the LHS of 
\eqref{sumcond} - and a ${\bf 2_M}$ in the antisymmetric basis. \eqref{tof4} is the only basis we will use for the ${\bf 3_A}$ representation.
\item Finally, if $f(t,u)$ obeys no symmetry property whatsoever, then the orbit of the permutation group on $f$ 
is six dimensional; we obtain the ${\bf 6_{\rm left}}$ 
representation. One basis for this representation is obtained follows. We can write $f(t,u)$ as a sum of a symmetric and an
antisymmetric function $$f(t,u)= \frac{f(t,u) + f(u,t)}{2}
+ \frac{f(t,u) - f(u,t)}{2} $$
The symmetric part of $f$ then generates a ${\bf 3}$ while the 
antisymmetric part generates a ${\bf 3_A}$. We can then use 
our standard basis for the ${\bf 3}$ and ${\bf 3_A}$. This 
is indeed the strategy we will adopt when we need to 
choose a basis for the ${\bf 6}_{\rm left}$ representation. We
emphasize, however, that unlike in every other case described 
above our basis for the ${\bf 6}_{\rm left}$ is ambiguous in 
the following manner. The decomposition of the ${\bf 6}_{\rm left}$
into the ${\bf 3}$ and the ${\bf 3}_{\bf A}$ is convention dependent. 
This can be understood in many ways. From a group theoretic 
viewpoint the grouping of the ${\bf 1_S}+ 2 \cdot {\bf 2_M} + {\bf 1_A}$ 
into a ${\bf 1_S}+ {\bf 2_M}$ and a ${\bf 1_A}$ + ${\bf 2_M}$ is inherently ambiguous, as there is no canonical way of choosing 
to associate one of the two ${\bf 2_M}$'s with the ${\bf 1_S}$ 
and the other ${\bf 2_M}$ with the ${\bf 1_A}$. Given one such grouping, we can as well replace the two ${\bf 2_M}$'s by any linearly independent linear combination of the two. From 
the practical constructive point of view the  procedure outlined above has a degree of arbitrariness because the basis 
we obtain depends not just on the orbit of functions obtained 
from $f(t,u)$ but also on a choice of a particular function
in the orbit (in this case $f(t, u)$ itself, rather than, 
for instance, $f(u, s)$ to get the procedure going). Different choices of starting functions differ in that they 
group different linear combinations of the two independent 
${\bf 2_M}$'s with the ${\bf 1_S}$ and ${\bf 1_A}$. When we need  
a basis for objects that transform in the  ${\bf 6}_{\rm left}$ below we will indeed decompose this representation into the 
${\bf 3}$ and ${\bf 3}_{\bf A}$, but will be very careful to explicitly spell out the particular conventions that give 
meaning to this decomposition. 
\end{itemize}

To end this subsection we present the explicit computation 
of a particular Clebsch-Gordon coefficient. Recall that 
the ${\bf 2_M} \times {\bf 2_M}= {\bf 1_S} + {\bf 1_A}+ {\bf 2_M}$.
Working with two copies of the ${\bf 2_M}$ in the symmetric 
basis we will now work out the explicit linear combination 
of products of original basis elements that transform in the 
${\bf 1_A}$. Let
\begin{equation} \label{toftd}
(N_1^{{\bf 2_M}, (1)}, N_1^{{\bf 2_M}, (2)}, N_1^{{\bf 2_M}, (3)}
), ~~~~(N_2^{{\bf 2_M}, (1)}, N_2^{{\bf 2_M}, (2)}, N_2^{{\bf 2_M}, (3)})
\end{equation}
represent two distinct vectors that transform in the
symmetric representation of the ${\bf 2_M}$  basis. Then
it is not difficult to verify that  
\begin{equation}\label{pcla}
N_1^{{\bf 2_M}, (1)} \left( N_2^{{\bf 2_M}, (2)} -N_2^{{\bf 2_M}, (3)}  \right) + N_1^{{\bf 2_M}, (2)} \left( N_2^{{\bf 2_M}, (3)} -N_2^{{\bf 2_M}, (1)}  \right) + 
N_1^{{\bf 2_M}, (3)} \left( N_2^{{\bf 2_M}, (1)} -N_2^{{\bf 2_M}, (2)}  \right)
\end{equation} 
transforms in the ${\bf 1_A}$ representation\footnote{One quick way to see this is to specialize to the 
particular case in which the two abstract triplets above 
are triplets of functions. Suppose the two ${\bf 2_M}s$ in symmetric basis be  
 $$ (f_1(t,u), f_1(s, u), f_1(u, t)), ~~~
 (f_2(t,u), f_2(s, u), f_3(u, t))$$
 (where all functions are symmetric in their arguments as 
 appropriate for the symmetric basis). It is trivial to verify 
 that the function 
\begin{equation}\label{antisym} 
f_1(u,t) \left( f_2(s, u)-f_2(t,s) \right) 
 +  f_1(s,u) \left( f_2(t, s)-f_2(t,u) \right)
 +  f_1(t,s) \left( f_2(t, u)-f_2(s,u) \right)
 \end{equation} 
is completely antisymmetric in its arguments and 
so transforms in the ${\bf 1_A}$ representation.

The function in \eqref{antisym} is antisymmetric.
	
	It follows
	that 
\begin{equation}\label{pclaa}
N_1^{{\bf 3}, (1)} \left( N_2^{{\bf 2_M}, (2)} -N_2^{{\bf 2_M}, (3)}  \right) + N_1^{{\bf 3}, (2)} \left( N_2^{{\bf 2_M}, (3)} -N_2^{{\bf 2_M}, (1)}  \right) + 
N_1^{{\bf 3}, (3)} \left( N_2^{{\bf 2_M}, (1)} -N_2^{{\bf 2_M}, (2)}  \right)
\end{equation}
and  
\begin{equation}\label{pclab}
N_1^{{\bf 3}, (1)} \left( N_2^{{\bf 3}, (2)} -N_2^{{\bf 3}, (3)}  \right) + N_1^{{\bf 3}, (2)} \left( N_2^{{\bf 3}, (3)} -N_2^{{\bf 3}, (1)}  \right) + 
N_1^{{\bf 3}, (3)} \left( N_2^{{\bf 3}, (1)} -N_2^{{\bf 3}, (2)}  \right)
\end{equation} 
also transform in the ${\bf 1_A}$ representation of $S_3$.}.

\subsection{Counting S-matrices}

As we have seen, four particle S-matrices are always 
characterized by an infinite number of parameters. However 
the number of parameters that appears at any given 
derivative order is  finite. For this reason it 
is useful to characterize the data needed to specify an 
$S$ matrix by a partition function 
\begin{equation}\label{Smpf}
Z_{ \text{S-matrix}}(x) = \sum_{m=0}^\infty n(m)x^m
\end{equation} 
where $n(m)$ is the number of free parameters that appear
in the most general polynomial S-matrix at $m$ derivative order\footnote{In computing the derivative order we assign $\epsilon_{\mu}$ dimension zero.}.

The local module structure presented above makes the 
computation of $Z_{ \text{S-matrix}}(x)$ rather simple. 
Let us first consider the case of both graviton and 
photon scattering in $D \geq 5$. As we have mentioned 
above, in this case the local modules are freely generated.
The generators of these modules, $E_J^\sigma$, are 
each characterized by their derivative dimension $\Delta_J$ and their $S_3$ representation ${\bf R_J}$. It follows from \eqref{cgi} that in this 
case $Z_{ \text{S-matrix}}(x)$ is 
given by
\begin{equation} \label{pfnnnn}
Z_{ \text{S-matrix}}(x) = \sum_{J} x^{\Delta_J} 
Z_{{\bf R_J}}(x)
\end{equation} 
where $Z_{\bf R_J}(x)$ are listed in \eqref{partfnsss}.

Let us now turn to the case $D \leq 4$. In this case the local Modules 
are not freely generated, but instead have relations. As we will see below 
it turns out that the relation modules are all themselves free (there are no 
relations for relations). Let the generators for the relations be labelled by their derivative dimension $\Delta_I$ and  $S_3$ representation ${\bf R_I}$. It follows that in $D \leq 4$ the formula \eqref{pfnn} is modified
to 
\begin{equation} \label{pfnn}
Z_{ \text{S-matrix}}(x) = \sum_{J} x^{\Delta_J} 
Z_{{\bf R_J}}(x) - \sum_{I} x^{\Delta_I} 
Z_{{\bf R_I}}(x)
\end{equation} 
where the sum over $J$ runs over all module generators, while the sum over 
$I$ runs over all relation generators. 

\subsection{Regge growth} \label{rgss}

Recall that the generators of 
the bare module are zero order in derivatives in the parity even sector, and also in the parity odd sector for even $D$. In these cases the generators are functions of $\alpha_i, \polo_i$ but are not separately functions of $s$, $t$ and $u$. On the other hand when $D$ is odd, the parity odd generators are
proportional to $\sqrt{stu}$ times functions of $\alpha_i, \polo_i$. In order to deal uniformly with all cases below, we introduce the variable $a$; $a=0$ for parity even S-matrices in every $D$  and parity odd S-matrices in even $D$. $a=3$ for parity odd S-matrices in odd $D$. 

We will now derive a lower bound for the Regge growth for 
local  S-matrices at $2n+a$ order in derivatives.  In order to do this we note that every such S-matrix is an $n^{\rm th}$ order descendant of some bare generator. The generator in question might transform in the 
${\bf 1_S}$, the ${\bf 1_A}$ or the ${\bf 2_M}$ representation, or a linear combination of these. We take these cases up in turn.

Consider any bare generator, say $|e_{\bf S}\rangle$, that transforms in the ${\bf 1_S}$. As explained under \eqref{partfnsss}, the most general 
$n^{\rm th}$ level descendant of this generator that itself 
transforms in the ${\bf 1_S}$ representation is given, for 
$n \geq 2$, by  
\begin{equation}\label{mjil}
\left( \sum_{k, m} a_{k,m} (stu)^k (s^2+t^2+u^2)^m \right) | e_{\bf S} \rangle 
\end{equation} 
where the sum runs over all terms with $3k+2m=n$. It is 
easy to convince oneself that all the S-matrices in 
\eqref{mjil} grow at least as fast as 
\begin{equation}\label{srg}
s^{ \left( 2 \left[\frac{n+2}{3}\right] + \frac{ a}{3} \right) } 
\end{equation}
in the Regge limit\footnote{When 
$n=3p$, we obtain the slowest growth when $a_{k,m}$  is non-zero only for $k=p$ and $m=0$. When $n=3p+1$ (and $p\geq 1$) the slowest growth is achieved when $a_{k,m}$ is non-zero only when $k=p-1$ and $m=2$. When $n=3p+2$ we get the slowest growth for the monomial with $k=p$ and $m=1$.}.
where $[m]$ represents the largest integer no smaller than $m$. 

Now consider a bare generator $|e_{\bf A}\rangle$ that transforms 
in the antisymmetric representation. The most general 
descendant at $2n+a$ order in derivatives is given by 
\begin{equation}\label{mjilas}
\left( s^2u-u^2s + t^2s-t^2u -s^2t+u^2 t \right) \left( \sum_{k, m} a_{k,m} (stu)^k (s^2+t^2+u^2)^m \right) | e_{\bf A}\rangle 
\end{equation} 
where $3k+2m=n-3$. For $n=3$ and $n \geq 5$ all terms in 
\eqref{mjilas} grow at least as fast in the Regge limit as\footnote{When 
	$n=3p$, we obtain the slowest growth when $a_{k,m}$  is non-zero only for $k=p-1$ and $m=0$. When $n=3p+1$ (and $p\geq 2$) the slowest growth is achieved when $a_{k,m}$ is non-zero only when $k=p-2$ and $m=2$. When $n=3p+2$ we get the slowest growth for the monomial with $k=p-1$ and $m=1$.}, 
\begin{equation}\label{srgas}
s^{ \left( 2 \left[\frac{n-1}{3}\right] + 3 + \frac{ a}{3} \right) }. 
\end{equation}

Finally consider a bare generator multiplet that transforms 
in the ${\bf 2_M}$ representation. Let the triplet of 
basis vectors 
$$ (|e_{\bf M}^{(1)}\rangle, |e_{\bf M}^{(2)}\rangle , |e_{\bf M}^{(3)} \rangle ) $$
transform in the ${\bf 2_M}$ representation in the symmetric basis (see around \eqref{tof1})\footnote{Note in particular that 
$$|e_{\bf M}^{(1)}\rangle +|e_{\bf M}^{(2)}\rangle +|e_{\bf M}^{(3)} \rangle=0.$$}.
The most general $2n+a$ derivative descendant of these basis 
vectors is given either by 
\begin{equation}\label{mjilmo}
\left( \sum_{k, m} a_{k,m} (stu)^k (s^2+t^2+u^2)^m \right) 
\bigg( s|e_{\bf M}^{(1)}\rangle +t|e_{\bf M}^{(2)}\rangle +u|e_{\bf M}^{(3)} \rangle \bigg)
\end{equation}
with $3k+2m=n-1$ 
or by 
\begin{equation}\label{mjilmt}
\left( \sum_{k, m} a_{k,m} (stu)^k (s^2+t^2+u^2)^m \right) 
\bigg( \left( t^2+u^2-2s^2 \right) |e_{\bf M}^{(1)}\rangle + 
\left( u^2+ s^2-2t^2 \right) |e_{\bf M}^{(2)}\rangle +
\left( s^2+ t^2-2u^2 \right)|e_{\bf M}^{(3)} \rangle \bigg)
\end{equation}
with $3k+2m=n-2$.
All the S-matrices in \eqref{mjilmo} and \eqref{mjilmt}  grow at least as fast in the Regge limit as\footnote{When 
	$n=3p$, we obtain the slowest growth from the term in  \eqref{mjilmo}  with $k=p-1$ and $m=1$. When $n=3p+1$ the slowest growth comes from the term in \eqref{mjilmo} with $k=p$ and $m=0$. When $n=3p+2$ we get the slowest growth for the monomial in \eqref{mjilmt} with $k=p$ and $m=0$.}  
\begin{equation}\label{srgm} \begin{split} 
&s^{ \alpha + \frac{ a}{3}  } \\
&\alpha = 2p +1 ~~~~~{\rm when } ~~~~{n=3p}\\
&\alpha = 2p+1 ~~~~~{\rm when } ~~~~{n=3p+1}\\
&\alpha=2p+2 ~~~~~{\rm when }~~~~{n=3p+2}\\
\end{split}
\end{equation}

Combining all the results above we conclude that every 
local S-matrix at $2n+a$ derivative order grows at least 
as fast in the Regge limit as 
\begin{equation}\label{srgtot} \begin{split} 
&s^{ \alpha(n) + \frac{ a}{3}  } \\
&\alpha(n) = 2p  ~~~~~~~~~{\rm when } ~~~~{n=3p}\\
&\alpha(n) = 2p+1 ~~~~~{\rm when } ~~~~{n=3p+1}\\
&\alpha(n) =2p+2 ~~~~~{\rm when }~~~~{n=3p+2}\\
\end{split}
\end{equation}
The bound in the first line in the first line 
in \eqref{srgtot} is saturated by the state $(stu)^p | e_{\bf S}\rangle$, the bound in the second line is saturated 
by the state 
$$(stu)^p \left( s|e_{\bf M}^{(1)}\rangle +t|e_{\bf M}^{(2)}\rangle +u|e_{\bf M}^{(3)}\right)$$
and the bound in the third line in \eqref{srgtot} 
is saturated both by  $(stu)^p (s^2+t^2+u^2)| e_{\bf S}\rangle$
and by 
$$(stu)^p \left( \left( t^2+u^2-2s^2 \right) |e_{\bf M}^{(1)}\rangle + 
\left( u^2+ s^2-2t^2 \right) |e_{\bf M}^{(2)}\rangle +
\left( s^2+ t^2-2u^2 \right)|e_{\bf M}^{(3)} \rangle \right).$$

As we have mentioned in the introduction, we 
are particularly interested in local S-matrices that grow no faster than $s^2$ in the Regge limit. We end this section with a complete listing of all module elements that 
have this feature. For parity even S-matrices - or parity odd S-matrices 
in even $D$ these possibilities are 
\begin{itemize} 
	\item At zero order in derivatives generators of the bare module in the 
	${\bf 1_S}$ representation yield S-matrices that grow like $s^0$ in the 
	Regge limit. 
	\item At the two derivative level we have 
	S-matrices of the form \eqref{mjilmo} with 
	$a_{k,m}$ non-zero only when $k=m=0$. These S-matrices grow like $s$ in the Regge limit. 
	\item At fourth order in derivatives 
	we have S-matrices of the form \eqref{mjil}
	with $a_{k,m}=0$ unless $k=0, m=1$. We also have S-matrices of the form \eqref{mjilmt} with $a_{k, m}=0$ unless $k=m=0$. These S-matrices grow like $s^2$ 
	in the Regge limit. 
	\item At six derivative order the unique such S-matrix is of the form \eqref{mjil}
	with $a_{k,m}=0$ unless $k=1$ and $m=0$. 
	\end{itemize} 
All other local S-matrices - in particular all 
S-matrices that are of 8 or higher order in 
derivatives - necessarily grow faster than 
$s^2$ in the Regge limit. 

In the case of parity odd S-matrices in odd 
$D$, the only Module elements that grow no faster than $s^2$ in the Regge limit are 
\begin{itemize}
		\item At 3 derivative order we have generators of the bare module in the 
	${\bf 1_S}$ representation. The corresponding S-matrices grow like $s$ 
	in the Regge limit. 
	\item At the five derivative level we have 
	S-matrices of the form \eqref{mjilmo} with 
	$a_{k,m}$ non-zero only when $k=m=0$. These S-matrices grow like $s^2$ in the Regge limit.  
	\end{itemize} 
All other S-matrices - in particular all S-matrices of 7 or higher order in derivatives - grow faster than $s^2$ in the Regge limit.

\section{Generators of the bare module}\label{index-structure}

In this section we will enumerate and explicitly construct $e_I(\alpha_i,\polo_i)$, the basis of the bare module defined in subsection \ref{fbm}, both for four photon scattering and for four 
graviton scattering in every spacetime dimension. We also keep track of the $S_3$ transformation properties of $e_I$. 
This section is devoted purely to the study of the bare module. We postpone the study of the embedding of the local module  into 
the bare module that we construct in this subsection to later in this paper, beginning with the next section in which we use  a systematic plethystic program  to enumerate the analytic index structures $E_J(p_i,\epsilon_i)$.

\subsection{Enumeration} \label{enumeration}

In this subsection we  count the rank of the bare module, \emph{i.e.} the number of linearly independent basis elements $e_I$. As explained 
in subsection \ref{fbm}, these are simply the set of $SO(D-3)$ 
and $\Z_2 \times \Z_2$ invariant polynomials of $\alpha_i$ and 
$\polo_i$ with the appropriate homogeneity properties. We separately enumerate parity odd and parity even generators of the bare module. As photons/gravitons have no propagating degrees of freedom when $D \leq 2$ and $D\leq 3$ respectively, 
we restrict our attention to $D\geq 3$ for photons $D\geq 4$ 
for gravitons. In the next subsection we will proceed to actually construct these basis elements $(e_I)$ and list their 
$S_3$ transformation properties.  

Under $SO(D-3)$, the effective polarization for photon takes values in the space $\rho =(\ts\oplus \tv)$ 
(for gravitons, $\rho=(\ts\oplus\tv\oplus \tt)$)\footnote{As we have explained above, photon $S$ matrices 
are separately linear in each of $(\polo_i, \alpha_i)$. Here $\polo_i$ is the $\tv$ while $\alpha_i$ is the $\ts$.   
On the other hand gravitational S-matrices are quadratic separately in each of $(\polo_i, \alpha_i)$; 
and are evaluated subject to the constraint 
$\polo_i \cdot \polo_i+\alpha_i^2=0$. The terms $\polo_i \polo_i$ is 
the $\tt$ above (this term is effectively traceless as the 
constraint \eqref{constrem} allows us to trade its trace for $\alpha_i^2$), 
the terms $\alpha_i \polo_i$ is the $\tv$ and the terms $\alpha_i^2$ are the $\ts$.}.
Here $\ts,\tv,\tt$ are scalar, vector and symmetric traceless tensor of $SO(D-3)$ respectively.  The number of index structures is the number of singlets in 
\begin{equation}\label{cgpg}
\rho^{\otimes 4}|_{\Z_2\times \Z_2}:  i.e. ~~~~~~ (\ts\oplus \tv)^{\otimes 4}|_{\Z_2\times \Z_2}~~ \rm{for\,\,photons},\qquad (\ts\oplus \tv\oplus \tt)^{\otimes 4}|_{\Z_2\times \Z_2}~~ \rm{for\, \,gravitons}
\end{equation}
where the notation $|_G$ stands for  projection onto $G$ invariants. 

In order to perform the necessary enumeration we use the formula
\begin{equation} \label{expzt}
\rho^{\otimes 4}|_{\Z_2\times \Z_2} = \rho^4 \ominus 3(S^2\rho\otimes \wedge^2 \rho).
\end{equation}
where $S^2\rho$ and $\wedge^2\rho$ stand for the symmetric and antisymmetric square of $\rho$ respectively.  \eqref{expzt} 
was derived and employed in \cite{Kravchuk:2016qvl} to study a closely  related problem, namely the enumeration of inequivalent 
tensor structures in CFT four point functions. We present a simple `physics' derivation of \eqref{expzt} in Appendix  
\ref{projection}.

 It follows from \eqref{expzt} that the dimensionality of 
  $\rho^{\otimes 4}|_{\Z_2\times \Z_2}$ can be enumerated by 
 employing the following simple algorithm. Let 
  $n^{\bf 1_S}_a$ and $n^{\bf 1_A}_a$ respectively represent the number of copies of the $SO(D-3)$ representation $a$ that appear in the symmetric and antisymmetric square of $\rho$. Then the number of $SO(D-3)$ singlets in the $\Z_2\times \Z_2$ invariant tensor product of 
 four copies of $\rho$ is given by,
\begin{equation}\label{sppnn}
\sum_a \left (   (n^{\bf 1_S}_a+n^{\bf 1_A}_a)^2 -3n_a^{\bf 1_S} n_a^{\bf 1_A} \right) 
= \sum_a \left (   (n^{\bf 1_S}_a-n^{\bf 1_A}_a)^2 + n_a^{\bf 1_S} n_a^{\bf 1_A} \right).
\end{equation} 
Using \eqref{sppnn} the enumeration of simultaneous $\Z_2\times \Z_2$ and $SO(D-3)$ invariants is straightforward. We have tabulated the result, in every dimension,  in Table \ref{counting-is}. Even though the counting based on 
\eqref{sppnn} only yields the total number of simultaneous $SO(D-3)$ and
$\Z_2 \times \Z_2$ singlets, in Table \ref{counting-is} we have anticipated the 
results of the rest of this section (see subsection \ref{constructionei} and Appendix \ref{appendix-is}) to 
separate this total into the separate contribution of even and odd structures.
\begin{table}
\begin{center}
\begin{tabular}{|l|l|l|}
\hline
photons & even & odd \\
\hline
$D\geq 8$ & 7& 0 \\
\hline
$D=7$ & 7 & 1 \\
\hline
$D=6$ & 7 & 1 \\
\hline
$D=5$ & 7 & 0 \\
\hline
$D=4$ & 5 & 2\\
\hline
$D=3$ &  1 & 1\\
\hline
\end{tabular}
\qquad \qquad \qquad 
\begin{tabular}{|l|l|l|}
\hline
gravitons & even & odd \\
\hline
$D\geq 8$ & 29& 0 \\
\hline
$D=7$ & 29 & 7 \\
\hline
$D=6$ & 28 & 9 \\
\hline
$D=5$ & 22 & 3 \\
\hline
$D=4$ & 5 & 2\\
\hline
$D=3$ & - & -\\
\hline
\end{tabular}
\end{center}
\caption{Number of parity even and parity odd index structures for 4-photon and 4-graviton S-matrix as various dimensions.}
\label{counting-is}
\end{table}

Note that the number of parity even structures do not depend on dimension for $D\geq 7$. 
This is because the most general configuration of four transverse polarizations $\polo_i$ can always be rotated to be in a specified $4$ dimensional hyperplane orthogonal to the scattering plane - say the hyperplane spanned by the directions 3, 4, 5, 6 if the scattering plane consists of the directions 0, 1, 2. It follows that the most general scattering kinematics can be achieved if the number of spacetime dimensions is $3+4=7$. Any additional dimensions merely plays the role of a spectator. 

The analysis of the previous paragraph does not apply to parity odd amplitudes as they depend on $\varepsilon$. The reason that the number of parity odd amplitudes stabilizes (to zero) for $D \geq 8$ may be understood as follows.  Parity odd S-matrices are linear in the  Levi-Civita tensor $\varepsilon$. Lorentz invariance requires that all free indices of $\varepsilon$ are contracted. As $\varepsilon$ is completely antisymmetric any given vector can contract with only one index of  $\varepsilon$. As four particle scattering amplitudes are functions of 7 independent vectors (the four polarization vectors $\epsilon_i$ and the three independent scattering momenta), parity odd  S-matrices do not exist in $D \geq 8$, but do exist in $D=7$, see table \ref{counting-is}.

As on-shell inequivalent $AdS$ bulk Lagrangians give rise to distinct boundary correlators, the counting of graviton four point structures in $D$ dimensions is the same as the counting of index structures for stress tensor four point functions in a  $D-1$ dimensional conformal field theory. This counting for parity even structures was performed in \cite{Dymarsky:2013wla}. It agrees with the counting presented here.

\subsection{Explicit construction of $e_I$ and 
$S_3$ transformations}\label{constructionei}

In this subsection we explicitly construct the basis elements $e_I(\alpha_i,\polo_i)$ and thereby obtain their $S_3$ transformation properties. Our construction is motivated by construction of index structure for CFT four point functions in \cite{Costa:2011dw}.
 In the main text we present the details of our construction only for $D \geq 7$ and simply
tabulate our final representation wise counting results in other 
dimensions. Detailed construction for $D < 7$ are presented in 
appendix \ref{appendix-is}. 

The parity even and odd generators for photons will be denoted by the letters $e$ and $o$ respectively. We will label the structures with the $S_3$ representation they transform under. For example, a parity even photon generator transforming in ${\bf 3}$ will be denoted as $e_{\bf 3}$. If there are multiple of them then we also include an arbitrarily assigned serial number in the subscript. We will only need to concern ourselves with ${\bf 1}_S$, ${\bf 1}_A$, ${\bf 3}$ and ${\bf 6}_{\rm left}={\bf 3}\oplus {\bf 3_A}$ representation. We always choose to work in standard representations \eqref{tof3} and \eqref{tof2} for ${\bf 3}$ and ${\bf 3_A}$. \footnote{ i.e,  the basis of structure transforming in ${\bf 3}$ (resp. ${\bf 3_A}$) such that state $(1)$ is $\Z_2$ symmetric (resp. antisymmetric) under the swap $1\leftrightarrow 2$. States $(2)$ and $(3)$ are obtained from $(1)$ by cyclic permutation $234\to 342$.} We will also include the space-time dimension in the superscript when it needs to be emphasized. This helps especially in the case of parity odd structures which crucially depend on space-time dimensions.

Let us first consider the case of parity even structures for photons. The structures will be labelled by the  $S_3$ representation they transform under.  

In this case  $\rho^{\otimes 4}|_{\Z_2\times \Z_2}$ has  $7$ distinct basis elements (see table \ref{counting-is}). Keeping in mind that $\rho=\ts + \tv$, it follows that these 7 structures each have their origin in one (and only one) of the 
tensor products\footnote{The tensor products $\ts^{\otimes 3}\tv$ and 
$\tv^{\otimes 3} \ts$ do not contribute as they contain no 
$SO(D-3)$ singlets.}
\be
\ts^{\otimes 4}|_{\Z_2\times \Z_2},\qquad \ts^{\otimes 2}\tv^{\otimes 2}|_{\Z_2\times \Z_2},\qquad \tv^{\otimes 4}|_{\Z_2\times \Z_2}.
\ee
 A slight generalization of the enumeration 
method described in the previous subsection and appendix \ref{projection} allows us to separately enumerate the basis 
elements in each of these sectors. We find that there is one element in $\ts^{\otimes 4}|_{\Z_2\times \Z_2}$ and 
three each in $\ts^{\otimes 2}\tv^{\otimes 2}|_{\Z_2\times \Z_2}$
and $\tv^{\otimes 4}|_{\Z_2\times \Z_2}$.

It is easy to explicitly construct these basis elements. Consider, for example, the sector $\tv^{\otimes 4}|_{\Z_2\times \Z_2}$. The $3$ basis elements in this sector are
\begin{equation}\label{exampstruct}
e_{{\bf 3},1}^{(1)}=(\polo_1\cdot\polo_2)(\polo_3\cdot\polo_4),\qquad e_{{\bf 3},1}^{(2)}=(\polo_3\cdot\polo_1)(\polo_2\cdot\polo_4)\qquad e_{{\bf 3},1}^{(3)}=(\polo_2\cdot\polo_3)(\polo_1\cdot\polo_4)
\end{equation}
It is easy to check that these structures are $\Z_2\times \Z_2$ invariant as desired. Also, each of the three elements 
listed in \eqref{exampstruct} happens to be invariant under a single  $\Z_2$ exchange transformation\footnote{Viewed as elements of 
$S_3=S_4/(\Z_2 \times \Z_2)$ and working in the `gauge' in which the fourth particle is fixed, the exchange elements that leave 
the three structures in \eqref{exampstruct} fixed  respectively are $(213)$, $(321)$ and $(132)$. }; moreover the three elements 
 are mapped to each other by the action of the cyclic elements of $S_3$ \footnote{See the discussion under \eqref{adjact} for a definition and listing of these cyclic elements.}. It follows from the discussion in the second last paragraph of 
subsection \ref{irrep} that these elements transform in the  $\bf 3$ representation of $S_3$ defined in and around \eqref{decomp3}. 
\begin{figure}
	\centering
	\includegraphics[scale=0.28]{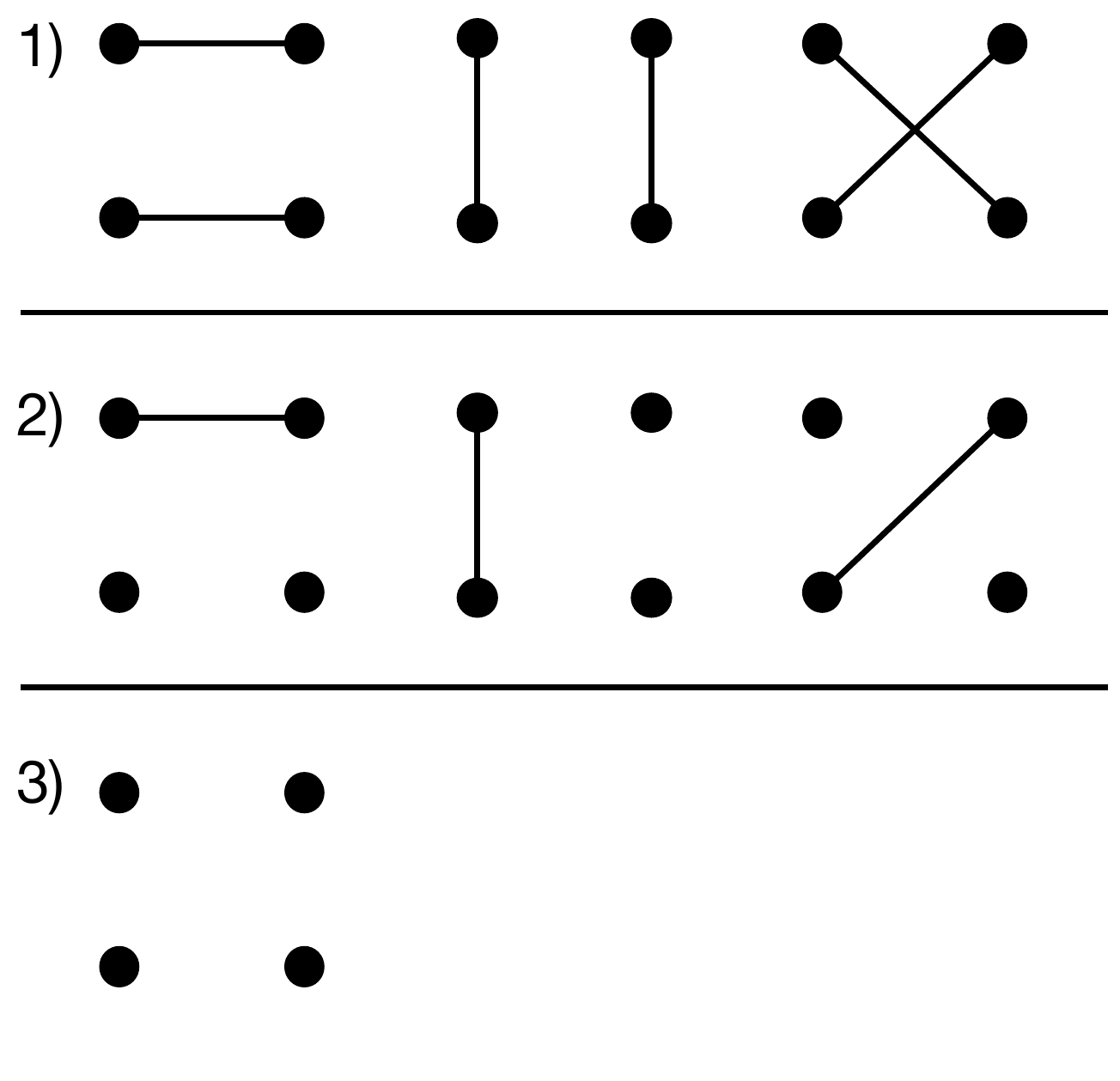}
	\caption{Parity even index structures of 4-photon scattering}
	\label{photon-structures}
\end{figure}

It is useful to introduce a graphical notation to denote 
structures such as those explicitly presented in \eqref{exampstruct}. Consider a graph with four vertices, 
one corresponding to each scattering particle. A line between any two vertices denotes the contraction of an $\polo$ at the corresponding vertices; for instance a line between vertex 1 and 
vertex 2 signifies the term $\polo_1 . \polo_2$; two lines 
between the same two vertices represents two factors of this dot product\footnote{Restated, 
	$\ts,\tv,\tt$ are denoted by a vertex with valency $0,1,2$ respectively. The contraction of indices means connecting pairs of vertices by lines such that all the valencies are saturated.}. Factors of $\alpha_i$ are not explicitly denoted in the graph, but are inserted into the corresponding expressions to saturate homogeneity. 
With these conventions, the three expressions  in \eqref{exampstruct} are represented by the three graphs on the first line of \ref{photon-structures}. On the other hand 
the three structures that lie in the $\ts^{\otimes 2}\tv^{\otimes 2}|_{\Z_2\times \Z_2}$ sector are the $\Z_2 \times \Z_2$ symmetrized versions of the graphs listed in the 
second line of figure \ref{photon-structures}. After $\Z_2 \times \Z_2$ symmetrization, the algebraic expressions corresponding 
to the figures in this second line are 
\begin{equation}\label{algexpsl} \begin{split} 
e_{{\bf 3},2}^{(1)}=\left( \polo_1 . \polo_2\, \alpha_3 \alpha_4 + \polo_3.\polo_4\,  \alpha_1 \alpha_2 \right), \quad
e_{{\bf 3},2}^{(2)}=\left( \polo_1 . \polo_3 \, \alpha_2 \alpha_4 + \polo_2.\polo_4 \, \alpha_1 \alpha_3 \right), \quad
e_{{\bf 3},2}^{(3)}=\left( \polo_2 . \polo_3 \,\alpha_1 \alpha_4+ 
\polo_1.\polo_4 \, \alpha_2 \alpha_3 \right)
\end{split}
\end{equation}
As in the case of \eqref{exampstruct}, the expressions in \eqref{algexpsl} are each invariant under a single  $\Z_2$ exchange element and are also mapped to each other by the action of $\Z_3$. It thus follows that the expressions in \eqref{algexpsl} - like those in \eqref{exampstruct} - transform in the ${\bf 3}$ representation of $S_3$.

Finally 
the single structure in the $\ts^{\otimes 4}|_{\Z_2\times \Z_2}$ is denoted by the very simple figure 
in the third line of figure \ref{photon-structures}. The 
corresponding expression is simply $\alpha_1\alpha_2\alpha_3\alpha_4$, and clearly transforms 
in the ${\bf 1_S}$ representation of $S_3$. We denote it as $e_{\bf S}$.
Using \eqref{decomp3} it follows that the seven parity even photon structures for $D \geq 7$ transform under $S_3$ as 
\begin{equation}\label{photonlisting}
7= 3 \cdot {\bf 1_S}+2 \cdot {\bf 2_M}
\end{equation}

\begin{figure}
	\centering
	\includegraphics[scale=0.35]{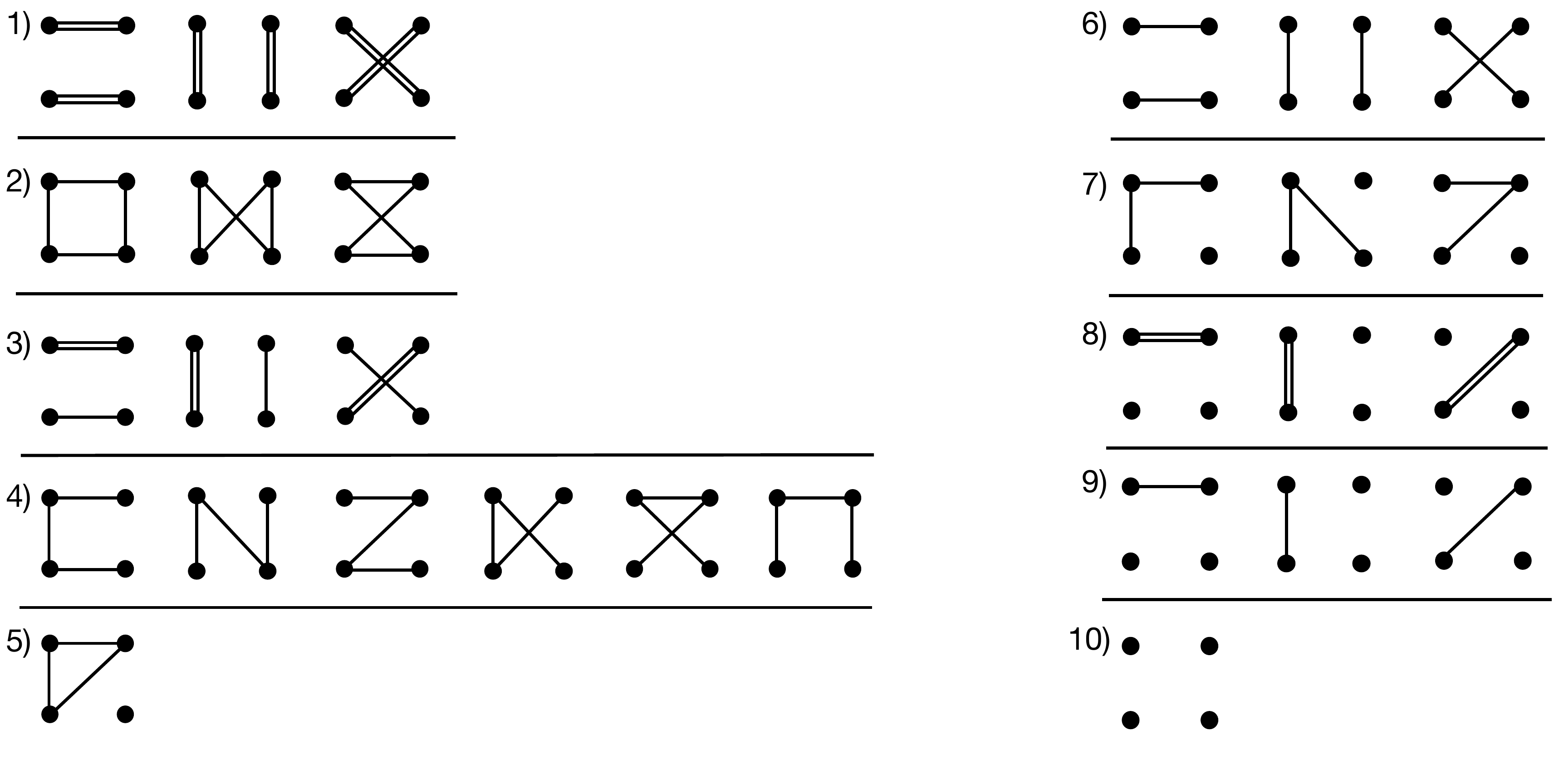}
	\caption{Parity even index structures of 4-graviton scattering.}
	\label{graviton-structures}
\end{figure}

We now turn to the explicit construction of the 29 parity even structures index structures for gravitational scattering  in $D \geq 7$ (see table \ref{counting-is} ). The structures in question are all listed in figure \ref{graviton-structures}. 
Several comments are in order. First, we re-emphasize that 
individual diagrams in figure \ref{graviton-structures} are 
not always $\Z_2\times \Z_2$ invariant but by the corresponding figure, we mean the object obtained after $\Z_2\times \Z_2$ symmetrization\footnote{In other words $\Z_2\times \Z_2$ images, i.e. images under double transpositions, are implicitly added in our diagrams when needed.}.
Second, we have arranged the graphs in figure \ref{graviton-structures} so that structures that lie in the same $\Z_3$ orbits all appear on the same line. More specifically, expressions denoted by the diagrams in lines 1), 2), 3), 6), 7), 8) and 9) all transform in the ${\bf 3}$ representation of $S_3$. Terms signified by the diagrams in line 4) transform in the ${\bf 6}_{\rm left}$ dimensional 
representation (see around \eqref{adjact}). And the terms denoted by the diagrams in lines 5) and 10) transform in the 
trivial ${\bf 1_S}$ representation of $S_3$. Using \eqref{decomp3},\eqref{sdr} it follows that the $S_3$ representation content of parity invariant graviton index structures is given by 
\begin{eqnarray} \label{gravstruct} 
29 &=& 10 \cdot {\bf 1_S}+ 9 \cdot {\bf 2_M}+1 \cdot {\bf 1_A}.
\end{eqnarray}

We now turn to a discussion of parity odd S-matrices in $D \geq 8$. Such S-matrices are linear in ${\tilde \varepsilon}^{D-3}$ (see \eqref{dmt} for a definition). For $D\geq8$ the number of free indices of ${\tilde \varepsilon}^{D-3}$ (or $N({\tilde \varepsilon})^{D-3}$) tensor is  $\geq 5$. As the only vectors available to contract with this tensor are the 4 $\polo_i$, there are non parity odd S-matrix structures in $ \geq 8$.

\begin{figure}
	\centering
	\includegraphics[scale=0.35]{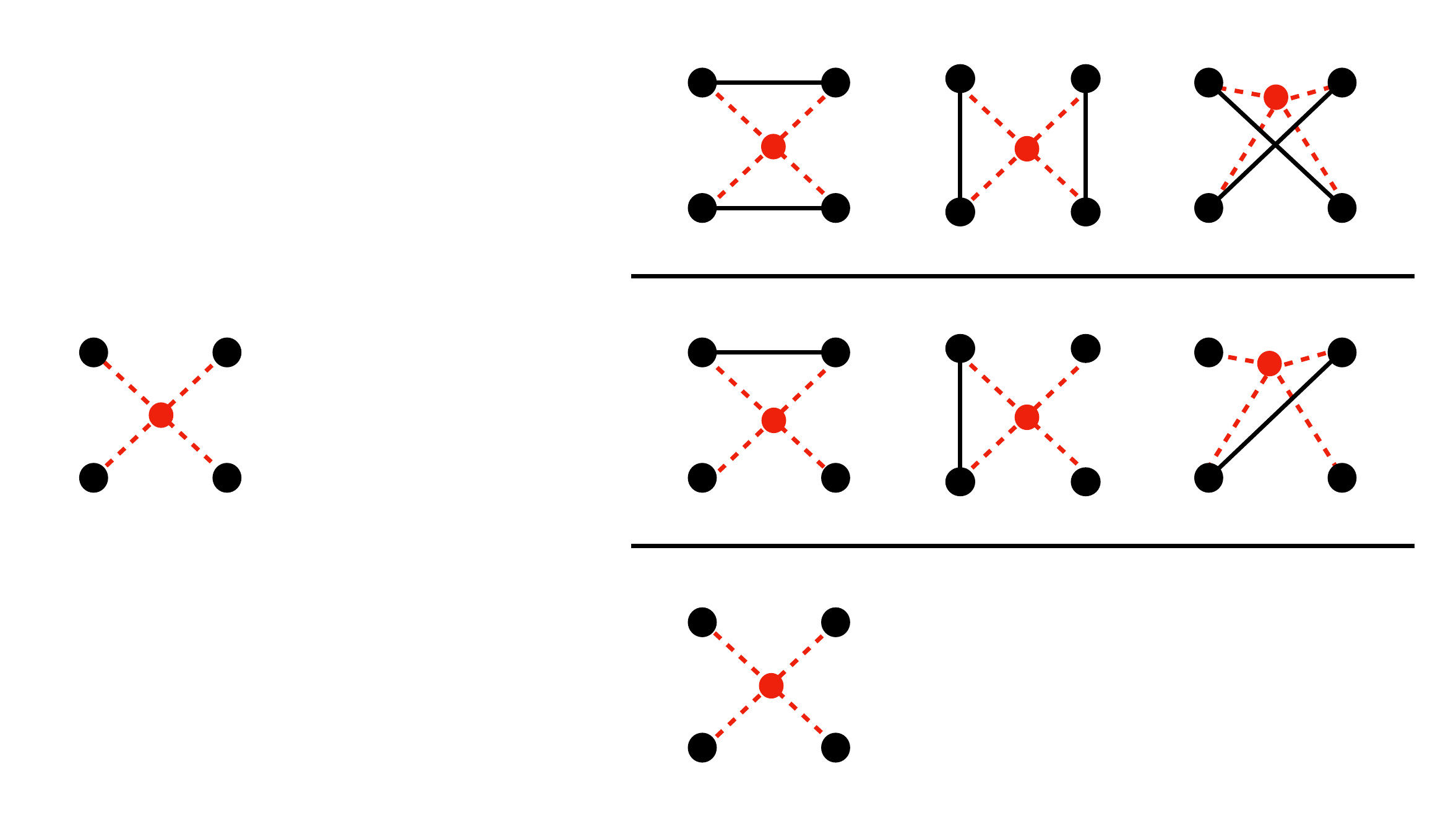}
	\caption{Parity odd index structures of 4-photon and 4-graviton scattering in $D=7$ respectively. The $\tilde \varepsilon$ symbol is denoted by a red dot with $D-3$ valency.}
	\label{parity-odd-7}
\end{figure}  
Let us now construct the parity odd structures for photon S-matrix in $D=7$. The tensor ${\tilde \varepsilon}^{4}$ has $4$ free indices so  it can be contracted with the 4 $\polo_i$ in a unique way. 
\begin{equation} \label{moreep}
o^{D=7}_{\bf S}={\tilde\varepsilon}^{4}_{\mu\nu\rho\sigma} \polo_1\,^\mu\polo_2\,^\nu\polo_3\,^\rho\polo_4\,^\sigma={\varepsilon}_{\alpha\beta\gamma\mu\nu\rho\sigma}\, p_1^\alpha \,p_2^\beta \,p_3^\gamma \,\polo_1\,^\mu\polo_2\,^\nu\polo_3\,^\rho\polo_4\,^\sigma.
\end{equation}
Consequently there is a single parity odd structure in
$\rho^{\otimes 4}|_{\Z_2\times \Z_2}$ for the case of photons in 
seven dimensions. This S-matrix transforms in the ${\bf 1}_S$ 
representation of $S_3$\footnote{In order to obtain the correct symmetry transformation property of this term it is important 
to permute the momenta that go into the definition of 
${\tilde \varepsilon}^4$ along with the factors of $\polo_i$. 
In order to avoid errors it is best to express ${\tilde \varepsilon}^4$ in term of $\varepsilon$ using \eqref{dmt} -
as has been done on the RHS of \eqref{moreep} - before 
performing permutations.}. In equations
\begin{equation}\label{phot8}
1= 1 \cdot {\bf 1_S}
\end{equation}

As for the case of parity even structures, it is useful to develop a graphical notation to denote parity odd contraction structures.  Our graphs now have 5 vertices; four for each of the scattering particles and a red dot for ${\tilde \varepsilon}^{D-3}$. The red vertex always has valency $D-3$ 
i.e. always has $D-3$ dotted lines emerging out of it.
A dotted line between the red dot and, say, the vertex $1$ 
denotes that $\polo_1$ has contracted with one of the 
free indices of ${\tilde \varepsilon}^{D-3}$. The meaning of lines between usual vertices is the same as for the parity even 
diagrams. With these conventions in place, the unique parity 
odd structure for 4 photon scattering in $D=7$ is denoted by the first diagram in figure \ref{parity-odd-7}. 

There are 7 parity odd structures for 4 graviton scattering in $D=7$. These structures are depicted in the second part of 
figure \ref{parity-odd-7}. The first two lines of this part of 
the figure depict structures that transform under $S_3$ in the ${\bf 3}$ representation. The single structure in the last line, 
of course, transforms in the ${\bf 1}_S$ representation.
In equations, the 7 parity odd graviton scattering structures in $D=7$ transform as 
\begin{eqnarray} \label{grav7}
7&=& 3 \cdot {\bf 1_S}+2 \cdot {\bf 2_M}.
\end{eqnarray}

\begin{table}
	\begin{center}
		\begin{tabular}{|c|c|c|c||c|c|c|}
			\hline
			\,\,photons\,  & \multicolumn{3}{c||}{even} & \multicolumn{3}{c|}{odd} \\
			\hline
			&   $n_{\bf 1_S}$  &   $n_{\bf 2_M}$ & $n_{\bf 1_A}$ &  $n_{\bf 1_S}$  &   $n_{\bf 2_M}$ & $n_{\bf 1_A}$   \\
			\hline
			$D\geq 8$ & 3 & 2 & 0 & 0 & 0 & 0 \\
			\hline
			$D= 7$ &3 &2 &0 &1 & 0& 0\\
			\hline
			$D= 6$ &3 &2 &0 &0 & 0& 1\\
			\hline
			$D= 5$ &3 &2 &0 &0 & 0& 0\\
			\hline
			$D= 4$ &3 &1 &0 &2 & 0 & 0\\
			\hline
			$D= 3$ &1 &0 &0 &0 & 0& 1\\
			\hline
		\end{tabular}
		\begin{tabular}{|c|c|c|c||c|c|c|}
			\hline
			gravitons    & \multicolumn{3}{c||}{even} & \multicolumn{3}{c|}{odd} \\
			\hline
			&   $n_{\bf 1_S}$  &   $n_{\bf 2_M}$ & $n_{\bf 1_A}$ &  $n_{\bf 1_S}$  &   $n_{\bf 2_M}$ & $n_{\bf 1_A}$   \\
			\hline
			$D\geq 8$ & 10 & 9 & 1 &0 &0 &0 \\
			\hline
			$D= 7$ & 10 & 9 & 1 &3 &2 &0 \\
			\hline
			$D= 6$ & 9 & 9 & 1 & 0 & 3 & 3 \\
			\hline
			$D= 5$ & 7 & 7 & 1 & 0& 1& 1 \\
			\hline
			$D= 4$ & 3 & 1 & 0 & 2& 0 &  0\\
			\hline
			$D= 3$ & - & - & - & -& -&-  \\
			\hline
		\end{tabular}
	\end{center}
	\caption{Number of parity even and parity odd index structures for 4-photon and 4-graviton S-matrix as various dimensions.}
	\label{s3-structures}
\end{table}

In appendix \ref{appendix-is} we have presented a detailed construction of all basis index structures - and their $S_3$ transformation properties for $D=6, 5, 4, 3$. 
Here we content ourselves with merely tabulating the $S_3$ transformation properties of the basis of the bare module 
in every dimension in table \ref{s3-structures}. Note of course  that, for photons and gravitons, both for parity even and odd, in any given dimension, $n_{\bf 1_S}+2n_{\bf 2_M}+n_{\bf 1_A}$ equals the total number of structures tabulated in \ref{counting-is}

\section{Counting local Lagrangians} \label{cll}

\subsection{Local Lagrangians and local S-matrices}
\label{llls}

As we have mentioned in section \ref{lsm}, one of the principal goals in this paper is the enumeration and explicit construction all local or polynomial S-matrices (recall Lorentz and gauge invariant S-matrix are said to be local, analytic or polynomial  if they are polynomial function of the variables $(p_i,\epsilon_i)$). The set of local four particle S-matrices is, of course, closely 
related to the set of all local  quartic Lagrangians\footnote{$L(x)$ is a local Lagrangian if is is a function only of fields and their derivatives evaluated at $x$, subject to the restriction that the number of derivatives acting on any field is bounded from above (i.e. is finite).}.  There is an obvious map from the set of local gauge invariant quartic 
vertices to the set of local 4 particle S-matrices. This map, however, is  many to one. Two  Lagrangians generate the same S-matrix if they differ only by total derivatives when evaluated on-shell (we will make this statement completely precise below). 
The map from equivalence classes of Lagrangians to S-matrices can also be inverted. Given polynomial S-matrix one can construct a local quartic Lagrangian vertex that is invariant under linearized gauge transformations that gives rise to that S-matrix\footnote{The map from S-matrices to Lagrangians  played an important role in \cite{Heemskerk:2009pn} for scalars.}. There exists, in other words, a one to one map from the space of local equivalence classes of Lagrangians and 
local S-matrices; the classification of local S-matrices is the 
same as the classification of equivalence classes of local Lagrangians.

In the rest of this subsection we pause to explore 
the space of inequivalent Lagrangians - and their connection with inequivalent four particle scattering 
- more carefully, separately for the case of scalars, photons and gravitons.

\subsubsection{Scalars}

In this subsubsection we closely follow the analysis of \cite{Heemskerk:2009pn}. Consider a theory of real massless scalars $\phi$ invariant under the $\Z_2$ transformations $\phi \rightarrow -\phi$.
We wish to study the most general local action for this theory, retaining only those terms that affect four scalar scattering. 
$\Z_2$ symmetry ensures that we need only consider terms in the Lagrangian that are quadratic and quartic. We require that our scalar propagator have a single massless pole at $p^2=0$. This 
constraint (plus a convenient choice of normalization) determines the quadratic term in the Lagrangian to be
\begin{equation}\label{cls}
S_2= -\frac12 \int d^Dx\, \partial_\mu \phi \partial ^\mu \phi.
\end{equation} 

The most general local quartic interaction Lagrangian takes the form
\begin{equation}\label{mjsa} \begin{split} 
&S_4=\int d^Dx L_4, ~~~~~~~~~L_4= \sum a_{m_1, m_2, m_3, m_4} \partial^{m_1} \phi~ \partial^{m_2}
\phi ~ \partial^{m_3} \phi ~\partial^{m_4} \phi
\end{split}
\end{equation} 
where the schematic summation in the last line of \eqref{mjsa} runs over both the number of derivatives $m_i$ on the fields $\phi$ as well as the distinct ways of contracting the various derivative indices. A tree diagram computation using the action \eqref{mjsa} 
yields a 4 scalar S-matrix. This procedure establishes a map from the space of local Lagrangians $L_4$ \eqref{mjsa} and the set of 
local S-matrices.  As mentioned in the introduction to this subsection, however, this map is many to one. Two Lagrangians $L_4$ 
yield the same S-matrix if 
\begin{itemize}
	\item They differ by a total derivative.
	\item They can be related to each other by a field redefinition. 
\end{itemize}
Consider a field redefinition of the schematic form 
\begin{equation}\label{frsf} \begin{split}
&\phi \rightarrow \phi+ \delta \phi\\
&\delta \phi = \left( \sum b_{m_1, m_2, m_3}  \partial^{m_1} \phi~ \partial^{m_2}
\phi ~ \partial^{m_3} \phi \right) \\
\end{split} 
\end{equation} 
Up to terms of sextic and higher order that we ignore, the 
field redefinition \eqref{frsf} shifts $L_4$ by 
\begin{equation}\label{deltas4}
\delta L_4=  \partial^2 \phi 
\left( \sum b_{m_1, m_2, m_3} \partial^{m_1} \phi~ \partial^{m_2}
\phi ~ \partial^{m_3} \phi \right) 
\end{equation} 
It follows that the space of quartic terms $L_4$ may be divided up into equivalence classes. Two local quartic terms lie in the same equivalence class either if they agree up to a total derivative  when we set $\partial^2 \phi=0$\footnote{In the introduction to this subsection we mentioned that Lagrangians that are `on-shell equivalent' generate the same S-matrices. 
In the context of the scalar theory we study in this subsubsection, the precise meaning of `on-shell equivalent'
is `obeys the equation $\partial^2 \phi=0$. In the case of 
the photon/graviton studied in subsequent sub subsections, 
`on-shell equivalent' means obeys the (free Maxwell) / (vacuum Einstein) equations respectively. }.

The map between equivalence classes of $L_4$ and  four scalar S-matrices is one to one. To see this it is useful to move to  momentum space. Let 
\begin{equation}\label{ftms} \begin{split}
\phi(x)&= \int \frac{d ^dk}{(2 \pi)^d} e^{i k.x } ~{\tilde \phi}(k) \\
L_4&= \int \prod_i \frac{d ^dk}{(2 \pi)^d} e^{i \left( \sum_{j} k_j x_j \right) } ~{\tilde L}_4(k_1, k_2 , k_3, k_4) ~\tilde\phi(k_1) \tilde\phi(k_2) \tilde\phi(k_3) \tilde\phi(k_4)\\
\end{split} 
\end{equation} 
It follows from the discussion above that ${\tilde L}^1_4(k_1, k_2 , k_3, k_4)$  and ${\tilde L}^2_4(k_1, k_2 , k_3, k_4)$ lie in the same equivalence class if and only if 
\begin{equation}\label{iandon}
{\tilde L}^1_4(k_1, k_2 , k_3, k_4)={\tilde L}^2_4(k_1, k_2 , k_3, k_4) ~~~{\rm when}~~~
\end{equation}
\begin{equation} \label{condits}
 \sum_{i=1}^4 k_i=0, ~~~~{\rm and}  ~~~~k_i^2=0, ~i=1 \ldots 4 
\end{equation} 
But ${\tilde L}_4(k_1, k_2, k_3, k_4)$, evaluated 
subject to \eqref{condits},  is precisely 
the tree level S-matrix evaluated using the Lagrangian $L$\footnote{Up to a universal normalization factor.}.
 It follows that the equivalence classes of quartic Lagrangian terms are in fact labelled by their tree level S-matrix. 
 Moreover any polynomial S-matrix $S(k_1, k_2, k_3, k_4)$ (defined on the space of momenta \eqref{condits}) can be extended to 
 a polynomial function of unconstrained variables $k_1$, $k_2$, $k_3$ and $k_4$ in many inequivalent ways. Choose any such 
 extension, name it ${\tilde L}_4(k_1, k_2, k_3, k_4)$. The equation \eqref{ftms} then 
 may be viewed as a map from polynomial S-matrices to (any particular representative of) the equivalence classes of local 
 Lagrangians.
 It follows that local 4 scalar S-matrices are in one to one correspondence with the equivalence classes $L_4$ described in this subsubsection. 
 
 Note that a complete classification of scalar field primaries transforming in any representation and  with arbitrary number of  $\phi$s has been carried out using algebraic methods in \cite{deMelloKoch:2017dgi, deMelloKoch:2018klm}.

\subsubsection{Photons} 

The discussion of electromagnetic Lagrangians parallels that of the previous subsubsection in many respects but also has some new elements. To ensure gauge invariance, in this subsection we study Lagrangians built only out of the 
field strengths $F_{\mu\nu}$ \footnote{This class covers almost all gauge invariant electromagnetic Lagrangians. The exceptions have to do with Chern Simons terms which we ignore in this 
subsection but which we will account for later in the paper.} 
and work order by order in powers of the field strength. We require our vector field in Lorentz gauge have a propagator 
proportional to $\eta_{\mu\nu}/p^2$. This condition together 
with a convenient choice of normalization determines 
the quadratic term in our Lagrangian to be
\begin{equation}\label{qlph}
S_2= -\frac14\int d^Dx \,F_{\mu \nu} F^{\mu \nu}.
\end{equation}
As we do not demand that our theory have a $\Z_2$ invariance, 
the Lagrangian could have cubic terms in the field strength. 
 In Appendix \ref{A}, however, we demonstrate that 
it is always possible to perform a field redefinition of 
the form 
\begin{equation}\label{frphf} \begin{split}
&A_\mu \rightarrow A_\mu+\delta A_\mu \\
&\delta A_\mu = \left( \sum d_{n_1, n_2 } 
\partial^{m_1} F~ \partial^{m_2}F ~  \right)_\mu \\
\end{split} 
\end{equation} 
to set the cubic action $S_3$ to zero. 

It follows that the 3 photon S-matrix vanishes. This also means that the most general classical S-matrix of a theory of photons (and no other fields) has no exchange poles and so is purely local in nature, as was the case of the $\Z_2$ invariant scalar 
theory of the previous subsubsection.

The most general quartic Lagrangian takes the schematic form
\begin{equation}\label{mjpha} \begin{split} 
&S_4=\int d^Dx L_4, ~~~~~~~~~L_4= \sum a_{m_1, m_2, m_3, m_4} \partial^{m_1} F~ \partial^{m_2}
F ~ \partial^{m_3} F ~\partial^{m_4} F
\end{split}
\end{equation} 
(where all indices on $\partial_\mu$ and $F_{\mu\nu}$ have been suppressed). A field redefinition of the schematic form 
\begin{equation}\label{frphf1} \begin{split}
&A_\mu \rightarrow A_\mu+\delta A_\mu \\
&\delta A_\mu = \left( \sum b_{n_1, n_2 n_3}  \partial^{m_1} F~ \partial^{m_2}F ~ \partial^{m_3} F \right)_\mu \\
\end{split} 
\end{equation} 
generates of shift of $L_4$ 
\begin{equation}\label{deltaph4}
\delta L_4=  \partial_\nu F^{\mu\nu}
\left( \sum b_{m_1, m_2, m_3} \partial^{m_1} F ~ \partial^{m_2} F
 ~ \partial^{m_3} F \right)_\mu.
\end{equation} 
It follows that Lagrangians $L_4$ that differ from each other by 
\begin{itemize}
	\item Total derivatives
	\item Terms that include $\partial_\nu F^{\nu \mu}$
	\item Terms that include $\partial_{[\alpha} F_{\beta \gamma]}$
	as a factor (such terms are, of course, identically zero when expressed in terms of $A_\mu$)
	\end{itemize}
generate the same S-matrix and should be treated as equivalent to each other. 

As in the case of scalars, equivalence classes of 
Lagrangians are labelled by their S-matrices. Moreover
we expect that there is a one to one map from classes of local Lagrangians to polynomial S-matrices.
See Appendix \ref{smatpho} for a discussion. 

\subsubsection{Gravitons} \label{gravlagfp}

In order to ensure diffeomorphism invariance, in this section we 
study gravitational Lagrangians constructed out of the Riemann 
tensor\footnote{This covers almost all diffeomorphism invariant
gravity Lagrangians. The exceptions to this rule are gravitational Chern Simons terms which we ignore in this subsubsection, but whose effects we account for later in this paper.} and work order by order in powers of the Riemann 
tensor\footnote{We define an action to be of $m^{\rm th}$ 
	order in Riemann tensors if there is no manipulation that allows us to express the same action as an expression 
	of higher orders in Riemann tensors in a local manner. 
	For instance, we count an expression containing $[\nabla_\mu, \nabla_\nu]$ acting on $m$ explicit copies of the Riemann tensor as being of degree $m+1$ as the antisymmetric combination of derivatives can be replaced by a Riemann tensor. An expression that is of $m^{\rm th}$ 
	order in Riemann tensors does not contribute to $n$ point
	scattering amplitudes of gravitons for $n<m$. Terms of $m^{\rm th}$ order typically do contribute to S-matrices for 
	$m$ and higher point S-matrices. There are exceptions to this last rule; it is sometimes possible for an object to be of $m^{\rm th}$ order in Riemann tensors but to contribute 
	to S-matrices only at order $m+1$ or higher. }. Before commencing our discussion we pause to define some terminology. Throughout this subsubsection 
the symbol $H^{(n)}_{\mu\nu}[R_{\alpha\beta\gamma\delta}]$ will 
denote the most general local functional that is rank 2 symmetric tensor and that is  $n^{\rm th}$ 
order in the Riemann tensor, but with arbitrary powers of the 
metric and arbitrary numbers of symmetrized derivatives\footnote{One example of an allowed functional is
	$$H^{(1)}_{\mu\nu} = a R g_{\mu\nu} 
	+ \nabla^2 R g_{\mu\nu} + c R_{\mu\nu} + d 
	\nabla^\alpha \nabla \beta R_{\alpha \mu \beta \nu} \ldots $$}. 

The unique diffeomorphism invariant action that is linear in Riemann tensors is, of course, the Einstein action 
\begin{equation} \label{einst}
S_E=\int \sqrt{-g} R.
\end{equation}
In Appendix \ref{A} we demonstrate that the field 
redefinition
\begin{equation}\label{frd} 
\delta g_{\mu \nu}=H^{(1)}_{\mu\nu}[R_{\alpha\beta\gamma\delta}]
\end{equation} 
may be used to cast the most general Lagrangian, quadratic 
in Riemann tensors,  into the form 
\begin{equation}\label{einstpgb}
S= S_E+S_{GB}+\int {\cal O}(R_{\alpha \beta \gamma \delta})^3, 
\end{equation} 
where,
\begin{equation} \begin{split}  \label{gblag}
S_{GB} &= \int \sqrt{-g} \,\, \delta_{[a}^{g}\delta_{b}^{h}\delta_{c}^{i}\delta_{d]}^{j}\,\, R_{ab}^{\phantom {ab}gh} R_{cd}^{\phantom{cd}ij}\\
& \propto
\int \sqrt{-g}\left(R^2 -4R^{\mu \nu}R_{\mu\nu}+R^{\mu\nu\rho\sigma}R_{\mu\nu\rho\sigma}\right).
\end{split}
\end{equation}
and ${\cal O}(R_{\alpha \beta \gamma \delta})^3$ denotes 
all terms that are of cubic or higher order in the 
Riemann tensor\footnote{In four dimensions the Gauss-Bonnet term vanishes identically.}. In other words Einstein-Gauss-Bonnet is the most general 
action quadratic in the  Riemann tensor up to total derivatives or terms  
terms that involve explicit factors of $R_{\mu\nu}$ and 
the Ricci scalar $R$\footnote{In particular  the Einstein equations $R_{\mu\nu}=0$ tell us  that we need only work with Riemann tensor - terms containing Ricci tensor or Ricci scalar are effectively trivial.}.

When evaluated in a spacetime of the form 
\begin{equation}\label{fs}
g_{\mu\nu}= \eta_{\mu\nu} + h_{\mu\nu}
\end{equation}
it turns out that the Gauss-Bonnet term in \eqref{einstpgb} starts out at order $h^3$ (up to total derivatives). It follows, 
in other words, that - despite the appearance - 
the Gauss-Bonnet term does not modify the Einstein propagator but does contribute 
to three point scattering of gravitons. This term is, of course, topological 
in $D=4$. 

In Appendix \ref{A} we next demonstrate that field redefinitions of the form   \begin{equation}\label{frdq} 
\delta g_{\mu \nu}=H^{(2)}_{\mu\nu}[R_{\alpha\beta\gamma\delta}]
\end{equation} 
can be used to cast the most general cubic  correction to the Einstein-Gauss-Bonnet action into the form 
\begin{equation}\label{einstpgbc}
S= S_E+S_{GB}+a S_{R^3}^{(1)}+b \chi_6 +\int \sqrt{-g} \left( {\cal O}(R_{\alpha \beta \gamma \delta})^4 \right) 
\end{equation} 
\begin{equation} \begin{split} 
\label{r31}
S_{R^3}^{(1)}&=\int \sqrt{-g}\left(R^{pqrs}R_{pq}^{\phantom{pq}tu}R_{rstu}+2R^{pqrs}R_{p\phantom{t}r}^{\phantom{p}t\phantom{r}u}R_{qtsu}\right)\\
\chi_6 &= \int \sqrt{-g} \left(  \frac{1}{8}\,\, \delta_{[a}^{g}\delta_{b}^{h}\delta_{c}^{i}\delta_{d}^{j}\delta_{e}^{k}\delta_{f]}^{l}\,\,R_{ab}^{\phantom {ab}gh} R_{cd}^{\phantom{cd}ij}R_{ef}^{\phantom{ef}kl} \right) \\
&= \int \sqrt{-g} \bigg( 4 R_{ab}^{\phantom{ab}cd}R_{cd}^{\phantom{cd}ef}R_{ef}^{\phantom{ef}ab}-8R_{a\phantom{c}b}^{\phantom{a}c\phantom{b}d}R_{c\phantom{e}d}^{\phantom{c}e\phantom{d}f}R_{e\phantom{a}f}^{\phantom{e}a\phantom{f}b}-24R_{abcd}R^{abc}_{\phantom{abc}e}R^{de}+3R_{abcd}R^{abcd}R\\
&\quad \quad \quad \quad +24 R_{abcd}R^{ac}R^{bd}+16R_{a}^{\phantom{a}b}R_{b}^{\phantom{b}c}R_{c}^{\phantom{c}a}-12R_{a}^{\phantom{a}b}R_{b}^{\phantom{b}a}R+R^3 \bigg)
\end{split}
\end{equation}   
In other words, Einstein-Gauss-Bonnet corrected by two specific cubic terms 
is the most general action cubic in Riemann tensors - up to total derivatives and terms that vanish by the Einstein equations\footnote{In four dimension another cubic term is present that is parity odd \cite{Maldacena:2011nz} $\varepsilon^{\alpha\beta\gamma\delta}R_{\alpha\beta a b}R_{\g \d c d}R_{a b c d}$.  This is dual to a parity violating three point function structure for stress tensors in three dimensions \cite{Giombi:2011rz, Costa:2011mg}. We will not consider exchange due to this term in this paper.}.

When evaluated on the metric \eqref{fs}, the term $\chi_6$ starts out at order $h_{\mu\nu}^4$ (up to total derivatives). It follows in particular that this 
term does not contribute to three graviton scattering. $\chi_6$
is field redefinition equivalent to the simpler looking expression 
\begin{equation}
\label{r32}
S_{R^3}^{(2)}=\int \sqrt{-g}\left(R^{pqrs}R_{pq}^{\phantom{pq}tu}R_{rstu}-2R^{pqrs}R_{p\phantom{t}r}^{\phantom{p}t\phantom{r}u}R_{qtsu}\right)
\end{equation}
(obtained by setting all terms involving $R_{\mu\nu}$ in $\chi_6$ to zero). 
The reason that we prefer to use $\chi_6$ rather than \eqref{r32} in our 
action is the following; when evaluated on the configuration \eqref{fs}, 
the expression $S_{R^3}^{(2)}$ is of order $h_{\mu\nu}^4$ only on-shell; 
when evaluated off-shell this expression is of order $h_{\mu\nu}^3$. 
As a consequence, while the actions $\chi_6$ and $S_{R^3}^{(2)}$ lead to the 
same polynomial graviton  4 point function, this scattering amplitude has its source 
purely in contact terms in the case of $\chi_6$, but in the more complicated  sum of contact and exchange diagrams (which are polynomial as on-shell 3 point functions vanish) in the case of $S_{R^3}^{(2)}$. Consequently $\chi_6$ is clearly 
dynamically simpler than $S_{R^3}^{(2)}$, even though it superficially looks 
more complicated. $\chi_6$ also has other interesting properties; it vanishes 
identically in less than six dimensions, and is topological in $d=6$. In fact
$\chi_6$ is sometimes called the `6 dimensional Euler density'. It is also the 
second in the sequence of Lovelock terms (the first is the Gauss-Bonnet term 
written above). 

In contrast to $\chi_6$,  $S_{R^3}^{(1)}$ does contribute to the three point functions\footnote{ The specific choice we have made for $S_{R^3}^{(1)}$
is a bit arbitrary; we could have added any multiple of $S_{R^3}^{(2)}$ to it without changing its essential features.}.
In fact the Einstein term, the Gauss-Bonnet term and $S_{R^3}^{(1)}$ each contribute 
to three graviton scattering. It follows that the most general 
3 graviton S-matrix is a linear sum of 3 independent structures. 
The Einstein action is quadratic in derivatives and leads to 
a 3 graviton S-matrix proportional to 
\begin{equation}\label{R1sm}
{\cal{A}}^{R}= (\epsilon_1.\epsilon_2 \epsilon_3.p_1 +\epsilon_1.\epsilon_3 \epsilon_2.p_3 +\epsilon_2.\epsilon_3 \epsilon_1.p_2 )^2
\end{equation}
The Gauss-Bonnet action is quartic in derivatives and leads 
to a 3 graviton S-matrix proportional to\footnote{As remarked in the earlier footnote, this structure vanishes in $D=4$ but a parity odd structure appears in its place.} 
\begin{equation}\label{R2sm}
{\cal{A}}^{R^2}=(\epsilon_1\wedge\epsilon_2\wedge\epsilon_3\wedge p_1 \wedge p_2)^2
\end{equation}
The Riemann cube term is sextic in derivatives and leads to a
3 graviton S-matrix proportional to 
\begin{equation}\label{R3sm}
{\cal{A}}^{R^3}= \left( {\rm Tr} F_1 F_2 F_3 \right)^2.
\end{equation}

As the 3-graviton S-matrix is non vanishing, 4-graviton S-matrices that follow from the Lagrangian \eqref{einstpgbc}
have contributions from Feynman diagrams with a single graviton 
exchange. Such exchange diagrams lead to S-matrices that 
are not polynomial in $s$, $t$ and $u$ but instead have poles. 
We have explicitly evaluated the 4 graviton S-matrix that follows from the action \eqref{einstpgbc}. Our results are 
presented in Section \ref{ec}. Notice that the three 
graviton scattering amplitudes ${\cal{A}}^{R^2}$ and ${\cal{A}}^{R^3}$
are gauge invariant\footnote{i.e. are invariant under the shift $\epsilon_i \rightarrow \epsilon_i + k_i$.} even off-shell (i.e. even if $k_i^2 \neq 0$). It follows that exchange diagrams that sew together two of these vertices are individually gauge invariant\footnote{Off shell gauge invariance of the three point function is relevant here because the intermediate graviton is off-shell in an exchange diagram.}. On the other hand ${\cal{A}}^{R}$ is not off-shell gauge invariant (thought it certainly is on-shell gauge invariant). The { \it sum} of all channels of exchange diagrams involving either one or two copies of 
${\cal{A}}^{R}$ is, therefore, gauge invariant only once we add the 
contribution of the relevant polynomial contact term (this 
is the explicit 4 $h_{\mu\nu}$ term in the Gauss-Bonnet action or 
in $S_{R^3}^{(1)}$ in the case of an Einstein Gauss-Bonnet or $S_{R^3}^{(1)}$ exchange diagram  or the explicit 4 $h_{\mu\nu}$ term in the Einstein action in the case of an Einstein - Einstein exchange diagram.)

We pause here to note that the discussion of the previous 
paragraph illustrates the interesting interplay between 
the on-shell invariance of S-matrices under {\it linearized} gauge transformations and the  {\it non-linear} off-shell gauge invariance  of the Lagrangians that generate them. It might at first seem that the requirement of linearized gauge invariance of S-matrices is a weaker condition than the off-shell non-linear gauge invariance of Lagrangians; this is not the case. Consider, for example, four point scattering in the pure Einstein gravity (the Einstein-Einstein case discussed in the previous paragraph). As we have mentioned above, the  exchange diagrams by themselves are not linearized-gauge invariant, but the four graviton contact structure in the Einstein-Hilbert Lagrangian is such that its linear-gauge transformations compensates the linear-gauge transformation of the exchange term to render the full S-matrix linear-gauge invariant. This works precisely because the non-linear gauge invariance of the Einstein Hilbert term relates cubic and quartic vertices (the gauge transformation of the cubic vertex linear in $h_{\mu\nu}$ cancels the linear gauge transformation of the quartic vertex {\emph i.e.} gauge transformation that is independent of $h_{\mu\nu}$). This relation between the vertices 
guarantees that the  sum of exchange and contact diagrams
generates a 4 graviton S-matrix that is  {\it on-shell, linearized } gauge invariant. 

As we have mentioned above, $\chi_6$ does not contribute to  3 graviton scattering, but does contribute (polynomiallally) to four graviton scattering. 
The four graviton S-matrix that follows from this term is proportional to 
\begin{equation}\label{4gsm}
\begin{split}
T_1&=(\epsilon_1 \wedge \epsilon_2 \wedge\epsilon_3\wedge\epsilon_4 \wedge k_1 \wedge k_2 \wedge k_3)^2\\ 
&\propto        \delta_{[a}^{i}\delta_{s}^{j}\delta_{d}^{k}\delta_{f}^{l}\delta_{g}^{m}\delta_{h}^{n}  \delta_{j]}^{p}  \,\,   \epsilon^1_{i}\epsilon^2_{j}\epsilon^3_{k}\epsilon^4_{l}p^1_m p^2_n p^3_p \epsilon^{1a}\epsilon^{2s}\epsilon^{3d}\epsilon^{4f}p^{1g} p^{2h} p^{3j}.
\end{split}
\end{equation} 
Finally we turn to local Lagrangians of quartic or higher order
in Riemann tensors. These terms, of course, do not contribute
to 3 graviton scattering, but all  give rise four graviton S-matrices that are polynomial in $\epsilon_i$ and $k_i$. As above field redefinitions of the form   \begin{equation}\label{frdc} 
\delta g_{\mu \nu}=H^{(3)}_{\mu\nu}[R_{\alpha\beta\gamma\delta}]
\end{equation} 
can be used to simplify the most general quartic 
correction to the Einstein-Gauss-Bonnet-Riemann-cube 
action. Even up to the simplification afforded by
the field redefinitions \eqref{frdc}, however, the 
most general action that is quartic in Riemann tensors, turns out to be characterized by an infinite (rather than a finite, as was the case at quadratic and cubic order) number of parameters. In subsection \eqref{pleth} below we turn to the problem of enumerating 
such Lagrangians. 

As in the previous subsubsection there is a close 
connection between gauge invariant local S-matrices and equivalence classes of quartic local Lagrangians built out of the Riemann tensor. To first approximation the relationship 
between these two structures is as for the gauge fields but there are some additional complications stemming from the non-linear nature of the gravitational field.
The map from Lagrangians to S-matrices continues to be the obvious one. When evaluated on-shell, Lagrangians that differ 
only by total derivatives { \it or terms of order $h^5$ or higher} 
yield the same S-matrix\footnote{The stipulation about terms of higher order is necessary because the non-linearity of gravity makes it possible for two terms built out of four Riemann tensors, that are distinct even on-shell at the non-linear level, to agree at ${\cal O}(h^4)$.}. There is also a complication 
in the reverse map: it is possible for local S-matrices 
to correspond to Lagrangians that are of lower than quadratic 
order in the Riemann tensor, as we have already seen in the 
example of the second Riemann cube term above. Later in this paper we will come to grips with all these complications in 
a quantitative manner.

\subsection{Plethystic program for scalars} \label{pleth}

In the previous subsection we have argued that enumeration of four scalar S-matrices is isomorphic to the enumeration equivalence classes of local Lagrangians, quartic in $\phi$. In this subsection we explicitly count 
these equivalence classes graded by number of derivatives. 

It is useful to define the single letter partition function i.e. partition function over all the operators that involve a single field, modulo the free equation of motion. The space of such operators for scalars is spanned by 
\be\label{scalar-desc}
\partial_{\mu_1} \partial_{\mu_2}\ldots  \partial_{\mu_l}\phi,\qquad \qquad {\rm for }\quad l=0,\ldots \qquad \qquad {\rm subjected \,\,to }\quad \partial_\mu\partial^\mu \phi=0.
\ee
 The single letter partition function is easily obtained; 
 (see e.g. \cite{Sundborg:1999ue, Aharony:2003sx}); it is given by 
\begin{eqnarray}\label{scalar-single}
i_\ts(x,y)&=&{\rm Tr}\,\,x^{\Delta} y_i^{H_i}= (1-x^2)\denom(x,y).\nonumber\\
\denom(x,y) &=&\Big(\prod_{i=1}^{D/2}(1-x y_i)(1-x y_i^{-1})\Big)^{-1}\qquad \qquad \qquad \qquad{\rm for \,\, D\,\, even}\nonumber \\
&=&\Big((1-x)\prod_{i=1}^{\lfloor D/2\rfloor}(1-x y_i)(1-x y_i^{-1})\Big)^{-1}\qquad \qquad \,\,{\rm for \,\, D\,\, odd}.
\end{eqnarray}
Here $H_i$ stands for the Cartan elements of $SO(D)$. The denominator factor $\denom(x,y)$ encodes the tower of derivatives on $\phi(x)$ keeping track of the degree and the charges under the Cartan subgroup of $SO(D)$. We have kept track of the Cartans of $SO(D)$ because we will eventually need to project polynomials built out of scalar letters to the space of  $SO(D)$ singlets below.

Equivalence classes of scalar Lagrangians are given by scalar quartic polynomials of the expressions \eqref{scalar-desc} modulo polynomials that are total derivatives. We will first  enumerate all quartic scalar polynomials we can make out of 
\eqref{scalar-desc} and then subtract those polynomials 
that are total derivatives. The partition function of polynomials of the expressions \eqref{scalar-desc} - the so 
called multi-letter partition function is given by the formula 
of Bose statistics 
\be\label{multi-particle}
\sum_{k=1}^{\infty}t^k i_\ts^{(k)}(x,y)=\exp\Big(\sum_{n=1}^{\infty}\frac{t^n }{n} i_\ts(x^n,y^n)\Big).
\ee
where $i_\ts^{(k)}$ to be the partition function over $k$-letter partition function, $i_\ts^{(1)}=i_\ts$.  

The four-letter partition function - relevant for counting quartic Lagrangians - is easily read off from equation \eqref{multi-particle}: 
\be\label{4-particle}
i_\ts^{(4)}(x,y)=\frac{1}{24}\Big(i^4_\ts(x,y) +6 i^2_\ts(x,y) i_\ts(x^2,y^2)+3i^2_\ts(x^2,y^2)+8i_\ts(x,y)i_\ts(x^3,y^3)+6i_\ts(x^4,y^4)\Big).
\ee
This partition function over four particle states includes operators that are total derivatives which we wish to remove.  In conformal field theory language, this means we want to count only primary scalar quartic operators. \emph{Assuming that there are no null states in the multiplet}, if the character of the conformal primary is $P(x,y)$ then the character over its entire multiplet is given by $P(x,y)\denom(x,y)$ where $\denom(x,y)$ encodes the contribution coming from the tower of derivatives. So in order to obtain only the partition function over primaries, we need to divide $i^{(4)}_\ts(x,y)$ by $\denom(x,y)$\footnote{As we will explain later, the assumption of not having null states is always valid for the scalar case but fails in $D=9$ both for photon and gravitons. In that case, we need to take care of the null state separately. }.
The partition function over polynomials of \eqref{scalar-desc}, modulo total derivatives is given by 
\be\label{Ddivide} i_\ts^{(4)}(x,y)/\denom(x,y). \ee

The partition function over scalar operators is now obtained simply by projecting onto $SO(D)$ invariant states. This is achieved by integrating $i_\ts^{(4)}(x,y)/\denom(x,y)$ over the Haar measure of the group. The resulting integral can be gauge fixed to the Cartan subgroup with the measure that is the $SO(D)$ version of the  Van der Monde determinant $\Delta(y_i)$ (see Appendix \ref{epi} for details) 
\be\label{singlet-proj}
I^D_\ts(x):=\oint \prod_{i=1}^{\lfloor D/2\rfloor}dy_i \,\Delta(y_i)\,  i_\ts^{(4)}(x,y)/\denom(x,y).
\ee  
This integral and its generalizations to higher points were used to enumerate $n$-point scalar S-matrices in $D=3$ and $D=4$ in \cite{Henning:2017fpj}.

It is somewhat cumbersome to evaluate this integral analytically for general dimensions. However it is easy to come up with a 
conjecture for the final answer. Parity invariant scalar scattering amplitudes are the same in every $D\geq 3$.  The argument is similar to the one made for photons and gravitons (about asymptotic dimensions being $7$) in the previous section.  For $D\geq 3$, the entire 4-particle scalar particle scattering can be chosen to lie in a 3-plane and the remaining dimensions are mere spectators (unlike photons and gravitons, there are no polarizations that can occupy transverse dimensions). Moreover there are no parity odd S-matrices for 
$D \geq 4$ while such structures do exist for $D=3$\footnote{The reason for this is simple. A parity 
odd S-matrix is proportional to $\epsilon_{\mu_1 \ldots \mu_D}$
All indices of this tensor have to contract with some vector. 
However there are only three independent vectors in 3 dimensional scattering; namely $k_1$, $k_2$ and $k_3$. It follows that parity odd S-matrices exist in $D=3$ but not for
$D \geq 4$.}. The correspondence between equivalence classes 
of Lagrangians and S-matrices leads us to conjecture that 
the result of the integral \eqref{singlet-proj} is independent 
of $D$ for $D \geq 4$. We have obtained some direct evidence
for this conjecture by using ${\mathtt {Mathematica}}$ to evaluate  \eqref{singlet-proj} in a power series in $x$ up to  ${\cal O}(x^{20})$ in $D=4, 5, 6, 7$, and verifying the results are 
independent of $D$ (and in fact equal to the results of the large 
$D$ computation, see below).

Given the conjecture described at the end of the previous paragraph, it is easy to evaluate \eqref{singlet-proj} for 
all $D\geq 4$. This is done analytically by evaluating the integral 
in \eqref{singlet-proj} in the large $D$ limit. The details of 
the computation are presented in Appendix \ref{epi}. 
Our final results - together with the results in $D=3,2$ (in these special cases the integral is easily separately evaluated analytically) are  tabulated results in \ref{scalar-plethystic} (with $\denom\equiv 1/((1-x^4)(1-x^6))$).
\begin{table}
\begin{center}
\begin{tabular}{|l|l|}
\hline
dimension & scalar partition function \\
\hline
$D\geq  4$ & $\denom$\\
\hline
$D=3$ & $(1+x^9)\denom$\\
\hline
$D=2$ & $(1-x^6)\denom$ \\
\hline
\end{tabular}
\end{center}
\caption{Partition function over the space of Lagrangians involving four $\phi$'s. This includes both parity even and parity odd Lagrangians. $\denom\equiv 1/((1-x^4)(1-x^6))$.}
\label{scalar-plethystic}
\end{table}

Note in particular that $I_\ts(x)$ agrees with the partition function over $S_3$ symmetric polynomials of $s,t$, $Z_{\bf 1_S}(x)$ for $D\geq 4$. This is entirely expected because the scalar S-matrices are precisely parameterized by $S_3$ symmetric polynomials of $s,t$. So this exercise serves as a check of something that we already know and expect. We will get new and more interesting results when we apply this machinery to Lagrangians of photons and gravitons.

\subsection{Photons and gravitons}\label{pleth-spin}
The single letter partition function for scalars $i_{\ts}(x,y)$ can be thought of as the following sum
\be
i_{\ts}(x,y)=1+x\,\chi_{\syng{1}}+x^2\, \chi_{\syng{2}}+x^3\, \chi_{\syng{3}}+x^4\, \chi_{\syng{4}}+x^5\, \chi_{\syng{5}}\ldots.
\ee
Here $\chi_R(y_i)$ is the character of representation $R$ of $SO(D)$. We have suppressed the arguments $y_i$ of $\chi_R$ in the above formula and will continue to do so when there is no danger of confusion. In the rest of the section, Young diagrams stand for $SO(D)$ representations unless otherwise mentioned. That only symmetric traceless representations of $SO(D)$ appear is clear from the explicit form of the descendants \eqref{scalar-desc}. These are also the representations of scalar spherical harmonics. This is also expected, because through state-operator map, the single-letter local operators are in one to one correspondence with single particle states on sphere $S^{D-1}$. The spectrum of states is precisely that of the scalar spherical harmonics.

For photons, the single letter partition function over gauge invariant operators is obtained by acting derivatives on the field strength $F_{\mu\nu}$, subjected to the equation of motion $\partial^\mu F_{\mu\nu}=0 $ and  the Bianchi identity $\partial_{[\rho}F_{\mu\nu]}=0$. 
The representations appearing after the action of a single derivative on $F_{\mu\nu}$ can be obtained by taking the direct product of rank 2 antisymmetric with a vector,
\be
{\tyng(1,1)}\,\otimes\,{\tyng(1)}\,=\, {\tyng(2,1)}\,\oplus {\tyng(1,1,1)} \,\oplus\, {\tyng(1)}.
\ee
The second and third representations stand for the Bianchi identity and equation of motion respectively, both of which vanish. So we get a single irreducible representation after the action of one derivative. Action of another derivative is obtained by the tensor product,
\be
{\tyng(2,1)}\,\otimes\,{\tyng(1)}\,=\,{\tyng(3,1)}\,\oplus {\tyng(2,2)} \,\oplus\, {\tyng(2,1,1)} \,\oplus\, {\tyng(2)}\, \oplus\,{\tyng(1,1)}. 
\ee
Again, the third and fourth representations vanish due to the Bianchi identity and equation of motion respectively. The last one vanishes because $\partial_\mu\partial^\mu=0$ (this follows from the Bianchi identity and the equation of motion), and the second vanishes because the two derivatives can not be anti-symmetrized. As a result, we again get the first one to be the only irreducible representation. This reasoning continues in general and we  get
\be
i_{\tv}(x,y)=x\,\chi_{\syng{1,1}}+x^2\, \chi_{\syng{2,1}}+x^3\, \chi_{\syng{3,1}}+x^4\, \chi_{\syng{4,1}}+x^5\, \chi_{\syng{5,1}}+x^6\, \chi_{\syng{6,1}}\ldots.
\ee
The first term is $F_{\mu\nu}$ and  has a single derivative, hence a single power of $x$. These are precisely the representations of vector spherical harmonics, which are the states of a gauge invariant vector field on $S^{D-1}$. This series is summed in appendix \ref{single-letter-sum}. The final 
result, which was also derived in e.g. \cite{Aharony:2003sx}, is given by 
\be\label{photon-sl}
i_\tv(x,y)=(((x-x^3)\chi_{\syng{1}}-(1-x^4))\denom(x,y)+1)/x.
\ee

For gravitons, the single letter partition function over gauge invariant operators is obtained by acting derivatives on the Riemann tensor $R_{\mu\nu\alpha\beta}$, subjected to the equation of motion $\partial^\mu R_{\mu\nu\alpha\beta}=0 $ and  the Bianchi identity $\partial_{[\rho}R_{\mu\nu]\alpha\beta}=0$. Recall that the Riemann tensor enjoys the symmetries,
\be
R_{\mu\nu\alpha\beta}=R_{\alpha\beta\mu\nu}=-R_{\nu\mu\alpha\beta}=-R_{\mu\nu\beta\alpha},\qquad R_{\mu[\nu\alpha\beta]}=0.
\ee
It belongs to the following representation of $SO(D)$,
\be
{\tyng(2,2)} \,\oplus\, {\tyng(2)} \,\oplus \, \cdot
\ee
The irreducible representations correspond to the Weyl tensor, Ricci tensor and Ricci scalar respectively. The last two representations vanish on-shell  and we are only left with the Weyl tensor. Note also that this representation is precisely the representation of the first tensor spherical harmonic. Just like the case of scalar and photon, we expect the graviton single letter partition function to be equal to the generating function for characters of  tensor spherical harmonics.
\be
i_{\tt}(x,y)=x^2\, \chi_{\syng{2,2}}+x^3\, \chi_{\syng{3,2}}+x^4\, \chi_{\syng{4,2}}+x^5\, \chi_{\syng{5,2}}+x^6\, \chi_{\syng{6,2}}+x^7\, \chi_{\syng{7,2}}\ldots.
\ee
The first term comes from  $R_{\mu\nu\alpha\beta}$ and  has a two derivatives, hence two powers of $x$.
This series is summed in appendix \ref{single-letter-sum}.
\be\label{graviton-sl}
i_\tt(x,y)=(((x^2-x^4)(1+\chi_{\syng{2}})-(x-x^5)\chi_{\syng{1}})\denom(x,y)+x^2\chi_{\syng{1,1}}+x\chi_{\syng{1}})/x^2.
\ee

\begin{table}
		\begin{center}
			\begin{tabular}{|l|l|}
				\hline
				dimension & photon partition function \\
				\hline
				$d\geq  10$ & $x^4({\color{mblue}2+3x^2+2x^4})\denom$\\
				\hline
				$d=9$ & $x^4({\color{mblue}2+3x^2+2x^4})\denom-{\color{myellow}x^5}$\\
				\hline
				$d=8$ & $x^4({\color{mblue}2+3x^2+2x^4})\denom+{\color{myellow}x^4}$ \\
				\hline
				$d=7$ & $x^4({\color{mred}x^{-1}}+{\color{mblue}2+3x^2+2x^4})\denom-{\color{myellow}x^3}$ \\
				\hline
				$d=6$ & $x^4({\color{mblue}2+3x^2+2x^4}+{\color{mred}x^8})\denom$  \\
				\hline
				$d=5$ & $x^4({\color{mblue}2+3x^2+2x^4})\denom$ \\
				\hline
				$d=4$ & $x^4({\color{mblue}2+3x^2+2x^4-(x^4+x^6)}+{\color{mred}1+2x^2+x^4-(x^4+x^6)})\denom$  \\
				\hline
				$d=3$ & $x^4({\color{mblue}1+x^2+x^4-(x^4+x^6)}+{\color{mred}x^5})\denom$    \\
				\hline
			\end{tabular}
		\end{center}
		\caption{Partition function over the space of Lagrangians involving four $F_{\mu\nu}$'s. This includes both parity even and parity odd Lagrangians. Recall $\denom\equiv 1/((1-x^4)(1-x^6))$.\newline
			{\color{mblue}Blue}: Parity {\color{mblue}Even}.\hspace{0.5cm}
			{\color{mred}Red}: Parity {\color{mred}Odd}.\hspace{0.5cm}
			{\color{myellow}Yellow}: Plethystic {\color{myellow}Miscount}}
		\label{photon-plethystic}
	\end{table}

\begin{table}
		\begin{center}
			\begin{tabular}{|l|l|}
				\hline
				dimension & graviton partition function \\
				\hline
				$d\geq 10$ & $x^{8}({\color{mblue}x^{-2}+6+9x^2+10x^4+3x^6})\denom-{\color{myellow}(x^6-x^8)}$ \\
				\hline
				$d=9$ & $x^{8}({\color{mblue}x^{-2}+6+9x^2+10x^4+3x^6})\denom-{\color{myellow}(x^6-x^8)}-{\color{myellow}2x^{9}}$ \\
				\hline
				$d=8$ & $x^{8}({\color{mblue}x^{-2}+6+9x^2+10x^4+3x^6})\denom-{\color{myellow}(x^6-x^8)}+{\color{myellow}2x^{8}}$ \\
				\hline
				$d=7$ & $x^{8}({\color{mblue}x^{-2}+6+9x^2+10x^4+3x^6}+{\color{mred}2x^{-1}+3 x+2x^3})\denom-{\color{myellow}(x^6-x^8)}$\\
				&$-{\color{myellow}(x^8)}-{\color{myellow}(2x^7)}$  \\
				\hline
				$d=6$ & $x^{8}({\color{mblue}6+9x^2+10x^4+3x^6}+{\color{mred}3x^{2}(x^2+x^4+x^6)})\denom$  \\
				\hline
				$d=5$ & $x^{8}({\color{mblue}4+7x^2+8x^4+3x^6}+{\color{mred}x^{3}(x^2+x^4+x^6)})\denom$  \\
				\hline
				$d=4$ & $x^{8}({\color{mblue}2+2x^2+3x^4-x^6-x^8}+{\color{mred}1+x^2+2x^4-x^6-x^8})\denom$ \\
				\hline
			\end{tabular}
		\end{center}
		\caption{Partition function over the space of Lagrangians involving four $R_{\mu\nu\rho\sigma}$'s (no $R_{\mu\nu}$, no $R$). This includes both parity even and parity odd Lagrangians. }
		\label{graviton-plethystic}
	\end{table}

Equipped with the single letter partition functions for photons \eqref{photon-sl} and gravitons \eqref{graviton-sl}, the partition function over quartic vertices is obtained in the same way as for  scalars. We simply change the subscript $\ts$ to $\tv$ and $\tt$ respectively in equation \eqref{multi-particle}, \eqref{4-particle} and \eqref{singlet-proj}.  Again, this computation difficult to perform analytically in general dimensions. 
As argued earlier, we naively expect the counting of S-matrices to be independent of $D$ for $D\geq 7$. As we will discuss in detail below, however, there are subtleties concerning parity odd operators and null states for $D=7, 8, 9$. For this reason 
we conjecture that the photon and graviton versions of 
\eqref{scalar-desc} are independent of $D$ for $D \geq 10$. 
Assuming this conjecture the photon and graviton analogues of \eqref{scalar-desc} for $D\geq 10$ are easily evaluated by performing the 
relevant large $D$ computation (see Appendix \ref{epi}). 
Our results are listed in Tables \ref{photon-plethystic} and \ref{graviton-plethystic} respectively. For $D \leq 9$ we have 
evaluated the relevant integrals to high order (up to ${\cal O}(x^{26})$ for photons and ${\cal O}(x^{28})$ for graviton partition functions); the results of these ${\mathtt {Mathematica}}$ experiments are all 
consistent with particular conjectures for the results for this 
integral that we present in every dimension in Tables \ref{photon-plethystic} and \ref{graviton-plethystic} respectively. In \cite{Henning:2017fpj}, photon S-matrix partition function was computed for $D=4$. Our result matches with theirs.

Of course, the conjectures for the results of the plethystic 
integrals in $D \leq 9$ were not obtained blind but were
motivated by physical considerations involving S-matrices 
which we describe in detail in subsequent sections. The  
power series evaluation of the plethystic performed in 
this subsection is  a direct computational 
evidence for the correctness of these physically motivated conjectures.

\subsection{Module generators and Lagrangians} \label{mgl} 

Earlier in this section we presented a detailed discussion of the correspondence between S-matrices and Lagrangians (up to equivalences). Note that the correspondence described so far relates two structures, both of which are  $S_4$ invariant. Lagrangians built out 
of identical bosonic fields are automatically $S_4$ invariant, 
while $S$ matrices are $S_4$ invariant by construction 
(see Section \ref{Permsfst} for a detailed discussion). 

In our general discussion about the structure of S-matrices in 
Section \ref{gsms} we found it useful to regard $S_4$ invariant 
$S$ matrices as special members of a larger family of $\Z_2 \times \Z_2$ 
invariant `quasi-invariant S-matrices' (see subsection \ref{lsm}). Recall, in particular, that it is  the  space of quasi-invariant polynomial S-matrices (rather than the space of 
fully $S_4$ invariant polynomial $S$ matrices) that form a module.
The space of physical (i.e. completely $S_4$ invariant) polynomial $S$ matrices is obtained by first enumerating the modules of 
quasi-invariant S-matrices and then projecting onto the subspace of $S_3$ singlets.

 As the module structure of local S-matrices plays a key role in their enumeration, it is somewhat unsatisfying to have Lagrangian structures `dual' only to fully $S_4$ invariant S-matrices. In particular, recall that S-matrix modules are labelled by their generators which, in general, transform
 in non-trivial representations of $S_3$. In this brief subsection we describe a procedure that allows us to associate
 Lagrangians with generators of the local module even when 
 the generators in question are not $S_3$ invariant. 

Any set of generators $M_a$ of the local module  (that transform in some representation of $S_3$) is naturally associated with an infinite class of genuine ($S_3$ invariant) S-matrices $S(M_a)$as follows. $S(M_a)$ is defined as the restriction of the span of $M_a$ to $S_3$ singlets, i.e. restriction to $S_3$ singlets of module elements of the form $r.M_a$ where $r$ is an element of the  ring (i.e. is a polynomial of $s$ and $t$). In other words $S(M_a)$ are all the $S_3$ invariant descendants of the generators.

Similarly any Lagrangian $L$ can be associated with an infinite
class of Lagrangians $C(L)$ defined as follows. $C(L)$ is defined as the set of Lagrangians obtained by taking derivatives the fields that appear in the Lagrangian and contracting the indices of these derivatives in pairs. 

We say that a Lagrangian $L$ is associated with the generators 
$M_a$ if the set of  S-matrices obtained from the Lagrangians $C(L)$ coincide with $S(M_a)$. This association allows us to use 
Lagrangians to label generators (and more generally elements)
of the local module. We will use this association in the next section.

As an example consider the photon Lagrangian ${\rm Tr}(F^2){\rm Tr}(F^2)$. The corresponding generators of the local Module are ${\rm Tr}(F_1 F_2){\rm Tr}(F_3 F_4)$, ${\rm Tr}(F_1 F_3){\rm Tr}(F_2 F_4)$ and ${\rm Tr}(F_1 F_4){\rm Tr}(F_3 F_2)$; this set of generators transforms in the $\bf 3$ of $S_3$. 

For another example consider the photon Lagrangian term $F^{ab}  \partial_a F^{\mu\nu} \partial_b F^{\nu\rho} F^{\rho \mu}$. In this case the generators corresponding to the given Lagrangian consist of 
the single element
\be \label{nteg}
\frac14(F^{ab}_{1}  \partial_a F_{2}^{\mu\nu} \partial_b F_{3}^{\nu\rho} F_{4}^{\rho \mu}+F^{ab}_{2}  \partial_a F_{1}^{\mu\nu} \partial_b F_{4}^{\nu\rho} F_{3}^{\rho \mu}+F^{ab}_{3}  \partial_a F_{4}^{\mu\nu} \partial_b F_{1}^{\nu\rho} F_{2}^{\rho \mu}+F^{ab}_{4}  \partial_a F_{3}^{\mu\nu} \partial_b F_{2}^{\nu\rho} F_{1}^{\rho \mu}).
\ee 
\eqref{nteg} had four terms rather than one because no single 
one of the terms above is $\Z_2\times \Z_2$ invariant. 
It is easy to see that the resultant expression \eqref{nteg} transforms in the completely symmetric representation of $S_3$.

\section{Polynomial photon S-matrices and corresponding Lagrangians} \label{mgls}

Recall that  quasi-invariant polynomial S-matrices form a module, called the local module. In this section we completely characterize this module by specifying the generators $E_J(p_i,\epsilon_i)$ for 4-photon S-matrices (the case of graviton S-matrices is the topic of the next section). We also present an explicit 
parameterization of the physical ($S_4$ invariant) 
S-matrices that are `descendants' of these generators and thereby 
present an explicit parameterization of the most general 
allowed polynomial four photon and four graviton S-matrix 
in every dimension. Finally we also present explicit 
expressions for the Lagrangians from which these S-matrices follow.

Before we dive into the analysis let us spare some time fixing up the notation and convention. In the case of photons, we denote the parity even generators of the local module as $E_{\bf R}$ and parity odd generators as $O_{\bf R}$. The subscript ${\bf R}$ is either ${\bf S},{\bf A}$ or ${\bf 3}$ denoting its $S_3$ representation ${\bf 1_S}, {\bf 1_A}$ or ${\bf 3}$ respectively.  When there are multiple generators transforming in the same representation are present, we assign them serial numbers which are also denoted in the subscript. For example, if there two symmetric parity even generators then they are denoted as $E_{{\bf S},1}$ and $E_{{\bf S},2}$. In the case when ${\bf R}={\bf 3}$ or ${\bf 3_A}$, we use a superscript to denote the specific state of the three dimensional representation (the transformation property of these generators are listed in \eqref{tof2} and \eqref{tof4}). By convention, we always choose $E_{\bf 3}^{(1)}$ (or $O_{\bf 3}^{(1)}$) to be invariant under the swap $1\leftrightarrow2$. The second and the third components are obtained by permuting $(234)\to (342)$. This means, the component $(2)$ is invariant under the swap $1\leftrightarrow3$ and the component $(3)$ is invariant under the swap $1\leftrightarrow4$. In one case we have to deal with the generator transforming in ${\bf 3}_A$ representation. Recall that this is the representation obtained by acting on a state $(1)$ that is antisymmetric in the exchange of $1\leftrightarrow 2$ by the cyclic permutation $(234)\to (342)$.  Sometimes we also include the space-time dimension in the superscript when it needs to be emphasized, e.g. $E_{\bf 3}^{D=4,(1)}$. 

For gravitons, we have the same notation except that the letters $G$ and $H$ are used, instead of $E$ and $O$, to denote the parity even and parity odd local module generators. In all cases, the corresponding bare module generators are denoted by lower-case letters i.e. the parity even and parity odd bare module generators for photons are denoted as $e$ and $o$ respectively  and for gravitons they are denoted as $g$ and $h$ respectively. 
In order to avoid excessive notation we use the photon notation $E, O$ and $e,o$ for scalars as well. Hopefully this does not cause any confusion as the discussion of the scalar case is very brief and serves as a warm up for the photon and graviton analysis.  

To construct the most general physical (i.e. $S_4$ invariant) S-matrix in the span of a quasi-invariant generator, say $E_{\bf R}$ we need to take the ``inner product" with a general polynomial of $(s,t)$ that transforms in the same representation ${\bf R}$. For example, if ${\bf R}={\bf S}$,
\be
\CS=\cf^{E_{\bf S}}(t,u)E_{\bf S}.
\ee
where the function $\cf^{E_{\bf S}}(t,u)$ is totally symmetric under $S_3$. When ${\bf R}={\bf A}$,
\be
\CS=\cf^{E_{\bf A}}(t,u)E_{\bf A}.
\ee
where the function $\cf^{E_{\bf A}}(t,u)$ is totally antisymmetric under $S_3$.

The S-matrix is more involved for quasi-invariant structure  that transforms in ${\bf 3}$ representation of $S_3$. It is given by
\be\label{3-general}
\CS=\cf^{E_{\bf 3}}(t,u) E_{\bf 3}^{(1)}+ \cf^{E_{\bf 3}}(u,s)E_{\bf 3}^{(2)}+\cf^{E_{\bf 3}}(s,t)E_{\bf 3}^{(3)}.
\ee
where $\cf^{E_{\bf 3}}(t,u)$ is a symmetric function in its two arguments (symmetry under the exchange of $t$ and $u$ is the same as the symmetry under the exchange of $1\leftrightarrow 2 $ which matches with the symmetry of $E_{\bf 3}^{(1)}$ and so on). Sometimes, it helps use the shorthand 
\be\label{example1}
\cf^{E_{\bf 3}^{(1)}}(t,u)\equiv \cf^{E_{\bf 3}}(t,u),\qquad \cf^{E_{\bf 3}^{(2)}}(t,u)\equiv \cf^{E_{\bf 3}}(u,s),\qquad \cf^{E_{\bf 3}^{(1)}}(t,u)\equiv \cf^{E_{\bf 3}}(s,t).
\ee
So that the above S-matrix can be written as 
\be
{\cal S}=\sum_{i=1,2,3}\cf^{E_{\bf 3}^{(i)}}(t,u) E_{\bf 3}^{(i)}.
\ee
The momenta functions \eqref{example1} transform as noted in \eqref{tof2}. 
The S-matrix corresponding to a generator transforming in ${\bf 3_A}$ representation is also given by the equation \eqref{3-general} except that the function $\cf$ is antisymmetric rather than symmetric in its two arguments (consequently the momenta functions transform as \eqref{tof4}). We will always label the function $\cf$ by the quasi-invariant structure that it multiplies.

 \subsection{Scalar polynomial S-matrices and corresponding Lagrangians}
 
As a warm up for the main analysis of this section let us 
first consider the case of four scalar scattering. 
As scalar S-matrices don't have any index structures, the local module and the bare module are identical (and so, in particular, are freely generated). In $D\geq 4$, they both are generated by a unique generator $E_{\bf S}=1$ which is clearly $S_3$ invariant (i.e. transforms in the ${\bf 1_S}$ representation of $S_3$). The Lagrangian corresponding to this generator is simply $\phi^4$. 

\subsubsection{Module generators and S-matrix partition functions}

The module analysis of scalar S-matrices (see \eqref{pfnn}) that the partition function $Z_\text{S-matrix}(x)$ over S-matrices in this case - which we denote by ${\cal I}_{\rm s}^{D \geq 4}(x)$  should simply be given by 
\begin{equation}\label{sc}
{\cal I}_{\rm s}^{D \geq 4}(x)=\denom
\end{equation} 
(see \eqref{partfnsss} for definitions). This prediction is precisely borne out by the partition function obtained via plethystic counting (see Table \ref{scalar-plethystic}).

All four scalar scattering amplitudes in $D \geq 4$ are parity invariant. The reason for this is easy to understand. 4-scalar scattering involves only $3$ independent vectors (which can be chosen to be any three of the four scattering momenta). It follows that no $D\geq 4$ parity odd S-matrix exists as the number of free indices in the Levi-Civita tensor exceeds the number of independent vectors. 

It is clear that the argument of the previous paragraph fails in $D=3$ however. In this case we have the following parity odd 
structure which is a second generator of the local Module (the first generator continues to be unity)
\begin{equation}\label{geno}
O_{\bf A}^{D=3}=\varepsilon_{\mu\nu \rho} k_1^\mu k_2^\nu k_3^\rho.
\end{equation} 
The generator \eqref{geno} is precisely ${\tilde \varepsilon}$ in \eqref{dmt} for $D=3$.
The `Lagrangian' associated with this generator (in the sense of subsection \ref{mgl})
is\footnote{The Lagrangian  \eqref{aslag} vanishes for symmetry reasons; however
its `descendants' (Lagrangians obtained by taking derivatives of the four 
$\phi$ fields in \eqref{aslag} and contracting the indices in pairs) do not, 
in general, vanish. Consequently the Lagrangian \eqref{aslag} - while trivial 
as a functional - is non-trivial as the Lagrangian that labels module generators
in the sense of subsection \ref{mgl}.}, 
\be\label{aslag}
\varepsilon_{\mu\nu \rho}\partial_\mu \phi \partial_\nu \phi \partial_\rho \phi \phi.
\ee

Clearly \eqref{geno} transforms in the anti-symmetric representation of $S_3$ (see \eqref{tmo} )and so it follows from \eqref{pfnn} that it's contribution to $Z_\text{S-matrix}(x)={\cal I}_{\rm s}^{D= 3}(x)$ is  $x^3 Z_{\bf 1_A}(x)$. It follows that the study of the 
local S-matrix in this case predicts that 
\begin{equation}\label{zsmd3}
{\cal I}_{\rm s}^{D= 3}(x)= Z_{\bf 1_S}(x)+ x^3 Z_{\bf 1_A}(x)= \left( 1+x^9 \right) \denom
\end{equation} 
in perfect agreement with the plethystic result reported in 
\ref{scalar-plethystic}.

In $D=2$, the scalar Lagrangians again are descendants of $\phi^4$. This would lead to the expectation that the partition function over Lagrangians should be $\denom$. But in two dimensions Mandelstam variables degenerate, either $t$ or $u$ vanish. Hence, the completely symmetric functions of $(s,t,u)$ are generated by only the four derivative generator $s^2+t^2+u^2$. The six derivative generator $stu$ is identically zero. The partition function over Lagrangians is then, $(1-x^4)^{-1}$. This precisely matches the plethystic counting.  

\subsubsection{Explicit Expressions for most general S-matrix 
	and corresponding Lagrangians}

For completeness we present a completely explicit parameterization of the 
most general 4 scalar S-matrix and associated Lagrangians. For $D\geq 4$, there is a unique quasi-invariant generator $E_{\bf S}=1$. The general S-matrix is
\be\label{mjlsc}
\CS=\cf^{E_{\bf S}}(t,u)
\ee
where this function is completely symmetric under the exchange of $s,t,u$. 

Recall that $\cf^{E_{\bf S}}(t,u)$ is a polynomial in $t$ and $u$
and so can be expanded as a finite sum of the form
\begin{equation}\label{expsca}
\cf^{E_{\bf S}}(t,u)= 
\Big(\cf^{E_{\bf S}}\Big)_{n,m}t^n u^m
\end{equation} 
It follows from the analysis of subsection  \ref{rgss} 
that the only S-matrices of the form \eqref{mjlsc} that grow no faster than 
$s^2$ in the Regge limit are 
\begin{equation}\label{fosm} 
\cf^{E_{\bf S}}(t,u)|_{<s^2}
=a_0 + a_4 (s^2+t^2+u^2) + a_6 (stu).
\end{equation} 

The Lagrangian from which the S-matrix \eqref{fosm} follows is proportional to 
\begin{equation} \label{lagscfol}
L^{D\geq 4}=\sum_{m, n} \Big(\cf^{E_{\bf S}}\Big)_{m,n} 2^{m+n}\left(\prod_{i=1}^m \prod_{j=1}^n  \left(\partial_{\nu_j} \partial_{\mu_i} \phi \right) \phi \partial_{\mu_i}
\phi \partial_{\nu_j} \phi \right). \\
\end{equation} 

Turning now to $D=3$, the most general scalar S-matrix is given by
\begin{equation}\label{mjlscthree}
\cf^{E_{\bf S}^{D=3}}(t,u) E_{\bf S}^{D=3} +\cf^{O_{\bf A}^{D=3}}(t,u)
O_{\bf A}^{D=3} 
\end{equation} 
where $E_{\bf S}^{D=3}=1$ and $O_{\bf A}^{D=3}=\varepsilon_{\mu\nu \rho} k_1^\mu k_2^\nu k_3^\rho|_{\Z_2\times \Z_2}$ as given in equation \eqref{geno}. The function  $\cf^{E_{\bf S}^{D=3}}(t,u)$
is a general polynomial that is completely symmetric under 
interchanges of $s, t, u$ while $\cf^{O_{\bf A}^{D=3}}(t,u)$ is 
a general polynomial that is completely antisymmetric under the 
same interchanges. We Taylor series expand the functions in \eqref{mjlscthree} 
in a manner completely analogous to \eqref{expsca}. The Lagrangians from
which the S-matrix \eqref{mjlscthree} follows is proportional to 
\begin{equation} \label{lagscfolth}
\begin{split}
L^{D=3}&=\sum_{m, n} \left(\cf^{E_{\bf S}^{D=3}}\right)_{m,n} 2^{m+n}\left(\prod_{i=1}^m \prod_{j=1}^n  \left(\partial_{\nu_j} \partial_{\mu_i} \phi \right) \phi \partial_{\mu_i}
\phi \partial_{\nu_j} \phi \right) \\
&-i\sum_{m, n} \left(\cf^{O_{\bf A}^{D=3}}\right)_{m,n} 2^{m+n}\left(\prod_{i=1}^m \prod_{j=1}^n  \epsilon_{\a\b\g}\left(\partial_{\nu_j} \partial_{\mu_i} \partial^\a\phi \right) \partial^\b\phi \partial_{\mu_i}
\partial^\g\phi \partial_{\nu_j} \phi \right). \\
\end{split}
\end{equation}

\subsection{Construction of all parity even photon S-matrices 
	for $D \geq 5$ } \label{photsm}

We now turn to the main topic of this section - the study of 
polynomial photon S-matrices. In this subsection we begin this analysis by presenting an explicit construction 
of all parity even S-matrices in $D \geq 5$.

In appendix \ref{Lag}, we have painstakingly shown that the most general parity even gauge invariant Lagrangian can be obtained by taking linear combinations of pairs of contracted derivatives on the three `generator' Lagrangians
\begin{equation}\label{lagstruct}
{\rm Tr}(F^2){\rm Tr}(F^2), ~~~~~ {\rm Tr}(F^4), ~~~~~~ -F^{ab} \partial_a F^{\mu\nu} \partial_b F^{\nu\rho} F^{\rho \mu}
\end{equation} 
The generators of the local module dual to these Lagrangians (in the sense of section \ref{mgl}) are given by

\bea\label{photon-lag}
E^{(1)}_{{\bf 3}, 1} &=& 8{\rm Tr}(F_1F_2){\rm Tr}(F_3F_4),~~~E^{(2)}_{{\bf 3}, 1} = 8{\rm Tr}(F_1F_3){\rm Tr}(F_2F_4),~~~E^{(3)}_{{\bf 3}, 1} =8{\rm Tr}(F_1F_4){\rm Tr}(F_3F_2),\nonumber\\
E^{(1)}_{{\bf 3}, 2} &=& 8{\rm Tr}(F_1F_3F_2F_4),~~~~~~~~E^{(2)}_{{\bf 3}, 2} = 8{\rm Tr}(F_1F_2F_3F_4),~~~~~~~~E^{(3)}_{{\bf 3}, 2}= 8{\rm Tr}(F_1F_3 F_4F_2),\nonumber\\
E_{\bf S}&\simeq& -6F^{ab}_{1}  \partial_a F_{2}^{\mu\nu} \partial_b F_{3}^{\nu\rho} F_{4}^{\rho \mu}|_{\Z_2\times \Z_2}\nonumber\\
&=& 6\left(-F^{ab}_{1}  \partial_a F_{2}^{\mu\nu} \partial_b F_{3}^{\nu\rho} F_{4}^{\rho \mu} -F^{ab}_{2}  \partial_a F_{1}^{\mu\nu} \partial_b F_{4}^{\nu\rho} F_{3}^{\rho \mu}-F^{ab}_{3}  \partial_a F_{4}^{\mu\nu} \partial_b F_{1}^{\nu\rho} F_{2}^{\rho \mu}-F^{ab}_{4}  \partial_a F_{3}^{\mu\nu} \partial_b F_{2}^{\nu\rho} F_{1}^{\rho \mu}\right). \nonumber\\
\eea

We have demonstrated that quantities listed in \eqref{photon-lag} are the generators of the parity even 
part of the local Module in every dimension. Note that 
there are two four derivative generators in the ${\bf 3}$ 
and one six derivative generators in the ${\bf 1_S}$ of $S_3$. The second subscript on $E_{\bf 3}$ is simply an arbitrarily assigned serial number.

It remains to check whether the parity even part of the  local module is freely generated. Note that the local module has 7 generators in every dimension. From Table \ref{counting-is} we see that the number of 
generators of the parity even part of the bare module is $5$ in $D=4$ and $1$ in $D=3$. 
As the number of generators of the local Module exceeds the number of generators of the bare module, it follows (see subsubsection \ref{elmbm}) that the local Module is {\it not} freely generated in these dimensions. 
In $D=4$ and $3$ the local parity even S-matrix module is not completely specified by their generators \eqref{photon-lag}; we also need to specify the relations obeyed within the modules 
generated by these generators. We will return to this point later in this subsection.  

In $D\geq5$, on the other hand, we see from 
Table \ref{counting-is} that the bare module has exactly as many generators as the local 
module (the representations of these generators also match those of the local module: two in the {\bf 3} and 
one in the ${\bf 1_S}$). It follows that the local module is freely generated unless the extremely stringent condition \eqref{freeconds} is satisfied. We will demonstrate below that \eqref{freeconds} is not satisfied so that the local module is also freely generated. In this case, therefore, parity even 
local S-matrix module - and so the space of all local parity even S-matrices - is completely specified by the generators 
\eqref{photon-lag}.

To proceed we express local module generators 
\eqref{photon-lag} in terms of the generators of the bare modules $e_I(\alpha_i,\polo_i)$ that are constructed in section \ref{constructionei},
\bea \label{expeei}
e_{{\bf 3},1}^{(1)}(\alpha_i,\polo_i)&=&(\polo_1\cdot\polo_2)(\polo_3\cdot\polo_4),~~
e_{{\bf 3},1}^{(2)}(\alpha_i,\polo_i)=(\polo_1\cdot\polo_3)(\polo_2\cdot\polo_4),\nonumber\\
e_{{\bf 3},1}^{(3)}(\alpha_i,\polo_i)&=&(\polo_1\cdot\polo_4)(\polo_3\cdot\polo_2)\nonumber\\
\nonumber\\
e_{{\bf 3},2}^{(1)}(\alpha_i,\polo_i)&=&( \polo_1 . \polo_2\, \alpha_3 \alpha_4 + \polo_3.\polo_4 \, \alpha_1 \alpha_2),~~
e_{{\bf 3},2}^{(2)}(\alpha_i,\polo_i)=( \polo_1 . \polo_3\, \alpha_2 \alpha_4 + \polo_2.\polo_4 \, \alpha_1 \alpha_3), \nonumber\\
e_{{\bf 3},2}^{(3)}(\alpha_i,\polo_i)&=&( \polo_1 . \polo_4\, \alpha_3 \alpha_2 + \polo_3.\polo_2 \, \alpha_1 \alpha_4)\nonumber\\
\nonumber\\
e_{{\bf S}}(\alpha_i,\polo_i)&=& \alpha_1 \alpha_2 \alpha_3 \alpha_4.
\eea
As explicitly indicated  $e_{{\bf 3},1}$ and $e_{{\bf 3},2}$ transform in $\bf 3$ of $S_3$.  We have\footnote{ If $\alpha^{(i)}$ and $\beta^{(j)}$ transform 
	in the ${\bf 3}$ representation, then $\theta^{(i)}= \alpha^{(i)} \beta^{(i)}$, 
	$\phi^{(1)}=\alpha^{(3)} \beta^{(2)} +\alpha^{(2)} \beta^{(3)}~ ~{\rm and ~cyclic}$, and 
	$\zeta^{(1)}=\alpha^{(1)} (\beta^{(2)} + \beta^{(3)}) ~ ~{\rm and ~ cyclic}$
all also transform in the ${\bf 3}$ representation. The reader can use 
these rules - together with the fact that $(s,t,u)$ transform in ${\bf 2_M}$ and the triplets of functions 
$(s^2, t^2, u^2)$, $(st,tu,us)$ transform in the ${\bf 3}$ 
- to show that the expressions that appear on the RHS of 
the definitions of  $E_1^{(i)}$ and $E_2^{(i)}$ transform in the ${\bf 3}$. }:
\bea\label{explicit-embeddingo}
E_{{\bf 3},1}^{(1)}&=&-8s^2 e_{{\bf 3},1}^{(1)}+8s^2e_{{\bf 3},2}^{(1)}-8s^2e_{\bf S},\qquad
E_{{\bf 3},1}^{(2)}=-8t^2 e_{{\bf 3},1}^{(2)}+8t^2e_{{\bf 3},2}^{(2)}-8t^2e_{\bf S},\nonumber\\
E_{{\bf 3},1}^{(3)}&=&-8u^2 e_{{\bf 3},1}^{(3)}+8u^2e_{{\bf 3},2}^{(3)}-8u^2e_{\bf S},\nonumber\\
\nonumber\\
E_{{\bf 3},2}^{(1)}&=&-2(u^2 e_{{\bf 3},1}^{(2)}+t^2 e_{{\bf 3},1}^{(3)})+2(u(s-t)e_{{\bf 3},2}^{(2)}+t(s-u)e_{{\bf 3},2}^{(3)})-2(t^2+u^2)e_{\bf S},\nonumber\\
E_{{\bf 3},2}^{(2)}&=&-2(s^2 e_{{\bf 3},1}^{(3)}+u^2 e_{{\bf 3},1}^{(1)})+2(s(t-u)e_{{\bf 3},2}^{(3)}+u(t-s)e_{{\bf 3},2}^{(1)})-2(u^2+s^2)e_{\bf S},\nonumber\\
E_{{\bf 3},2}^{(3)}&=&-2(t^2 e_{{\bf 3},1}^{(1)}+s^2 e_{{\bf 3},1}^{(2)})+2(t(u-s)e_{{\bf 3},2}^{(1)}+s(u-t)e_{{\bf 3},2}^{(2)})-2(s^2+t^2)e_{\bf S},\nonumber\\
\nonumber\\
E_{\bf S}&=&3\,stu\,(e_{{\bf 3},2}^{(1)}+e_{{\bf 3},2}^{(2)}+e_{{\bf 3},2}^{(3)}-2e_{\bf S}).
\eea

With the explicit expressions \eqref{explicit-embeddingo} in hand  it is 
not difficult to check that \eqref{freeconds} is indeed not obeyed and the local module is generated freely by the above generators (from \eqref{explicit-embeddingo}, we find, Det $\left[ p_{IJ}(s,t) \right] = 393216 s^5 t^5 u^5 $). It follows that \eqref{itv} is indeed the expected answer 
for the partition function over S-matrices.

For completeness we present an explicit parameterization of the most general parity even S-matrix for four photon scattering in $D \geq 5$ and also of the Lagrangians that generate these S-matrices. The most general S-matrix 
is parameterized by two $\Z_2$ symmetric functions of $t$ and $u$, and one 
completely $s$, $t$, $u$ symmetric function. Explicitly we have
\begin{eqnarray}  \label{expparphdte} 
&{\cal S}^{D\geq 5}_{\rm even}=4\bigg(\cf^{E_{{\bf 3},1}}(t,u)\left( p^1_\mu \epsilon^1_\nu - p^1_\nu \epsilon^1_\mu  
\right) \left( p^2_\mu \epsilon^2_\nu - p^2_\nu \epsilon^2_\mu  \right)
\left( p^3_\alpha \epsilon^3_\beta - p^3_\beta \epsilon^3_\alpha  
\right) \left( p^4_\alpha \epsilon^4_\beta - p^4_\beta \epsilon^4_\alpha  \right) \nonumber\\
&+\cf^{E_{{\bf 3},1}}(u,s)\left( p^1_\mu \epsilon^1_\nu - p^1_\nu \epsilon^1_\mu  
\right) \left( p^3_\mu \epsilon^3_\nu - p^3_\nu \epsilon^3_\mu  \right)
\left( p^2_\alpha \epsilon^2_\beta - p^2_\beta \epsilon^2_\alpha  
\right) \left( p^4_\alpha \epsilon^4_\beta - p^4_\beta \epsilon^4_\alpha  \right) \nonumber\\
&+\cf^{E_{{\bf 3},1}}(s,t)\left( p^1_\mu \epsilon^1_\nu - p^1_\nu \epsilon^1_\mu  
\right) \left( p^4_\mu \epsilon^4_\nu - p^4_\nu \epsilon^4_\mu  \right)
\left( p^3_\alpha \epsilon^3_\beta - p^3_\beta \epsilon^3_\alpha  
\right) \left( p^2_\alpha \epsilon^2_\beta - p^2_\beta \epsilon^2_\alpha  \right)\bigg) \nonumber\\
&+4\bigg( \cf^{E_{{\bf 3},2}}(t,u)\left( p^1_\mu \epsilon^1_\nu - p^1_\nu \epsilon^1_\mu  
\right) \left( p^3_\nu \epsilon^3_\alpha - p^3_\alpha \epsilon^3_\nu  \right)
\left( p^2_\alpha \epsilon^2_\beta - p^2_\beta \epsilon^2_\alpha  
\right) \left( p^4_\beta \epsilon^4_\mu - p^4_\mu \epsilon^4_\beta  \right) \nonumber\\
&+\cf^{E_{{\bf 3},2}}(u,s)\left( p^1_\mu \epsilon^1_\nu - p^1_\nu \epsilon^1_\mu  
\right) \left( p^2_\nu \epsilon^2_\alpha - p^2_\alpha \epsilon^2_\nu  \right)
\left( p^3_\alpha \epsilon^3_\beta - p^3_\beta \epsilon^3_\alpha  
\right) \left( p^4_\beta \epsilon^4_\mu - p^4_\mu \epsilon^4_\beta  \right) \nonumber\\
&+\cf^{E_{{\bf 3},2}}(s,t)\left( p^1_\mu \epsilon^1_\nu - p^1_\nu \epsilon^1_\mu  
\right) \left( p^3_\nu \epsilon^3_\alpha - p^3_\alpha \epsilon^3_\nu  \right)
\left( p^4_\alpha \epsilon^4_\beta - p^4_\beta \epsilon^4_\alpha  
\right) \left( p^2_\beta \epsilon^2_\mu - p^2_\mu \epsilon^2_\beta  \right)\bigg) \nonumber\\
&+\cf^{E_{{\bf S}}}(t,u)\nonumber\\
& \left( \left( p_a^1 \epsilon_b^1-p_b^1 \epsilon_a^1 \right)
p^2_a\left( p_\mu^2 \epsilon_\nu^2-p_\nu^2 \epsilon_\mu^2 \right) p^3_b\left( p_\nu^3 \epsilon_\alpha^3-p_\alpha^3 \epsilon_\nu^3 \right) \left( p_\alpha^4 \epsilon_\mu^4- p_\mu^4 \epsilon_\alpha^4 \right) \right.\nonumber\\
&\left.+\left( p_a^2 \epsilon_b^2-p_b^2 \epsilon_a^2 \right)
p^1_a\left( p_\mu^1 \epsilon_\nu^1-p_\nu^1 \epsilon_\mu^1 \right) p^4_b\left( p_\nu^4 \epsilon_\alpha^4-p_\alpha^4 \epsilon_\nu^4 \right) \left( p_\alpha^3 \epsilon_\mu^3- p_\mu^3 \epsilon_\alpha^3 \right)\right.\nonumber\\
&\left.+\left( p_a^3 \epsilon_b^3-p_b^3 \epsilon_a^3 \right)
p^4_a\left( p_\mu^4 \epsilon_\nu^4-p_\nu^4 \epsilon_\mu^4 \right) p^1_b\left( p_\nu^1 \epsilon_\alpha^1-p_\alpha^1 \epsilon_\nu^1 \right) \left( p_\alpha^2 \epsilon_\mu^2- p_\mu^2 \epsilon_\alpha^2 \right)\right.\nonumber\\
&\left.+\left( p_a^4 \epsilon_b^4-p_b^4 \epsilon_a^4 \right)
p^3_a\left( p_\mu^3 \epsilon_\nu^3-p_\nu^3 \epsilon_\mu^3 \right) p^2_b\left( p_\nu^2 \epsilon_\alpha^2-p_\alpha^2 \epsilon_\nu^2 \right) \left( p_\alpha^1 \epsilon_\mu^1- p_\mu^1 \epsilon_\alpha^1 \right)\right).
\end{eqnarray}

The functions $\cf^{E_{{\bf 3},1}}(t,u), \cf^{E_{{\bf 3},2}}(t,u)$ are each arbitrary functions that are symmetric in their two arguments. These functions with permuted arguments transform in the ${\bf 3}$ of $S_3$ 
(see the discussion around \eqref{tof1}) explaining the superscript ${\bf 3}$ on these functions. On the other hand 
$\cf^{E_{{\bf S}}}(t,u)$ is a function that is 
completely symmetric under interchange of $s$, $t$ and $u$.

It is not difficult to verify (see Appendix 
\ref{phorb}) that the most general S-matrix 
of the form \eqref{expparphdte} that grows 
no faster than $s^2$ in the Regge limit is
given by the four parameter set 
\begin{equation}\label{mjppph}
\cf^{E_{{\bf 3},1}}(t,u)= c_1, ~~~
\cf^{E_{{\bf 3},2}}(t,u)=c_2 + c_3(u+t), ~~~\cf^{E_{{\bf S}}}(t,u)
=c_4.
\end{equation} 
The S-matrices parameterized by $c_1$ and $c_2$
are both four derivative. The S-matrices 
parameterized by $c_3$ and $c_4$ are both 
6 derivatives. All 4 S-matrices corresponding to $c_i$, $i=1,2,3,4$ grow like
$s^2$ in the Regge limit.

The three functions $\cf^{E_{{\bf 3},1}}(t,u), \cf^{E_{{\bf 3},2}}(t,u)$ and $\cf^{E_{{\bf S}}}(t,u)$ can be Taylor expanded in a manner completely 
analogous to \eqref{expsca}. The Lagrangians that generates the S-matrix 
\eqref{expparphdte} is given by 
\begin{equation} \label{lagphten}
\begin{split} 
L^{D\geq 5}_{\rm even}=&\sum_{m, n} \left(\cf^{E_{{\bf 3},1}}\right)_{m,n} 2^{m+n}\left(\prod_{i=1}^m \prod_{j=1}^n 
{\rm Tr} \left(  
\partial_{\nu_j} \partial_{\mu_i} F 
F \right) {\rm Tr}  \left( \partial_{\mu_i}
F \partial_{\nu_j}F \right)  \right) \\
&+\sum_{m, n} \left(\cf^{E_{{\bf 3},2}}\right)_{m,n}2^{m+n}
\left( \prod_{i=1}^m \prod_{j=1}^n 
{\rm Tr} \left(  
\partial_{\nu_j} \partial_{\mu_i} F 
\partial_{\mu_i} F    F\partial_{\nu_j}F \right) \right)\\
&+\sum_{m, n} \left(\cf^{E_{{\bf S}}}\right)_{m,n}2^{m+n}
\left(- \prod_{i=1}^m \prod_{j=1}^n 
\partial_{\mu_i} \partial_{\nu_j}F_{ab} {\rm Tr} \left(  
\partial_{\mu_i}  \partial_a F 
\partial_{\nu_j} \partial_b F  F \right) \right). \\
\end{split} 
\end{equation}

As mentioned above \eqref{expparphdte} and \eqref{lagphten}
describe the most general polynomial S-matrix (and corresponding
local Lagrangian) for parity even four photon scattering in 
dimensions $D \geq 5$. In these dimensions the three functions  
label polynomial S-matrices in a one to one manner; every distinct choice of these functions yields a distinct S-matrix, and every polynomial S-matrix corresponds to some choice of these 
functions. 

In fact the expressions \eqref{expparphdte} and \eqref{lagphten}
also apply to $D=4$ and $D=3$. In this case, however, the 
map between the functions $\cf^{E_{{\bf 3},1}}(t,u), \cf^{E_{{\bf 3},2}}(t,u)$ and $\cf^{E_{{\bf S}}}(t,u)$ and polynomial S-matrices is many to one. While 
every S-matrix continues to correspond to some choice of these
three functions, many different choices of these functions 
yield the same local S-matrix (this is another way of saying 
that the parity odd local S-matrix module in these dimensions 
is not freely generated but has relations).

We now turn to a brief discussion of parity odd S-matrices, i.e. S-matrices that use a single copy of the Levi-Civita tensor. As this tensor has a different numbers of indices in different dimensions, the structure of the parity 
odd local module tends to be very specific to dimension. However there is one universal statement about parity odd 
S-matrices that is easy to make, namely that no such 
S-matrices exist for $D\geq 8$. This simple fact follows 
from the observation that in these the Levi-Civita tensor has 8 or more indices but only 7 independent vectors - three momenta
and four polarizations -for these indices to contract with. 
It follows that all four photon S-matrices are parity even 
in $D \geq 8$ (this fact is also clear from Table \ref{counting-is}).

In the rest of this subsection we will use the discussion 
above to understand the detailed structure of the local S-matrix module, the partition function over S-matrices, and 
also provide a completely explicit parameterization of S-matrices and their corresponding Lagrangians dimension 
by dimension. In order to do this we will construct the 
parity odd local S-matrix modules in the dimensions in which 
they exist. We will also completely characterize the 
relations in the local parity even S-matrix module in 
$D=4$ and $D=3$. Finally we will reconcile our results 
with the explicit results of plethystic counting presented 
in Table \eqref{photon-plethystic}.

Finally a notational remark: In the rest of this subsection we use the notation
$I_{\tv}^{D=m}(x)$ for the partition function $Z_\text{S-matrix}$
(see \eqref{Smpf}) for the case of 4 photon scattering in $m$ 
dimensions. 

\subsection{$D\geq 8$:}

In these dimensions all S-matrices are parity even. The 
local module of parity even S-matrices is freely generated. 
The module (see \eqref{photon-lag}) has two 4 derivative 
generators both in the ${\bf 3}$ and one 6 derivative generator 
in the ${\bf 1_S}$. 

It follows from \eqref{pfnnnn} that the partition function 
\eqref{Smpf} over S-matrices in these dimensions is given by 
\begin{equation}\label{itv} 
I_{\tv}^{D\geq 8}(x) = 2 x^4  Z_{\bf 3}(x) +x^6 Z_{\bf 1_S}(x). 
\end{equation} 

It remains to compare the prediction \eqref{itv} against 
the explicit results of plethystic counting presented in Table \ref{photon-plethystic}. For $D\geq 10$ the results reported in \ref{photon-plethystic} match exactly with the prediction \eqref{itv}; we view this matching as a highly non-trivial confirmation of \eqref{itv}.

It is at first puzzling, however, that the result of Plethystic \ref{photon-plethystic} 
differs from \eqref{itv} by $-x^5$ in $D=9$ and $+x^4$ in $D=8$. The resolution to this apparent contradiction is that  the plethystic procedure we adopted in \ref{pleth-spin}  slightly miscounts the S-matrices in $D=9$ and $D=8$ as 
we now explain.

Recall that the plethystic counting procedure of section \ref{pleth-spin} proceed in three steps. 
\begin{itemize} 
	\item Step 1: We computed the partition function over operators built out of four (gauge invariant and on-shell) letters.
    \item Step 2: We organized four letter operators into derivatives of primaries. Assuming descendants were freely generated we obtained a partition function over  primaries (non total derivatives ) by multiplying the result of Step 1 by $\denom^{-1}$.
    \item Step 3: We integrated the result of Step 2 over the $SO(D)$ group with the Haar measure to isolate $SO(D)$ singlet primaries. 
    \end{itemize}

In $D=9$ the error in the plethystic procedure lies in the assumption in step 2 that derivative descendants are freely generated for 
all primaries. This assumption fails for one primary, namely 
 $J=*F\wedge F\wedge F\wedge F$. $J^\mu$ is an identically conserved current (this follows from use of the 
Bianchi identity) so that $\partial_\mu J^\mu=0$. Of course it is also true that $\partial_ {\alpha_1} \partial_{\alpha_2} \ldots \left( \partial_\mu J^\mu \right)=0 $. Borrowing language from the representation theory of the conformal algebra, $J_\mu$ is the primary of a `short representation' and  $\partial_\mu J^\mu$ is a primary null state. It follows that the contribution derivatives of $J_\mu$ to the final result of Step 2 
is 
\begin{equation}\label{shortrep}
x^4 \chi_{\syng{1}} -x^5
\end{equation} 
where $\chi_V$ is the $SO(9)$ character of the vector and 
$x^5$ multiplies unity, the character of the scalar. 
It follows that the contribution of this primary to the integral over the $SO(9)$ gauge group in Step 3 is $-x^5$.  Removing this fake contribution (by adding $x^5$) turns the $D=9$ entry of Table \ref{photon-plethystic} into the correct module prediction \eqref{itv}. 

In $D=8$ the plethystic counting makes the opposite error; 
it omits to recognize that the quantity $F\wedge F \wedge F 
\wedge F$ is a total derivative. The reason for this failing 
is that the plethystic procedure (Step 2 above) only removes total derivatives of quartic polynomials built out of field 
strength letters. However $F\wedge F \wedge F \wedge F$ is the total derivative of the 7 dimensional Chern Simons form which is not a polynomial in gauge invariant letters. In order to
get the correct S-matrix partition function in $D=8$ we must, 
consequently, remove this total derivative by hand, i.e. 
subtract $x^4$ from the $D=8$ entry of Table \ref{photon-plethystic}. Once again this procedure yields the correct 
module prediction \eqref{itv}\footnote{Operators formed out of $F\wedge F\wedge F\wedge F$ by taking derivatives of the four field strengths and contracting indices in pairs are also 
total derivatives. however it is not difficult to check that 
such operators can be written as total derivatives of 
operators that are quartic in letters and so are correctly 
subtracted out by the plethystic procedure. For example 
consider $\epsilon^{abcdefgh}\partial_\mu F_{ab}\partial^\mu F_{cd}F_{ef}F_{gh}$. We write this as,
\be
\begin{split}
\epsilon^{abcdefgh}\partial_\mu F_{ab}\partial^\mu F_{cd}F_{ef}F_{gh} &= -2 \epsilon^{abcdefgh}\partial_a F_{b \mu}\partial^\mu F_{cd}F_{ef}F_{gh} \\
&=-2 \epsilon^{abcdefgh} \partial_a\left(F_{b \mu}\partial^\mu F_{cd}F_{ef}F_{gh}\right)+2\epsilon^{abcdefgh}F_{b \mu}\partial^\mu \partial_a F_{cd}F_{ef}F_{gh}\\
&+ 2\epsilon^{abcdefgh}F_{b \mu}\partial^\mu  F_{cd}\partial_a F_{ef}F_{gh} + 2\epsilon^{abcdefgh}F_{b \mu}\partial^\mu  F_{cd} F_{ef}\partial_aF_{gh}\\
&=-2 \epsilon^{abcdefgh} \partial_a\left(F_{b \mu}\partial^\mu F_{cd}F_{ef}F_{gh}\right)
\end{split}
\ee
where in going from the first line to the second line we have used Bianchi identity.}.
	
In summary, the correct partition function over S-matrices is given by \eqref{itv} for all $D \geq 8$.  The most 
general S-matrix in these dimensions is given by  \eqref{expparphdte} and the Lagrangian that generates this S-matrix continues to be given by \eqref{lagphten}.

\subsection{$D=7$:} 

In this case the plethystic partition function reported in 
Table \ref{photon-plethystic} can be recast as
\begin{equation}\label{itvsevp} 
 2 x^4  Z_{\bf 3}(x) +x^6 Z_{\bf 1_S}(x)
+ x^3 Z_{\bf 1_S}(x) -x^3.
\end{equation} 
Once again the plethystic counting in $D=7$ makes a small 
error; it omits to count the 3 derivative gauge invariant 
Lagrangian, 
\begin{equation} \label{cssevdef}
O_{\bf S}^{D=7}={\rm CS}_7= *\left( A\wedge F\wedge F \wedge F \right) 
\end{equation} 
the 7 dimensional Chern Simons form,  as this expression is not a polynomial 
in gauge invariant letters\footnote{On the other hand Lagrangians formed out of $A\wedge F\wedge F\wedge F$ by taking derivatives and contracting indices in pairs {\it can} be written entirely out of field strengths - up to total derivatives - and so are correctly 
	counted by the plethystic procedure. For example 
	consider $\epsilon^{bcdefgh}\partial_\mu A_{b}\partial^\mu F_{cd}F_{ef}F_{gh}$
\begin{eqnarray}
\epsilon^{bcdefgh}\partial^\mu A_{b}\partial_\mu F_{cd}F_{ef}F_{gh}&=& -2\epsilon^{bcdefgh}\partial^\mu A_{b}\partial_c F_{d \mu}F_{ef}F_{gh} \nonumber\\
&=&-2\epsilon^{bcdefgh}\partial_c\left(\partial^\mu A_{b} F_{d \mu}F_{ef}F_{gh}\right) +2\epsilon^{bcdefgh}\partial_\mu F_{cb} F_{d \mu}F_{ef}F_{gh}\nonumber\\
&&+2\epsilon^{bcdefgh}\partial^\mu A_{b} F_{d \mu}\partial_c F_{ef}F_{gh}+2\epsilon^{bcdefgh}\partial^\mu A_{b} F_{d \mu}F_{ef}\partial_c F_{gh}\nonumber\\
&&=-2\epsilon^{bcdefgh}\partial_c\left(\partial^\mu A_{b} F_{d \mu}F_{ef}F_{gh}\right) +2\epsilon^{bcdefgh}\partial_\mu F_{cb} F_{d \mu}F_{ef}F_{gh}
\end{eqnarray}	
where in going from the first to second step we have used Bianchi identity and we have removed the last two terms in the second line using Bianchi identity again.}.
This error is corrected for by adding $x^3$ to \eqref{itvsev} 
and so the corrected Plethystic result predicts that 
\begin{equation}\label{itvsev} 
I_{\tv}^{D=7}(x)=
2 x^4  Z_{\bf 3}(x) +x^6 Z_{\bf 1_S}(x)
+ x^3 Z_{\bf 1_S}(x) 
\end{equation}
Note that in $D=7$ (and all odd dimensions), contributions to the S-matrix partition function even in $x$ count parity even S-matrices while contributions that are odd in $x$ count parity 
odd structures\footnote{This follows immediately from the 
	fact that the Levi-Civita symbol has an odd number of 
	indices when $D$ is odd. When $D$ is even, on the other hand, parity odd and parity even S-matrices both yield even (in $x$) contributions to the S-matrix partition function.}
It follows that \eqref{itvsev} can be refined into 
\begin{equation}\label{oepf}
I_{\tv}^{D=7~\rm even}=x^4(2+3x^2+2x^4)\denom ,\qquad I_{\tv}^{\rm D=7~odd}=x^3\denom.
\end{equation}

The parity even part of the prediction \eqref{oepf} is precisely the partition function of a freely generated 
module with generators \eqref{photon-lag}. As described 
earlier in this subsection, this is the expected structure 
of the local module for parity even S-matrices in this dimension. This agreement is non-trivial confirmation of 
the module prediction that the most general local  parity even 
S-matrix in $D=7$ continues to be given by \eqref{expparphdte} and the Lagrangian that generates this S-matrix continues to 
be given by \eqref{lagphten}.

We now turn to the study of the parity odd S-matrix module. 
In this case the parity odd bare module has a single 
generator in the ${\bf 1_S}$ which in fact coincides with 
the single parity odd generator of the local module; 
the generator in question is given by 
\begin{equation}\label{genbar}
 O_{\bf S}^{D=7}=\epsilon_1 \wedge \epsilon_2 \wedge \epsilon_3 \wedge \epsilon_4 \wedge k_1 \wedge k_2 \wedge k_3
 \end{equation} 
and the corresponding Lagrangian is the 7 dimensional 
Chern Simons form that we have already encountered above. 
As this generator has derivative dimension $3$ its contribution
to the S-matrix partition function precisely agrees with
\eqref{oepf}. We thus have a complete `module' explanation
for the corrected plethystic result \eqref{oepf}.

The most general parity odd S-matrix in $D=7$, is given by 
\begin{equation}\label{mjposm}
{\cal S}^{D=7}_{\rm odd} = -i4\cf^{O_{\bf S}^{D=7}}(t,u)(8*(\epsilon^1\wedge  p^2 \wedge\epsilon^2 \wedge p^3 \wedge\epsilon^3\wedge p^4 \wedge\epsilon^4))
\end{equation} 
where, as ${\bf S}$ in superscript  suggests, the function 
$\cf^{O_{\bf S}^{D=7}}(t,u)$ is an arbitrary 
completely symmetric function of $s$, $t$, $u$. 
This function $\cf^{O_{\bf S}^{D=7}}(t,u)$ can be expanded as in \eqref{expsca}. The parity odd Lagrangian 
from which  \eqref{mjposm} follows takes the form 
\begin{equation}\label{mjposml}
L^{D=7}_{\rm odd}= \sum_{m, n} \left(\cf^{O_{\bf S}^{D=7}}\right)_{m,n} 2^{m+n}\left(\prod_{i=1}^m \prod_{j=1}^n 
\left(  
\partial_{\nu_j} \partial_{\mu_i} A\wedge \partial_{\mu_i} F\wedge \partial_{\nu_j}F\wedge F\right)\right),
\end{equation} 

There is exactly one parity odd photonic 
S-matrix that grows no faster than $s^2$ 
in the Regge limit; this is the S-matrix 
with $\cf^{O_{\bf S}^{D=7}}(t,u)= 
{\rm const}$. Explicitly the S-matrix 
is given by \eqref{genbar} - the generator 
of the local module which, in this case, also 
happens to be the generator of the bare 
module. This momentum dependence of this 
S-matrix is $\sqrt{stu}$ and so it scales like 
$s$ in the Regge limit.

\subsection{$D=6$:}

In this case the plethystic partition function reported in 
Table \ref{photon-plethystic} can be recast as
\begin{equation}\label{itvsixp} 
2 x^4  Z_{\bf 3}(x) +x^6 Z_{\bf 1_S}(x)
+ x^6 Z_{\bf 1_A}(x)
\end{equation} 
In this case it turns out that the plethystic counting 
makes no errors, and \eqref{itvsixp} is the correct formula
for the partition function over S-matrices. The analysis 
of parity even module structures presented earlier in this 
subsection predicts that the parity even part of the 
partition function over S-matrices is given by 
\eqref{itv} for all $D\geq 5$, and so, in particular, for 
$D=6$. Comparing with \eqref{itvsixp} it follows that 
\begin{equation}\label{deqsix}
I_{\tv}^{D=6,\rm even}=2 x^4  Z_{\bf 3}(x) +x^6 Z_{\bf 1_S}(x),\qquad I_{\tv}^{D=6, \rm odd}=x^6 Z_{\bf 1_A}(x).
\end{equation}
The even part of the S-matrix is no different from higher 
dimensions; in particular the most general parity even polynomial S-matrix continues to be given by \eqref{expparphdte} and the Lagrangian that generates this S-matrix continues to 
be given by \eqref{lagphten}.

We now turn to the parity odd S-matrix module. 
Once again the bare and local module each have a single 
generator, from which it follows immediately that the local 
module is freely generated. The generator of the local module 
is proportional to
\begin{equation}\label{golm}
O^{D=6}_{\bf A}=stu \cdot o^{D=6}_{\bf A}
\end{equation} 
(see \eqref{pophdsix})
an expression that is dual, in the sense of subsection 
\ref{mgl}, to the Lagrangian\footnote{While this Lagrangian vanishes as an expression 
	it is non-trivial as the Lagrangian corresponding to 
	the local Module generator, in the sense of subsection 
	\ref{mgl}.}  
\be\label{parity-odd-6d}
O^{D=6}_{\bf A}=F^{ab}*(\partial_a F\wedge \partial_b F \wedge F).
\ee
As this generator has derivative dimension $6$ and - like $o^{D=6}_{\bf A}$ 
(see \eqref{pophdsix}) transforms in the ${\bf 1_A}$ of $S_3$, 
its contribution
to the S-matrix partition function precisely agrees with
\eqref{oepf}. We thus have a complete `module' explanation
for the plethystic result \eqref{deqsix}. 
	
The most general parity odd S-matrix in $D=6$ is given by 
\begin{equation}\label{mjposmsix}
\begin{split}
{\cal S}^{D=6}_{\rm odd} &=  \cf^{O_{\bf A}^{D=6}}(t,u) \left( 8\left( p^1_a \epsilon^1_b - p^1_b \epsilon^1_a\right)p^2_ap^3_b *(\epsilon^2\wedge k^2\wedge\epsilon^3\wedge k^3\wedge\epsilon^4\wedge k^4) \right.\\
&\left.+8\left( p^2_a \epsilon^2_b - p^2_b \epsilon^2_a\right)p^1_a p^4_b *(\epsilon^1\wedge k^1\wedge\epsilon^3\wedge k^3\wedge\epsilon^4\wedge k^4)\right. \nonumber\\
&\left.+8\left(p^3_a \epsilon^3_b - p^3_b \epsilon^3_a\right)p^4_a p^1_b *(\epsilon^2\wedge k^2\wedge\epsilon^1\wedge k^1\wedge\epsilon^4\wedge k^4) \right.\nonumber\\
&\left. +8\left(p^4_a \epsilon^4_b - p^4_b \epsilon^4_a\right)p^3_a p^2_b *(\epsilon^2\wedge k^2\wedge\epsilon^3\wedge k^3\wedge\epsilon^1\wedge k^1) \right).
\end{split}
\end{equation} 
where the function $ \cf^{O_{\bf A}^{D=6}}(t,u)$ is a completely antisymmetric function of $s$, $t$ and $u$. 
 The parity odd Lagrangian 
from which  \eqref{mjposmsix} follows takes the form 
\begin{equation}\label{mjposmlsix}
L^{D=6}_{\rm odd} = -\sum_{m, n} \left(\cf^{O_{\bf A}^{D=6}}\right)_{m,n} 2^{m+n}\left(\prod_{i=1}^m \prod_{j=1}^n 
\left(  
\partial_{\nu_j} \partial_{\mu_i} F_{ab}  ( \partial_{\mu_i}\partial_a F\wedge \partial_{\nu_j}\partial_b F \wedge F)\right)\right).
\end{equation} 

Recall that the generator \eqref{golm} 
transforms in the ${\bf 1_A}$ representation. 
Even though the momentum dependence of the  generator is $stu$ (and so scales like $s^2$ in the Regge limit) the first descendant of this generator 
that is completely symmetric occurs at 12 derivative order (and scales like $s^5$ in 
the Regge limit). None of the parity odd S-matrices in $D=6$ grow like $s^2$ or slower 
in the Regge limit.

\subsection{$D=5$:}
Remarkably enough the plethystic partition functions in $D=5$ 
is identical to that for $D \geq 10$. As in $D=6$, 
the $D=5$ plethystic result has no errors that need correction, so we conclude that the correct S-matrix partition function is given by 
\be \label{phodf}
I_{\tv}^{D=5 ~\rm even}=2 x^4  Z_{\bf 3}(x) +x^6 Z_{\bf 1_S}(x),\qquad I_{\tv}^{D=5~\rm odd}=0.
\ee
Every aspect of \eqref{phodf} is easy to understand from 
our module analysis. The fact that there are no parity 
odd S-matrices is a consequence of the fact that the parity 
odd bare module vanishes (see Table \ref{counting-is}). 
And the parity even local module of S-matrices is freely 
generated starting with the generators \eqref{photon-lag}. 
As for $D\geq 8$ most general polynomial $D=4$  S-matrix and corresponding Lagrangian is given by \eqref{expparphdte} and  \eqref{lagphten}.

\subsection{$D=4$:}
For all $D\geq 5$, the number of  generators of the local module  agreed with the number of generators of the bare module. Also, the condition \eqref{freeconds} is not satisfied. As a result the local module is also freely generated. 
In $D=4$, we see a new phenomenon. The number of generators of the local module is more than the number of generators of the bare module. Hence the local generators obey certain relations. This happens both in the parity even as well as parity odd sector. We discuss this case in detail below.

\subsubsection{Parity even}
The rank of the free bare module reduces from $7$ to $5$ in four dimensions as discussed in appendix \ref{appendix-is} while the generators of the local module $E_J$ in equation \eqref{photon-lag} all remain non-zero.
In order to characterize the relations obeyed by $E_J$, we focus on the embedding of the local module into bare module \eqref{explicit-embeddingo}. As discussed in appendix \ref{appendix-is}, in four dimensions, the bare structures $e_{{\bf 3},1}^{D=4,(i)}$ for $i=1,2,3$ become identical. Hence, $e_{{\bf 3},1}^{D=4}$ transforms in ${\bf 1_S}$ - rather than $\bf 3$ as in higher dimensions (i.e. the ${\bf 2_M}$ part of $e_{{\bf 3},1}$ trivializes in $D=4$).  We denote this generator as $e_{{\bf 3}\to {\bf S}}^{D=4}$.
 As $E_{\bf S}^{D=4}$ doesn't have $e_{{\bf 3},1}^{D=4,(i)}$ it is unaffected by this change. The local generators $E_{{\bf 3},1}^{D=4}$ and $E_{{\bf 3},2}^{D=4}$ become,  
\bea \label{simpgen}
E_{{\bf 3},1}^{(1)}&=&-8s^2 e_{{\bf 3}\to {\bf S}}+8s^2e_{{\bf 3},2}^{(1)}-8s^2e_{\bf S},\nonumber\\
E_{{\bf 3},2}^{(1)}&=&-2(u^2+t^2 )e_{{\bf 3}\to {\bf S}}+2(u(s-t)e_{{\bf S},2}^{(3)}+t(s-u)e_{{\bf 3},2}^{(2)})-2(t^2+u^2)e_{\bf S}.
\eea
Here and in the rest of the subsection we drop the superscript $D=4$  to avoid clutter (the $(2)$ and $(3)$ components of \eqref{simpgen} follow by cyclicity as in 
\eqref{expeei}).
We have relations in the modules generated by $E_{{\bf 3},1}^{(i)}$ and $E_{{\bf 3},2}^{(i)}$ whenever there are non-trivial solutions to 
the equations 
\be\label{4d-relation}
\big(\sum_{i=1,2,3}~ \frac12 \cf^{E_{{\bf 3},1}^{(i)}}(t,u) E_{{\bf 3},1}^{(i)}\big)+\big(\sum_{i=1,2,3} ~ \frac12\cf^{E_{{\bf 3},2}^{(i)}}(t,u) E_{{\bf 3},2}^{(i)}\big)=0.
\ee

It is not difficult to see that there are two \emph{independent} families of  solutions, 
\bea \label{defouramb}
\cf_1^{E_{{\bf 3},1}}(t,u)&=&\frac18(s^2+t^2+u^2)f(t,u),\nonumber\\  
\cf_1^{E_{{\bf 3},2}}(t,u)&=& t u f(t,u).
\eea
and 
\bea
\cf_2^{E_{{\bf 3},1}}(t,u)&=&\frac98 s tu \,g(t,u),\nonumber\\
\cf_2^{E_{{\bf 3},2}}(t,u)&=&-\frac12(2(t^3+u^3)-stu) \,g(t,u).
\eea
where $f(t,u)$ and $g(t,u)$ are arbitrary functions that are symmetric in the two arguments.

Completely explicitly, the most general 
S-matrix is specified a completely symmetric polynomial $\cf^{E_{{\bf S}}}(t,u)$ along with 
two variable symmetric polynomials 
$\cf^{E_{{\bf 3},1}}(t,u)$
and $\cf^{E_{{\bf 3},2}}(t,u)$
that are subjected to the equivalence relations 
\bea
\cf^{E_{{\bf 3},1}}(t,u)&\sim & \cf^{E_{{\bf 3},1}}(t,u) +\frac98 s tu \,g(t,u) + \frac18(s^2+t^2+u^2)f(t,u),\nonumber\\
\cf^{E_{{\bf 3},2}}(t,u)&\sim& \cf^{E_{{\bf 3},2}}(t,u) -\frac12(2(t^3+u^3)-stu) \,g(t,u) + t u \,f(t,u)
\eea
where $f(t,u)$ and $g(t,u)$ are arbitrary 
functions symmetric in $t$ and $u$.

The relation module is thus a rank two free module with  one basis element at 8-derivative (the first 
solution in \eqref{4d-relation} - recall that $E_{{\bf 3},1}$ and $E_{{\bf 3},2}$ themselves start at 4 derivative order)  and the other at 10-derivative 
(the second solution in \eqref{4d-relation}). Both generators of the relation module transform in the symmetric representation. Their contribution to the partition function is given by 
\be \label{contplethm} 
-(x^8+x^{10}) Z_{{\bf 1_S}}(x)=-(x^8+x^{10})\denom.
\ee
It follows that module considerations lead us to predict that the 
partition function over parity even S-matrices in $D=4$ is given by 
\be \label{contpleth} 
I_{\tv}^{D=4~\rm even}=2 x^4Z_{{\bf 3}}(x) + x^6 Z_{{\bf 1_S}}(x) -(x^8+x^{10}) Z_{{\bf 1_S}}(x)
\ee
We will compare the prediction \eqref{contpleth}  against the results of plethystic counting 
after incorporating contribution of parity odd S-matrices below. 

Interestingly, both the relations \eqref{defouramb}  can be thought of as a consequence of a certain $6$ derivative quasi-invariant structure reducing from ${\bf 3}$ to ${\bf 1_S}$ \emph{i.e.} vanishing of the mixed representation ${\bf 2_M}$ at $6$ derivative. This quasi-invariant structure is,
\be
\tilde E^{(1)}\equiv\frac12(u \,E_{{\bf 3},2}^{(2)}+t \,E_{{\bf 3},2}^{(3)})+\frac{s}{8}(E_{{\bf 3},1}^{(1)}+E_{{\bf 3},1}^{(2)}+E_{{\bf 3},1}^{(3)}).
\ee
It is easy to check that $\tilde E^{(1)}=\tilde E^{(2)}=\tilde E^{(3)}$. The first and the second relation \eqref{4d-relation} are simply consequences of 
\be
s\tilde E^{(1)}+t\tilde E^{(2)}+u\tilde E^{(3)}=0, \qquad (s^2+2ut)\tilde E^{(1)}+(t^2+2us)\tilde E^{(2)}+(u^2+2st)\tilde E^{(3)}=0
\ee
respectively, which hold true for any symmetric structure.

Note that the most general polynomial S-matrix - and the Lagrangian that 
generates it - continues to be given by the equations \eqref{expparphdte}
and \eqref{lagphten}. The subtlety in this case is that distinct choices of 
the three functions $\cf^{E_{{\bf 3},1}^{D=4}}(t,u)$, $\cf^{E_{{\bf 3},2}^{D=4}}(t,u)$ and $\cf^{E_{{\bf S}}^{D=4}}(t,u)$ do not all generate distinct 
S-matrices; choices for these functions that differ by \eqref{defouramb}
yield the same S-matrix (and same corresponding Lagrangian).

\subsubsection{Parity odd}
It is not difficult to verify that the local module of parity odd S-matrices 
has two sets of generators, one at $4$ derivative order and the other at $6$ derivative order; in the sense of 
subsection \ref{mgl} these generators are 
`dual' to the Lagrangians 
\be \label{Lagrangiansdfpo}
 *(F\wedge F){\rm Tr}(F^2),\qquad \qquad \varepsilon_{\mu\nu\rho\sigma}F^{\mu\nu} \partial^\rho F^{ab}\partial^\sigma F^{bc} F^{ca}.
\ee
The corresponding generators are 
\be\label{podfgen} 
\begin{split}
O_{\bf 3}^{D=4,(1)}&\equiv 2*(F_1\wedge F_2){\rm Tr}(F_3 F_4)|_{\Z_2\times \Z_2}=4*(F_1\wedge F_2){\rm Tr}(F_3 F_4)+4*(F_3\wedge F_4){\rm Tr}(F_1 F_2),\nonumber\\
 O_{\bf S}^{D=4}&\equiv 6\varepsilon_{\mu\nu\rho\sigma}F_1^{\mu\nu} \partial^\rho F_2^{ab}\partial^\sigma F_3^{bc} F_4^{ca}|_{\Z_2\times \Z_2}=6\bigg(\varepsilon_{\mu\nu\rho\sigma}F_1^{\mu\nu} \partial^\rho F_2^{ab}\partial^\sigma F_3^{bc} F_4^{ca}+ \varepsilon_{\mu\nu\rho\sigma}F_2^{\mu\nu} \partial^\rho F_1^{ab}\partial^\sigma F_4^{bc} F_3^{ca}\nonumber\\
 &+ \varepsilon_{\mu\nu\rho\sigma}F_3^{\mu\nu} \partial^\rho F_4^{ab}\partial^\sigma F_1^{bc} F_2^{ca}+\varepsilon_{\mu\nu\rho\sigma}F_4^{\mu\nu} \partial^\rho F_3^{ab}\partial^\sigma F_2^{bc} F_1^{ca}\bigg).
\end{split}
\ee
While we have not carefully checked that there are no additional generators 
of the parity odd local module at higher than six derivatives we strongly believe 
this to be the case (the matching of our final result with plethystic counting 
can be thought of as extremely non-trivial evidence in favor of this guess).

Here $O_{\bf 3}^{D=4}$ transforms in $\bf 3$ under $S_3$ while $O_{\bf S}^{D=4}$ transforms in ${\bf 1_S}$. Naively, the contribution of these terms to the partition function would have been 
\begin{equation} \label{ncpopf} 
x^4 Z_{{\bf 3}} + x^6 Z_{{\bf 1_S}} 
\end{equation} 
but just like in the case of parity even structures, $O_{\bf 3}^{D=4}$'s do not generate the local module freely. This is because the bare module is of rank $2$ as described in appendix \ref{appendix-is}. Its generators are
\be 
o_{{\bf S},1}^{D=4}\equiv i N(\tilde \varepsilon)_\mu {\polo_{4\mu}}\alpha_1\alpha_2\alpha_3|_{\Z_2\times \Z_2},\qquad 
o_{{\bf S},2}^{D=4}\equiv i N(\tilde \varepsilon)_\mu {\polo_{4\mu}}\polo_{1\nu}\polo_{2\nu}\alpha_3|_{\Z_2\times \Z_2}.
\ee
Note that $o_{{\bf S},1}^{D=4}$ and $o_{{\bf S},2}^{D=4}$ both transform in the ${\bf 1_S}$ 
representation under $S_3$. The embedding of the local module into the bare module is given below. In the rest of the subsection, again we will suppress $D=4$ superscript on the local and bare module generators  to avoid clutter.
\begin{eqnarray} \label{oot}
O_{\bf 3}^{(1)}&=&8s^2 \left(o_{{\bf S},1}-o_{{\bf S},2} \right),~ O_{\bf 3}^{(2)}=8t^2 \left(o_{{\bf S},1} -o_{{\bf S},2} \right),~O_{\bf 3}^{(3)}=8u^2 \left(o_{{\bf S},1} -o_{{\bf S},2} \right)\nonumber\\
O_{\bf S}&=&  6stu \left(o_{{\bf S},1}- 3 o_{{\bf S},2} \right) .
\end{eqnarray}
Note that $O_{\bf 3}$ in \eqref{oot}  transform in the ${\bf 3}$ even 
thought $o_{{\bf S},1}$ and $o_{{\bf S},2}$ transform in the ${\bf 1_S}$ simply 
because the triplet of functions $(s^2, t^2, u^2)$ 
transforms in the ${\bf 3}$. 

As we have mentioned above, $O_{\bf 3}$ and $O_{\bf S}$ do not generate the  local module freely. As every state in 
the $O_{\bf S}$ module is proportional to $o_{{\bf S},1}- 3 o_{{\bf S},2}$ while 
every state in the $O_{\bf 3}$ module is proportional to 
$o_{{\bf S},1}-  o_{{\bf S},2}$, there can be no relations that involve both 
states  generated by $O_{\bf 3}$  and states generated by $O_{\bf S}$; 
all relations that exist have to work module by module. 
As $stu \cf^{O_{\bf S}}(t,u)$ vanishes only 
when $\cf^{O_{\bf S}}(t,u)=0$ it is clear that the module generated by $O_{\bf S}$ has no relations. 

On the other hand the three distinct generators $O_{\bf 3}^{(i)}$ are all `descendants' of the same primary 
state $o_{{\bf S},1}-  o_{{\bf S},2}$ so clearly there exist relations between 
these states.  More generally the  non-trivial solution to the equations 
\be\label{poin}
\sum_{i=1,2,3} ~\frac12\cf^{O_{\bf 3}^{(i)}}(t,u)O_{\bf 3}^{(i)}=0.
\ee
are given by the two families 
\be \label{solpoin} 
\cf_1^{O_{\bf 3}}(t,u) =2 t u f(t,u), \qquad \cf_2^{O_{\bf 3}}(t,u) = (2(t^3+u^3)+(t+u)tu)g(t,u).
\ee
where $f(t,u)$ and $g(t,u)$ are arbitrary functions that are symmetric in the two arguments. 
It follows that the relation module is a rank two free module with  one basis element at 8-derivative (first solution in \eqref{solpoin} ) and the other at 10-derivative (second solution in \eqref{solpoin}). The generators of the relation module are explicitly given by 
\bea
&&2tu O_{\bf 3}^{(1)}+2us O_{\bf 3}^{(2)}+2st O_{\bf 3}^{(3)} \nonumber\\
&&(2(t^3+u^3)-stu) O_{\bf 3}^{(1)}+(2(u^3+s^3)-stu) O_{\bf 3}^{(2)}+(2(s^3+t^3)-stu) O_{\bf 3}^{(3)}.
\eea
Both these generators transform in the ${\bf 1_S}$ representation. The contribution of the relations  to the S-matrix partition function is given by 
\be \label{contpf} 
-(x^8+x^{10}) Z_{{\bf 1_S}}(x)=-(x^8+x^{10})\denom.
\ee 
Combining \eqref{ncpopf}, \eqref{contpf} and \eqref{contpleth}, the 
module prediction for the $D=4$ partition function over S-matrices is 
given by 
\be\label{fpfdf}
\begin{split}
I_{\tv}^{\rm even}&=x^4(2+3x^2+x^4-x^6)\denom=2x^4Z_{{\bf 3}}+x^6Z_{{\bf 1_S}}-(x^8+x^{10})Z_{{\bf 1_S}}\nonumber\\
I_{\tv}^{\rm odd}&=x^4(1+2x^2-x^6)\denom=x^4Z_{{\bf 3}}+x^6Z_{{\bf 1_S}}-(x^8+x^{10})Z_{{\bf 1_S}}.
\end{split}
\ee
Remarkably enough the prediction \eqref{fpfdf} matches exactly with the 
results of plethystic counting in $D=4$ (see Table \ref{photon-plethystic}). 
We view this match as an extremely non-trivial check of the completeness 
of our understanding of photon S-matrices in $D=4$.

The most general parity four photon parity odd S-matrix is given by 

\begin{eqnarray}\label{mnjposm} 
S^{D=4}_{\rm odd}&=& 2 \cf^{O_{\bf 3}}(t,u)   \left(4*(p_1\wedge \epsilon_1\wedge p_2 \wedge \epsilon_2 \left( p^3_\mu \epsilon^3_\nu - p^3_\nu \epsilon^3_\mu)  
\right)\left( p^4_\mu \epsilon^4_\nu - p^4_\nu \epsilon^4_\mu  
\right)+\left(1\rightarrow 3, 2\rightarrow 4\right)\right)\nonumber\\
&&+2\cf^{O_{\bf 3}}(u,s)  \left(4*(p_1\wedge \epsilon_1 \wedge  p_3\wedge \epsilon_3 \left( p^2_\mu \epsilon^2_\nu - p^2_\nu \epsilon^2_\mu)  
\right)\left( p^4_\mu \epsilon^4_\nu - p^4_\nu \epsilon^4_\mu  
\right)+\left(1\rightarrow 2, 3\rightarrow 4\right)\right)
\nonumber\\
&&+2\cf^{O_{\bf 3}}(s,t)  \left(4*(p_1\wedge \epsilon_1\wedge p_4 \wedge \epsilon_4\left( p^3_\mu \epsilon^3_\nu - p^3_\nu \epsilon^3_\mu)  
\right)\left( p^2_\mu \epsilon^2_\nu - p^2_\nu \epsilon^2_\mu  
\right)+\left(1\rightarrow 3, 4\rightarrow 2\right)\right)\nonumber\\
&& -\cf^{O_{\bf S}}(t,u) \times
 \nonumber\\
&&\left( 2*(p^2\wedge p^3\wedge p^1\wedge\epsilon^1) 
\left(p^2_c \epsilon^2_d - p^2_d \epsilon^2_c\right)\left(p^3_d \epsilon^3_e - p^3_e \epsilon^3_d\right)\left(p^4_e \epsilon^4_c - p^4_c \epsilon^4_e\right)\right.\nonumber\\
&&\left.+2*(p^1\wedge p^3\wedge p^2\wedge\epsilon^2) 
\left(p^1_c \epsilon^1_d - p^1_d \epsilon^1_c\right)\left(p^3_d \epsilon^3_e - p^3_e \epsilon^3_d\right)\left(p^4_e \epsilon^4_c - p^4_c \epsilon^4_e\right)\right.\nonumber\\
&&\left.+ 2*(p^2\wedge p^1\wedge p^3\wedge\epsilon^3 )
\left(p^2_c \epsilon^2_d - p^2_d \epsilon^2_c\right)\left(p^1_d \epsilon^1_e - p^1_e \epsilon^1_d\right)\left(p^4_e \epsilon^4_c - p^4_c \epsilon^4_e\right)\right.\nonumber\\
&&\left.+2*(p^2\wedge p^3\wedge p^4\wedge\epsilon^4) 
\left(p^2_c \epsilon^2_d - p^2_d \epsilon^2_c\right)\left(p^3_d \epsilon^3_e - p^3_e \epsilon^3_d\right)\left(p^1_e \epsilon^1_c - p^1_c \epsilon^1_e\right)
\right)
\end{eqnarray} 
where $\cf^{O_{\bf S}}(t,u)$ is a general polynomial that is completely symmetric under $S_3$ and $\cf^{O_{\bf 3}}(t,u)$ is a polynomial that is symmetric in the two arguments and is only defined up to the addition of specific functions $\cf_1^{O_{\bf 3}}(t,u)$ and $\cf_2^{O_{\bf 3}}(t,u)$ given in equation \eqref{solpoin}.

Recall that all the S-matrices discussed 
in this section are descendants of the two 
completely symmetric bare module generators
$o_{{\bf S},1}-o_{{\bf S},2}$ and $o_{{\bf S},1}-3 o_{{\bf S},2}$. From the analysis
of subsection \eqref{rgss} we know that the six descendants of these bare generators 
that grow no faster than $s^2$ in the Regge 
limit are 
\begin{eqnarray}
&&(o_{{\bf S},1}-o_{{\bf S},2}), ~~~(s^2+t^2 +u^2) (o_{{\bf S},1}-o_{{\bf S},2}), ~~~
stu(o_{{\bf S},1}-o_{{\bf S},2}), \nonumber\\
&&(o_{{\bf S},1}-3o_{{\bf S},2}), ~~~(s^2+t^2 +u^2) (o_{{\bf S},1}-3o_{{\bf S},2}), ~~~
stu(o_{{\bf S},1}-3o_{{\bf S},2})
\end{eqnarray}
Of these six structures only $(s^2+t^2 +u^2)(o_{{\bf S},1}-o_{{\bf S},2})$ and $stu(o_{{\bf S},1}-3o_{{\bf S},2})$  are elements 
of the local module (see \eqref{oot}). 
It follows that the only local parity odd 
photonic S-matrices in $D=4$ that grow 
no slower than $s^2$ in the Regge limit are
those of the form \eqref{mnjposm} with 
$\cf^{O_{\bf 3}}(t,u)$ and 
$\cf^{O_{\bf S}}(t,u)$ both 
constant.

The functions that appear in \eqref{mnjposm} can be Taylor expanded as in 
\eqref{expsca}. The Lagrangians that generate the S-matrices \eqref{mnjposm} are given by 
\begin{eqnarray}\label{mnjposml} 
L^{D=4}_{\rm odd}&=& \sum_{m, n} \left(\cf^{O_{\bf 3}}\right)_{m.n} 2^{m+n}\left(\prod_{i=1}^m \prod_{j=1}^n 
{\rm Tr}  
\partial_{\nu_j} \partial_{\mu_i} (F \wedge F) \text{Tr}( \partial_{\mu_i} F\partial_{\nu_j}F )\right) \nonumber\\
&&+\sum_{m, n} \left(\cf^{O_{\bf S}}\right)_{m.n} 2^{m+n}\left(\prod_{i=1}^m \prod_{j=1}^n 
\partial_{\nu_j} \partial_{\mu_i} (*F_{ab}) \partial_a \partial_{\mu_i}F_{cd} \partial_b \partial_{\nu_j} F_{de}F_{ec} \right), \nonumber\\
\end{eqnarray} 

Distinct choices of the two functions $ $ and $ $ do not all yield inequivalent 
S-matrices; choices of functions that differ by shifts of the form \eqref{solpoin} yield 
the same S-matrix and same effective Lagrangian. 

\subsection{$D=3$}

\subsubsection{Parity even}
As we have noted in Appendix \ref{appendix-is} the parity even part of the  bare module in $D=3$ is freely generated by the single generator $e_{\bf S}$. The fact that there are no transverse direction to the momenta plane - and so no transverse polarizations -  ensure that  $e_{{\bf 3},1}$ and $e_{{\bf 3},2}$ simply vanish in this dimension.  

The generators  $E_{{\bf 3},1}^{D=3},E_{{\bf 3},2}^{D=3}$ and $E_{\bf S}^{D=3}$ are easily evaluated 
in terms of $e_{\bf S}$, the generator of the bare module. Suppressing the $D=3$ superscript, we find
\bea \label{simpgen3}
E_{{\bf 3},1}^{(1)}&=&-8s^2e_{\bf S},\qquad E_{{\bf 3},1}^{(2)}=-8t^2e_{\bf S},\qquad E_{{\bf 3},1}^{(3)}=-8u^2e_{\bf S},\nonumber\\
E_{{\bf 3},2}^{(1)}&=&-2(t^2+u^2)e_{\bf S},\qquad E_{{\bf 3},2}^{(2)}=-2(u^2+s^2)e_{\bf S},\qquad E_{{\bf 3},2}=-2(s^2+t^2)e_{\bf S},\nonumber\\
E_{{\bf S}}&=& -6\,stu\,e_{\bf S}.
\eea
Each of the generators transform in the ${\bf 3}$ representation. It is easy to see that the ${\bf 1_S}$ 
part and the ${\bf 2_M}$ parts of these two generators 
are both proportional to each other so that the generators $E_{{\bf 3},1}^{(i)}$ and $E_{{\bf 3},2}^{(i)}$ are simply linear combinations of each other. The precise relationship between them is 
\begin{equation} \label{releoei} 
E_{{\bf 3},2}^{(i)} =-\frac14 E_{{\bf 3},1}^{(i)}
+ \frac{\sum_{j=1}^3 E_{{\bf 3},1}^{(j)}}{4} 
\end{equation}
For this reason we can simply ignore the generators $E_{{\bf 3},2}$ and work only with $E_{{\bf 3},1}$. 

Now it is also easy to see that the singlet $E_{{\bf S}}$
is a `level one' descendant of $E_{{\bf 3},1}$. The precise
relationship is 
\begin{eqnarray}
 E_{{\bf S}}&=& \frac14 (sE_{{\bf 3},1}^{(1)}+t E_{{\bf 3},1}^{(2)}+u E_{{\bf 3},1}^{(3)})
\end{eqnarray}
Consequently, $E_{{\bf 3},1}^{(i)}$ are the only independent local module generators and they transform in the ${\bf 3}$.
The Lagrangian `dual' to these generators is simply  
$$({\rm Tr} F^2)^2.$$  

It is completely clear that the three $E_{{\bf 3},1}$ are 
simply level two descendants of the single generator $e_{\bf S}$ of the bare module. It follows that the local module 
is not freely generated but is instead subject to relations. In particular the relations 
\begin{eqnarray}\label{3d-relation}
\sum_{i=1,2,3}~ \frac12 \cf^{E_{{\bf 3},1}^{(i)}}(t,u) E_{{\bf 3},1}^{(i)}=0
\end{eqnarray}
hold whenever 
\be \label{dethreeamb}
 \cf_1^{E_{{\bf 3},1}}(t,u) =2tu \,f(t,u),\qquad    \cf_2^{E_{{\bf 3},1}}(t,u) =(2(t^3+u^3)+(t+u)tu) \,g(t,u).
\ee
where $f(t,u)$ and $g(t,u)$ are arbitrary functions that are symmetric in the two arguments. 
The functions $g$ and $f$ parameterize null modules. 
The generators of these modules are both in the ${\bf 1_S}$
representation and occur at 10 and 8 derivative order 
respectively. It follows that their contribution to the partition function over S-matrices is given by 
\be 
-(x^8+x^{10}) Z_{{\bf 1_S}}(x)=-(x^8+x^{10})\denom.
\ee

The full parity even part of the partition function, after taking into account the module relations, is thus given by 
\be  
I_{\tv}^{D=3~\rm even}= x^4Z_{{\bf 3}}(x)  -(x^8+x^{10}) Z_{{\bf 1_S}}(x)
\ee
in agreement with the restriction of the $D=3$ part of the results of Table \ref{photon-plethystic} to even powers of $x$. 

It is easy to check that 
\be \label{contpleth3} 
I_{\tv}^{D=3~\rm even}= Z_{{\bf 1_S}}(x)-1
\ee
\eqref{contpleth3} expresses the fact that the the 
restriction to singlets of the module generated by $E_1^{(i)}$ is the same as the restriction to singlets 
of the bare module minus the contribution of the dimension 
zero generator $e_3$ itself.  

The most general parity even S-matrix and corresponding  Lagrangian are given by \eqref{expparphdte} and \eqref{lagphten} by setting  $\cf^{E_{{\bf 3},2}^{D=3}}(t,u)=0$ and  $\cf^{E_{{\bf S}}^{D=3}}(t,u)=0$ and $\cf^{E_{{\bf 3},1}^{D=3}}(t,u)$ is a polynomial symmetric in $(t,u)$ subject to the equivalence relation \eqref{dethreeamb}.

\subsubsection{Parity odd}
The local module of parity odd S-matrices is completely
generated by the module elements `dual' to the $5$ derivative Lagrangian
\be \label{Lagrangiansdtpo}
 \epsilon_{\a\b\g}F^{\a\b}\partial^\g F_{ab}F_{bc}F_{ca}
\ee
The quasi-invariant local generators coming from this Lagrangian transform in representation ${\bf 2_M}$. They are given by 
\begin{eqnarray} \label{symgen}
&&O_{\bf M}^{D=3,(1)}=\frac13 (2\epsilon^{\a\b\g}F^1_{\a\b}\partial^\g F^2_{ab}F^3_{bc}F^4_{ca}-\epsilon^{\a\b\g}F^1_{\a\b}\partial^\g F^3_{ab}F^4_{bc}F^2_{ca} -\epsilon^{\a\b\g}F^1_{\a\b}\partial^\g F^4_{ab}F^2_{bc}F^3_{ca}) \nonumber\\
&&O_{\bf M}^{D=3,(2)}= \frac13(2\epsilon^{\a\b\g}F^1_{\a\b}\partial^\g F^3_{ab}F^4_{bc}F^2_{ca} -\epsilon^{\a\b\g}F^1_{\a\b}\partial^\g F^4_{ab}F^2_{bc}F^3_{ca}-\epsilon^{\a\b\g}F^1_{\a\b}\partial^\g F^2_{ab}F^3_{bc}F^4_{ca})  \nonumber\\
&&O_{\bf M}^{D=3,(3)}= \frac13(2\epsilon^{\a\b\g}F^1_{\a\b}\partial^\g F^4_{ab}F^2_{bc}F^3_{ca} -\epsilon^{\a\b\g}F^1_{\a\b}\partial^\g F^2_{ab}F^3_{bc}F^4_{ca}-\epsilon^{\a\b\g}F^1_{\a\b}\partial^\g F^3_{ab}F^4_{bc}F^2_{ca})
\end{eqnarray}  
The generators $O_{\bf M}^{D=3,(i)}$ are antisymmetric under the swap $1\leftrightarrow2, 1\leftrightarrow3 $ and $1\leftrightarrow4$ respectively. The fact that they transform in ${\bf 2_M}$ means that $\sum_i O_{\bf M}^{D=3,(i)}=0$ (this last statement is a direct consequence of the fact that the generators 
\eqref{symgen} are a level one descendant of the bare generator 
$\epsilon_{abc} p_1^a p_2^b p_3^c$ and so transform in the ${\bf 2_M}$ 
representation, simply because the triplet of functions $s, t, u$ transform in the ${\bf 2_M}$ representation). 

As the parity bare module has a single generator (see Appendix \ref{appendix-is}) it follows immediately that the parity odd 
 local module is not freely generated. The explicit expression for the 
generator of the bare module is  
$$
o_{\bf A}^{D=3}=4\varepsilon_{abc} p_1^{a}p_2^{b}p_3^{c} \,\,\alpha_1\alpha_2\alpha_3\alpha_4.
$$
 
 The generators $O_{\bf M}^{D=3,(i)}$ above are given simply in terms of 
 $o_{\bf A}^{D=3}$  by 
 \begin{equation}\label{henum}
 (O_{\bf M}^{(1)}, O_{\bf M}^{(2)}, O_{\bf M}^{(3)} )= (s \,o_{\bf A}, t \,o_{\bf A}, u \,o_{\bf A}) .
 \end{equation} 
 We have dropped the superscript $D=3$ to avoid clutter. 
It follows immediately from \eqref{henum} that 
\begin{equation} \label{nst} 
(u-t)O_{\bf M}^{(1)} + (s-u) O_{\bf M}^{(2)} +  (t-s) O_{\bf M}^{(3)}=0
\end{equation} 
so that the LHS of \eqref{nst} is a null state. Notice that our 7 derivative null state  transforms in the ${\bf 1_S}$. It follows that the partition function over parity odd S-matrices is given by 
\begin{equation}\label{contofpod}
x^5 Z_{{\bf 2_M}}(x) -x^7 Z_{{\bf S }}(x)=\left(x^7+x^9-x^7\right)  \denom= x^9 \denom
\end{equation} 
to the partition function over S-matrices, in agreement with 
the restriction of the results of the $D=3$ part of Table 
\ref{photon-plethystic} to odd powers of $x$.

The most general S-matrix coming from \eqref{Lagrangiansdtpo} is given by 
\begin{eqnarray}\label{smatsdtpo}
S_{\rm odd}^{D=3} &=& i\left(\cf^{O_{\bf M}^{D=3}} (t,u) \left(2*(p_1\wedge \epsilon_1 \wedge p_2)\left(p^2_c \epsilon^2_d - p^2_d \epsilon^2_c\right)\left(p^3_d \epsilon^3_e - p^3_e \epsilon^3_d\right)\left(p^4_e \epsilon^4_c - p^4_c \epsilon^4_e\right) \right.\right.\nonumber\\
&&\left.\left. + (1 \leftrightarrow 2, 3 \leftrightarrow 4) + (1 \leftrightarrow 3, 2 \leftrightarrow 4) +(1 \leftrightarrow 4, 2 \leftrightarrow 3) \right)\right.\nonumber\\
&& +\left.\cf^{O_{\bf M}^{D=3}}(u,s) \left(2*(p_1\wedge \epsilon_1 \wedge p_3)\left(p^3_c \epsilon^3_d - p^3_d \epsilon^3_c\right)\left(p^2_d \epsilon^2_e - p^2_e \epsilon^2_d\right)\left(p^4_e \epsilon^4_c - p^4_c \epsilon^4_e\right) \right.\right.\nonumber\\&&\left.\left.
+ (1 \leftrightarrow 2, 3 \leftrightarrow 4) + (1 \leftrightarrow 3, 2 \leftrightarrow 4) +(1 \leftrightarrow 4, 2 \leftrightarrow 3) \right)\right.\nonumber\\
&& +\left.\cf^{O_{\bf M}^{D=3}}(s,t) \left(2*(p_1\wedge \epsilon_1 \wedge p_4)\left(p^4_c \epsilon^4_d - p^4_d \epsilon^4_c\right)\left(p^3_d \epsilon^3_e - p^3_e \epsilon^3_d\right)\left(p^2_e \epsilon^2_c - p^2_c \epsilon^2_e\right) \right.\right.\nonumber\\&&\left.\left.
+ (1 \leftrightarrow 2, 3 \leftrightarrow 4) + (1 \leftrightarrow 3, 2 \leftrightarrow 4) +(1 \leftrightarrow 4, 2 \leftrightarrow 3) \right)\right)
\end{eqnarray}
As their notation indicates,
$\cf^{O_{\bf M}^{D=3}}(t,u)$ is the most general antisymmetric  polynomial of its arguments satisfying $\sum_i\cf^{O_{\bf M}^{{D=3},(i)}}(t,u)=0$. Furthermore, it is subjected to 
\begin{equation}\label{equivrel}
\cf^{O_{\bf M}^{D=3}}(t,u) \sim \cf^{O_{\bf M}^{D=3}}(t,u) +  (-t+u)f(t,u),
\end{equation} 
where $f(t,u)$ is any function antisymmetric in its arguments. The equivalence relation is a consequence of 
\eqref{nst}. 

All parity odd S-matrices are descendants 
of the generator $o_{\bf A}^{D=3}$ that transforms in 
the ${\bf 1_A}$. It follows immediately 
from subsection \ref{rgss} that there are 
no S-matrices in the module of $o_{\bf A}^{D=3}$ that 
yield S-matrices whose Regge growth is slower than $s^2$ (the first S-matrix in this module occurs at 9 derivative order and grows like 
$s^4$ in the Regge limit). The most general descendant Lagrangian that gives rise to  S-matrix \eqref{smatsdtpo} is given by
\begin{eqnarray}
L^{D=3}_{\rm odd}&=& \sum_{m, n} \left(\cf^{O_{\bf M}^{D=3}}\right)_{m,n} 2^{m+n}\left(\prod_{i=1}^m \prod_{j=1}^n 
 \epsilon_{\a\b\g}\partial_{\nu_j} \partial_{\mu_i} F^{\a\b} \text{Tr}(\partial^{\g} F  \partial_{\mu_i} F\partial_{\nu_j}F )\right).\nonumber\\
\end{eqnarray}

\subsection{Tree level 4-photon S-matrix in string theory}
In this section, we apply our technology to tree level 4-photon S-matrix, first in Type I string theory then in Bosonic string theory. We will simply express S-matrix in terms of our local structures $E_{{\bf 3},1}, E_{{\bf 3},2}$ and $E_{{\bf S}}$. 

\subsection*{Type I Superstring}
The 4-photon scattering amplitude (\cite{Schwarz:1982jn}) is given by,
\be
\CS_{{\rm Type \,I}}=f_{{\rm Type \,I}}(t,u)\sum_{i=1,2,3} (E_{{\bf 3},2}^{(i)}-\frac14 E_{{\bf 3},1}^{(i)}).
\ee
where $f_{{\rm Type \,I}}(t,u)$ is specific $S_3$ symmetric function. In terms of Lagrangians, this S-matrix is obtained by considering specific derivative contractions on 
\be
L_{{\rm Type \,I}} \propto \text{Tr} (F^4) -\frac{1}{4} (\text{Tr} (F^2))^2. 
\ee     
This specific contraction of field-strength is sometimes called $t_8 F^4$, see for example \cite{Deser:2000xz}, where $t_8$ is a tensor such that
\be\label{t8tensor}
\text{Tr} (F^4) -\frac{1}{4} (\text{Tr} (F^2))^2= t^{\mu_1\ldots \mu_8} F_{\mu_1\mu_2}\ldots F_{\mu_7,\mu_8}.
\ee
    
\subsection*{Bosonic string}
The 4-photon scattering amplitude (\cite{Schwarz:1982jn}) in the case of bosonic strings is given by \eqref{expparphdte}, with
\bea
\cf^{E_{{\bf 3},1}}(t,u)&=&f_{{\rm Bosonic}}(t,u)\Big(\frac{1}{32}\left(\frac{-(t+u)}{2}+1\right)\left(\frac{t}{2}+1\right)\left(\frac{u}{2}+1\right) + \frac{1}{256}(4tu+t^2u^2+2(t^2+u^2))\Big)\nonumber\\
\cf^{E_{{\bf 3},2}}(t,u)&=&f_{{\rm Bosonic}}(t,u)\Big(\frac{-1}{8}\left(\frac{-(t+u)}{2}+1\right)\left(\frac{t}{2}+1\right)\left(\frac{u}{2}+1\right)+\frac{1}{64} (t+2) (u+2)\nonumber\\
&& \left(t^2+2 t (u-1)+(u-2) u\right)\Big)\nonumber\\
\cf^{E_{{\bf S}}}(t,u)&=&f_{{\rm Bosonic}}(t,u)\Big(\frac{1}{96} \left((t+u)^2 (-(2 t+3))+2 (t+u) t^2-3 t^2+16\right)\Big)
\eea

where $f_{{\rm Bosonic}}(t,u)$ is specific $S_3$ symmetric function.

\section{Polynomial graviton S-matrices and corresponding Lagrangians} \label{gravsm} 

We now turn to a study of Gravitational S-matrices and 
corresponding Lagrangians.

\subsection{$D \geq 8$} 

As in the case of photon scattering, there are no parity 
odd gravitational S-matrices for $D \geq 8$. In the rest of
this subsubsection we will provide a detailed description of the (automatically parity even) local S-matrix module in $D\geq 8$.

\subsubsection{Modules generated by Lagrangians with 8 or fewer derivatives} \label{mgef}

No gravitational Lagrangian that is linear or quadratic in $R_{\mu\nu \alpha \beta}$ produces a polynomial 4 graviton S-matrix (see  subsubsection \ref{gravlagfp}).  $G_{{\bf S},1}\equiv \chi_6$ is the unique 3 Riemann Lagrangian that produces a polynomial S-matrix (see \eqref{r31} and \eqref{4gsm}).  All other parity even Lagrangians that generate polynomial S-matrices can be written as the sum of products of derivatives of four Riemann tensors.

The simplest four Riemann Lagrangians are those with eight derivatives. These are constructed from contractions of four Riemann tensors (no derivatives). All inequivalent contractions of four  Riemann tensors have been enumerated in \cite{Fulling:1992vm}. Excluding those structures that involve $R$ and $R_{\mu\nu}$ and so can be removed by field redefinitions (see  subsubsection \ref{gravlagfp}), the authors of \cite{Fulling:1992vm} find 7 inequivalent contractions in $D \geq 8$.

Five of the seven transform as ${\bf 3}$ and are labeled as $G_{{\bf 3},i}$, $i=1,\ldots, 5$. One generator transforms as ${\bf 6}_{\rm left}$, it is labeled as $G_{\bf 6}$. It is convenient to decompose ${\bf 6}_{\rm left}$ into ${\bf 3}\oplus {\bf 3_A}$. We label these pieces as $G_{{\bf 3}, 6}$ and $G_{{\bf 3_A}}$.
 All these are listed in \eqref{lablaggravgen} and the associated Lagrangians are in \eqref{elo}. The remaining 8-derivative generator $G_{{\bf 3}, 9}$ transforms in the ${\bf 3}$. The Lagrangian associated to it is,
\begin{equation}\label{sevecha}
{G}_{{\bf 3},9}=R_{pqrs}R_{ptru}R_{tvqw}R_{uvsw}.
\end{equation}

Finally there is one additional subtlety that needs to be 
taken into account. Each of the seven generators  $G_{{\bf 3},1} \ldots G_{{\bf 3},6}$ and $G_{{\bf 3},9}$  have a single generator in the ${\bf 1_S}$ \footnote{In each case 
this generator is simply the S-matrix that follows constructed from tree diagrams using the Lagrangians `dual' the module elements above - i.e. the Lagrangians listed in \cite{Fulling:1992vm}.}. One linear combination of these 
seven ${\bf 1_S}$ structures is simply the third Lovelock term \cite{Deser:2000xz}
\be\label{chiei}
\begin{split}
\chi_8&= \epsilon^{abcdefgh}\epsilon^{\a\b\g\d\mu\nu\rho\sigma}R_{ab\a\b}R_{cd\g\d}R_{ef\mu\nu}R_{gh\rho\sigma}\\
&\propto \left( G_{{\bf 3},1} +2 G_{{\bf 3},2} +16 G_{{\bf 3},3} +32 G_{{\bf 3},4} +8 G_{{\bf 3},5} -16 G_{{\bf 3},6} -64 G_{{\bf 3},9}
\right)|_{\bf S}.
\end{split}
\ee  
When expanded to fourth order in fluctuations, $\chi_8$ and all its `descendants' simply vanish on-shell. It follows that $\chi_8$ corresponds to no module element and plays no role in the discussion that follows. When studying S-matrices, 
 therefore, one of the seven ${\bf 1_S}$ structures above - lets 
 say the ${\bf 1_S}$ in $G_{{\bf 3},9}$ - can be re-expressed as a linear
 combination of the other six, and so is not an independent 
 module generator. As we remove the completely symmetric part from $G_{{\bf 3},9}$, let us relabel it as $G_{\bf 2_M}$ to reflect its correct transformation properties.

In the rest of this  subsubsection we focus on the submodule - lets call it 
$M_8$ -  of the local 
gravitational module that is generated by Lagrangians with 
at most 8 derivatives, i.e. the (independent terms in) descendants of $G_{{\bf S},1}$ plus $G_{{\bf 3},1} \ldots G_{{\bf 3},6}$, $G_{{\bf 3_A}}$ and $G_{\bf 2_M}$\footnote{In the next subsubsection we will continue to describe the rest of the local module (the part of the module generated by terms with 10 or more derivatives).}. 
  It turns out that the submodule of interest to this subsubsection is freely generated (the same holds true for the full local module). This statement - which is simply an unproved 
assertion at this stage - will effectively be demonstrated 
later in this subsection by comparison of the `module' and plethystic partition functions.

Proceeding with the assumption that it is freely generated, we  now provide two different but equivalent descriptions of the submodule $M_8$.
The simpler of the two descriptions of our sub module goes as follows. $G_{{\bf S},1}$ is clearly a generator of our submodule. The module generated by this
6 derivative element has exactly two 8 derivative `descendants' which together transform in a single copy of in the ${\bf 2_M}$. This 8 derivative ${\bf 2_M}$ 
is a linear combination of the 8 eight derivative 
${\bf 2_M}$'s present in the generators dual to the eight derivative generators $G_{{\bf 3},1} \ldots G_{{\bf 3},6}$, $G_{\bf 3_A}$ and $G_{\bf 2_M}$. The relation can be expressed as follows. Let us define $(r^{(1)} , r^{(2)}, r^{(3)})=(s,t,u)$. Then, for example\footnote{ Each of $G_{{\bf 3},1} \ldots G_{{\bf 3},5}$ transforms as ${\bf 1_S}$ + ${\bf 2_M}$, $G_{\bf 6}$ transforms as ${\bf 1_S}$ + ${\bf 1_A}$ 
+ $2 \cdot {\bf 2_M}$ and $G_{{\bf 2_M}}$ transforms as ${\bf 2_M}$. It 
follows that there are a total of 8 ${\bf 2_M}$'s.}, 
\be\label{nyny}
r^{(i)}\, G_{{\bf S},1}=4\Big( -G_{{\bf 3},1}^{(i)}-2G_{{\bf 3},2}^{(i)}-16 G_{{\bf 3},3}^{(i)}+16 G_{{\bf 3},4}^{(i)}-2G_{{\bf 3},5}^{(i)}+10 G_{{\bf 3},6}^{(i)}+16 G_{{\bf 2_M}}^{(i)}+(4G_{\bf 3_A}^{(i+1)}-4G_{\bf 3_A}^{(i+2)})\Big).
\ee

where  $(i+1)$ and $(i+2)$ are defined cyclically; 
- for instance when $i=2$, $(i+1)=(3)$ and $(i+2)=(1)$.

This means, the LHS of \eqref{nyny} (i.e. the  8 derivative descendant of $G_{{\bf S},1}$) together with the generators
 $G_{{\bf 3},1} \ldots G_{{\bf 3},6}$ and $G_{\bf 3_A}$ span the space of 8 derivative module elements. Note that $G_{\bf 2_M}$ does not appear in this list 
 of generators; we have \eqref{chiei} to eliminate the 
${\bf 1_S}$ part of $G_{{\bf 3},9}$ and have used \eqref{nyny} 
 to eliminate the ${\bf 2_M}$ part of this generator.

 As $G_{{\bf S},1}$, $G_{\bf 3_A}$ and $G_{{\bf 3},1} \ldots G_{{\bf 3},6}$ are the  generators of our (assumed freely generated) submodule. It follows that the  partition function over singlets of this submodule is given by 
\begin{equation}\label{pfzm}
Z(x)=5 x^8 Z_{{\bf 3}}(x) + x^8
Z_{{\bf 6}}(x) +  x^6 Z_{{\bf 1_S}}(x). 
\end{equation} 

While the description of the submodule $M_8$ presented above is 
completely adequate for the purpose of listing and counting S-matrices, it is not sufficient to allow us to find a Lagrangian dual to 
each of these S-matrices. The reason for this is as follows. 
While $G_{{\bf S},1}=\chi_6$ itself is a perfectly good  Lagrangian, the 
S-matrices in the module generated by the module element 
dual to $G_{{\bf S},1}$ do not have obvious non linearly gauge 
invariant Lagrangian descriptions.  This is because  $G_{{\bf S},1}$ is the product of only 3 Riemann tensors, while in order to build 
the most general descendant of a generator, we need to be able to 
act derivatives on four independent objects.

As all 8 derivative terms are given by products of four Riemann tensors, the descendants of all such terms have obvious Lagrangian
descriptions.  For this reason we will now present a second 
construction of $M_8$  in terms of descendants of only 8 derivative elements. This description allows us to associate 
a Lagrangian with every element of $M_8$.

The second, slightly more complicated description of the 
submodule  $M_8$ goes as follows. Let us study a slightly
modified submodule - lets call it $M_8'$ - which is defined 
to equal the set of all elements in $M_8$ that have eight or more  derivatives. Since the only element in $M_8$ that had 
fewer than $8$ derivatives is only the generator  $G_{{\bf S},1}$, $M_8'$ equals $M_8$ minus (the module 
element dual to) $G_{{\bf S},1}$.

Clearly {\it all} 8 derivative Lagrangians, i.e. $G_{{\bf 3},1} \ldots G_{{\bf 3},6}$, $G_{\bf 3_A}$  {\it and} $G_{{\bf 2_M}}$ are generators of $M_8'$. Unlike $M_8$, however, $M_8'$ is not freely generated. This follows from the fact that one of the 8 derivative ${\bf 2_M}$ generators  of 
$M_8'$ is a descendant of $\chi_6$, and so takes the form (see \eqref{nyny}) 
\begin{equation}\label{tripe}
(|s \rangle, |t \rangle, |u \rangle ) \equiv (s |G_{{\bf S},1}\rangle , t |G_{{\bf S},1}\rangle, u |G_{{\bf S},1} \rangle)
\end{equation} 
As $|G_{{\bf S},1}\rangle$ is not a state in our modified 
submodule, \eqref{tripe} is does not express a relationship between two states within the modified submodule. However we can use \eqref{tripe} to deduce relations within the submodule 
as follows. The most general state generated by the triplet \eqref{tripe} 
is  
\begin{equation}\label{sede}
|\chi \rangle= A_s |s\rangle + A_t|t \rangle+A_u |u \rangle
\end{equation} 
where the otherwise arbitrary functions of $s, t, u$, 
$A_s$, $A_t$ and $A_u$ obey 
\begin{equation}\label{constra}
A_s+A_t+A_u=0.
\end{equation} 
Now the choice 
\begin{equation}\label{coa}
A_s= (u-t) \chi , ~~~A_t=(s-u) \chi, ~~~A_u=(t-s) \chi
\end{equation}
(where $\chi$ is an arbitrary function of $s, t, u$) 
clearly obeys \eqref{constra} and so gives a legitimate 
descendant of the ${\bf 2_M}$ \eqref{sede}. At the level 
of \eqref{sede} this descendant is non-trivial. But  
once we substitute \eqref{tripe} into \eqref{sede} we 
immediately find that the state generated by \eqref{coa} 
vanishes. In other words the vanishing \eqref{sede} 
with the choice \eqref{coa} is a non-trivial relation
within $M_8'$ (even though this relationship is automatic 
when viewed within $M_8$). 

As $\chi$ is an arbitrary function, the relations 
\eqref{coa} form a freely generated relation (or null state) module whose primary (or generator) is \eqref{sede} with 
$\chi=1$, i.e. is 
\begin{equation}\label{chigen}
(u-t) |s \rangle + (s-u) |t \rangle + 
(t-s) |u\rangle
\end{equation}
As the triplets $(|s \rangle , |t \rangle , |u\rangle)$
and $(s, t, u)$ both transform like \eqref{tof1} under 
$S_3$,  it follows  from  \eqref{pcla} that the 10 derivative state \eqref{chigen} transforms in the 
${{\bf 1_A}}$. It follows that the partition function over the module  $M_8'$ is 
\begin{equation}\label{pfzmc}
Z(x)= 5 x^8 Z_{{\bf 3}}(x) + x^8
Z_{{\bf 6}}(x) + 
\left( x^8 Z_{{\bf 2_M}}(x) -x^{10} Z_{{\bf 1_A}}(x) \right)
\end{equation} 
Adding in the contribution of the six derivative state, 
it then follows that the partition function over $M_8$ is 
\begin{equation}\label{pfzmcn}
Z(x)= 5 x^8 Z_{{\bf 3}}(x) + x^8
Z_{{\bf 6}}(x) + 
\left( x^6 +x^8 Z_{{\bf 2_M}}(x) -x^{10} Z_{{\bf 1_A}}(x) \right)
\end{equation} 
Using the explicit expressions \eqref{partfnsss} it is easy to verify that  
\begin{equation}\label{smoi}
x^6 + x^8 Z_{{\bf 2_M}}(x) -x^{10} Z_{{\bf 1_A}}(x)=x^6 Z_{{\bf 1_S}}(x)
\end{equation} 
so that \eqref{pfzmc} and \eqref{pfzm} are actually the same equation.

\subsubsection{The rest of the local submodule} \label{rmg}

In the previous  subsubsection we have constructed the submodule 
of the local gravitational module that is generated by 6 and 8 derivative terms. In this subsubsection we will construct the rest
of the local module. We proceed by guided guesswork; our
guesses will be validated by comparison with plethystic counting.

As we have already accounted for the contribution of $G_{{\bf S},1}$, all remaining polynomial S-matrices are produced by Lagrangians quartic in the Riemann tensor.  In order to capture the contribution  of such terms to four graviton S-matrices, it is sufficient to linearize each of the four  Riemann tensors and also to work on-shell. On-shell and at linearized order 
\be \label{reimf}
R_{\mu\nu\rho\sigma} \propto (p_{\mu}\epsilon_{\nu}-p_{\nu}\epsilon_{\mu})(p_{\rho}\epsilon_{\sigma}-p_{\sigma}\epsilon_{\rho})
\propto  F_{\mu\nu} F_{\rho \sigma} 
\ee
Note that the RHS of \eqref{reimf} is quadratic in $\epsilon$ as expected. At fixed momentum the Riemann tensor is - formally- the second symmetric power of field strengths
\be
R_{\mu\nu\rho\sigma}(p)=  \frac12 F_{\mu\nu}(p)\otimes F_{\rho\sigma}(p).
\ee

One simple (but not necessarily exhaustive) way to construct polynomial graviton S-matrices is to take the second symmetric tensor power of photon S-matrices. The set of all polynomial gravitational 
S- matrices that can be constructed in this manner 
clearly form a submodule of the complete local gravitational module. The symmetric products of the three generators 
of the local photon module\footnote{Recall that the module of parity even 
	polynomial photon S-matrices in $D \geq 5$ was generated 
	by two four derivative generators $E_1$ and $E_2$ 
	(both of which transform in the ${\bf 3}$)and 
	one six derivative generator $E_3$ (which transforms
	in the ${\bf 1_S}$)} are a special set of states 
within this submodule. The products of generators may be decomposed  into familiar representations of $S_3$ as follows 
\bea
&&\frac{1}{16}S^2 E_{{\bf 3},1}=G_{{\bf 3},1}\oplus G_{{\bf 3},2}, \quad \frac{1}{16}S^2 E_{{\bf 3},2}=G_{{\bf 3},3}\oplus G_{{\bf 3},4},  \nonumber\\
&&\frac{1}{16}E_{({\bf 3},1}\otimes E_{{\bf 3},2)}=G_{{\bf 3},5}\oplus G_{\bf 6}=G_{{\bf 3},5}\oplus G_{{\bf 3},6} \oplus G_{{\bf 3_A}}, \nonumber\\
 && \frac{1}{16}E_{({\bf 3},1}\otimes E_{{\bf S})}=G_{{\bf 3},7}, \quad \frac{1}{16}E_{({\bf 3},2}\otimes E_{{\bf S})}=G_{{\bf 3},8},\quad \frac{1}{16}S^2 E_{\bf S}=G_{{\bf S},2}.
\eea
where $S^2$ represents the symmetric square of an 
$S_3$ representation. The new (with $10$ or higher number of derivatives) generators are $G_{{\bf 3},7}, G_{{\bf 3},8}$ and $G_{{\bf S},2}$. They are labeled by their $S_3$ transformation properties as per the convention.More explicitly the 
generators so obtained are given by 
\bea \label{lablaggravgen}
G_{{\bf 3},1}^{(1)}&\equiv& \frac{1}{16}E_{{\bf 3},1}^{(1)}\otimes E_{{\bf 3},1}^{(1)}=R^1_{abpq}R^2_{abpq}R^3_{cdrs}R^4_{cdrs}\nonumber\\
G_{{\bf 3},2}^{(1)}&\equiv& \frac{1}{16}E_{{\bf 3},1}^{(1)}\otimes E_{{\bf 3},1}^{(2)}|_S=R^1_{abpq}R^3_{abrs}R^4_{cdpq}R^2_{cdrs}+R^1_{abpq}R^4_{abrs}R^3_{cdpq}R^2_{cdrs}\nonumber\\
G_{{\bf 3},3}^{(1)}&\equiv& \frac{1}{16}E_{{\bf 3},2}^{(1)}\otimes E_{{\bf 3},2}^{(1)}=R^1_{abpq}R^2_{cdrs}R^3_{bcqr}R^4_{dasp}\nonumber\\
G_{{\bf 3},4}^{(1)}&\equiv& \frac{1}{16}E_{{\bf 3},2}^{(1)}\otimes E_{{\bf 3},2}^{(2)}|_S=R^1_{abpq}R^3_{cdqr}R^4_{bcrs}R^2_{dasp}+R^1_{abpq}R^4_{cdqr}R^3_{bcrs}R^2_{dasp}\nonumber\\
G_{{\bf 3},5}^{(1)}&\equiv& \frac{1}{16}E_{{\bf 3},1}^{(1)}\otimes E_{{\bf 3},2}^{(1)}=R^1_{abpq}R^2_{abrs}R^3_{cdqr}R^4_{cdsp}\nonumber\\
G_{{\bf 3},6}^{(1)}&\equiv& \frac{1}{16}E_{{\bf 3},1}^{(1)}\otimes E_{{\bf 3},2}^{(2)}|_S=R^1_{abpq}R^2_{abqr}R^3_{cdrs}R^4_{cdsp}+R^1_{abpq}R^2_{abqr}R^4_{cdrs}R^3_{cdsp}\nonumber\\
G_{{\bf 3_A}}^{(1)}&\equiv& \frac{1}{16}E_{{\bf 3},1}^{(1)}\otimes E_{{\bf 3},2}^{(2)}|_A=R^1_{abpq}R^2_{abqr}R^3_{cdrs}R^4_{cdsp}-R^1_{abpq}R^2_{abqr}R^4_{cdrs}R^3_{cdsp}\nonumber\\
G_{{\bf 3},7}^{(1)}&\equiv& \frac{1}{16}E_{{\bf 3},1}^{(1)}\otimes E_{\bf S}|_{\Z_2\otimes \Z_2}=R^1_{abpq}\partial_p R^2_{abrs}\partial_q R^3_{cdst}R^4_{cdtr}|_{\Z_2\otimes \Z_2}\nonumber\\
G_{{\bf 3},8}^{(1)}&\equiv& \frac{1}{16}E_{{\bf 3},2}^{(1)}\otimes E_{\bf S}|_{\Z_2\otimes \Z_2}=R^1_{abpq}\partial_p R^2_{cdrs}\partial_q R^3_{bcst}R^4_{datr}|_{\Z_2\otimes \Z_2}\nonumber\\
G_{{\bf S},2} &\equiv& \frac{1}{16}E_{\bf S}|_{\Z_2\otimes \Z_2} \otimes E_{\bf S}|_{\Z_2\otimes \Z_2}=R^1_{abpq}\partial_a\partial_p R^2_{cdrs}\partial_b\partial_q R^3_{dest}R^4_{eatr}|_{\Z_2\otimes \Z_2}.
\eea
In \eqref{lablaggravgen} we have explicitly only listed the $(1)$ components of the generators that transform in ${\bf 3}$ and in one case in ${\bf 3_A}$ and their transformation properties are given in \eqref{tof2} and \eqref{tof4}.

The Lagrangians corresponding to all these generators 
are given by 
\begin{equation}\label{elo}
\begin{split}
G_{{\bf 3},1} &=R_{abpq}R_{baqp}R_{cdrs}R_{dcsr}\\
G_{{\bf 3},2}&=R_{pqrs}R_{pqtu}R_{tuvw}R_{rsvw}\\
G_{{\bf 3},3}&=R_{pqrs}R_{ptru}R_{tvuw}R_{qvsw}\\
G_{{\bf 3},4}&=-R_{pqrs}R_{ptuw}R_{tvws}R_{qvru}\\
G_{{\bf 3},5}&=R_{pqrs}R_{pqtu}R_{rtvw}R_{suvw}\\
G_{{\bf 6}}&=G_{{\bf 3},6}\oplus G_{{\bf 3_A}}=R_{pqrs}R_{pqrt}R_{uvwt}R_{uvws}\\
G_{{\bf 3},7}&=R_{pqab}\partial_a R_{qp \mu\nu} \partial_b R_{rs\nu \alpha} R_{sr \alpha \mu}\\
G_{{\bf 3},8}&=R_{pqab}\partial_a R_{qr \mu\nu} \partial_b R_{rs\nu \alpha}R_{sp\alpha\mu}\\
G_{{\bf S},2}&=R_{abpq} \partial_p \partial_a R_{\mu\nu\beta\gamma}\partial_q\partial_b R_{\nu \alpha\gamma \delta}R_{\alpha\mu\delta\beta}.
\end{split}
\end{equation}

In our discussion above we have already encountered the 
6 eight derivative module elements, $G_{{\bf 3},1} \ldots G_{{\bf 3},6}$ and $G_{{\bf 3_A}}$ that arise in the symmetric product $S^2 (E_{{\bf 3},1}+E_{{\bf 3},2})$. Recall that these elements, together with 
$G_{{\bf S},1}=\chi_6$, were the generators of the submodule $M_8$ discussed 
in subsubsection \ref{mgef}. 
The set of seven $8$-derivative Lagrangians constructed out of Riemann tensors has also appeared in \cite{Bellazzini:2015cra}.
Using ${\mathtt {Mathematica}}$ we have verified
that this result has the following extension to the full local gravitational module for $D \geq 7$.
{\it The set of module elements $G_{{\bf S},1},G_{{\bf S},2}$ and $G_{{\bf 3},1} \ldots G_{{\bf 3},8}$ and $G_{\bf 3_A}$ are all independent generators of the the 
parity even part of the local gravitational module. In other 
words no one of these objects can be written as a linear sum over descendants of the others.}

Note that we have only demonstrated that the module elements  
listed above are  all generators; we have not algebraically 
demonstrated that there are no other generators of the module.   
Nonetheless a comparison of the generators we have already 
identified with the generators of the bare module suggests 
that this is indeed the case. Note that the generators of the local module identified in the previous paragraph consist of one ${\bf 1_S}$ at $6$ derivative order, five ${\bf 3}s$ and one ${\bf 6}$ at $8$ derivative order, two ${\bf 3}$s at $10$ derivative order  and one ${\bf 1_S}$ at $12$ derivative order.  The number (total of 29) and representation content of this collection of generators precisely matches the number and state content of the bare gravitational module (see Table \eqref{counting-is} and  Fig.\ref{graviton-structures}). 

It will turn out - and we will proceed under the assumption that - the list of generators described above is exhaustive; i.e. that $G_{{\bf S},1},G_{{\bf S},2}$ and $G_{{\bf 3},1} \ldots G_{{\bf 3},8}$ and $G_{\bf 3_A}$ generate the local gravitational module. This statement can be taken to be a guess at this stage, which 
will be verified below by comparison with  explicit results of 
plethystic counting below. 
As the number of local generators matches the number of bare generators, it is of importance to know whether the stringent condition \eqref{freeconds} is obeyed. It turns out it is not. It follows that $G_{{\bf S},1},G_{{\bf S},2}$ and $G_{{\bf 3},1} \ldots G_{{\bf 3},8}$ and $G_{\bf 3_A}$ generate the local gravitational module freely.

\subsubsection{Explicit form of S-matrices and Lagrangians}
\label{efsm}

We use the notation $I_{\tt}^{D}(x)$ to 
denote the partition function over gravitational 
S-matrices in $D$ dimensions. Our analysis of the module 
structure presented above amounts to the prediction that  
\begin{equation}\label{ittdgs}
I_{\tt}^{D \geq 8,~ {\rm even}}(x)= x^6 Z_{{\bf 1_S} }+ 
5x^8 Z_{{\bf 3} } + x^8 Z_{{\bf 6} } + 2x^{10} Z_{{\bf 3} }
+ x^{12} Z_{{\bf 1_S} }
\end{equation} 
The explicit form of the S-matrices (and corresponding 
local Lagrangians) summed over by \eqref{ittdgs} 
are, respectively, listed in section \ref{localgravsmatrix} (see \eqref{b0smat}, \eqref{b1smat}, \eqref{b2smat}, \eqref{b3smat}, \eqref{b4smat}, \eqref{b5smat}, \eqref{b6smat}, \eqref{b8smat} and \eqref{b9smat}).

\subsubsection{Comparison with Plethystic}

As there are no parity odd S-matrices when $D\geq 8$, 
it is particularly straightforward to  compare the parity even module predictions, \eqref{ittdgs}, with the results of plethystic counting, Table \ref{graviton-plethystic}. 

Let us first consider the case $D \geq 10$. In this case 
it is easy to check that the plethystic partition function 
listed in Table \ref{graviton-plethystic} exceeds the module prediction \eqref{ittdgs} by $x^8-x^6$. This difference 
is easy to understand. The term $-x^6$ reflects the fact that the plethystic counting omits to count the Lagrangian $G_{{\bf S},1}=\chi_6$ (recall that the plethystic procedure only counts operators built out of the products of four Riemann letters). On the other hand the extra $x^8$ reflects the fact that 
the plethystic partition function incorrectly counts $\chi_8$, defined in \eqref{chiei}, as a non-trivial Lagrangian term. $\chi_8$ does not actually generate a non-trivial S-matrix as its restriction to fourth order in amplitudes is a total derivative. The plethystic procedure does not recognize this fact, as the object that $\chi_8$ is a total derivative of is not gauge invariant and 
so is not itself a quartic polynomial of non-trivial letters.

Let us now turn to $D=9$. In this case the difference between 
the plethystic result and \eqref{ittdgs} is $-x^6+x^8 -2 x^9$
The explanation for the  $x^8 -x^6$ is the same as for 
$D \geq 10$. The explanation for the additional $-2x^9$ 
is very similar to the explanation for the difference between 
plethystic and module predictions for the $D=9$ photon S-matrix
(see above \eqref{shortrep}). In the gravitational case 
we have two conserved currents (that play the role that 
$J=*F\wedge F\wedge F\wedge F$ played in the analysis for 
$D=9$ photons, see  above
\eqref{shortrep}). These conserved currents are 
\be
*(F_1\wedge F_2\wedge F_3\wedge F_4){\rm Tr}(F_1 F_2){\rm Tr}(F_3 F_4),\qquad  *(F_1\wedge F_2\wedge F_3\wedge F_4){\rm Tr}(F_1 F_3 F_2 F_4).
\ee
As in the photon case, each of these conserved currents 
contributes $-x^9$ to the plethystic counting of singlets. These terms do   not correspond to genuine scalar Lagrangians. They represent
an error of the plethystic procedure. Once they are removed, the plethystic result reported in Table \ref{graviton-plethystic} matches perfectly with the correct 
S-matrix partition function \eqref{ittdgs}. 

In $D=8$ the plethystic result differs from \eqref{ittdgs}
by $-x^6 +x^8 + 2 x^8$. The explanation for $-x^6+ x^8$ is the 
same as in $D \geq 10$. The explanation for the additional $2x^8$ 
is, again, similar to the explanation of the excess of $x^4$
in the $D=8$ photon plethystic partition function: the plethystic procedure fails to recognize that some Lagrangian terms are total derivatives (and so incorrectly counts them) because they 
are not total derivatives of objects formed from the products
of four gauge invariant letters. In this case the total 
derivatives that are incorrectly counted by the plethystic 
procedure are the Lagrangians corresponding to the generators 
\be
*(F_1\wedge F_2\wedge F_3\wedge F_4){\rm Tr}(F_1 F_2){\rm Tr}(F_3 F_4),\qquad  *(F_1\wedge F_2\wedge F_3\wedge F_4){\rm Tr}(F_1 F_3 F_2 F_4).
\ee
Explicitly the corresponding Lagrangians are 
\be
\begin{split}
\epsilon^{abcdefgh}R_{ab\a\b}R_{cd\b\a}R_{ef\g\d}R_{gh\d\g} \sim *({\rm Tr}(R \wedge R){\rm Tr}(R \wedge R)), \\
\qquad \epsilon^{abcdefgh}R_{ab\a\b}R_{cd\b\g}R_{ef\g\d}R_{gh\d\a} \sim *({\rm Tr}(R \wedge R \wedge R \wedge R))
\end{split}
\ee

The Hodge dual of these Lagrangians can be expressed as total derivatives of non-gauge invariant objects \cite{Chowdhury:2016cmh} (see appendix A)
\begin{eqnarray}
\textrm{Tr}(R \wedge R\wedge R\wedge R) &=&d[\textrm{Tr} (\omega \wedge d\omega\wedge d\omega\wedge d\omega) + \frac{8}{5}\textrm{Tr} (d\omega \wedge d\omega \wedge \omega \wedge \omega \wedge \omega) \nonumber\\
&&+\frac{4}{5}\textrm{Tr} (d\omega \wedge \omega \wedge d\omega  \wedge \omega \wedge \omega) + \frac{4}{7} \textrm{Tr} (\omega \wedge \omega \wedge \omega \wedge \omega \wedge \omega \wedge \omega \wedge \omega) \nonumber\\
&&+ 2\textrm{Tr} (d\omega \wedge \omega\wedge \omega\wedge \omega\wedge \omega\wedge \omega)].
\end{eqnarray}     
Similarly,
\begin{eqnarray}
\textrm{Tr}(R \wedge R) \textrm{Tr} (R \wedge R) &=& d[ \textrm{Tr}(\omega \wedge d\omega +\frac{2}{3} \omega \wedge \omega \wedge\omega)\wedge \textrm{Tr}(R\wedge R)]. 
\end{eqnarray}    
where $\omega_{\mu ab}$ is the spin connection and in notations,
$${\rm Tr}(R \wedge R){\rm Tr}(R \wedge R) \sim R_{a \phantom{a}cd}^{ \phantom{a}b}  R_{b \phantom{a}ef}^{ \phantom{a}a} R_{\a  \phantom{a}gh}^{ \phantom{a}\b}  R_{\b  \phantom{a}ij}^{ \phantom{a}\a} 
~dx^c \wedge dx^d \wedge dx^e \wedge dx^f \wedge dx^g \wedge dx^h \wedge dx^i \wedge dx^j$$ 
$${\rm Tr}(R \wedge R \wedge R \wedge R) \sim R_{a \phantom{a}cd}^{ \phantom{a}b}  R_{b \phantom{a}ef}^{ \phantom{a}\a} R_{\a  \phantom{a}gh}^{ \phantom{a}\b}  R_{\b  \phantom{a}ij}^{ \phantom{a}a}~ dx^c \wedge dx^d \wedge dx^e \wedge dx^f \wedge dx^g \wedge dx^h \wedge dx^i \wedge dx^j$$ 
Once again the extra $2x^8$ is simply an error of the plethystic procedure; once it is removed the plethystic 
partition function agrees perfectly with \eqref{ittdgs}.

In summary, for all $D \geq 8$ the difference  between the plethystic result and \eqref{ittdgs} is entirely a consequence of the fact that the
plethystic procedure counts S-matrices only up to minor errors. 
Once we correct the plethystic result to account for these 
errors we do indeed recover 
\eqref{ittdgs}, the correct partition function over local 
S-matrices. The  agreement between the (corrected) plethystic result and \eqref{ittdgs} may be viewed as confirmation of the assumption made in \ref{rmg}, namely that Lagrangians \eqref{elo} 
plus $\chi_6$ are the full set of generators of the local graviton
module, and that this module is freely generated.

\subsection{$D=7$} 

As far as the parity even terms are concerned  
the only difference between $D=7$ and $D \geq 8$ is that 
the expression $\chi_8$ defined in \eqref{chiei} vanishes 
identically in $D=7$. From the point of view of enumerating 
and listing S-matrices and their corresponding Lagrangians 
this makes absolutely no difference, as, being a total 
derivative at four graviton order, $\chi_8$ did not contribute 
to four graviton scattering anyway even when it did not 
identically vanish. It follows, in particular, that 
\eqref{ittdgs} and all the results of subsubsection \eqref{efsm} 
apply unchanged to the parity even part of the $D=7$ S-matrix. 

The fact that $\chi_8$ vanishes identically in $D=7$ (and so 
is not counted by the plethystic procedure) does, however, 
impact the comparison of \eqref{ittdgs} with plethystic, as
we see immediately below. 

Even though parity odd S-matrices exist and are counted by 
the plethystic procedure in $D=7$, it is easy to 
compare the parity even module prediction \eqref{ittdgs} with the results of plethystic counting reported in Table \ref{graviton-plethystic} 
The reason for this 
is that the plethystic partition function of Table \ref{graviton-plethystic} is easily decomposed into a part 
that receives contributions only from the even part of the 
S-matrix and a part that receives contributions only from the 
odd part of the S-matrix. The decomposition is as follows: even powers of 
$x$ in Table \ref{graviton-plethystic} count parity even structures while odd 
powers of $x$ count parity odd structures. 

The difference between the parity even part of the result reported in Table \ref{graviton-plethystic} and \eqref{ittdgs} is $-x^6$. As above this term reflects the fact that $G_{{\bf S},1}=\chi_6$, which 
continues to be well defined in $D=7$, is not captured by 
plethystic counting\footnote{The difference is simply $-x^6$ rather than $x^8 -x^6$ (as was the case in $D\geq 10$) because $\chi_8$ is not counted - and so is not over-counted - by the plethystic procedure in these dimensions. }.

Once we correct the plethystic result (by adding $x^6$ to it 
to account for $\chi_6$), the even part of the result reported 
in Table \ref{graviton-plethystic} agrees perfectly with 
the correct parity even partition function \eqref{ittdgs}.

\subsubsection{Parity odd S-matrices}

We now turn to a study of the parity odd S-matrices in $D=7$. 
As in subsubsection \ref{rmg} we proceed by guided guesswork
and then use the results of plethystic counting to validate 
our guess. The key insight of subsubsection \ref{rmg} is that 
the symmetric square of photon module generators play a 
special role in the gravitational module. In subsubsection 
\ref{rmg} we only considered the symmetric square of the 
parity even generators. Extending our considerations to 
parity odd generators we obtain three parity odd elements 
of the gravitational module given by 
\begin{equation}\label{toog}
H_{{\bf 3},1}^{D=7}=O_{\bf S}^{D=7}\otimes E_{{\bf 3},1},\qquad H_{{\bf 3},2}^{D=7}=O_{\bf S}^{D=7} \otimes E_{{\bf 3},2}, \qquad H_{{\bf S}}^{D=7}=O_{\bf S}^{D=7} \otimes E_{{\bf S}}.
\end{equation}
Recall that $O_{\bf S}^{D=7}$ is the unique parity odd photon local module generator in $D=7$. It is simply equal to the seven dimensional  Chern-Simons term. The first two expressions in \eqref{toog} are of 7th order 
in derivatives, and transform in the ${\bf 3}$. The last 
expression is of 9th order in derivatives and transforms in 
the ${\bf 1_S}$. We have checked that the third expression 
in \eqref{toog} is not a linear combination of `descendants' 
of the first two. Consequently all three expressions in 
\eqref{toog} are generators of the odd part of the local 
gravitational module. Note that the number and representation
content of the generators \eqref{toog} matches the number 
and representation content of the odd part of the bare 
module (see Table \ref{counting-is} and appendix \ref{appendix-is}). As in 
subsubsection \ref{rmg} it is thus natural to guess that 
\eqref{toog} exhausts the space of odd generators in $D=7$. 
We proceed on this assumption - the correctness of this guess 
will be demonstrated by the agreement with plethystic. 
If there are indeed no more generators, it is easy to check that the stringent condition \eqref{freeconds} is not obeyed so that the odd part of the local 
module is freely generated. We thus tentatively predict that 
\begin{equation}\label{oppfse}
I_{\tt}^{D=7,~{\rm odd}}= 2x^7 Z_{{\bf 3}}(x)+ x^9 Z_{{S}}(x)
\end{equation}

The odd part of the plethystic partition function reported 
in Table \ref{graviton-plethystic} almost agrees with 
\eqref{oppfse}; the difference between the plethystic result 
and \eqref{oppfse} is -$2 x^7$ and is easy to understand. It is
a consequence of the fact that the first two generators 
in \eqref{toog} cannot be written as products of four 
Riemann tensor letters (even though all their `descendants' can) and so these expressions are not counted by the plethystic procedure. Once we correct the plethystic partition function by 
adding in the contribution of these terms, it agrees perfectly 
with \eqref{oppfse}.  This demonstrates that  \eqref{oppfse} is indeed the correct partition function over odd S-matrices, and validates our guess that the expressions 
in \eqref{toog} are the only odd generators of the $D=7$ module. 

The explicit form of the most general parity odd $D=7$ 
S-matrix is given in \eqref{parityodd7smatrix}.

To end this subsection we explain how the non-linearly gauge invariant 
Lagrangians that generate \eqref{parityodd7smatrix} may be determined. 
It is not difficult to obtain the Lagrangians dual to the generators  \eqref{toog}. In order to obtain these quantities  
we note that working on-shell and to linear order in 
fluctuations, in the vierbein formulation of general 
relativity
\begin{equation}\label{osvf} 
\left( k_\mu \omega_\nu(k) - k_\nu \omega_\mu(k)
\right)_{ab} \propto R_{\mu\nu ab}(k) \propto 
F_{\mu\nu}(k) F_{ab}(k)
\end{equation} 
\eqref{osvf} can be used to find a linearization of 
the quantity $(\omega_\mu)_{ab}$. Up to gauge transformations 
\begin{equation} \label{ohgf} 
(\omega_\mu)_{ab}(k) = A_{\mu}(k) F_{ab}(k)
\end{equation} 
(it is easy to check that \eqref{ohgf} implies \eqref{osvf}; at least locally it follows that 
\eqref{osvf} in turn implies that \eqref{ohgf} holds 
up to gauge transformations). Using \eqref{ohgf} and 
\eqref{osvf} it is easy to check that the following linearizations hold 
\begin{equation}\label{lagno} \begin{split} 
&
{\rm Tr} ( \omega \wedge d\omega +\frac{2}{3} \omega \wedge \omega \wedge\omega)\wedge \textrm{Tr}(R\wedge R)] \\
&\propto*(A_1\wedge F_2\wedge F_3\wedge F_4){\rm Tr}(F_1 F_2){\rm Tr}(F_3 F_4)\propto H_{{\bf 3},1}^{D=7} \\
&\\
&\textrm{Tr} (\omega \wedge d\omega\wedge d\omega\wedge d\omega) + \frac{8}{5}\textrm{Tr} (d\omega \wedge d\omega \wedge \omega \wedge \omega \wedge \omega) \nonumber\\
&+\frac{4}{5}\textrm{Tr} (d\omega \wedge \omega \wedge d\omega  \wedge \omega \wedge \omega) + \frac{4}{7} \textrm{Tr} (\omega \wedge \omega \wedge \omega \wedge \omega \wedge \omega \wedge \omega \wedge \omega) \nonumber\\
&+ 2\textrm{Tr} (d\omega \wedge \omega\wedge \omega\wedge \omega\wedge \omega\wedge \omega)\\
& \propto*(A_1\wedge F_2\wedge F_3\wedge F_4){\rm Tr}(F_1 F_3 F_2 F_4) \propto H_{{\bf 3},2}^{D=7}\\
&\\
& \epsilon^{abcdefg}R_{ab\a\b}R_{c\a \mu\nu} \partial_\b R_{de\nu \a}R_{fg\a \mu}\\
&
\propto *(A_1\wedge F_2\wedge F_3\wedge F_4)F^{\a\b}_1{\rm Tr}(\partial_\a F_3 \partial_\b F_2 F_4) \sim H_{{\bf S}}^{D=7}
\end{split} 
\end{equation}

Note that the third expression in \eqref{lagno} is a product of (derivatives 
of) four Riemann tensors. The Lagrangians that generate the third generator 
listed in \eqref{toog} are given by contracting indices of the derivatives 
that act on each of these four Riemann tensors (as in, for instance, the previous subsection). 

On the other hand the situation with descendants of the first two (multiplets of)
generators of \eqref{toog} is different. As the gauge invariant expressions
for the Lagrangians dual to these generators are {\it not} given as the products 
of four gauge invariant `letters (e.g. four Riemann tensors)', the prescription 
for obtaining (non-linearly) gauge invariant Lagrangians dual to the descendants 
of these generators is less clear. In some ways this `puzzle' is similar to that encountered in  subsubsection \ref{mgef}, and the resolution to this puzzle is also similar 
to that of subsubsection \ref{mgef}. It is easy to find all the first level (i.e. two derivative) descendants of the first two generators in \eqref{toog}. 
In each case - $H_{{\bf 3},1}^{D=7}$ and $H_{{\bf 3},2}^{D=7}$ - there are 6 such 
descendants. It is not difficult to recast the explicit expressions for each 
of these descendants in terms entirely of  products of field strengths (i.e. not 
involving explicit factors of the gauge non invariant $A_\mu$). It is thus possible to find a non-linearly gauge invariant Lagrangian dual to of each of these two groups of level 1 descendants; the corresponding Lagrangians are quartic 
polynomials in (derivatives of) $R_{\mu\nu \alpha \beta}$. Consequently, non-linearly gauge invariant Lagrangians that give rise to descendants of these level one descendants - i.e. any descendants of level 2 or higher in the original module - are obtained, as usual, by contracting the indices of derivatives on these 
four Riemann terms in pairs. 

That is not, however, the end of the story. In each case the original module 
transformed in a ${\bf 3}$ and so had three generators. In each case 
the set of level one descendants transform in the $2\cdot {\bf 2_M}$ + ${\bf 1_S}$ 
+ ${\bf 1_A}$ and so are six in number. It follows that the modules 
generated by these new generators (which were level one descendants of the  
original generator) is not freely generated but has relations. The generators of these relation modules occur at level two above the generator of the original primary - or level one above the generators of the new module (the analogue of 
$M_8'$ in subsection \ref{mgef} ) - and are easily seen to transform in the ${\bf 1_A} + {\bf 2_M}$. 

In summary, as in subsection \ref{mgef} there are two different descriptions 
of the module generated by each of the first two terms in \eqref{toog}. The first 
simple description is that of a freely generated module generated by the expressions in \eqref{toog}. The second consist of the special seven derivative
states listed in \eqref{toog} plus the module generated by a 9 derivative primary 
in the ${\bf 6_{\rm left}}$ (the new module is the analogue of $M_8'$ in 
subsection \ref{mgef}; its primaries are the level one descendants of the 
primaries of the original module) minus the states in the relation module which is freely generated by 11 derivative primaries in the ${\bf 2_M}$ plus the ${\bf 1_A}$. The fact that these two descriptions
are the same is captured by the true partition function identity 
\begin{equation}\label{pfident} 
x^7 Z_{{\bf 3}}(x)= x^7 + x^9 Z_{{\bf 6_{\rm left}}}(x) - x^{11} 
\left( Z_{{\bf 2_{M}}}(x) + Z_{{\bf 1_A}}(x) \right) 
\end{equation} 
The second, more complicated, description for the module allows for a 
straightforward transcriptions to Lagrangians.

While the algorithm for obtaining the (non-linearly gauge invariant) Lagrangians 
dual to each of the generators of \eqref{toog} is now clear, the final 
expressions are moderately lengthy, and we do not provide an explicit 
listing of the corresponding Lagrangians, in the hope that the interested reader would be able to re create these expressions without too much trouble.

\subsection{D=6} 

\subsubsection{Parity even}

In $D=6$ there are two changes in the space of 
$6$ and $8$ derivative generators of the even part of the 
local graviton scattering module (see \cite{Fulling:1992vm}). 
The obvious change is that $G_{{\bf S},1}$ - which was nonzero 
for all $D\geq 7$ - vanishes identically for $D \leq 6$. 

The less obvious change in $D=6$ goes as follows. Recall
that at the algebraic level $G_{{\bf 3},1} \ldots G_{{\bf 3},6}$, $G_{\bf 3_A}$ and $G_{{\bf 3},9}$ 
were all completely independent for $D\geq 8$. As explained above in $D=7$ \eqref{chiei} together with the vanishing of $\chi_8$, implied a linear equation between the Lagrangians $G_{{\bf 3},1} \ldots G_{{\bf 3},6}$, $G_{\bf 3_A}$ and $G_{{\bf 3},9}$, and so between the ${\bf 1_S}$ components of the generators
`dual' to these Lagrangians in the sense of subsubsection 
\ref{mgl}. In $D\leq 6$ this relationship between the ${\bf 1_S}$ 
components of the generators  $G_{{\bf 3},1} \ldots G_{{\bf 3},6}$, $G_{\bf 3_A}$ and $G_{{\bf 3},9}$ is enhanced
to a similar relationship between {\it all} components of the 
corresponding generators. In other words in $D\leq 6$ the 
full ${\bf 3}$ generator $G_{{\bf 3},9}$ is no longer 
independent of the generators $G_{{\bf 3},1} \ldots G_{{\bf 3},6}$, $G_{\bf 3_A}$, but 
can now be written as a sum over the generators 
$G_{{\bf 3},1} \ldots G_{{\bf 3},6}$, $G_{\bf 3_A}$  - see \eqref{nyny} for a formula. In  six dimensions the ${\bf 2_M}$ part of the generator of $G_{{\bf 3},9}$ can be expressed as ${\bf 2_M}$ of $G_{{\bf 3},1} \ldots G_{{\bf 3},6}$, $G_{\bf 3_A}$ due to vanishing of the LHS of \eqref{nyny} and the ${\bf 1_S}$ part was already related to the ${\bf 1_S}$ of the $G_{{\bf 3},1} \ldots G_{{\bf 3},6}$, $G_{\bf 3_A}$ due to vanishing of \eqref{chiei}.

In subsubsection \ref{mgef} we had two ways of generating 
the submodule of the local part of the gravitational module 
generated by terms with 8 or fewer derivatives. The module 
$M_8$ had $G_{{\bf S},1}$ as one of its generators and was freely 
generated. On the other hand the module $M_8'$ did not 
include $G_{{\bf S},1}$ as a generator and consequently was not 
freely generated. For $D \leq 6$ $M_8$ and $M_8'$ are simply 
identical as $G_{{\bf S},1}$ vanishes identically. 
In $D \geq 7$ the module $M_8'$ was not freely generated because 
there was a `secret' relationship between the 8 derivative ${\bf 2_M}s$ (see \eqref{tripe}). This relationship led to 
relationships between the descendants of these ${\bf 2_M}s$. 
For $D \leq 6$ the fact that $|G_{{\bf S},1} \rangle$ vanishes identically turns \eqref{tripe} into the simpler equation
\begin{equation}\label{tripem}
(|s \rangle, |t \rangle, |u \rangle ) =0
\end{equation} 
(the LHS is the linear combination of 4 R structures that would 
have been a descendant of $\chi_6$ for $D\geq 7$). We can use 
\eqref{tripem} to simply remove $G_{{\bf 3},9}$ from the list of 
generators of $M_8'$. The new module $M_8$ is now {\it freely} generated by $G_{{\bf 3},1} \ldots G_{{\bf 3},6}$, $G_{\bf 3_A}$. 

The end result of this slightly complicated discussion is simply 
that we continue to expect the parity even part of the Graviton 
module to be freely generated in $D=6$, but now with one generator
- namely $\chi_6$ - removed.  This expectation is strengthened by the fact that we see precisely the same reduction in bare structures (Table \ref{counting-is} and Appendix \ref{appendix-is}). Our (slightly tentative and to be conformed by comparison with plethystic) prediction 
for the partition function over the parity even part of the
 $D=6$ S-matrix 
\begin{equation}\label{ittdgsi}
I_{\tt}^{D =6,~ {\rm even}}(x)= 
5x^8 Z_{{\bf 3} } + x^8 Z_{{\bf 6} } + 2x^{10} Z_{{\bf 3} }
+ x^{12} Z_{{\bf 1_S} }
\end{equation}
The most general S-matrix and Lagrangian continue to be given by  \eqref{Dgeq7expsmatrix} and \eqref{Dgeq7explag} but with the function $\cf^{G_{{\bf S},1}}(t,u)$ set equal to zero 

\subsubsection{Parity odd} 

In $D=6$ (and every even dimension)  parity 
odd and parity even S-matrices both contribute terms even 
in $x$ to the plethystic partition function. The contribution 
of parity odd and parity even terms to the results of Table \ref{graviton-plethystic} cannot thus be easily disentangled. 
For this reason we need to first analyze odd S-matrices 
before comparing our results with plethystic. 

As in subsubsection \ref{rmg} we can generate a submodule of 
the parity odd part of the $D=6$ S-matrix by taking the 
direct product of the parity even and parity odd parts of the 
local photon S-matrix modules. As in subsubsection \ref{rmg} 
the direct product of generators of these photon modules 
play a distinguished role in the corresponding product submodule. Recall that the parity odd 
photon module has a single generator (see \eqref{parity-odd-6d})  
\be \label{resi}
O_{{\bf A}}^{D=6}= F_1^{ab}*(\partial_a F_2\wedge \partial_b F_3 \wedge F_4).
\ee
As the term in \eqref{resi} is totally anti-symmetric, the tensor products of $E_{{\bf 3},1}$ and $E_{{\bf 3},2}$ with this term transform in the  ${\bf 3_A}\equiv {\bf 2_M}\oplus {\bf 1_A}$. The explicit Lagrangians dual to these generators 
are given by 
\begin{eqnarray} \label{pods}
&&H_{{\bf 3_A},1}^{D=6}\equiv  O_{{\bf A}}^{D=6}\otimes E_{{\bf 3},1}=\epsilon^{abcdef}R^1_{\mu\nu\a\b } \partial_\a  R^2_{\nu\mu ab} \partial_\b R^3_{\g\d cd} R^4_{\d\g ef}\nonumber\\
&&H_{{\bf 3_A},2}^{D=6}\equiv  O_{{\bf A}}^{D=6}\otimes E_{{\bf 3},2} =\epsilon^{abcdef}R^1_{\mu\nu\a\b } \partial_\a  R^2_{\nu\g ab} \partial_\b R^3_{\g\d cd} R^4_{\d\mu ef}.
\end{eqnarray}
In addition to \eqref{pods} there is a third set of 10 derivative 
generators that transform in the ${\bf 3_A}$; these generators 
are given by 
\begin{equation}\label{onj}
H_{{\bf 3_A},3}^{D=6}=\epsilon^{abcdef}F^1_{ab}(\partial_cF^2_{\mu\nu}\partial_d F^3_{\nu\alpha}F^{4}_{\a \mu})(F^1_{\g\d}F^2_{e\rho}F^3_{f\rho}F^4_{\d\g})
\end{equation} 
and are generated by the Lagrangian 
\begin{equation} \label{onjj}
\epsilon^{abcdef}R_{ab\g\d}\partial_cR_{\mu\nu e\rho}\partial_d R_{\nu\alpha f\rho}R_{\a \mu \d\g}.
\end{equation} 
We conjecture that the parity odd part of the local module 
is freely generated by the three generators $H_{{\bf 3_A},1},H_{{\bf 3_A},2}$ and $H_{{\bf 3_A},3}$. 
Note that our guess matches the fact that the bare module
in this dimension also consists of 3 copies of the ${\bf 3_A}$. 
Note that the Lagrangians in  \eqref{pods} and  \eqref{onj} themselves vanish as numbers because it they have 
no ${\bf 1_S}$ part. However they do not vanish as module generators in the sense of subsection \eqref{mgl}.

Note that the tensor product of $E_{\bf S}$ with the generator $O_{{\bf A}}^{D=6}$ also yields a parity odd module element. 
This element transforms in the ${\bf 1_A}$; it is the 
completely antisymmetric part of the generator dual to the 12 derivative Lagrangian 
\begin{equation}\label{nlag}
\epsilon^{abcdef}R_{\mu\nu\a\b } \partial_\mu\partial_\a  R_{\d \rho ab} \partial_\nu \partial_\b R_{\rho\g cd} R_{\g\d ef}.
\end{equation} 
Assuming our conjecture is correct \eqref{nlag} 
must be a descendant of \eqref{onjj} and \eqref{pods}. We 
have not explicitly checked this; it would be useful to do so. 

The module prediction for the 
partition function over parity odd S-matrices is thus 
\begin{equation}\label{mpdesi}
I_{\tt}^{D=6~{\rm odd}}= 3 x^{10}  
Z_{{\bf 3_A}}(x)  
\end{equation}
Remarkably enough the sum of the partition functions 
\eqref{mpdesi} and \eqref{ittdgsi} exactly match the 
$D=6$ part of the the plethystic results reported in 
Table \ref{graviton-plethystic}. This agreement is a consistency check for all the guesses in our analysis of both the parity even and parity odd parts of the $D=6$ local module. 

The explicit forms of the most general $D=6$ parity odd S-matrices are given in \eqref{oddsmatd6grav}, and the Lagrangians that generate these 
S-matrices are listed in \eqref{oddlagd6grav}. 

\subsection{$D = 5$} \label{dles}

\subsubsection{Parity even} 

We have noted above that for $D\leq 6$  generator  $G_{{\bf 3},9}$ are not independent of $G_{{\bf 3},1} \ldots G_{{\bf 3},6}$, $G_{\bf 3_A}$. In $D=5$ 
there are further reductions (see \cite{Fulling:1992vm}). 
It turns out, in particular, that the five eight derivative  Lagrangians $G_{{\bf 3},1} \ldots G_{{\bf 3},5}$ which lead to generators 
in the ${\bf 3}$   are no
longer independent of each other but obey two linear 
relations (see \eqref{5dgravrelation1} and \eqref{5dgravrelation2} for details). These relations 
may be used  to remove two of the module elements - say 
$G_{{\bf 3},3}$ and $G_{{\bf 3},4}$ from the list of generators. Remarkably 
enough we see a similar reduction in the number and representation 
content of generators 
of the bare module in this dimension (see Table \ref{counting-is} and appendix \ref{appendix-is}). We thus expect that 
the module continues to be freely generated (once again 
this guess is verified by the results of plethystic 
counting below). In other words we tentatively predict that 
\begin{equation}\label{ittdgfi}
I_{\tt}^{D =5,~ {\rm even}}(x)= 
3x^8 Z_{{\bf 3} } + x^8 Z_{{\bf 6} } + 2x^{10} Z_{{\bf 3} }
+ x^{12} Z_{{\bf 1_S} }
\end{equation} 
It is easy to check that \eqref{ittdgfi} exactly matches 
the part of the $D=5$ result reported in Table \ref{graviton-plethystic} that is even in $x$, confirming 
the guess that led to \eqref{ittdgfi}. 

The most general parity even part of the  S-matrix and Lagrangian continue to be given by \eqref{Dgeq7expsmatrix} and \eqref{Dgeq7explag} but with the functions $\cf^{G_{{\bf 3},3}}(t,u)=0$ and $\cf^{G_{{\bf 3},4}}(t,u)=0$ set equal to zero.

\subsubsection{Parity odd}

There were are no parity odd photonic S-matrices in $D=5$; 
as a consequence no parity odd graviton S-matrices are 
generated by the direct product of photonic structures. 
Nonetheless parity odd gravitational S-matrices do exist. 
We have verified that the lowest (derivative) dimension generator
for the parity odd part of the local module is 
\be \label{podefive}
H_{{\bf 3_A}}^{D=5}=\epsilon^{abcde}R_{\a\b cd}\partial_\a\partial_e R_{\mu\nu fg} \partial_\b R_{\nu\mu gh} R_{abhf}=\epsilon^{abcde}\left( F^1_{\a\b} \partial_\a F^2_{\mu\nu}\partial_\b F^3_{\nu\mu} F^4_{ab}\right)\left( F^1_{cd}\partial_e F^2_{fg}F^3_{gh}F^4_{hf}\right).
\ee
This 11 derivative term transforms in ${\bf 3_A}={\bf 2_M}\oplus {\bf 1_A}$. If \eqref{podefive} were to exhaust the set of 
parity odd generators of the local module, then the number 
of generators and representation content of the local module 
would match that of the bare module (see Table \ref{counting-is}
and appendix \ref{appendix-is}). This leads us to guess that the parity 
odd part of the local module has no other generators, and 
that this module is freely generated. This guess yields the 
following prediction
\begin{equation} \label{poyu}
 I_{\tt}^{D=5~{\rm odd}}=x^{11} Z_{{\bf 3_A}}(x) 
\end{equation}
It is easily verified that \eqref{poyu} agrees exactly with 
the part of the $D=5$ result reported in Table \ref{graviton-plethystic} that is odd in $x$, verifying our 
guess for the structure of the odd part of the local module. The most general parity odd S-matrix is listed in 
\eqref{oddsmatd5grav} and the corresponding Lagrangian is listed in 
\eqref{oddlagd5grav}. 

\subsection{$D=4$}

\subsubsection{Parity even}

In $D=5$ the independent generators of the parity even part of 
the local module were $G_{{\bf 3},1}$, $G_{{\bf 3},2}$, $G_{{\bf 3},5}$, $G_{{\bf 3},6}$, $G_{{\bf 3},7}$, $G_{{\bf 3},8}$, $G_{{\bf 3_A}}$ and $G_{{\bf S},2}$.
The list of parity even generators is further reduced in 
$D=4$ as follows.  First, as noted in \cite{Fulling:1992vm}, 
$G_{\bf 6}=G_{{\bf 3},6}\oplus G_{{\bf 3_A}}$ vanishes in $D=4$. Next, as again noted in \cite{Fulling:1992vm}, in $D=4$ there is a new linear relationship
between $G_{{\bf 3},1}$, $G_{{\bf 3},2}$ and $G_{{\bf 3},5}$  
(see \eqref{parityevenreld4} for the exact relation). This relationship can be used to eliminate $G_{{\bf 3},2}$ from our list of generators.
Finally, and most subtly, in $D=4$ the 10 derivative 
generators $G_{{\bf 3},7}$ and and $G_{{\bf 3},8}$ both turn out to be `descendants' of $G_{{\bf 3},1}$ and $G_{{\bf 3},5}$ (i.e. 
$G_{{\bf 3},7}$ and $G_{{\bf 3},8}$ lie in the module generated by 
$G_{{\bf 3},1}$ and $G_{{\bf 3},5}$) (see \eqref{parityevenreld43} and \eqref{parityevenreld42}). 

It follows that the parity even part of the S-matrix
module is generated by  $G_{{\bf 3},1},G_{{\bf 3},5}$ and $G_{{\bf S},2}$ (a total of 7 generators). We have verified that there are no further relations between these generators themselves.  However in this case the reduction in the number of generators of the bare module
is even greater (the bare module has 5 independent structures; one in the ${\bf 3}$ and two in the ${\bf 1_S}$ see Table 
\ref{counting-is} and appendix \ref{appendix-is}). As the local module has more generators than the bare module  it follows that it is not freely generated. This is not surprising.  Recall $G_{{\bf 3},1}^{(1)}=E_{{\bf 3},1}^{(1)}\otimes E_{{\bf 3},1}^{(1)}$ and $G_{{\bf 3},5}^{(1)}=E_{{\bf 3},1}^{(1)}\otimes E_{{\bf 3},2}^{(1)}$ and from the discussion near equation \eqref{4d-relation} we know that $E_{{\bf 3},1}$ and $E_{{\bf 3},2}$ are not independent but obey the relation \eqref{4d-relation}. Indeed \eqref{4d-relation} can be used to show that the condition
\be
\sum_{i=1,2,3}\Big(\cf^{G_{{\bf 3},1}^{(i)}}(t,u) G_{{\bf 3},1}^{(i)} +\cf^{G_{{\bf 3},5}^{(i)}}(t,u) G_{{\bf 3},5}^{(i)}\Big)=0
\ee
has two independent families of solution 
\bea
\cf_1^{G_{{\bf 3},1}}(t,u)&=&  (-s t u-\frac{2}{5}\left(t^3+u^3\right))f(t,u),\qquad \cf_1^{G_{{\bf 3},5}}(t,u) = -4\cf_1^{G_{{\bf 3},1}}(t,u)\nonumber\\
\cf_2^{G_{{\bf 3},1}}(t,u)&=& (2 u^2 t^2)g(t,u),\qquad\qquad \qquad \qquad  \cf_2^{G_{{\bf 3},5}}(t,u) = -4\cf_2^{G_{{\bf 3},1}}(t,u).
\eea
where $g(t,u)$ and $f(t,u)$ are  arbitrary functions symmetric in the two arguments.
It follows that the relation module has two generators, 
both in the ${\bf 1_S}$, the first at 14 derivative order 
and the second at 16 derivative order. The module prediction 
for the parity even part of the 4 Graviton S-matrix is thus
\begin{equation}\label{ittdgfo}
I_{\tt}^{D =4,~ {\rm even}}(x)= 
2x^8 Z_{{\bf 3} } + x^{12} Z_{{\bf 1_S} } -x^{14}Z_{{\bf 1_S} }
-x^{16} Z_{{\bf 1_S} }
\end{equation} 

The most general parity invariant 4 graviton S-matrix in 
4 dimensions (and corresponding Lagrangian) can now be
written down; 
	\begin{equation}
		\begin{split} 
			&S^{E}_{T,4}=\frac14\bigg(\cf^{G_{{\bf 3},1}}(t,u)\left[\left( p^1_p \epsilon^1_q - p^1_q \epsilon^1_p  
			\right) \left( p^2_p \epsilon^2_q - p^2_q \epsilon^2_p  \right)
			\left( p^3_r \epsilon^3_s - p^3_s \epsilon^3_r  
			\right) \left( p^4_r \epsilon^4_s - p^4_s \epsilon^4_r  \right)\right.\\
			&\left.\left( p^1_a \epsilon^1_b - p^1_b \epsilon^1_a  
			\right) \left( p^2_a \epsilon^2_b - p^2_b \epsilon^2_a  \right)
			\left( p^3_c \epsilon^3_d - p^3_d \epsilon^3_c  
			\right) \left( p^4_c \epsilon^4_d - p^4_d \epsilon^4_c  \right)\right] \\
			&+\cf^{G_{{\bf 3},1}}(u,s)\left[3\leftrightarrow 2\right]+\cf^{G_{{\bf 3},1}}(s,t)\left[2\leftrightarrow 4\right]\bigg)\\
			&
	+\frac14\bigg(\cf^{G_{{\bf 3},5}}(t,u)\left[\left( p^1_p \epsilon^1_q - p^1_q \epsilon^1_p  
			\right) \left( p^2_p \epsilon^2_q - p^2_q \epsilon^2_p  \right)
			\left( p^3_v \epsilon^3_w - p^3_w \epsilon^3_v  
			\right) \left( p^4_v \epsilon^4_w - p^4_w \epsilon^4_v  \right)\right.\\
			&\left.\left( p^1_r \epsilon^1_s - p^1_s \epsilon^1_r  
			\right) \left( p^2_t \epsilon^2_u - p^2_u \epsilon^2_t  \right)
			\left( p^3_r \epsilon^3_t - p^3_t \epsilon^3_r  
			\right) \left( p^4_s \epsilon^4_u - p^4_u \epsilon^4_s  \right)\right] \\
			&+\cf^{G_{{\bf 3},5}}(u,s)\left[3\leftrightarrow 2\right]+\cf^{G_{{\bf 3},5}}(s,t)\left[2\leftrightarrow 4\right]\bigg)\\
	&+\frac{1}{16}\left(\cf^{G_{{\bf S},2}}(t,u)\right) \times \\
		&	\left[ \left( p_a^1 \epsilon_b^1-p_b^1 \epsilon_a^1 \right)
			p^2_a\left( p_\mu^2 \epsilon_\nu^2-p_\nu^2 \epsilon_\mu^2 \right) p^3_b\left( p_\nu^3 \epsilon_\alpha^3-p_\alpha^3 \epsilon_\nu^3 \right) \left( p_\alpha^4 \epsilon_\mu^4- p_\mu^4 \epsilon_\alpha^4 \right) \right.\\
		&	\left.\left( p_p^1 \epsilon_q^1-p_q^1 \epsilon_p^1 \right)
			p^2_p\left( p_\beta^2 \epsilon_\gamma^2-p_\gamma^2 \epsilon_\beta^2 \right) p^3_q\left( p_\gamma^3 \epsilon_\delta^3-p_\delta^3 \epsilon_\gamma^3 \right) \left( p_\delta^4 \epsilon_\beta^4- p_\beta^4 \epsilon_\delta^4 \right) \right.\\
			&+\left.(1\leftrightarrow 2)+(1\leftrightarrow 3)+ (1\leftrightarrow4)\right].
		\end{split}
	\end{equation}
The most general descendant Lagrangian which gives rise to this is given by,	
\begin{equation}
\begin{split}
L^{4,E}
&=\sum_{m, n} \left(\cf^{G_{{\bf 3},5}}\right)_{m,n} 2^{m+n}\left(\prod_{i=1}^{m}\prod_{j=1}^{n} \left(\partial_{\nu_j}\partial_{\mu_i}R_{pqrs}\right)R_{pqtu}\left(\partial_{\mu_i}R_{rtvw}\right)\left(\partial_{\nu_i}R_{suvw}\right)\right)\\
	&+\sum_{m, n} \left(\cf^{G_{{\bf S},2}}\right)_{m,n} 2^{m+n}\left(
	\prod_{i=1}^{m}\prod_{j=1}^{n}\left(\partial_{\mu_i}\partial_{\nu_j}R_{abpq}\right) \left(\partial_p \partial_a R_{\mu\nu\beta\gamma}\right)\left(\partial_q \partial_b R_{\nu\alpha\gamma\delta}\right)R_{\alpha \mu\delta\beta}\right)\\
	&+
	\sum_{m, n} \left(\cf^{G_{{\bf 3},1}}\right)_{m,n} 2^{m+n}\left(\prod_{i=1}^m\prod_{j=1}^n\left(\partial_{\mu_i}\partial_{\nu_j} R_{abpq}\right)R_{baqp}\left(\partial_{\mu_i}R_{cdrs}\right)\left(\partial_{\nu_j}R_{dcsr}\right)\right).
		\end{split}
		\end{equation}
	
\subsubsection{Parity odd} 

In $D=4$, there are also two parity odd generators for the 
local module of S-matrices. One of these be thought of as the tensor product $H_{\bf 3}^{D=4,(i)}=O_{\bf 3}^{D=4,(i)}\otimes E_{{\bf 3},1}^{(i)}$. This term transforms in ${\bf 3}$ of $S_3$. In addition, there is a 12 
derivative primary
\be
H_{\bf S}^{D=4}=\epsilon^{abmn}R_{mngh}\partial_a \partial_g R_{cdij}\partial_b \partial_h R_{dejk} R_{ecki}=(*F^1_{ab})\text{Tr}(\partial_a F^2\partial_b F^3F^4))((F^1_{\a\b})\text{Tr}(\partial_\a F^2\partial_\b F^3F^4))
\ee
which transforms as ${\bf 1_S}$. We have $4$ local generators while, as discussed in appendix \ref{appendix-is}, the bare module is of rank $2$. There must exist relations in this case as well. It turns out that the relations involve only the 3 components of  $H_{\bf 3}^{D=4,(i)}$ and are independent of $H_{\bf S}^{D=4}$. The relations take the form 
\be
\Big(\sum_{i=1,2,3} \cf^{H_{\bf 3}^{D=4,(i)}}(t,u) H_{\bf 3}^{D=4,(i)}\Big)=0
\ee
with two families of solutions
\bea
\cf_1^{H_{\bf 3}^{D=4}}(t,u)&=&(-2(t^3+u^3)-stu)f(t,u)\nonumber\\
\cf_2^{H_{\bf 3}^{D=4}}(t,u)&=&(-4(t^3u+u^3t)+2t^2u^2)g(t,u).
\eea
where $g(t,u)$ and $f(t,u)$ are arbitrary functions symmetric in the two arguments.

The relation module parameterized by the function $f$ is headed 
by a 14 derivative primary and transforms in the ${\bf 1_S}$ representation. The relationship module parameterized by the 
function $g$ is headed by a 16 derivative primary and 
also transforms in the ${\bf 1_S}$. It follows that the partition 
function over the parity even part of S-matrices is given by 
\be \label{pfpis4}
 I_{\tt}^{D=4~\rm odd}=x^8 Z_{{\bf 3}}(x) + x^{12}Z_{{\bf 1_S}}(x) 
 -x^{{14}} Z_{{\bf 1_S}}(x) - x^{{16}}Z_{{\bf 1_S}}(x).
\ee
Remarkably enough the sum of the partition functions \eqref{pfpis4} and \eqref{ittdgfo} exactly matches the 
$D=4$ plethystic partition function reported in Table \ref{graviton-plethystic}, verifying the local S-matrix 
module predicted above. 
The most general parity odd S-matrix is given by \eqref{oddsmatd4grav} and 
the corresponding Lagrangian is given by 
\eqref{oddlagd4grav}.

\subsection{Regge growth of gravity S-matrices} \label{rggsm}

In subsection \ref{rgss} we have demonstrated
that local S-matrices with 7 or more derivatives all grow faster than $s^2$ in 
the Regge limit. Of all the gravitational 
S-matrices that appear in the lists 
provided in this section (in every dimension, 
and ranging over both parity even and parity 
odd structures) there is only one that occurs at less than 7 derivative order. This is the 
S-matrix that follows from $\chi_6$ \eqref{b0smat} for $D \geq 7$.
 This S-matrix is takes the 
schematic form $stu \times |b \rangle$ 
where $|b \rangle$ is a linear combination of the symmetric part of the bare generators 
listed in lines 1) and 2) of Fig. \ref{graviton-structures}. This S-matrix 
grows like $s^2$ in the Regge limit. 
 
It follows, in summary, that there are 
no local gravitational S-matrices that 
scale like $s^2$ or slower for $D \leq 6$. 
For $D \geq 7$ there is a unique S-matrix 
of this form, namely the S-matrix 
\eqref{b0smat}.

\subsection{Tree level 4-graviton S-matrix in string theory} 
\subsection*{Einstein gravity}

In this section we reproduce the S-matrix coming from Einstein gravity in our basis. The scattering amplitude for Einstein gravity is given by \cite{Sannan:1986tz},  

\begin{eqnarray}
A^{EG}_{4h} &=& \frac{-4\kappa^2}{stu}\left( \frac{1}{2} \epsilon _2.\epsilon _3 \left(s \epsilon _1.k_3 \epsilon _4.k_2+t \epsilon _1.k_2 \epsilon _4.k_3\right)+\frac{1}{2} \epsilon _1.\epsilon _4 \left(s \epsilon _2.k_4 \epsilon _3.k_1+t \epsilon _2.k_1 \epsilon _3.k_4\right)\right.\nonumber\\
&&\left. +\frac{1}{2} \epsilon _2.\epsilon _4 \left(s \epsilon _1.k_4 \epsilon _3.k_2+u \epsilon _1.k_2 \epsilon _3.k_4\right) +\frac{1}{2} \epsilon _1.\epsilon _3 \left(s \epsilon _2.k_3 \epsilon _4.k_1+u \epsilon _2.k_1 \epsilon _4.k_3\right)\right.\nonumber\\
&&\left. +\frac{1}{2} \epsilon _3.\epsilon _4 \left(t \epsilon _1.k_4 \epsilon _2.k_3+u \epsilon _1.k_3 \epsilon _2.k_4\right) +\frac{1}{2} \epsilon _1.\epsilon _2 \left(t \epsilon _3.k_2 \epsilon _4.k_1+u \epsilon _3.k_1 \epsilon _4.k_2\right)\right. \nonumber\\
&&\left.-\frac{1}{4} s t \epsilon _1.\epsilon _4 \epsilon _2.\epsilon _3-\frac{1}{4} s u \epsilon _1.\epsilon _3 \epsilon _2.\epsilon _4-\frac{1}{4} t u \epsilon _1.\epsilon _2 \epsilon _3.\epsilon _4\right)^2
\end{eqnarray}         
In terms of notation introduced above,
\be
\CS=\frac{-4\kappa^2}{stu}(\frac{1}{32} G_{{\bf 3},1}-\frac12 G_{{\bf 3},6}+\frac{1}{16} G_{{\bf 3},2}-\frac14 G_{{\bf 3},5}+G_{{\bf 3},4}+\frac12 G_{{\bf 3},3})|_{\bf 1_S}.
\ee
This implies that the following Lagrangian term reproduces the tensor structure of Einstein gravity S-matrix. More precisely $stu \times A^{EG}_{4h}$ is reproduced by the following Lagrangian. 

\begin{eqnarray}
stu\, L^{EG}_{4h} &\propto & \frac{1}{32}(R_{pqrs}R_{pqrs})^2 -\frac{1}{2}R_{pqrs}R_{pqrt}R_{uvws}R_{uvwt}+\frac{1}{16}R_{pqrs}R_{pqtu}R_{tuvw}R_{rsvw}\\
&&-\frac{1}{4}R_{pqrs}R_{pqtu}R_{rtvw}R_{suvw}-R_{pqrs}R_{p tru}R_{tvws}R_{qvuw}+ \frac{1}{2}R_{pqrs}R_{ptru}R_{tvuw}R_{qvsw}\nonumber
\end{eqnarray} 
Sometimes this Lagrangian is written as $t_{16}R^4$ \cite{Deser:2000xz}, where $t_{16}$ is a sixteen index tensor which is the square of eight index tensor $t_8$ defined in \eqref{t8tensor}.
\be
L^{EG}_{4h} \propto t^{\mu_1\ldots \mu_8} t^{\nu_1\ldots \nu_8} R_{\mu_1\mu_2\nu_1\nu_2}\ldots R_{\mu_7\mu_8\nu_7\nu_8}.
\ee

\subsubsection*{Type II}
The $4$-graviton amplitude in Type II  superstring theory is given by \cite{Schwarz:1982jn}
\begin{eqnarray}
A^{ss}_{4h} &=& h(s,t,u,\alpha')A^{EG}_{4h}
\end{eqnarray} 
Hence this index structure is reproduced by the same Lagrangian term which reproduces Einstein gravity up to momentum factors.

\section{Exchange contributions} \label{ec}

The most general classical (i.e. tree level) four particle
S-matrix that follows from a local Lagrangian is given 
as the sum of two kinds of terms. These are 
\begin{itemize} 
	\item Local S-matrices (i.e. S-matrices that are 
	polynomials in the variables $\epsilon_i, k_i)$. 
	These S-matrices, which have their origin in local contact type interactions in the Lagrangian, have been the focus of this paper so far.
	\item Pole terms that come from the exchange of an intermediate particle.
\end{itemize}
In the earlier sections of this paper we have performed 
an exhaustive classification of local S-matrices for 
four scalar, four photon and four graviton scattering. 
In this section we will present a preliminary discussion 
of the pole contributions. 

Consider for instance, a four graviton (gggg) scattering amplitude. Consider the pole contribution to this amplitude from the exchange of a particle $P$ of mass $m$ that transforms in the representation ${\cal P}$ of the massive Lorentz little group $SO(D-1)$. The most important thing about this amplitude is that the residue of its pole is 
completely fixed by the on-shell three particle S-matrix 
$ggP$\footnote{Let $s=-(k_1+k_2)^2$ denote the exchange momentum. The full exchange diagram involves an intermediate off-shell $P$ particle of squared mass $s$- and so is completely specified only once we are given a `generalized' three point amplitude in which the gravitons are on-shell but the particle $P$ is off-shell. However 
	all off-shell extensions of the same on-shell amplitude 
	agree when $s=m^2$. Moreover these three point amplitudes are polynomial in momenta, so the difference between 
	the numerators exchange diagram built out of any two distinct off-shell extensions of the same on-shell 3 point function contains at least one factor of $(s-m^2)$. This overall propagator cancels the 
	pole originating from the exchange propagator, and 
	we are left with a polynomial S-matrix.}. It follows 
that the most general S-matrix that comes from a local 
Lagrangian is characterized by the masses and spins of the
exchange particles $P$ together with the three point 
$(ggP)$ couplings - in addition to the data that specifies 
polynomial S-matrices. 

In order to complete the classification of polynomial (e.g. 4 graviton) S-matrices presented earlier in this paper into a complete classification of all S-matrices that could possibly originate in local Lagrangians, all we need to do is to work out all possible $ggP$ couplings, and stitch two of these couplings together through the propagator for the 
particle $P$. Every element in this program is straightforward to carry through.
 It is easy to list 
the representations ${\cal P}$ of $P$ that can have nonzero 
on-shell three point functions with our scattering 
particles\footnote{For instance for scalar $P$ scattering can be nonzero only if $P$ transforms in the traceless 
symmetric representation with an even number of indices. }.
It is also not difficult to enumerate the most 
general kinematically allowed on-shell three point functions. The spin $P$  propagator is given 
simply by the projection - in index space -  onto the representation space ${\cal P}$ (in the $D-1$ dimensional space orthogonal to $k_1+k_2$ ) divided by $s-m^2$. Sewing 
these elements together allows us to explicitly 
construct the most general pole contributions to S-matrices\footnote{When $P$ is itself a graviton there 
	are some minor additional complications, as we will see below.}. 

In this paper we will not systematically carry through 
the program outlined in the previous paragraph; we leave this exercise for future work. In this section and Appendix \ref{sampexch} we first present a few sample computations of tree level four point exchange 
S-matrices. In particular in Appendix \ref{sampexch}
we present \begin{itemize} 
	\item The most general pole contribution to four scalar scattering. 
	\item One family of photon photon spin S exchange contributions to four photon scattering, including, in particular the unique scalar exchange contribution to this process.
		\item  One family of graviton graviton spin S exchange contributions to four graviton scattering, including, in particular the unique scalar exchange contribution to this process.
		\end{itemize} 
In the main text below we also compute and present results for 
\begin{itemize}	
	\item The most general graviton exchange contribution to four graviton scattering. 
	\item The most general massive scalar exchange contribution to four graviton scattering.
	\item The most general massive spin two exchange contribution to four graviton scattering. 
	\end{itemize} 

The main focus of our discussion in this section is 
the Regge growth of exchange contributions to S-matrices. As in the case of contact interactions 
discussed earlier in this paper, we are particularly interested in classifying those exchange contributions 
to 4 particle scattering that grow no faster than 
$s^2$ in the Regge limit. \footnote{More generally 
we would like to classify those exchange contributions which grow no faster than $s^2$ in the Regge limit after being combined with suitable polynomial S-matrices of the sort we have enumerated earlier in this paper. Any such combination of an exchange S-matrix plus a `local counterterm subtraction' 
reflects an addition to four particle scattering that 
is not ruled out by the CRG conjecture.} It is very 
easy to see that the exchange of a massive particle of 
spin $J$  \footnote{Any massive exchanged particle transforms under some representation of the little group $SO(D-1)$. There representations can be labelled 
by Young Tableaux. We say that a particle has spin $J$ 
if the length of the largest row in the Young Tableaux 
labelling that particle is $J$. } in the $t$ channel 
yields a contribution to scattering that cannot 
grow faster than $s^J$. \footnote{The reason for this 
is as follows. In the $t$ channel, the scattering particles are grouped into  those with momenta 
$k_1, ~k_3$ and those with momenta $k_2, ~k_4$. 
Contraction of momenta within a group - e.g. the 
dot products $k_1.k_3$ - produces factor of $t$ but 
never of $s$. Moreover the unique contraction of momenta between two groups - which happens through the propagator of the exchanged particle - is  $(k_1-k_3). (k_2-k_4)$. If the exchanged particle has no more than
$J$ symmetrized indices, there cannot be more than 
$J$ factors of $(k_1-k_3). (k_2-k_4)$, simply because the original three point function between 
two scattering particles and the exchanged particle 
could not have had any vector - in this case $k_1-k_2$ 
- contract with more than $J$ indices of the exchanged 
particle. See Appendix \ref{sampexch} for examples 
and more details.} Moreover we  expect that this inequality is generically saturated - i.e. that spin $J$ exchange in the $t$ channel will grow like $s^J$ 
in the Regge limit (see Appendix \ref{sampexch} for 
examples). We thus expect that the exchange of spin 
$J$ particles with $J>2$ will always violate the 
CRG conjecture; note that this violation is non polynomial in $t$ and so cannot be canceled by a local counterterm. This discussion applies equally 
well to the scattering of scalars, photons and gravitons. 

Let us now turn to the exchange of particles with 
spin $\leq 2$. The $t$ channel contributions to such exchange processes are always consistent with the
CRG conjecture. However the question of whether the  $s$ and $u$ channel contributions to exchange contributions violates the CRG conjecture depends 
on the nature of the external particle. In Appendix 
\ref{sampexch} we demonstrate that spin zero and 
spin 2 exchange contributions to four scalar and four 
photon scattering are both consistent with the CRG 
conjecture even in the $s$ and $u$ channels. Quite 
remarkably, however, the explicit computations presented in subsections \ref{4gm} and \ref{4ge} demonstrate that the same 
is not true for four graviton scattering. In four graviton scattering all  possible 
(non Einstein) exchange of massless gravitons, 
massive scalars and massive spin 2 particles violate 
the CRG conjecture in a way that cannot be fixed by a local counterterm.

Why did the sample low spin exchange contributions that we have explicitly computed violate the CRG conjecture for the case of external gravitons? 
The key point here is that three point $gg P$ 
S-matrices appear always to be generated by 
Lagrangian couplings of (derivatives of) two factors of the Riemann tensor to the particle $P$; consequently the three point 
couplings are always at least 4 derivative order in derivatives. \footnote{The assumption of this section - namely that  $ggp$ couplings are always 4 derivative or higher - 
	can, easily be verified by algebraic means - i.e. by simply constructing all gauge invariant $ggP$ three point scattering amplitudes. We leave this straightforward (but potentially lengthy) exercise 
	to future work. }
Assuming this to be the case,  in subsection \ref{ecgsr}, we have given an argument that demonstrates that such exchange contributions 
{\it always} violate the CRG conjecture in a way 
that cannot be canceled by local counterterms, 
at least in $D \leq 6$. The argument of subsection 
\ref{ecgsr} applies to every exchange contribution
including those that we have not explicitly computed.

\subsection{4 Graviton scattering from graviton exchange} \label{4ge}

In this section we construct and study all possible graviton exchange contributions to four point scattering amplitudes of gravitons. As we have explained above, the pole contributions to these exchange diagrams is given 
by sewing on-shell three point functions through 
graviton propagators. The kinematically allowed on-shell
3 point functions for gravitons have been listed in 
\eqref{R1sm} , \eqref{R2sm}, \eqref{R3sm} (in the exceptional case $d=4$ the amplitude \eqref{R2sm} is replaced by a parity odd six derivative structure - we will not separately consider this special case).
For the convenience of the reader we reproduce the 
relevant expressions here\footnote{The $R^2$ and $R^3$ three point functions 
	are sometimes quoted as 
	$$A^{R^2}=  2(\epsilon_{1}.\epsilon_{2}\epsilon_{3}.p_{1}+\epsilon_{1}.\epsilon_{3}\epsilon_{2}.p_{3}+\epsilon_{2}.\epsilon_{3}\epsilon_{1}.p_{2})(\epsilon_{1}.p_{2}\epsilon_{2}.p_{3}\epsilon_{3}.p_{1}) $$
and
$$A^{R^3} = ~6(\epsilon_{1}.p_{2}\epsilon_{2}.p_{3}\epsilon_{3}.p_{1})^2 $$
On-shell these expressions agree with those listed in 
\eqref{onshellthreepoint}. However the form of the 
expressions in \eqref{onshellthreepoint} has the added 
advantage that they are off-shell gauge invariant (i.e. 
gauge invariant without needing to use the conditions 
$k_i^2=0.$}\footnote{As remarked in the footnote of section \ref{gravlagfp}, the Gauss-Bonnet term vanishes in $D=4$ but a new parity odd term appears that contributes to graviton three point function.}.
\begin{align}\label{onshellthreepoint}
A^R ~ =&~(\epsilon_{1}.\epsilon_{2}\epsilon_{3}.p_{1}+\epsilon_{1}.\epsilon_{3}\epsilon_{2}.p_{3}+\epsilon_{2}.\epsilon_{3}\epsilon_{1}.p_{2})^2  \\
A^{R^2} ~ =&(\epsilon^1\wedge\epsilon^2\wedge\epsilon^3\wedge k^1\wedge k^2)^2\\
A^{R^3} ~=&~F^1_{ab}F^{2}_{ab}F^{2}_{cd}F^3_{cd}F^3_{ef}F^1_{ef} + \text{perm.}
\end{align}
The graviton propagator is simple; it is 
\begin{equation} \label{gravprop}
G_{\mu\nu, \rho\sigma}=\frac{1}{k^2}\left( \frac{1}{2}(\eta_{\mu\rho}\eta_{\nu\sigma} +\eta_{\nu\rho}\eta_{\mu\sigma} )~-\frac{1}{D-2}~\eta_{\mu\nu}\eta_{\rho\sigma}\right). 
\end{equation}
We will now use \eqref{onshellthreepoint} and \eqref{gravprop} to obtain  
the graviton exchange contribution for the four point functions in a theory whose  (ggg) three point function is given by 
\begin{equation} \label{gtp} 
A=\alpha_R A^{R} + \alpha_{R^2} A^{R^2}  + \alpha_{R^3} A^{R^3}.
\end{equation}

The general exchange S-matrix takes the form 
\begin{equation} \label{gsm} 
{\cal A}= \alpha_R^2 {\cal A}_{R-R} + \alpha_{R} \alpha_{R^2} {\cal A}_{R-R^2} + \alpha_{R} \alpha_{R^3} {\cal A}_{R-R^3}+ \alpha_{R^2}^2  {\cal A}_{R^2-R^2}
+ \alpha_{R^2} \alpha_{R^3} {\cal A}_{R^2-R^3}
+ \alpha_{R^3}^2 {\cal A}_{R^3-R^3}
\end{equation}
Note that the structures $A^{R^2}$ and $A^{R^3}$ are 
gauge invariant off-shell (i.e. without using $k_i^2=0$). It follows that exchange diagrams that 
sew two of these vertices together - i.e. 
${\cal A}_{R^2-R^2}$, ${\cal A}_{R^2-R^3}$ and 
		${\cal A}_{R^3-R^3}$ - 
are automatically gauge invariant separately in each channel.
In other words these three amplitudes are can be evaluated 
using the same sewing process utilized when the exchanged 
particle is not a graviton but another particle\footnote{In particular each of the amplitudes ${\cal A}_{R^2-R^2}$, ${\cal A}_{R^2-R^3}$ and ${\cal A}_{R^3-R^3}$ can be 
 decomposed admits a gauge invariant decomposition into a piece that has an $s$ pole, a piece that has a $t$ pole and a piece that has a $u$ pole.}\footnote{ On the other hand exchange pieces that involve one or two $R$ vertices (i.e. terms 
proportional to one or two powers of $\alpha_R$) cannot, 
in general, be decomposed as described above in a gauge 
invariant manner. This term is best written in the form 
$\frac{B}{stu}$ where $B$ is a gauge invariant polynomial. }.

The sewing process is easily 
performed in each channel: we find

\begin{eqnarray} \label{fftp}
{\cal A}_{R^2-R^2}&=&\begin{cases}S^{G_{\bf 6}} + S^{G_{{\bf 3},1}}, \nonumber\\
 \cf^{G_{{\bf 6}}}(t,u)=\frac{1}{8s}, \cf^{G_{{\bf 3},1}}(t,u)=\frac{-D}{32(D-2)s}  \end{cases} \nonumber\\
 \nonumber\\  \nonumber\\
{\cal A}_{R^2-R^3}&=& \begin{cases}S^{G_{{\bf 3},6}} + S^{G_{{\bf 3},5}}+S^{G_{{\bf 3},1}}+S^{G_{{\bf 3},2}} + S^{G_{{\bf 3},7}}+S^{G_{\bf 3_A}}, \nonumber\\  
\cf^{G_{{\bf 3},6}}(t,u)=\frac{3}{8},~\cf^{G_{{\bf 3},5}}(t,u)=\frac{-3}{16}, ~\cf^{G_{{\bf 3},1}}(t,u)=\frac{-3(D+2)}{64(D-2)},\nonumber\\
~\cf^{G_{{\bf 3},2}}(t,u)=\frac{3}{32},~ \cf^{G_{{\bf 3},7}}(t,u)=\frac{3}{2s} ,~\cf^{G_{\bf 3_A}}(t,u)=\frac{-12(t-u)}{64s}.  
\end{cases}\nonumber\\
\nonumber\\  \nonumber\\
{\cal A}_{R^3-R^3}&=& \begin{cases}
S^{G_{{\bf 3},6}}+S^{G_{{\bf 3},5}} +S^{G_{{\bf 3},1}} +S^{G_{{\bf 3},2}}+S^{G_{{\bf 3},7}}+S^{G_{\bf 3_A}},\nonumber\\
\cf^{G_{{\bf 3},6}}(t,u)=\frac{9s}{32}, \cf^{G_{{\bf 3},5}}(t,u)=\frac{-9s}{32}, \cf^{G_{{\bf 3},1}}(t,u)=\frac{-9(t^2+u^2+D ut)}{64(D-2)s}, \cf^{G_{{\bf 3},2}}(t,u)=\frac{-9s}{128}, \nonumber\\
\cf^{G_{{\bf 3},7}}(t,u)=\frac{-9}{4}, ~\cf^{G_{\bf 3_A}}(t,u)=\frac{-9(t-u)}{32}. 
\end{cases}
\end{eqnarray}
These S-matrices \eqref{fftp} are formally generated by the 
non local Lagrangians
\begin{equation} \label{lagtt}
(stu){\cal A}_{R^2-R^2}\propto \Bigg(-\frac{D}{4(D-2)}\nabla_\mu \nabla_\nu R_{pqrs}R_{pqrs} \nabla^\mu R_{abcd}\nabla^\nu R_{abcd}+2\nabla_\mu \nabla_\nu R_{pqrs}R_{pqrt}\nabla^\mu R_{uvwt}\nabla^\nu R_{uvws}\Bigg)
\end{equation}
\begin{equation}
\begin{split}
&(stu){\cal A}_{R^2-R^3}\\
&\propto\Bigg(6~\nabla_x\nabla_y\nabla_{\mu}R_{pqrs}\nabla^{\mu}R_{pqrt} \nabla^x R_{uvwt} \nabla^y R_{uvws}-12~\nabla_x\nabla_y\nabla_{\mu}R_{pqrs}R_{pqrt}\nabla^x\nabla^{\mu}R_{uvwt}\nabla^y R_{uvws}\\&+6~\nabla_x\nabla_y\nabla^{\mu}R_{pqrs}R_{pqtu}\nabla^x\nabla_{\mu}R_{rtvw}\nabla^yR_{suvw} + \frac{3(D+2)}{2(D-2)}\nabla_x\nabla_y\nabla^{\mu}R_{pqrs}R_{pqrs}\nabla^x\nabla_{\mu}R_{pqrs}\nabla^y R_{pqrs} \\& +\frac{3}{2}~\nabla_x\nabla_y\nabla^{\mu}R_{pqrs}\nabla_{\mu}R_{pqtu}\nabla^xR_{tuvw}\nabla^yR_{rsvw} +12~\nabla_x\nabla_yR_{\mu\nu a b}\nabla_{a}R_{\nu\mu m n}\nabla_{b}\nabla^xR_{\alpha \beta n p}\nabla^yR_{\beta \alpha p m}\Bigg)
\end{split}
\end{equation}

\begin{equation}
\begin{split}
&(stu){\cal A}_{R^3-R^3}\\
&\propto \Bigg(-36\nabla^a\nabla^b\nabla^\mu\nabla^\nu R_{pqrs}\nabla_\mu R_{pqrt}\nabla_a\nabla_\nu R_{uvwt}\nabla_b R_{uvws} \\
&-18\nabla^a\nabla^b \nabla^\mu\nabla^\nu R_{pqrs} R_{pqtu}\nabla_a\nabla_\mu\nabla_\nu R_{rtvw} \nabla_b R_{suvw}-18\nabla^b\nabla^a\nabla^\mu\nabla^\nu R_{pqrs} R_{pqtu}\nabla_a\nabla_\mu R_{rtvw}\nabla_\nu \nabla_b R_{suvw}\\
&-\frac{9}{D-2}\nabla^a\nabla^b\nabla^\mu\nabla^\nu R_{\a\b cd}R_{\a\b cd}\nabla_aR_{pqrs}\nabla_b\nabla_\mu\nabla_\nu R_{pqrs}\\
&-\frac{9D}{2(D-2)}\nabla^a\nabla^b\nabla^\mu\nabla^\nu R_{\a\b cd}R_{\a\b cd}\nabla_a\nabla_\mu R_{pqrs}\nabla_\nu \nabla_b R_{pqrs}+\frac{9}{2}\nabla_b \nabla_a \nabla^{\mu}\nabla^\nu R_{pqrs}\nabla_{\mu}\nabla_\nu R_{pqtu}\nabla^b R_{tuvw}\nabla^a R_{rsvw}\\
&-72\nabla_l \nabla_k \nabla_\g R_{\mu\nu a b}\nabla_{a}R_{\nu\mu m n}\nabla^\g \nabla_{b}\nabla^k R_{\alpha \beta n p}\nabla^l R_{\beta \alpha p m}\Bigg)
\end{split}
\end{equation}

We now turn to the evaluation of the remaining three amplitudes; ${\cal A}_{R-R}$, ${\cal A}_{R-R^2}$ and 
${\cal A}_{R-R^3}$. These amplitudes are distinguished by 
the fact that they sew diagrams including at least one copy 
of the amplitude $A^{R}$, which is gauge invariant on-shell but 
not off-shell. As the exchange diagram includes an off-shell 
propagator, the corresponding diagrams are not gauge invariant.
Note that this complication is a direct consequence of the 
fact that the exchanged particle is, itself, a graviton - rather than some completely different particle. This is why  the three point functions are not automatically gauge invariant
when the exchanged particle is off-shell. It follows that 
the  three exchange diagrams discussed in this paragraph cannot be computed simply by  sewing the corresponding three point functions with the graviton propagator. In order to recover gauge invariance we must also add in the contribution of contact 4 point terms from the Einstein action (in the case of ${\cal A}_{R-R}$), the contribution of the 
contact term of  the Gauss-Bonnet action 
(in the case of ${\cal A}_{R-R^2}$) and the contribution of the contact term from Riemann cube 
action (in the case of ${\cal A}_{R-R^3}$)\footnote{For example, consider the $R-R^{2}$ exchange which occurs at ${\cal O}(\alpha_R\alpha_{R^2})$. There is a polynomial contribution to this exchange diagram from the Gauss bonnet term to fourth order in perturbation. Together the exchange diagram and the contact piece are gauge invariant.}.
A direct computation of four graviton tree level scattering matrix starting with the  Lagrangian 
\begin{equation}
S= \int \sqrt{g}~ ( \a_R R + \alpha_{R^2}(R^2 - 4 R_{\mu\nu}R^{\mu\nu}+R_{\mu\nu\rho\sigma}R^{\mu\nu\rho\sigma}) + \alpha_{R^3}R_{\mu\nu\rho\sigma}R^{\mu\nu a b}R_{a b}~^{\rho\sigma}~),
 \end{equation}
is algebraically intensive. We use a different method.

We first note that the full gauge invariant result for each of  ${\cal A}_{R-R}$, ${\cal A}_{R-R^2}$ and ${\cal A}_{R-R^3}$
is given by the sum of a term with an $s$ pole, a term with a 
$t$ pole, a term with a $u$ pole and a polynomial contact term. 
Every such S-matrix can be manipulated into the form
\begin{equation}\label{formofsm}
\frac{\sum_i \beta_i S_i}{stu}
\end{equation}
Here $S_i$ are the most general local gauge invariant 
S-matrices at 8 derivative order (in the case of ${\cal A}_{R-R}$), 10 derivative order (in the case of ${\cal A}_{R-R^2}$), and 12 derivative order (in the case of ${\cal A}_{R-R^3}$). Recall that we have already explicitly constructed a basis of all such local S-matrices in section \ref{gravsm} above. $\beta_i$ are the as yet unknown constant coefficients of these basis structures. 

In order to determine the as yet unknown constants $\beta_i$ 
we now impose the following conditions. The expression 
\eqref{formofsm} is a meromorphic function $\epsilon_i$ and $k_i$. Holding $\epsilon_i$ and all other components of $k_i$ constant,  for a moment, we note that \eqref{formofsm} has a pole in the variable $s$. We impose the condition that the 
residue of this pole is the residue of the s channel exchange diagram obtained by sewing the relevant 3 point functions 
through the graviton propagator (this residue is gauge invariant, because it only samples the 3 point functions 
when all participating particles are on-shell). This condition
unambiguously determines all $\beta_i$ coefficients 
in the case of the amplitudes ${\cal A}_{R-R}$ and ${\cal A}_{R-R^2}$. The fact that these
two amplitudes are unambiguously determined by their poles is easy to understand. These amplitudes are, respectively, of homogeneity 
2 and 4 in derivatives. An ambiguity in these amplitudes would be a gauge invariant 2 or 4 derivative polynomial S-matrix, and we have demonstrated above that no such S-matrix exists. 

On the other hand the amplitude ${\cal A}_{R-R^3}$ is of 
homogeneity 6 in derivatives, and so is determined by its poles only up to the addition of the unique 6 derivative local gauge 
invariant 4 graviton S-matrix \eqref{b0smat}. Algebraically we do
indeed find that $\beta_i$ are determined only up to this ambiguity\footnote{The fact that $\beta_i$ are determined only up to 
	this ambiguity is very natural from a Lagrangian viewpoint. 
	While the Lagrangians that gave rise to the Einstein and 
	Gauss-Bonnet 3 point functions were unique, the Lagrangians that give rise to the $R^3$ 3 point function have a one parameter ambiguity, parameterized by the coefficient 
	of the second Lovelock terms - which is an $R^3$ term 
	whose contribution to the 3 graviton S-matrix vanishes.
	It is thus clear that the 4 point function that follows from the exchange of such a vertex has a contribution from the  2nd Lovelock term with an undetermined coefficient.}.
In reporting our answer below we make an arbitrary choice to fix this ambiguity. Our final results are

\begin{eqnarray} \label{fftp2}
{\cal A}_{R-R}&=&\begin{cases}S^{G_{{\bf 3},6}} +S^{G_{{\bf 3},5}} +S^{G_{{\bf 3},3}}+ S^{G_{{\bf 3},4}}+S^{G_{{\bf 3},1}} +S^{G_{{\bf 3},2}}, \nonumber\\
\cf^{G_{{\bf 3},6}}(s,t)=\frac{-1}{64stu},~\cf^{G_{{\bf 3},5}}(t,u)=\frac{-1}{32stu},~\cf^{G_{{\bf 3},3}}(t,u)=\frac{-1}{32stu},\nonumber\\
\cf^{G_{{\bf 3},4}}(t,u)=\frac{-1}{8stu}, \cf^{G_{{\bf 3},1}}(t,u)=\frac{1}{256stu},\cf^{G_{{\bf 3},2}}(t,u)=\frac{1}{128stu}, \nonumber\\
 \end{cases}\nonumber\\
 \nonumber\\  \nonumber\\
 {\cal A}_{R-R^2}&=&\begin{cases}S^{G_{{\bf 3},6}} +S^{G_{{\bf 3},5}} +S^{G_{{\bf 3},3}}+ S^{G_{{\bf 3},4}}+S^{G_{{\bf 3},8}} +S^{G_{{\bf 3},7}}+S^{G_{\bf 3_A}}, \nonumber\\
\cf^{G_{{\bf 3},6}}(s,t)=\frac{1}{32tu},~\cf^{G_{{\bf 3},5}}(t,u)=\frac{-1}{16tu},~\cf^{G_{{\bf 3},3}}(t,u)=\frac{1}{4tu},\nonumber\\
\cf^{G_{{\bf 3},4}}(t,u)=\frac{1}{4tu}, \cf^{G_{{\bf 3},8}}(t,u)=\frac{1}{stu}, \cf^{G_{{\bf 3},7}}(t,u)=\frac{-1}{4stu}, \cf^{G_{\bf 3_A}}(t,u)=\frac{-(t-u)}{32stu} \nonumber\\
 \end{cases}\nonumber\\
 \nonumber\\  \nonumber\\
 {\cal A}_{R-R^3}&=&\begin{cases}S^{G_{{\bf 3},6}} +S^{G_{{\bf 3},5}} +S^{G_{{\bf 3},3}}+ S^{G_{{\bf 3},4}}+ S^{G_{{\bf 3},1}}+ S^{G_{{\bf 3},2}}+S^{G_{{\bf 3},8}} +S^{G_{{\bf S},2}}+S^{G_{\bf 3_A}}, \nonumber\\
\cf^{G_{{\bf 3},6}}(t,u)=\frac{-3s^2+2t^2+2u^2}{128stu},~\cf^{G_{{\bf 3},5}}(t,u)=\frac{3}{64s},~\cf^{G_{{\bf 3},3}}(t,u)=\frac{-2tu-10s^2}{64stu},\nonumber\\
\cf^{G_{{\bf 3},4}}(t,u)=\frac{5}{16s}, \cf^{G_{{\bf 3},1}}(t,u)=\frac{3tu+s^2}{512stu}, \cf^{G_{{\bf 3},2}}(t,u)=\frac{-tu-s^2}{256stu}, \cf^{G_{{\bf 3},8}}(t,u)=\frac{-1}{tu}, \nonumber\\
\cf^{G_{{\bf S},2}}(t,u)=\frac{-1}{3stu}, \cf^{G_{\bf 3_A}}(t,u)=\frac{-s(t-u)}{128stu}.
 \end{cases}\nonumber\\
 \nonumber\\
\end{eqnarray}
The non-local effective Lagrangians that generate these amplitudes are
\begin{equation}
\begin{split}
(stu){\cal A}_{R-R}\propto &\Bigg(\frac{1}{32}(R_{pqrs}R_{pqrs})^2 -\frac{1}{2}R_{pqrs}R_{pqrt}R_{uvws}R_{uvwt}+\frac{1}{16}R_{pqrs}R_{pqtu}R_{tuvw}R_{rsvw}\nonumber\\
&-\frac{1}{4}R_{pqrs}R_{pqtu}R_{rtvw}R_{suvw}-R_{pqrs}R_{p tru}R_{tvws}R_{qvuw}+ \frac{1}{2}R_{pqrs}R_{ptru}R_{tvuw}R_{qvsw}\Bigg)
\end{split}
\end{equation}
\begin{equation}
\begin{split}
(stu){\cal A}_{R-R^2}\propto &\Bigg(-2~(\nabla_\mu R_{pqrs}R_{pqrt} \nabla^\mu R_{uvwt}R_{uvws}) +2 (\nabla_\mu R_{pqrs}R_{pqtu}\nabla^\mu R_{rtvw}R_{suvw})~\\
&-8~(\nabla_\mu R_{pqrs}\nabla^\mu R_{ptru}R_{tvuw}R_{qvsw})-8~(\nabla_\mu R_{pqrs}\nabla^\mu R_{ptuw}R_{tvws}R_{qvru})\\
&-8R_{\alpha\beta a b} \nabla_a R_{\beta\gamma cd} \nabla_b R_{\gamma\delta de} R_{\delta \alpha ec}
+2 R_{\a\b ab}\nabla_a R_{\b\a cd}\nabla_bR_{\g\d de}R_{\d\g e c} \Bigg)
\end{split}
\end{equation}
\begin{equation}
\begin{split}
(stu){\cal A}_{R-R^3}\propto &\Bigg((-\nabla^\mu\nabla^\nu R_{pqrs}\nabla_\mu\nabla_\nu R_{pqrt}R_{uvwt}R_{uvws}+2\nabla^\mu\nabla^\nu R_{pqrs} R_{pqrt}\nabla_\mu\nabla_\nu R_{uvwt}R_{uvws}\\&+\nabla^\mu\nabla^\nu R_{pqrs} \nabla_\nu R_{pqrt}\nabla_\mu R_{uvwt}R_{uvws})+ \frac{3}{2}\nabla^\mu\nabla^\nu R_{pqrs} R_{pqtu}\nabla^\mu R_{rtvw}\nabla^\nu R_{suvw}\\
&-10\nabla_\mu\nabla_\nu R_{pqrs}\nabla^\mu\nabla_\nu R_{ptru}R_{tvuw}R_{qvsw}-11\nabla_\mu\nabla_\nu R_{pqrs}\nabla_\nu R_{ptru}R_{tvuw}\nabla^\mu R_{qvsw}\\
&+10\nabla_\mu\nabla_\nu R_{pqrs}\nabla_\mu R_{ptuw}\nabla_\nu R_{tvws}R_{qvru}+\frac{1}{8}\nabla^\mu \nabla^\nu R_{abcd}R_{abcd}R_{pqrs}\nabla_\mu \nabla_\nu R_{pqrs}\\&+\frac{5}{16}\nabla^\mu \nabla^\nu R_{abcd}R_{abcd}\nabla_\nu R_{pqrs}\nabla_\mu R_{pqrs}-\frac{1}{4} \nabla_\mu \nabla_\nu R_{pqrs}\nabla^\mu \nabla^\nu R_{pqtu}R_{tuvw}R_{rsvw}\\&-\frac{3}{8} \nabla_\mu \nabla_\nu R_{pqrs} \nabla^\nu R_{pqtu}R_{tuvw}\nabla^\mu R_{rsvw} -32 \nabla_\mu R_{\alpha\beta a b} \nabla^\mu \nabla_a R_{\beta\gamma cd} \nabla_b R_{\gamma\delta de} R_{\delta\a e c}\\
&-\frac{8}{3} R_{abcd}\nabla_a\nabla_c R_{\a\b\g\d}\nabla_b\nabla_d R_{\b\mu \d \nu} R_{\mu \a \nu \g}\Bigg)
\end{split}
\end{equation}
While it is not manifest from the expressions  above, we have checked that all scattering amplitudes involving the Gauss-Bonnet 3 point function vanishes for $D = 4~$, as expected (recall the Gauss-Bonnet Lagrangian is topological in 4 dimensions; in particular its contribution to 3 graviton scattering vanishes). 
\begin{table}
	\begin{center}
		\begin{tabular}{|l|l|l|}
			\hline
			Exchange & Regge behavior (large $s$, fixed $t$) & Regge behavior after subtraction \\
			\hline
			${\cal A}_{R-R}$&$s^2/t$ &- \\
			\hline
			${\cal A}_{R-R^2}$&  $s^2$& -\\
			\hline
			${\cal A}_{R-R^3}$& $s^2 t$ &-\\
			\hline
			${\cal A}_{R^2-R^2}$& $s^3$&- \\
			\hline
			${\cal A}_{R^2-R^3}$&$ s^4$& $s^3t$ \\
			\hline
			${\cal A}_{R^3-R^3}$& $s^5$ & $s^4t$ \\
			\hline
		\end{tabular}
	\end{center}
	\caption{Regge behavior of exchange diagrams}
	\label{reggeexchange}
\end{table}

\subsubsection*{Regge growth}
 The Regge behavior of the amplitudes constructed in this section is easily determined\footnote{As usual one obtains the Regge behavior by explicitly decomposing the polarizations into transverse and parallel components using \eqref{perpp} and \eqref{exppol} and evaluating the resulting S-matrix at large $s$, keeping t fixed. }. In every case the `$t$ channel contributions' (i.e. the 
 terms in the S-matrix that are non polynomial in $t$ when expressed as functions 
 of particle momenta and $\epsilon_i$)  grow no faster than $s^2$, consistent with the fact that we are studying  the exchange of a spin 2 particle. 
 All other contributions to the $S$ matrix are analytic in $t$. It follows 
 from dimensional analysis that these remaining contributions can grow no 
 faster than $s$ (in the case of ${\cal A}_{R-R}$), or $s^2$ (in the case of 
 ${\cal A}_{R-R^2}$). Dimensional analysis would have allowed  ${\cal A}_{R-R^3}$
 $s^3$ growth but we find that the amplitude actually grows more slowly 
 like $s^2t$. 
 
 In the case of ${\cal A}_{R^2-R^2}$ the sum of $s$ and $u$ channel 
 exchanges gives rise to an S-matrix that is 6th order in derivatives 
 and grows like $s^3$ - and so faster than $s^2$ - in the Regge limit. 
 It is easy to see that this faster than $s^2$ growth cannot be canceled by a local counter-term. This can be seen in two equivalent ways. First, in our exhaustive classification of local 
 counter-terms earlier in this paper there is only one S-matrix that is of 
 sixth order in derivatives, and this S-matrix grows like $s^2t$ rather 
 than like $s^3$ in the Regge limit. Equivalently, we have explicitly 
 constructed the Lagrangian that gives rise to the ${\cal A}_{R^2-R^2}$
 S-matrix (see \eqref{lagtt}) and it simply is not local, even in 
 the Regge limit. 
 
 In the case of ${\cal A}_{R^3-R^3}$ the S-matrix is 10th order in derivatives  and grows like $s^5$ - and so  considerably faster than $s^2$ - in the Regge limit. 
 This growth can be slightly ameliorated by counter-term 
 subtractions.  The explicit S-matrix for this term is listed in \eqref{fftp}. Notice that in \eqref{fftp} the functions  
 $\cf$ are all polynomials. It follows that 
 all the contributions from these functions can be 
 canceled by local counter-terms. The only piece in  
 ${\cal A}_{R^3-R^3}$ that cannot be cancelled by a local 
 counter-term is the part of the  S-matrix parameterized 
 by $\cf^{G_{{\bf 3},1}}(t,u)=\frac{-9(t^2+u^2+D ut)}{64(D-2)s}$. After a further local counter-term subtraction we are left with $\cf^{G_{{\bf 3},1}}(t,u) \propto \frac{tu}{t+u}$. In both the $u$ 
 and the $s$ channels the subtracted $\cf^{G_{{\bf 3},1}}(t,u)$ is now proportional to $t$ in the 
Regge limit, resulting in a (maximally subtracted ) scattering amplitude that scales like $s^4 t$ in the Regge limit. 
 
 Finally, in  case of ${\cal A}_{R^2-R^3}$ the explicit S-matrix (see \eqref{fftp}) is of 8 derivative order and grows 
 like $s^4$. Once again counter-term subtractions can be used to reduce this growth down to $s^3 t$. In particular, the contribution to this S-matrix from  $\cf^{G_{{\bf 3},7}}(t,u)=\frac{3}{2s}$ - a term which clearly cannot be cancelled by a local counter-term - grows like 
 $s^3t$. It follows that local counter-terms cannot be used to further reduce the Regge growth of this S-matrix.
 
 Our final results are summarized in Table \ref{reggeexchange}. Plugging the 
 results of Table \ref{reggeexchange} into \eqref{gsm} 
 we conclude that the only graviton exchange contributions 
 \eqref{gsm} that grow no faster than $s^2$ in the Regge limit  are those with $\alpha_{R^2}=\alpha_{R^3}=0$\footnote{While ${\cal A}_{R-R}$,${\cal A}_{R-R^2}$ and ${\cal A}_{R-R^3}$ grow like $s^2$ the remaining 3 amplitudes grow faster than $s^2$. The fact that the coefficient of  ${\cal A}_{R^2-R^2}$ must vanish forces $\alpha_{R^2}$ to vanish. 
The fact that the  coefficient of  ${\cal A}_{R^3R^3}$ must also vanish forces $\alpha_{R^3}$ to vanish.}\footnote{Also note that the six derivative contact term ambiguity that one encounters  in ${\cal A}_{R-R^3}$, scales as $stu$ and hence is Regge allowed.}.

The uniqueness of Einstein gravity three point function has also been argued in \cite{Kologlu:2019bco} on account of violation of the ``superconvergence sum rule" for non-zero $\alpha_{R^2}$ or non-zero $\alpha_{R^3}$.

\subsection{4 graviton scattering from exchanges of 
massive particles} \label{4gm}

In this section we compute the pole contributions to four graviton scattering from 
\begin{itemize}
	\item The exchange of a massive scalar.
	\item The exchange of a massive spin 2 particle.
	\item The exchange of a massive spin l particle, assuming that the graviton - graviton - spin $l$ three point S-matrix takes one of three kinematically allowed forms. 
\end{itemize}

The chief qualitative observation we will make is that (in contrast with the scalar and photon case)
all of the exchange contributions to four graviton scattering that we compute grow faster than $s^2$ 
in the Regge limit even after accounting for possible 
local counter-term subtractions.

\subsubsection{Massive scalar exchange }
A three point scattering amplitude between two gravitons and 
a scalar is a polynomial in $\epsilon_1$, $\epsilon_2$, $k_1$ and $k_2$ that is of second order in $\epsilon_1$ and $\epsilon_2$  and is also invariant under 
the Bose flip $1 \leftrightarrow 2$ of the external gravitons.
We consider three point functions for which the external gravitons are on-shell but the higher spin particle is allowed 
to be off-shell (or, equivalently, is of indeterminate mass). 
In other words the expressions we construct will not be functions of $p_1^2$ and $p_2^2$ (because these vanish) but 
will be allowed to depend on $p_1.p_2$. 
The most general expression that meets these conditions is 
given by 
\begin{align}
A (\epsilon_{1}.p_{2})^2(\epsilon_{2}.p_{1})^2 + B (\epsilon_{1}.\epsilon_{2})^2+ C (\epsilon_{1}.\epsilon_{2})(\epsilon_{1}.p_{2})(\epsilon_{2}.p_{1})
\end{align}
Where $A$, $B$ and $C$ are arbitrary functions of $p_1.p_2$. 
The requirement that the S-matrix also enjoys invariance under 
$\epsilon_i \rightarrow \epsilon_i + c_ik_i$ fixes the three 
point function to take the form 
\begin{equation}\label{msc}
A(\epsilon_{1}.p_{2}~\epsilon_{2}.p_{1} -  p_{1}.p_{2}~\epsilon_{1}.\epsilon_{2})^2
= A R_1^{\mu\nu\a\b} R_2^{\mu\nu\a\b}
\end{equation}
where, once again, $A$ is an arbitrary function of $p_1.p_2$. 
\eqref{msc} is off-shell gauge invariant for every choice of the 
function $A$. Different choices of $A$ yield four graviton exchange 
amplitudes that differ only in polynomial pieces. For simplicity 
we choose $A$ to be a constant. The Lagrangian that generates 
the scattering amplitude \eqref{msc} is given by 
\begin{equation}\label{massive scalar Lagrangian}
R_{abcd}R^{abcd}\phi
\end{equation}
where $\phi$ is the massive scalar field. 

The S-matrix that follows by stitching together two copies 
of \eqref{msc} through a scalar propagator is
proportional to $S^{G_{{\bf 3},1}}$ (see \eqref{b5smat}) with 
\begin{equation} \label{smatsc}
\cf^{G_{{\bf 3},1}}(t,u)=\frac{1}{s-m^2}. 
\end{equation} 
The S-matrix described above is formally generated by the non 
local Lagrangian 
\begin{equation}
{\cal A} = \frac{1}{s-m^2}\Big(R_{abcd}R_{abcd}R_{pqrs}R_{pqrs}\Big).
\end{equation}
\subsubsection*{Regge growth}
In the Regge limit i.e. large $s$ fixed $t$, it is  
easily verified that the amplitude \eqref{smatsc} grows like 
$s^3$. To be precise, we analyze the Regge behavior from each channel :
\begin{align}
{\CS}_s \to &s^3\Big((-1+\epsilon_1^\perp\cdot \epsilon_2^\perp)^2(-1+\epsilon_3^\perp\cdot \epsilon_4^\perp)^2\Big)  \\ 
{\CS}_t \to &t^{4}\Big((-1+\epsilon_1^\perp\cdot \epsilon_3^\perp)^2(-1+\epsilon_2^\perp\cdot \epsilon_4^\perp)^2\Big)  \\
{\CS}_u \to & - s^{3}\Big((-1+\epsilon_1^\perp\cdot \epsilon_4^\perp)^2(-1+\epsilon_2^\perp\cdot \epsilon_3^\perp)^2\Big). 
\end{align}
Notice that the $t$ channel contribution grows like $s^0$
(in agreement with general expectations for the $t$ channel growth in the exchange of a spin zero particle).,
The leading Regge growth is from the s and u channel, but they arise with opposite signs, and naively we may think that they might cancel. However, their growth comes with different polarization vectors, and hence cannot cancel for generic choice of the polarization vectors. Hence, the Regge growth for the amplitude is $s^{3}$, which arises from s and u channels. In the Regge limit $\cf^{G_{{\bf 3},1}}(t,u)=\frac{1}{s}$. The only available 6 derivative local counter-term is $G_{{\bf S},1}$, and the S-matrix 
it gives rise to does not cancel that coming from \eqref{smatsc}. It follows that the $s^3$ Regge growth 
from scalar exchange is irreducible (i.e. cannot be 
further reduced by local counter-term subtraction).

\subsubsection{Massive spin 2 exchange}
We now consider the contribution of an exchange of a massive 
spin two particle to four graviton scattering. The massive spin 2 field $S_{\mu\nu}$ is symmetric and traceless. We take its polarization tensor to be $\epsilon_3^\mu \epsilon_3^\nu$ 
and its momentum to be $p_3$. Note that 
$\epsilon_3. \epsilon_3=0$ and $p_3. \epsilon_3=0$. 

The independent data for the scattering amplitude is given by 
five vectors; $p_1$, $p_2$, $\epsilon_1$, $\epsilon_2$ and 
$\epsilon_3$. Recall that the final S-matrix is a scalar. 
Any scalar built out of a set of vectors is a polynomial in the 
seven independent dot products of these vectors 
We start by writing the Lorentz invariant building blocks involving the three particles :
\begin{eqnarray} \label{bbs}
\nonumber A_{1} &=& \epsilon_{1}.p_{2} ~~A_{2}=\epsilon_{2}.p_{1}~~A_{3}=\epsilon_{3}.(p_{1}-p_{2})\\
b_{12}&=&\epsilon_{1}.\epsilon_{2}~~b_{23}=\epsilon_{2}.\epsilon_{3}~~b_{13}=\epsilon_{1}.\epsilon_{3}
\end{eqnarray}

Note that $A_3$ is antisymmetric under $1 \leftrightarrow 2$. 
The most general polynomial built out of \eqref{bbs} that is of 
homogeneity two in each of $\epsilon_1$, $\epsilon_2$ and 
$\epsilon_3$ and also has the $1 \leftrightarrow 2$ Bose exchange 
symmetry is a linear combination of the structures 
\begin{align}\label{Building blocks}
A~=&~b_{12}b_{23}b_{13} \\ \nonumber
B~=&~b_{13}b_{23}A_{1}A_{2} \\ \nonumber
C~=&~b_{13}^2 A_{2}^2+b_{23}^2 A_{1}^2 \\ \nonumber
D~=&~b_{23}b_{12}A_{1}A_{3} - b_{13}b_{12}A_{2}A_{3}\\ \nonumber
E~=&~  A_{3}^2b_{12}^2 \\ \nonumber
F~=&~A_{3}b_{23}A_{2}A_{1}^2 - A_{3}b_{13}A_{1}A_{2}^2 \\ \nonumber
G~=&~  A_{3}^2A_{1}A_{2}b_{12}\\ \nonumber
H~=&~ A_{3}^2 A_{1}^2A_{2}^2
\end{align}
where the coefficients of the linear combination are arbitrary 
functions of $p_1.p_2$. 
It is not difficult to verify that only two linear combinations of \eqref{Building blocks} are invariant under  $\epsilon_1 \rightarrow \epsilon_2 + c_1 p_1$ and 
$\epsilon_2 \rightarrow c_2 p_2$. One of the two gauge  invariant S-matrices occurs at four derivative order and 
is given by  $(p_{1}.p_{2})^2 A-(p_{1}.p_{2})B-\frac{1}{2}(p_{1}.p_{2})D-\frac{1}{4}(p_{1}.p_{2})E+\frac{1}{2} F+\frac{1}{4} G$, or more explicitly 
by
\begin{equation}\label{4 derivative 3 point function}
\Bigg(p_{1}.\epsilon_{2}p_{2}.\epsilon_{1}-p_{1}.p_{2}\epsilon_{1}.\epsilon_{2}\Bigg)\Bigg(\epsilon_{1}.\epsilon_{3}\Big(p_{1}.\epsilon_{2}p_{2}.\epsilon_{3}-p_{1}.p_{2}\epsilon_{2}.\epsilon_{3}\Big)+p_{1}.\epsilon_{3}\Big(p_{2}.\epsilon_{1}\epsilon_{2}.\epsilon_{3} -\epsilon_{1}.\epsilon_{2}p_{2}.\epsilon_{3}\Big)\Bigg)
\end{equation}
The 3 particle S-matrix \eqref{4 derivative 3 point function} may 
be rewritten in terms of field strengths as
\begin{equation}
F_{1}^{ab}F_{2}^{ab}F_{1}^{c\mu}F_{2}^{c\nu}S^{\mu \nu}
\end{equation}
The Lagrangian that yields this 3 point function is :
\begin{equation}\label{Lagrangian A}
 R_{abc\mu}R^{abc}~_{\nu}S^{\mu\nu} 
\end{equation}
It is not difficult to check that the 3 point function \eqref{4 derivative 3 point function} may, in fact, be 
rewritten as 
\begin{equation}\label{R2smn}
(\epsilon_1\wedge\epsilon_2\wedge\epsilon_3\wedge p_1 \wedge p_2)^2
\end{equation}
and so is identical in structure to the 3 graviton 
scattering amplitude from the Gauss-Bonnet term 
\eqref{R2sm}.

There is another gauge invariant combination of the building blocks, at the level of six derivatives, which is $\frac{1}{2}(p_{1}.p_{2})^2 E-(p_{1}.p_{2})G+\frac{1}{2}H$, 
or more explicitly by 
\begin{equation}\label{6 derivative 3 point function}
A_6 =(p_{1}.\epsilon_{3}~p_{2}.\epsilon_{3})(p_{1}.\epsilon_{2}p_{2}.\epsilon_{1} - p_{1}.p_{2}\epsilon_{1}.\epsilon_{2})^2
\end{equation}
In terms of the Field strength , the above structure becomes :
\begin{equation}
-p_{1}^{\mu}F_{1}^{ab}F_{2}^{ab}p_{2}^{\nu}F_{1}^{cd}F_{2}^{cd}S^{\mu\nu}
\end{equation}
The Lagrangian structure which yields this 3 point function is :
\begin{equation}\label{Lagrangian B}
B =\nabla_{\mu}R^{abcd}\nabla_{\nu}R_{abcd}S^{\mu\nu}
\end{equation}
\newline
Formally defining $F_{3}^{\mu\nu}$ it is not difficult to 
verify that $A_6$ may be rewritten as

\begin{equation}\label{R3smn}
A_6 \propto \left( {\rm Tr} F_1 F_2 F_3 \right)^2
\end{equation}
and so is the analogue of the 3 graviton scattering amplitude \eqref{R3sm}.

Note that the three point functions from \eqref{Lagrangian A} and \eqref{Lagrangian B} 
are both gauge invariant even off-shell (i.e. without using 
$p_1^2=p_2^2=0$) \footnote{In fact the amplitudes are 
even formally off-shell gauge invariant under fictitious 
gauge transformations for the particle $3$. This observation is not really physical as, on-shell, the 
transformation $\epsilon_3 \rightarrow \epsilon_3 + p_3$ 
does not preserve the condition $\epsilon_3. p_3=0$ 
as particle $3$ is massive.}

In summary, there are exactly two graviton- graviton 
- massive spin-2 3-point functions. The first is 
the analogue of the Gauss-Bonnet 3 point coupling for 
three gravitons. The second is the analogue of the Riemann
cube 3 point coupling for 3 gravitons. In the case 
that particle 3 is massive, there is no analogue of the 
Einstein 3 point coupling for 3 gravitons. At least at the 
algebraic level the lack of a structure analogous to the 
Einstein 3 graviton coupling is a consequence of the fact 
that that \eqref{R1sm} is gauge invariant only on-shell; 
the manipulations that ensure this make crucial use of the 
fact that $p_3^2=0$ and have no extension to $p_3^2 \neq 0$\footnote{ Under the gauge transformation $\epsilon_1 \rightarrow \epsilon_1+ \lambda p_1$, this is proportional to 
$$(p_1\cdot p_2 \epsilon_2 \cdot \epsilon_3)$$
which vanishes only on-shell. Hence there are no possible gauge-invariant massive deformations of the three point function coming from Einstein gravity.  }.

In summary, the most general 3 point function is given by 
\begin{equation}\label{mjtpf}
\alpha_4 A_4 + \alpha_6 A_6
\end{equation}
The most general massive spin two exchange contribution to 
four graviton scattering is thus given by by stitching these 
two three point functions with a massive spin-2 propagator 
\begin{equation}\label{massivepropagator}
P_{\mu_{1}\mu_{2} , \nu_{1}\nu_{2}} (p)= \frac{1}{p^2 + m^2}\Bigg(\frac{1}{2}\Big(\theta_{\mu_{1}\nu_{1}}\theta_{\mu_{2}\nu_{2}}+\theta_{\mu_{1}\nu_{2}}\theta_{\mu_{2}\nu_{1}}\Big) - \frac{1}{D-1}\theta_{\mu_{1}\mu_{2}}\theta_{\nu_{1}\nu_{2}}\Bigg)
\end{equation}
and yields a 4 graviton scattering amplitude of the form 
\begin{equation} \label{mjspit}
{\cal A}= \alpha_4^2 {\cal A}_{4,4} + \alpha_4 \alpha_6 
{\cal A}_{4,6} + \alpha_6^2 {\cal A}_{6, 6}
\end{equation}
The explicit expressions for ${\cal A}_{4,4}$, ${\cal A}_{4,6}$ 
and ${\cal A}_{6, 6}$ are easy to obtain and are given by 
\begin{eqnarray} \label{fftp1}
{\cal A}_{A-A}&=&\begin{cases}S^{G_{{\bf 3},6}} + S^{G_{{\bf 3},1}}, \\
\cf^{G_{{\bf 3},6}}(s,t)=\frac{1}{4(s-m^2)}, ~\cf^{G_{{\bf 3},1}}(t,u)=\frac{8+D}{16(1-D)(s-m^2)}  \end{cases}\nonumber\\
\nonumber\\
{\cal A}_{A-B}&=& \begin{cases} S^{G_{{\bf 3},5}}+S^{G_{{\bf 3},1}}+S^{G_{{\bf 3},2}} + S^{G_{{\bf 3},7}}+S^{G_{\bf 3_A}}, \\  
 \cf^{G_{{\bf 3},5}}(t,u)=\frac{s}{2(s-m^2)}, ~ \cf^{G_{{\bf 3},1}}(t,u)=\frac{3s}{8(-1+D)(s-m^2)},\\
~\cf^{G_{{\bf 3},2}}(t,u)=\frac{s}{8(s-m^2)},~ \cf^{G_{{\bf 3},7}}(t,u)=\frac{4}{s-m^2},~ \cf^{G_{{\bf 3_A}}}(t,u)=\frac{2(t-u)}{4(s-m^2)}   
\end{cases}\nonumber\\
\nonumber\\
{\cal A}_{B - B}&=& \begin{cases}
S^{G_{{\bf 3},1}} ,\\
\cf^{G_{{\bf 3},1}}(t,u)=\frac{(-2+D)(u^2+t^2)-2Dut}{16(-1+D)(s-m^2)}. 
\end{cases}\nonumber\\
\end{eqnarray}
The non-local Lagrangians which generate these amplitudes are quite non-trivial functions of the mass $m$.
\begin{align} \label{Aaalag}
\begin{split} 
&(s-m^2)(t-m^2)(u-m^2){\cal A} _{A - A}=\Bigg(\frac{1}{2} \nabla_\mu\nabla_\nu R_{pqrs}R_{pqrt}\nabla^\mu R_{uvwt}\nabla^\nu R_{uvws} \\
&-\frac{m^2}{2} \nabla_\mu R_{pqrs}R_{pqrt}\nabla^\mu R_{uvwt} R_{uvws} -\frac{m^2}{2} \nabla_\nu R_{pqrs}R_{pqrt} R_{uvwt} \nabla^\nu R_{uvws}+\frac{m^4}{2} R_{pqrs}R_{pqrt} R_{uvwt}R_{uvws}\\
&+\frac{(D+8)}{32(1-D)}\nabla_{\mu}\nabla_{\nu}R_{abcd}R_{abcd}\nabla^{\mu}R_{pqrs}\nabla^{\nu}R_{pqrs} - \frac{m^2(D+8)}{32(1-D)}\nabla_{\mu}R_{abcd}R_{abcd}\nabla^{\mu}R_{pqrs}R_{pqrs}\\
&-\frac{m^2(D+8)}{32(1-D)}\nabla_{\nu}R_{abcd}R_{abcd}R_{pqrs}\nabla^{\nu}R_{pqrs} +\frac{m^4(D+8)}{32(1-D)}R_{abcd}R_{abcd}R_{pqrs}R_{pqrs}\Bigg)  \\
\end{split}
\end{align}
 The reader can check the Lagrangian \eqref{Aaalag} indeed generates the S-matrix of ${\cal A} _{A - A}$ in the first line of \eqref{fftp1}. The product factor $(t-m^2)(u-m^2)$ cancels with the contribution of the $m^2$ and $m^4$ terms in \eqref{Aaalag} to give the  required momenta functions in \eqref{fftp1}.  To avoid clutter, we suppress the $O(m^2,m^4)$ terms in writing down the non-local Lagrangians for  ${\cal A}_{A- B}$ and 
 ${\cal A}_{B - B}$
 
\begin{align}
\begin{split}
&(s-m^2)(t-m^2)(u-m^2){\cal A} _{A - B} =\Bigg(\nabla_\g \nabla_\d \nabla^\mu R_{pqrs}\nabla_\mu R_{pqrt}\nabla^\g R_{uvwt}\nabla^\d R_{uvws}\\
&+2 \nabla_\g\nabla_\d\nabla^\mu R_{pqrs}R_{pqrt}\nabla_\mu\nabla^\g  R_{uvwt}\nabla^\d R_{uvws}- \nabla_\g \nabla_\d \nabla^\mu R_{pqrs}R_{pqtu}\nabla_\mu \nabla^\g R_{rtvw}\nabla^\d R_{suvw}\\& +\frac{3}{4(D-1)}\nabla_\g\nabla_\d\nabla^\mu R_{pqrs}R_{pqrs}\nabla_\mu\nabla^\g R_{abcd}\nabla^\d R_{abcd}  -\frac{1}{4}\nabla_\g\nabla_\d\nabla_\mu R_{pqrs}\nabla^\mu R_{pqtu}\nabla^\g R_{tuvw}\nabla^\d R_{rsvw}\\& 
-2\nabla_\g\nabla_\d R_{\mu\nu a b} \nabla_{a}R_{\nu\mu m n}\nabla_{b}\nabla^\g R_{\alpha \beta n p}\nabla^\d R_{\beta \alpha p m}+{\CO (m^2)}\Bigg) \\
&(s-m^2)(t-m^2)(u-m^2){\cal A} _{B - B} =\Bigg(\frac{D-2}{4(D-1)} \nabla^\mu\nabla^\nu\nabla^\g\nabla^\d R_{abcd}R_{abcd}\nabla_{\g}R_{mnop}\nabla_{\mu}\nabla_{\nu}\nabla_{\d}R_{mnop} \\&+\frac{D}{4(1-D)}\nabla^\mu\nabla^\nu\nabla^\g\nabla^\d R_{abcd}R_{abcd}\nabla_\mu \nabla_\nu R_{mnop}\nabla_\g \nabla_\d R_{mnop}+\CO(m^2)\Bigg) 
\end{split}
\end{align}
\subsubsection*{Regge growth}
The analysis (and final results) of Regge behaviors of the amplitudes 
${\cal A}_{A - A}$, ${\cal A}_{A- B}$ and 
${\cal A}_{B - B}$ is essentially identical in structure 
to the analysis of the Regge growth in ${\cal A}_{R^2-R^2}$, ${\cal A}_{R^2-R^3}$ and 
${\cal A}_{R^3-R^3}$.
The Regge behavior of the different amplitudes is listed in table \ref{reggeexchange for massive exchanges}.
\begin{table}
	\begin{center}
		\begin{tabular}{|l|l|l|}
			\hline
			Exchange & Regge behavior (large $s$, fixed $t$) &  Regge behavior after subtraction\\
			\hline
			${\cal A}_{RR\phi}$&$s^3$ &  -\\
			\hline
			${\cal A}_{A - A}$&  $s^3$ & -\\
			\hline
			${\cal A}_{A- B}$& $s^4$ &  $s^3t$\\
			\hline
			${\cal A}_{B - B}$& $s^5$ & $s^4t$\\
			\hline
		\end{tabular}
	\end{center}
	\caption{Regge behavior of exchange diagrams for massive particles}
	\label{reggeexchange for massive exchanges}
\end{table}

\subsection{Exchange contribution to gravitational scattering and Regge growth}\label{ecgsr}

Above we have computed the contributions to graviton 
scattering from the exchange of massive scalars and 
massive spin two particles (we have also computed one series of exchange contributions from the exchange of massive spin
$2l$ particles). In each case we have seen that the 
exchange contributions grow faster than $s^2$ in the Regge 
limit, and also that this growth cannot be sufficiently 
tamed (i.e. brought down to growth like $s^2$ or slower)
by the subtraction of local counter-term contributions. 
In this subsection we argue (under a plausible but not completely justified assumption) that this feature is general: it applies to the contribution to four graviton scattering 
from the exchange of any particle, atleast when $D\leq 6$.

\subsubsection{Regge growth of general exchange contributions}

The contribution to gravitational scattering of the exchange of a massive particle of any spin takes the form
\be \label{smatmodf}
{\cal S}=\frac{|\alpha_1\rangle }{s-m^2}+\frac{|\alpha_2\rangle}{t-m^2}+\frac{|\alpha_3 \rangle}{u-m^2}.
\ee
Here $|\alpha_i\rangle$ are elements of the local Module 
(this follows from the fact that 3 point functions are local and gauge invariant).

Let us suppose that $|\alpha_i \rangle$ are of 
$2n^{\rm th}$ order in derivatives. It follows that 
\begin{equation}\label{ttmm}
(s-m^2)(t-m^2) (u-m^2) {\cal S} 
\end{equation} 
is local, and of degree $2n+4$ in derivatives. We note 
for later use that the part of \eqref{ttmm} that is of order $2n+4$ in derivative is given by 
\begin{equation} \label{ttmmm}
stu {\cal S} 
\end{equation} 

It follows that the growth of \eqref{ttmm} is no slower 
that $s^{\alpha(n+2) + \frac{a}{3} }$
where $\alpha(n)$ is listed in \eqref{srgtot}. It then 
follows that the growth of ${\cal S}$ in the Regge limit 
is at least as fast as $s^{\alpha(n-1) + \frac{a}{3}}$. In the special case that $a=0$ the exchange contributions always grow faster than $s^2$ whenever $n-1 > 3$, i.e. 
for $n >  4$. 

Let us now focus on the borderline `dangerous' case 
$n=4$. In this case we obtain an exchange S-matrix ${\cal S}$ that grows like $s^2$ in the Regge limit only when the 
quantity in \eqref{ttmmm} is of the form 
$$ 3 (stu)^2 |g_{\bf S} \rangle$$
where $|g_{\bf S} \rangle$ is a symmetric generator of the bare 
module (see subsection \ref{rgss}) \footnote{The factor of 3 is inserted for later algebraic convenience.}. When this is the case 
the S-matrix is given by 
\begin{equation}\label{smatrixfor}
{\cal S}= 3(stu) | g_{\bf S} \rangle
\end{equation} 
Comparing \eqref{smatrixfor} and \eqref{smatmodf} we conclude that an exchange S-matrix can have the `dangerous' $s^2$ 
growth only if and only if the module elements $|\alpha_i \rangle$
take the form 
\begin{equation} \label{solalpo} 
|\alpha_1 \rangle = s(stu) |g_{\bf S} \rangle, ~~~
|\alpha_2 \rangle = t(stu) |g_{\bf S} \rangle, ~~~
|\alpha_3 \rangle = u(stu) |g_{\bf S} \rangle.
\end{equation} 

\subsubsection{Structure of $|\alpha_i\rangle$ for the 
	case of gravitational scattering.}

Let us now specialize to the special case of exchange contributions to four graviton scattering by a particle of general spin, $\chi_{mn..}$. We expect - and assume - that  the three point function between $\chi_{mn..}$ and two gravitons to take the schematic form 
\begin{equation}\label{gfcou}
\chi_{mn..} R_{abcd} R_{efgh}
\end{equation} 
where all indices are appropriately contracted and the three 
point functions may also involve extra derivatives\footnote{Note that a coupling of the schematic form 
	$\chi_{mn..}  R_{abcd}$ induces mixing between 
	$\chi_{mn..}$ and the graviton at quadratic order. Such 
couplings are eliminated by field redefinitions. The 
lowest order couplings that survive after field redefinitions 
render the Lagrangian diagonal at quadratic level are those 
of the form \eqref{gfcou}.}.
The contribution of the exchange of $\chi_{mn..}$ to four 
graviton scattering thus leads to an S-matrix of the form
\eqref{smatmodf} with $|\alpha_i \rangle$ given by descendants 
of four Riemann structures, i.e. by elements of the module 
described in subsubsection \ref{mgef}. 

We have argued in the previous subsubsection that such a 
contribution can grow like $s^2$ or slower only $|\alpha_i\rangle$ 
are 8 derivative objects (i.e. are linear combinations of the 
four Riemann generators described in subsubsection \ref{mgef})
and additionally if \eqref{solalpo} holds. However 
the only multiplet of four R structures that 
is of the form \eqref{solalpo} are the descendants of 
$|G_{{\bf S},1} \rangle$ (see subsubsection \ref{mgef}, in particular 
see \eqref{nyny})\footnote{The fact that no other multiplet of four Riemann 
	structures are of the form \eqref{solalpo} follows 
	from the fact that the module $M_8$ described in 
	subsubsection \ref{mgef} is freely generated. Had 
	another relation like \eqref{nyny} existed, there would 
	have been null states in $M_8$ - of exactly the same 
	form as the null states of $M_8'$.}.
We conclude that the only possible exchange contribution that 
grows no faster than $s^2$ in the Regge limit is one 
proportional to the S-matrix from $G_{{\bf S},1}$. It follows, in 
particular, that all exchange contributions to graviton scattering in $D \leq 6$  grow faster than $s^2$ in the 
Regge limit. 

Note that while we have not been able to rule out the possibility 
of an exchange contribution proportional to $G_{{\bf S},1}$ in 
$D \geq 7$, it is entirely possible that such a term is never
actually generated\footnote{Such a term can only be generated
$|\alpha_1 \rangle=s|G_{{\bf S},1}\rangle$ is a sum of `perfect squares'; it is entirely possible that this is not the case. We hope to address this issue in the future}.
We leave the careful investigation of this point to the future.

\subsubsection{Counter-term cancellation}

Say we have an exchange contribution that grows faster than 
$s^2$ in the Regge limit. In this section we investigate whether its growth can be cancelled by a local counter-term. 

Let us once again focus on S-matrices of the form 
\eqref{smatmodf}, and focus on the part of $|\alpha_i\rangle$
that is of 8th order in derivatives.  We have just argued 
that all such terms grow faster than $s^2$ in $D \leq 6$. 
The denominator in \eqref{smatmodf} turns the 8 derivative numerator into a six derivative S matrix. It is immediately clear in $D \leq 6$ that this six derivative term 
cannot be cancelled by local counter-terms, simply because 
we have carefully enumerated all available counter-terms
earlier in this paper, and all these counter-terms are of 
8 or higher order in derivatives when $D \leq 6$.

It follows that it is impossible to use local counter-terms 
to cancel the offending large $s$ behavior of exchange diagrams 
unless the 8 derivative part of $|\alpha_i \rangle$ vanish. 
It seems extremely unlikely that this can happen unless 
$|\alpha_i \rangle $ itself vanishes\footnote{Exchange contributions are not homogeneous in 
	derivatives. An $|\alpha_i \rangle$ that, for instance, 
	starts out at 10th order in derivatives also has a 
	piece at $8th$ order in derivatives obtained by Taylor 
	expanding the answer in $m^2$.}. 

\subsubsection{Cancellation between exchange diagrams} 

The reader may wonder whether the offending Regge behavior 
in exchange contributions to gravitational scattering can 
cancel between themselves. Could, for instance, the 
contribution from the exchange of a particle at mass $m_1$ 
in some representation cancel offending part from the 
exchange of a particle of mass $m_2$ in the same representation?
We believe this cannot happen for the reasons we now describe.

When the particle exchanged lies in a representation with four 
or more symmetrized Lorentz indices, it is kinematically 
obvious that cancellation cannot sufficiently improve 
Regge behavior. This is because the exchange of such particles 
lead to violation of $s^2$ growth even in the $t$ channel.
The violating contribution in this channel scales like 
$$ \frac{s^{l}}{t-m^2}.$$ 
As the functional form of this amplitude is a function of 
$t$ with complicated $m^2$ dependence, it is obvious that the 
Regge growths of particles of different mass cannot cancel each 
other. 

When the particle exchanged lies in a representation with three 
or fewer symmetrized indices, the faster than $s^2$ Regge 
growth appears in the $s$ and $u$ channels. The dependence 
of these violations on $m^2$ are relatively simple. Even though 
this is the case, two different exchange contributions cannot 
cancel against each other, simply because each exchange contribution is a perfect square; contributions that are proportional to each other are all of the same sign, and so 
can only add and never cancel. The positivity demanded 
above follows from the requirement that all exchange particles 
have the right sign kinetic term, and that all three point 
couplings are real - these are both constraints that any 
sensible classical theory should clearly have.

\section{Discussion} \label{conc} \label{c}  

Even though this paper and related appendices runs over more than a hundred and fifty pages a great deal remains to be done. 

The principal technical accomplishment of this paper 
is the detailed classification of all polynomial four photon and four graviton S-matrices\footnote{Somewhat unrelated to the main theme of the paper, our classification of polynomial S-matrices can be thought of the classification of counter-terms that contribute to four photon and four photon scattering. We thank R. Loganayagam for this observation.}. The exhaustive classification presented in this paper has been arrived at with the aid of some guided guesswork. While our final results have passed several non-trivial consistency checks and so are very likely correct, it would be useful re-derive our classification in a mathematically rigorous manner. The methods of 
\cite{Henning:2015daa} and \cite{Henning:2017fpj} may prove 
useful in this regard. 

It would be useful - and should not be difficult - to complete the detailed classification of exchange contributions to four photon and four graviton scattering initiated in section \ref{ec}. In particular it would be 
useful to convert the arguments presented in section \ref{ec} for the absence of exchange contributions to four graviton S-matrices consistent 
with the CRG bound into a tight mathematical proof for the same 
result. 

Assuming the validity of the CRG conjecture, the 
analysis of this paper comes very near to establishing 
Conjectures 1 and 2 (see the introduction) for the 
case of four graviton scattering at least for $D \leq 6$. Note that the CRG conjecture is also violated 
by exchange contributions using the three point scattering 
amplitudes from the Gauss-Bonnet and three Riemann terms 
(see \eqref{r31}). In other words the CRG conjecture, in addition to constraining four graviton scattering, 
also gives an alternate derivation of the results for 
three graviton scattering obtained in 
\cite{Camanho:2014apa}.

Apart from tightening up the arguments presented in this paper 
there are two major directions in which it would be 
interesting to move forward. 

First it is unsatisfactory that progress towards 
establishing Conjectures 1 and 2 accomplished in this 
paper itself relies on a conjecture, namely the CRG 
conjecture. It would be very satisfying to prove 
this conjecture - and more generally to understand how once 
can systematically analyze the constraints of causality, positivity of energy, stability etc imposed on classical scattering matrices. We believe that such an analysis may not be too difficult to undertake. We hope to report on 
the results of such an analysis in future work. 

Next, in this paper we have focused on the study 
of four point functions. In order to complete 
a classification of classical theories of gravity
it is important that we are able to generalize our 
analysis to the study of $5$ and higher point scattering amplitudes as well. Such a study might
require a generalization of the CRG conjecture 
to higher point scattering, a result that would 
be easiest to obtain once (and if) we are able to 
prove the CRG conjecture for four particle scattering. 
The generalization of the analysis of this paper to 
the study of five and higher point gravitational 
scattering may also backreact on the study of 
four graviton scattering. Note, in this connection 
that even assuming the validity of the CRG conjecture 
we have not been able to demonstrate 
Conjectures 2 and 3 for four point scattering in $D \geq 7$.
In these dimensions the Lagrangian \eqref{sllintro} 
- when added to the Einstein Lagrangian - generates 
a four graviton S-matrix consistent with the CRG
conjecture. It is conceivable, however, that the 
Lagrangian \eqref{sllintro} is ruled out because it
generates unacceptable contributions to 5 or higher 
graviton scattering, in much the same way that the 
Gauss-Bonnet coupling is ruled because of its 
effect of four graviton scattering. A generalization of the CRG conjecture to $5$ (and higher) points would be needed to constrain $5$ (and higher) point graviton scattering. An inelastic bound on chaos might be relevant for these considerations \cite{Turiaci:2019nwa}.

Through this paper we have focused our attention on S matrices in flat space. What implications do our results have for consistent classical asymptotically $AdS$ solutions of gravity? It seems very likely to us that conjectures 2 and 3 have clear AdS analogues. The AdS analogue of 
conjecture 3 is the following claim. Consider any large $N$ CFT in which the stress tensor and its multitraces are closed under the OPE at leading nontrivial order in $1/N$. The AdS analogue of conjecture  3 asserts that the leading order large $N$ stress tensor correlators in any such theory 
must be those generated by classical Einstein gravity in the bulk. It seems quite likely that this result is correct. It has already been demonstrated \cite{Afkhami-Jeddi:2016ntf, Afkhami-Jeddi:2017rmx} that the single trace stress tensor exchange contributions to four point functions in such theories necessarily have the structure that would follow from a dual bulk Einstein action (this is the AdS version of the causality results of \cite{Camanho:2014apa}). 
As in the case of external scalars \cite{Turiaci:2018dht}, the inversion formula \cite{Caron-Huot:2017vep} for stress tensor correlators at large $N$  should allow us to determine  double trace exchange contributions 
\footnote{And hence local bulk four graviton - roughly speaking ${\rm Weyl}^4$ interaction terms.} to the same process almost uniquely in terms of single trace exchange. 
If it turns out that the words `almost uniquely' in the 
previous sentence can be replaced by `uniquely'
\footnote{i.e. if the 3 parameter ambiguity of the large 
	$N$ scalar inversion formula listed in \cite{Turiaci:2018dht} is absent in the case of stress tensor four point functions.} then 
the AdS analogue of conjecture 3 would have been established at the level of 4 point functions. This will be the case provided that the four point function computed from {\it any}
local bulk interaction violates the chaos bound; we expect
this to be the case at least in $D \leq 6$ because of the close relationship between Regge growth and the chaos 
bound.

The AdS analogue of conjecture 2 would be the claim that large $N$ stress tensor correlators with a finite number of single trace exchange contributions are all also generated by the classical Einstein action in the bulk\footnote{In other words there is no consistent theory in which the four point function of the stress tensor receives contributions 
from a finite number of single trace exchanges in addition 
to stress tensor exchange.}.
 At the level of 4 point functions, it may be possible to establish this conjecture  by generalizing the arguments of \cite{Afkhami-Jeddi:2016ntf, Afkhami-Jeddi:2017rmx} to demonstrate that any non stress tensor single trace exchange contribution to the four stress tensor correlator (plus arbitrary double trace contributions to the same correlator) always violates causality (equivalently the chaos bound). It seems entirely possible to us that this result is true; this is already been studied to some extent in \cite{Meltzer:2017rtf, Afkhami-Jeddi:2018own}.

We note that there is no simple AdS analogue of conjecture 1. Stress tensor correlators in ${\cal N}=4$ Yang Mills theory are nontrivial functions of the single continuous parameter $\lambda$. This lack of universality does not necessarily indicate a problem with conjecture 1, but may instead have its roots in the limited nature of the universality on $R^p \times M_{10-p}$ postulated in conjecture 1 (universality applies only within a consistent truncation and not in solutions - like presumably $AdS$ compactifications - that lie outside this truncation).

Returning to flat space, once several issues discussed 
in this section have been cleared up, and the status 
of Conjectures 2 and 3 is completely clarified, it may be possible to begin a meaningful study of the 
utterly fascinating possibility that string theory is the unique consistent classical extension of Einstein 
gravity, as effectively asserted in Conjecture 1. We leave these for future endeavors.

\section*{Acknowledgments}

We would like to thank O. Aharony, N. Arkani Hamed, T. Dumitrescu, R. Gopakumar,  T. Hartman, V. Hubeny, S. Caron-Huot, J. David, R. D. Koch, A. Maharana,   J. Maldacena, G. Mandal, D. Meltzer, M. Mirbabayi, P. Nayak,  K. Papadodimas, J. Penedones, R. Poojary, M. Rangamani, H. S. Reall, A. Rudra,  A. Sen, A. Sever, A. Sinha, R. Sinha,  R. Soni,  D. Simmons-Duffin, D. Stanford, A. Strominger, S. Trivedi and  A. Zhiboedov for very useful discussions. We would also like to thank J. David, T. Hartman,   A. Maharana,  P. Nayak, J. Penedones,  R. Poojary, A. Rudra, A. Sen,  D. Simmons-Duffin, A. Sinha, R. Soni, D. Stanford, S. Trivedi and  A. Zhiboedov for comments on the manuscript. 
The work of all authors was supported by the Infosys Endowment for the study of the Quantum Structure of Spacetime.
S.M. would like to thank ICTP Trieste for hospitality while this work was being carried out.  The work of A.G. is also supported by SERB Ramanujan fellowship. A.G. would like to acknowledge that part of this work was performed at the Aspen Center for Physics, which is supported by National Science Foundation grant PHY-1607611. The work of T.G. is  supported by DST Inspire fellowship. We would all also like to acknowledge our debt to the people of India for their steady support to the study of the basic sciences.

\appendix

\section{Discussion of Conjecture 1} \label{cono}

\subsection{Universality of the consistent truncation}\label{uct} 

The type II genus $g$ $n$ graviton scattering amplitude on $R^p\times M_{10-p}$ is 
schematically given by the formula 
\begin{equation} \label{scatamp} 
{\cal A}= \int dz d\tau ~\langle V_1(z_1) V_2(z_2)  \ldots V_n(z_n) \rangle _{S_\tau}^{{R^p\times M_{10-p} +{\rm ghosts}}}
\end{equation}
where $V_i$ is the vertex operator for the $i^{th}$ graviton, $z_i$ is 
the insertion point of this operator and the superscript of $\langle \rangle$ denotes the CFT in which the expectation value is taken, and the subscript indicates 
the manifold on which this expectation is computed.  $S_\tau$ is a genus $g$ Riemann 
surface with modulus $\tau$. The integral is over $\tau$ as well as (some of - i.e. 
the non fixed subset of) the $z_i$. 

As graviton vertex operators lie entirely in the $R^p+{\rm ghosts}$ part of the CFT 
\eqref{scatamp} can be rewritten as 
\begin{equation} \label{scatampt} 
{\cal A}= \int d\tau Z_{M_{10-p}}(S_\tau)  \left( \int  dz ~\langle V_1(z_1) V_2(z_2)\ldots  V_n(z_n) \rangle_{S_\tau}^{R^p +{\rm ghosts}} \right)
\end{equation}
where $Z_{M_{10-p}}(S_\tau)$ is the partition function on the Riemann surface 
$S_\tau$ of the sigma model on $M_{10-p}$. For $g \geq 1$ \eqref{scatamp} is a detailed probe of the manifold $M_{10-p}$. In the special case $g=0$, however, 
the Riemann surface - a sphere - has no moduli so \eqref{scatampt} further simplifies to 
\begin{equation} \label{scatampth} 
{\cal A}= Z_{M_{10-p}}(S^2)  \int  dz ~\langle V_1(z_1) V_2(z_2) \ldots  V_n(z_n) \rangle_{S^2}^{R^p +{\rm ghosts}}
\end{equation}
Note that the only dependence of \eqref{scatampth} on $M_{10-p}$ is through 
a single multiplicative constant $Z_{M_{10-p}}(S^2)$ which sets the effective value 
of the $p$ dimensional Newton constant.  

The correlator that appears on the RHS of \eqref{scatampth} can, in principle, be computed as follows. We take the two $V_i(z_i)$ operators that are nearest to each other and use the OPE to expand the product of these two operators in a series 
of single operators. For each of the operators that appears in this series we 
can then once again replace the product of the two nearest $V_i$ with a sequence of 
other operators and so on. We can continue this process until the only operator 
insertion left is the identity. At this stage the correlator on the RHS of \eqref{scatampth} is a number equal to the $S^2$ partition function of the $R^p + {\rm ghost}$ sigma model 

As the $V_i$ are separately invariant under $(-1)^{F_L}$, $(-1)^{F_R}$ and $\Omega$, all operators that appear on the RHS of the OPEs described above also have this property. It follows from the discussion of the previous paragraph that the graviton scattering amplitude is the same in type IIA, type IIB and also in type I theory. Moreover it also follows that the exchange poles that appear in the relevant amplitudes all correspond to particles whose vertex operators are simultaneously 
invariant under $(-1)^{F_L}$, $(-1)^{F_R}$ and $\Omega$.

\subsection{Warped Compactifications}  

The universality of graviton scattering 
amplitudes described in subsection \ref{uct} was a consequence of the fact 
that the sigma model on $R_p\times M_{10-p}$ is the sum of two completely  non 
interacting sigma models; namely the sigma model on $R_p$ and the sigma model on 
$M_{10-p}$. The argument of subsections  
\ref{uct} does not go through for warped compactifications, i.e. compactifications 
on a spacetime of the schematic form
\begin{equation}\label{wcsf} 
ds^2= e^{\chi(y)} dx_\mu^2 + g_{ij}(y) dy_i dy_j
\end{equation} 
(here $x_\mu$ are coordinates on $R_{p}$ 
while $y_i$ are coordinates on $M_{10-p}$).

If it is possible to find consistent backgrounds of string theory of the form 
\eqref{wcsf} in which $g_s$ can parametrically be taken to zero while 
the volume (and nature) of the compact 
manifold  $M_{10-p}$ is held fixed then 
graviton scattering on such compactifications would likely deviate from 
type II or Heterotic versions of the Virasoro Shapiro amplitude. Such examples would thus violate conjecture 1 - and would be of great interest as they would generate new examples of consistent classical gravitational scattering amplitudes. 

It is not, however, clear to us if examples of 
the form \eqref{wcsf} do exist. In the 
rest of this subsection we provide a brief discussion of this point\footnote{We thank A. Maharana and S. Trivedi for very useful discussions on these issues.}.

One class of examples of warped compactifications is obtained by turning 
on RR fluxes (see e.g. \cite{Giddings:2001yu}) on cycles of $M_{10-p}$. 
However the dilaton is often a fixed scalar 
in such compactifications, and so cannot be taken to be parametrically small. When this is the case the corresponding  flux compactifications do not contradict Conjecture 1. 

Models with a flat dilaton potential may be obtained by turning off fluxes but placing $R_p$ 
space filling branes and orientifold planes at points in $M_{10-p}$. In Type IIB theory with $p=4$, for instance, these branes would be $D3$ branes. The total number of 
branes in such situations is fixed to be an order one number - by the requirement that the dilaton tadpole vanish (equivalently from the $D_3$ brane charge Gauss law). The dilaton potential is now completely flat and the dilaton can be made parametrically small.
As the number of branes is held fixed while 
taking $g_s \to 0$, however, the cylinder, Klein bottle and Mobius strip contributions 
to $4d$ gravity scattering amplitudes are 
a factor of $g_s$ smaller than the contribution of closed string tree level exchange\footnote{Reflecting the fact that open strings, for instance, only run in loops in closed string scattering.} and can be ignored in the $g_s\to 0$ limit. Thus graviton scattering amplitudes in this case once again reduce to the type II Virasoro Shapiro formula, and 
Conjecture 1 is not violated. 

If $M_{10-p}$ is non-compact - e.g. is $R_6$ when $p=4$ - then there is no constraint on the number $N$ of $D3$ branes. 
In this the limit $g_s \to 0$, $N \to \infty$ with $g_s N$ fixed, 
the contribution of open string loops to closed string scattering is of the same order as the contribution of closed string tree exchanges. As the AdS/CFT correspondence has emphasized, the dual description of these open string loop diagrams is summarized by closed string propagation a geometry that is non-trivially warped in the neighborhood of the $N$ branes. As $M_{10-p}$ is non-compact, however, in this example the function $\chi(y)$ (see \eqref{wcsf}) is non-constant only on an infinitesimal fraction of the internal space. For this reason the  $p$ dimensional gravitational S matrices in such a manifold are again given by the IIB Virasoro Shapiro amplitude\footnote{Moreover with a value of Newton's constant that is strictly zero.}. Once again conjecture 1 is not violated. 

In summary, while it remains possible that the $g \to 0$ limit of  compactifications of the form \eqref{wcsf} exist and supply conterexaxmples to Conjecture 1, we have not been able to come up with a clear instance of such a counterexample.  

\subsection{Constructing counter examples}

One could also try to come up with counter examples for conjecture 1 by simply guessing a  gravity scattering amplitude that meets all consistency requirements. Such an attempt was made (for scalars), for instance, in  
\cite{Veneziano:2017cks}. Without the clear identification of a genuine string compactification that gives rise to the relevant amplitude, however, it would be  difficult to be sure that a gravitational version of something like \cite{Veneziano:2017cks} is a genuine counter example to Conjecture 1. In order to establish this we would have to have an ansatz for all point scattering amplitudes 
of all the fields that lie within the minimal consistent truncation that contains gravity. This is, of course, a tall order, 
one that would be difficult to come up with.  
\footnote{In order to be at all convincing, 
an ansatz construction would at least need to specify the four particle scattering amplitudes external particles taken to be any of the fields in the minimum consistent 
truncation. This also sounds like a tall order. }
\footnote{We thank J. Penedones, S. Trivedi and  S. Zhiboedov for discussions on this point.}

 \section{Action of $S_3$ on functions of $(s,t,u)$}\label{s3permutations}
 
 In this section we study polynomials of $s$, $t$ and $u$ 
 graded by their degree. We will be particularly interested 
 in decomposing the space of such functions into distinct 
 representations of the permutation group $S_3$ which 
 permutes the three variables. We will perform our study both 
 for unconstrained functions of $s$, $t$ and $u$, as well as 
 for `constrained' functions, i.e. functions that are 
 regarded as identical if they agree when $s+t+u=0$.  
 We start with a brief discussion of the permutation group 
 $S_3$ and its representations. 
 
 \subsection{$S_3$ and its irreducible representations}
 
 Let us first recall that the group $S_3$ 
 has six elements. An element of the permutation group is said to be odd or even 
 depending on whether it is built out of an 
 odd or even number of permutations. We label 
 an element of $S_3$ by the result of the action of that element on $(1,2,3)$, 
 Thus $(1,2,3)$, the identity element $I$ is 
 even. The other two even elements are the 
 cyclical permutations $C=(2,3,1)$  and $C^{-1}=(3,1,2)$. The set of even elements of $S_3$ form the abelian subgroup $Z_3$. 
 The odd elements of this group are the
 three permutations $P_{12}=(2,1,3)$, 
 $P_{13}=(3,2,1)$ and $P_{23}=(1,3,2)$. 
 
 If we think of $1$, $2$ and $3$ as basis elements of a three dimensional vector space then the action above yields a representation 
 of $S_3$ in terms of $ 3 \times 3 $ matrices. 
 The representation is clearly reducible: 
 all permutation elements act as identity on 
 the basis vector $(1+2+3)$. This one dimensional representation is the completely 
 symmetrical representation of $S_3$; this is the representation
 labelled by three boxes in the first row of the Young Tableaux.

 On the 
 other hand the two dimensional set of 
 vectors with $1+2+3=0$ mix only among 
 themselves under the permutation group, and 
 so transform under a 2 dimensional representation of this group. A convenient 
 basis for this space is found by diagonalizing 
 $C$. Let 
 \begin{equation}\label{basisnd}
 B_1=e^{- \frac{2 \pi i}{3}} |1\rangle + |2\rangle + e^{\frac{2 \pi i}{3}} |3 \rangle, ~~~B_2=e^{ \frac{2 \pi i}{3}} |1 \rangle + |2 \rangle + e^{-\frac{2 \pi i}{3}} |3 \rangle
 \end{equation}
 Then
 \begin{equation}\label{conb} 
 C \left( {\begin{array}{c}
 	B_1 \\
 	B_2 \\
 	\end{array} } \right) =
 \left( {\begin{array}{cc}
 	e^{-\frac{2 \pi i}{3}}&0 \\
 	0&e^{\frac{2 \pi i}{3}} \\
 	\end{array} } \right)
 \left( {\begin{array}{c}
 	B_1 \\
 	B_2 \\
 	\end{array} } \right)
 \end{equation}
 The action of the permutations on the same 
 basis is given by 
 \begin{equation} \label{ponb} \begin{split}
 &P_{12} 
 \left( {\begin{array}{c}
 	B_1 \\
 	B_2 \\
 	\end{array} } \right)=\left( {\begin{array}{cc}
 	0&e^{-\frac{2 \pi i}{3}} \\
 	e^{\frac{2 \pi i}{3}} &0 \\
 	\end{array} } \right) \left( {\begin{array}{c}
 	B_1 \\
 	B_2 \\
 	\end{array} } \right) \\
 &P_{23} 
 \left( {\begin{array}{c}
 	B_1 \\
 	B_2 \\
 	\end{array} } \right)=\left( {\begin{array}{cc}
 	0&e^{\frac{2 \pi i}{3}} \\
 	e^{\frac{-2 \pi i}{3}} &0 \\
 	\end{array} } \right) \left( {\begin{array}{c}
 	B_1 \\
 	B_2 \\
 	\end{array} } \right) \\
 &P_{13} 
 \left( {\begin{array}{c}
 	B_1 \\
 	B_2 \\
 	\end{array} } \right)=\left( {\begin{array}{cc}
 	0& 1 \\
 	1 &0 \\
 	\end{array} } \right) \left( {\begin{array}{c}
 	B_1 \\
 	B_2 \\
 	\end{array} } \right) \\
 \end{split}
 \end{equation}
 Note that the phases that appear in the top right corner of the 
 three matrices \eqref{ponb} are, respectively, $e^{-\frac{2\pi i}{3}}$, $1$ and $e^{\frac{2 \pi i}{3}}$. Of course the 
 redefinition  $B^1 \rightarrow  \alpha  B^1$ changes each of 
 these phases by $\alpha$. It follows that while the actual 
 value of each phase is convention dependent, the ratios of 
 the phases are physical (convention independent). We will encounter this fact below.

 The equations \eqref{conb} and \eqref{ponb} give a complete 
 characterization of this two dimensional irreducible 
 representation of $S_3$ (this is the representation labelled by 
 the Young Tableaux with two boxes in its first row and one in its second row, or equivalently two boxes in the first column
 and one in the second column). From the fact that $P_{ij}^2=1$
 it follows that the eigenvalues of the operator $P_{ij}= \pm 1$ 
 in every representation of $S_3$. In this particular 2 dimensional representation it is easily verified that 
 the two eigenvalues of $P_{ij}$ are plus one and minus one 
 for every choice of $i$ and $j$. 
 
 Though it does not show up in the decomposition described 
 above, there is a third irreducible representation of $S_3$. 
 This is the completely antisymmetric representation labelled 
 by a Young tableaux with three boxes in the first column. 
 In this one dimensional representation, every even element of 
 the permutation element acts as unity (identity) while 
 every odd element as $-1$ (negative identity). 
 
 \subsection{Left action of $S_3$ on itself}
 
 Now consider the left action of $S_3$ on itself. This clearly 
 generates a 6 dimensional reducible representation of the 
 permutation group. Clearly the basis vector 
 \begin{equation}\begin{split}\label{symbv}
 &(123) +(231) +(312) + (213) + (321) + (132)\\
 & I+S+S^{-1} +P_{12} +P_{13}+ P_{23} \\
 \end{split}
 \end{equation}
 transforms in the one dimensional symmetric representation. 
 Similarly the  basis vector 
 \begin{equation}\begin{split}\label{asymbv}
 &(123) +(231) +(312) - (213) - (321) - (132)\\
 & I+S+S^{-1} -P_{12} -P_{13}- P_{23} \\
 \end{split}
 \end{equation}
 transforms in the one dimensional antisymmetric representation. 
 What remains is the four dimensional vector space of elements
 $$ A(123) +B(231) +C(312) + p (213) + q  (132) + r (321)$$
 with $A+B+C=0$ and $p+q+r=0$. It is not difficult to decompose 
 this four dimensional vector space into a direct sum of two 
 copies of the two dimensional irreducible representation
 defined in the previous subsection. Define
 \begin{equation} \label{defbbeta} \begin{split}
 &b_1^\pm= e^{- \frac{2 \pi i}{3}} [(123)\pm(132)] + [(231)\pm(213)] + e^{\frac{2 \pi i}{3}} [(312)\pm(321)]\\
 &b_2^\pm= e^{ \frac{2 \pi i}{3}} [(123)\pm(132)] + [(231)\pm(213)] + e^{-\frac{2 \pi i}{3}} [(312)\pm(321)].\\
 \end{split}
 \end{equation}
 With this definition it is easy to see that $b^+_{1,2}$ and $b^-_{1,2}$ transform in representation ${\bf 2_M}$.

 \subsection{Clebsch-Gordon rules}\label{cg}
 
 Let us use the symbols ${\bf 1_S}$, ${\bf 1_A}$ and ${\bf 2_M}$ 
 to denote the symmetric, antisymmetric and 2 dimensional 
 irreducible representations of $S_3$.  Clearly
 \begin{equation}\label{cgoeff} 
 {\bf 1_S} \times {\bf R}={\bf R} 
 \end{equation}
 where ${\bf R}$ denotes any of the irreps   ${\bf 1_S}$, ${\bf 1_A}$ or ${\bf 2_M}$.
 On the other hand we have\footnote{The first two relations in \eqref{cgoeff1} are obvious. In order to see the last relation note that ${\bf 1_A} \times {\bf 2_M}$ is a two dimensional vector space. Let $a$ be the vector that transforms in the ${\bf 1_A}$ and let $(B_1, B_2)$ be the doublet of vectors that transform under $S_3$ according to \eqref{conb} and \eqref{ponb}. It is easily verified that the doublet of tensor product vectors 
 	$$\left( {\begin{array}{c}
 		-a B_1 \\
 		a B_2 \\
 		\end{array} } \right)
 	$$
 	also transforms under the permutation group precisely as listed in \eqref{conb} and \eqref{ponb}. It follows that the two dimensional vector space
 	${\bf 1_A} \times {\bf 2_M}$ transforms in the ${\bf 2_M}$ 
 	representation of $S_3$.}. 
 \begin{equation}\label{cgoeff1} 
 {\bf 1_A} \times {\bf 1_S}={\bf 1_A},~~~
 {\bf 1_A} \times {\bf 1_A}={\bf 1_S}, ~~~
 {\bf 1_A} \times {\bf 2_M}={\bf 2_M} 
 \end{equation}
 
 Finally we have 
 \begin{equation} \label{ttt}
 {\bf 2 } \times {\bf 2_M}= {\bf 1_S} + {\bf 1_A} + {\bf 2_M}
 \end{equation}
 To see how \eqref{ttt} works, let us use the tensor product 
 vectors $B_1B_1$, $B_1B_2$, $B_2B_1$ and $B_2B_2$ as a basis 
 for the four dimensional space ${\bf 2 }\times {bf 2}$. 
 The vector 
 \begin{equation}\label{singvec1}
 B_1B_2+B_2B_1
 \end{equation}
 transforms in the singlet ${\bf 1_S}$ representation. 
 The vector 
 \begin{equation}\label{singvec}
 B_1B_2-B_2B_1
 \end{equation}
 transforms in the ${\bf 1_A}$ representation. Finally the action of $S_3$ on the doublet of vectors 
 \begin{equation}\label{doubtwo}
 \left( {\begin{array}{c}
 	B_2 B_2 \\
 	B_1 B_1 \\
 	\end{array} } \right)
 \end{equation}
 is given precisely by the matrices listed in \eqref{conb} 
 and \eqref{ponb}. It follows that this doublet of vectors 
 transforms in the ${\bf 2_M}$ representation of $S_3$. 
 
 \subsection{Functions of 3 variables and the permutation group.}
 
 Consider a function of three variables $s$, $t$ and $u$. 
 Let the permutation group $S_3$ act on these three variables. 
 Given any particular function $f(s, t, u)$, the action of 
 the permutation group generates up to 5 new functions.
 
 If the original function was invariant under a subgroup of the 
 permutation group then we would obtain fewer than 5 new functions. Let us first suppose that this is not the case. In 
 this case the six dimensional linear vector space of the obtained functions transforms in a six dimensional representation of $S_3$. In fact the representation we find is 
 identical to that of the previous subsection (representation of 
 $S_3$ by left action on itself). So we obtain one copy of the 
 symmetric representation, one copy of the antisymmetric 
 representation and two copies of the two dimensional 
 representation. 
 
 Consider a general function  $f(s,t,u)$. Given any such function it is easy to break it up into a part that is completely symmetric, a part that is completely antisymmetric 
 and a part that lies somewhere in the (generically 4 dimensional) representation vector space of the two dimensional representations.  We have  
 \begin{equation}\label{fdecomp}\begin{split}
 &f(s,t,u)= f^{\rm sym}(s,t,u) + f^{\rm as}(s,t,u) + f^{\rm mixed}(s,t,u) \\
 &f^{\rm sym}(s,t,u)= P^{\rm sym} f =\frac16 (f(s,t,u) + f(t, u, s)+ 
 f(u, s, t) + f(t, s, u) + f(u, t, s) + f(s, u, t))\\
 &f^{\rm as}(s,t,u)= P^{\rm as} f= \frac16 (f(s,t,u) + f(t, u, s)+ 
 f(u, s, t) - f(t, s, u) - f(u, t, s) - f(s, u, t))\\
 &f^{\rm mixed}(s,t,u)= P^{\rm mixed}f = \frac13(2f(s,t,u) - f(t,u,s) -f(u,s,t))
 \end{split}
 \end{equation}
 It is easy to verify that $P^{\rm sym}$, $P^{\rm as}$ and
 $P^{\rm mixed}$ all square to themselves and so are  projectors.
 Moreover they project onto orthogonal subspaces, so that the 
 product of two non equal projectors vanishes. Finally, these 
 projectors commute with the action of the permutation group.
 The last equation in \eqref{fdecomp} asserts that the 
 polynomials that transform in the mixed representations 
 vanish under $\Z_3$ symmetrization as well as under complete
 symmetrization (these two facts imply these functions also 
 vanish under complete anti-symmetrization).

\section{$\Z_2\times \Z_2$ invariance}\label{projection}

In this Appendix we present a simple physics derivation of the 
formula \eqref{expzt}.
 
Consider any `single particle Hilbert space' with a basis 
single particle eigenstates $|i\rangle$ with definite values of the commuting charges $J_m$ and a single particle 
partition function 
\begin{equation}\label{sppf}
{\rm Tr}_{\rho} \left( \prod_m y_m^{J_m} \right) 
= \sum_i \langle i | \prod_m y_m^{J_m} | i \rangle = z(y_m)
\end{equation}

Next consider the Hilbert space of two identical bosons/fermions, each of whose single particle Hilbert space is $\rho$. Let the corresponding Hilbert spaces be denoted by $S^2\rho$ and $\wedge^2 \rho$ respectively. $S^2 \rho$ and 
$\wedge^2 \rho$ may, respectively, be thought of as the projection of the two (distinguishable) particle Hilbert spaces
onto the subspaces on which the permutation operator $P_{(12)}$ (exchange of the two particles) has eigenvalue $\pm 1$. It follows that  
\begin{equation}\label{spptt1}
\begin{split}
&{\rm Tr}_{S^2\rho} \left( \prod_m y_m^{J_m} \right)
=\sum_{i_1,i_2,} \langle i_i i_2  | \left( \prod_m y_m^{J_m} 
\right)
\left(\frac{ 1 + P_{(12)}}{2} \right)  | i_1 i_2 \rangle =\frac{z^2(y_m) + z(y_m^2)}{2}\\
&{\rm Tr}_{\Lambda^2\rho} \left( \prod_m y_m^{J_m} \right)
=\sum_{i_1,i_2,} \langle i_i i_2  | \left( \prod_m y_m^{J_m} 
\right)
\left(\frac{ 1 - P_{(12)}}{2} \right)  | i_1 i_2 \rangle =\frac{z^2(y_m) - z(y_m^2)}{2}\\
\end{split}
\end{equation}
where we have used the fact that  
\begin{equation}\label{maniporb}
\langle i_1 i_2 |\left( \prod_m y_m^{J_m} 
\right) P_{(12)} | i_1 i_2 \rangle=
\langle i_1 i_2 |\left( \prod_m y_m^{J_m} 
\right) | i_2 i_1 \rangle =
\delta_{i_1, i_2} \langle i_1 |\left( \prod_m (y^2_m)^{J_m} 
\right) | i_1 \rangle 
\end{equation}

Next consider the Hilbert space $\rho$  of four distinguishable particles, each of whose single particle state space is spanned by $|i\rangle$. The partition function over this Hilbert space is, of course, given by 
\begin{equation}\label{sppff1}
{\rm Tr}_{\rho^{\otimes 4}} \left( \prod_m y_m^{J_m} \right)
=\sum_{i_1,i_2, i_3, i_4} \langle i_i i_2 i_3 i_4 | \prod_m y_m^{J_m} | i_1 i_2 i_3 i_4 \rangle =z^4(y_m)
\end{equation}

Finally consider the same partition function but now over the four distinguishable particle Hilbert Space projected 
onto the subspace of  $\Z_2 \times \Z_2$ invariants, i.e. the 
space $\rho^{\otimes 4}|_{\Z_2\times \Z_2}$. We find 
\begin{equation}\label{sppff} \begin{split} 
&{\rm Tr}_{\rho^{\otimes 4}|_{\Z_2\times \Z_2}} \left( \prod_m y_m^{J_m} \right)=\sum_{i_1,i_2, i_3, i_4} \langle i_i i_2 i_3 i_4 | \left( \prod_m y_m^{J_m} \right) 
\left(\frac{ 1 + P_{(2143)} + P_{(3412)} + P_{(4321)}  }{4} 
\right)  | i_1 i_2 i_3 i_4 \rangle\\
=& \frac{1}{4} \sum_{i_1,i_2,i_3,i_4} \bigg( \langle i_1 i_2 i_3 i_4 | \prod_m y_m^{J_m} | i_1 i_2 i_3 i_4 \rangle 
+ \langle i_1 i_2 i_3 i_4 | \prod_m y_m^{J_m} | i_2 i_1 i_4 i_3 \rangle \\
&~~~~~~~~~~~ +\langle i_1 i_2 i_3 i_4 | \prod_m y_m^{J_m} | i_3 i_4 i_1 i_2 \rangle + \langle i_1 i_2 i_3 i_4 | \prod_m y_m^{J_m} | i_4 i_3 i_2 i_1 \rangle \bigg) \\
=& \frac{z^4(y_m) + 3z^2(y_m^2)}{4}\\
=& z^4(y_m) -3 \left(\frac{z^2(y_m)+z(y_m^2)}{2} \right) \times  \left(\frac{z^2(y_m)-z(y_m^2)}{2} \right)  
\end{split} 
\end{equation}
(In going from the second to the third line of \eqref{sppff} 
we have used manipulations similar to \eqref{maniporb}.)

Comparing \eqref{sppff}, \eqref{sppff1} and \eqref{spptt1} 
we conclude that
\begin{equation}\label{finres} 
{\rm Tr}_{\rho^{\otimes 4}|_{\Z_2\times \Z_2}} \left( \prod_m y_m^{J_m} \right)={\rm Tr}_{\rho^{\otimes 4}} \left( \prod_m y_m^{J_m} \right)-3{\rm Tr}_{S^2\rho} \left( \prod_m y_m^{J_m} \right){\rm Tr}_{\Lambda^2\rho} \left( \prod_m y_m^{J_m} \right)
\end{equation} 
an equation that can schematically be written in the form 
\eqref{expzt} i.e. 
\begin{equation}\label{finressimp}
\rho^{\otimes 4}|_{\Z_2\times \Z_2}
=\rho^{\otimes 4}- 3 S^2 \rho  \otimes \Lambda^2 \rho
\end{equation} 

In the context of the discussion around \eqref{expzt} the 
single particle Hilbert Space $\rho$ is the space of photon 
or graviton polarizations. $J_m$ are the $SO(D-3)$ Cartan 
charges. The projection onto $SO(D-3)$ singlets could be 
achieved by integrating \eqref{finres} over the $SO(D-3)$ 
Haar measure.

\section{Bare index structures in low dimensions}\label{appendix-is}
In this appendix we will complete the discussion of subsection \ref{constructionei} by constructing parity even and parity odd generators of the bare module for photons (called $e$'s and $o$'s respectively) and for gravitons for $D<7$. 

\subsection{Photons}
\subsection*{$D=6$:} In this case, the parity even bare structures 
are the same as those in \emph{i.e} $D\geq 7$ (i.e. continue to be given by 
the construction depicted in Fig. \ref{photon-structures} and so continue 
to transform in the representations listed in \eqref{photonlisting}). The parity odd structure which transforms in the  ${\bf 1_A}$ representation. This structure is simply 
\begin{equation} \label{fposi}
o_{\bf A}^{D=6}=N({\tilde\varepsilon}^3)_{\mu\nu\rho} \polo_1\,^\mu\polo_2\,^\nu\polo_3\,^\rho\,|_{\Z_2\times \Z_2}= \frac{{\varepsilon}_{\alpha\beta\gamma\mu\nu\rho}\, p_1^\alpha \,p_2^\beta \,p_3^\gamma }{\sqrt{stu}} \,\polo_1\,^\mu\polo_2\,^\nu\polo_3\,^\rho\,|_{\Z_2\times \Z_2}.
\end{equation}
Writing out the sum over $\Z_2 \times \Z_2$ orbits explicitly (and using 
$p_4=-p_1-p_2 -p_3$ to simplify the expressions), this structure is proportional to 
\begin{equation} \label{pophdsix} 
o_{\bf A}^{D=6}=\frac{{\varepsilon}_{\alpha\beta\gamma\mu\nu\rho}\, p_1^\alpha \,p_2^\beta \,p_3^\gamma}{\sqrt{stu}}\left( \polo_1\,^\mu\polo_2\,^\nu\polo_3\,^\rho\ \a_4 + \polo_2\,^\mu\polo_1\,^\nu\polo_4\,^\rho\ \a_3 + 
\polo_3\,^\mu\polo_4\,^\nu\polo_1\,^\rho\ \a_2+ \polo_4\,^\mu\polo_3\,^\nu\polo_2\,^\rho\ \a_1  \right) 
\end{equation}  
According to the rules spelt out in subsubsection \ref{etp},  \eqref{pophdsix} transforms in the completely antisymmetric 
representation of $S_4$ (and so, in particular, in the ${\bf 1_A}$ representation  $S_3= S_4/(\Z_2 \times \Z_2)$. 
\subsection*{$D=5$:} 

As above, the parity even bare structures 
are the same as those in $D\geq 7$ (i.e. continue to be given by 
the construction depicted in Fig. \ref{photon-structures} and so continue 
to transform in the representations listed in \eqref{photonlisting}).

We now turn to parity odd structures.  A priori, one could have tried to make parity odd structures using ${\tilde \varepsilon}^2_{\mu\nu}$. We need to be either in subsector  $\tv^{\otimes 4}$ or $\tv^{\otimes 2}\ts^{\otimes 2}$. We consider these two cases in turn. 

 In the first case, when we contract one pair of vectors using ${\tilde \varepsilon}$, the other needs to be contracted with Kronecker $\delta$. 
An example of such a term is the $Z_ 2 \times \Z_2$ symmetrized version of 
$$ \left( {\tilde \varepsilon}^2_{\mu\nu} 
\polo_1\,^\mu \polo_2\,^\nu \right) \left( \polo_3 . \polo_4 \right) $$
Explicitly performing the $\Z_2 \times \Z_2$ of this term we obtain 
\begin{equation}\label{algedem1}
{\tilde \varepsilon}^2_{\mu\nu}\Bigg( \polo_1\,^\mu \polo_2\,^\nu  \left( \polo_3 . \polo_4 \right)  + \polo_2\,^\mu \polo_1\,^\nu  \left( \polo_4 . \polo_3 \right)  + \polo_3\,^\mu \polo_4\,^\nu  \left( \polo_1 . \polo_2 \right)  
+ \polo_4\,^\mu \polo_3\,^\nu  \left( \polo_2 . \polo_1 \right)
\Bigg) =0
\end{equation} 
Consequently there are no $\Z_2 \times \Z_2$ symmetric expressions in  $\tv^{\otimes 4}$. 

The analysis for the second case is very similar. One can write down 
singlets in the sector $\tv^{\otimes 2}\ts^{\otimes 2}$: for example 
$$ \left( {\tilde \varepsilon}^2_{\mu\nu}
\polo_1\,^\mu \polo_2\,^\nu \a_3\a_4\right)$$
$\Z_2 \times \Z_2$ symmetrization kills these structures - the algebra that 
demonstrates this is very similar to \eqref{algedem1}. 

It follows that there are no parity odd photon scattering structures in 
$D=5$.

\subsection*{$D=4$:} In this case, there is a reduction in the number of parity even structures compared to the higher dimensions and also there are two parity odd structures. Let us first discuss the parity even case. In $D=4$, $\polo_i$ are vectors of $SO(1)$, they can simply be thought of as numbers. Given this, the three states of $e_{{\bf 3},1}$,
\be
(\polo_1\cdot\polo_2)(\polo_3\cdot\polo_4),\qquad (\polo_2\cdot\polo_3)(\polo_1\cdot\polo_4)\qquad (\polo_3\cdot\polo_1)(\polo_2\cdot\polo_4)
\ee
are now 
indistinguishable from each other and so transform in the ${\bf 1_S}$ 
representation. We denote this structure as $e_{{\bf 3}\to {\bf S}}$.

On the other hand the structure $e_{{\bf 3},2}$, 
which transformed in the ${\bf 3}$ for $D \geq 5$, continue to transform in  ${\bf 3}$ even in $D=4$. 
Similarly the symmetric  structure $e_{\bf S}$  continues to transform in the 
${\bf 1_S}$ in $D=4$ as it did for $D \geq 5$. 
In summary, the parity even bare photon structures in $D=4$ transform in the 
\begin{equation}\label{pebsdf}
{\bf 3} +2 \cdot {\bf 1_S}.
\end{equation}

We now turn to parity odd structures. All such structures
lie either in the $\tv \ts^{\otimes 3}$ or 
in the $\tv^{\otimes 3}\ts$ and  
are given by the $\Z_2 \times \Z_2$ symmetrization
\be \label{ztzt}
o_{{\bf S},1}^{D=4}=\left( N({\tilde \varepsilon})_{\mu} {\polo_{4\mu}} \right) \alpha_1\alpha_2\alpha_3|_{\Z2\times \Z2} \qquad
{\rm and} \qquad o_{{\bf S},2}^{D=4}=\left( N({\tilde \varepsilon})_{\mu} {\polo_{4\mu}} \right) \polo_{1\nu}\polo_{2\nu} \alpha_3||_{\Z2\times \Z2}. 
\ee
Explicitly performing the $\Z_2 \times \Z_2$ symmetrization we obtain 
\begin{equation}\label{algedem}
\begin{split} 
&o_{{\bf S},1}^{D=4}=N({\tilde {\varepsilon}})_{\mu}  \Bigg( \polo_1\,^\mu  
\alpha_2 \alpha_3 \alpha_4 + \polo_2\,^\mu \alpha_1 \alpha_3 \alpha_4  + \polo_3\,^\mu  \alpha_1 \alpha_2 \alpha_4  + \polo_4\,^\mu \alpha_1 \alpha_2 \alpha_3
\Bigg) \\
& o_{{\bf S},2}^{D=4}=N({\tilde {\varepsilon}})_{\mu} \Bigg( \polo_3\,^\mu  
\polo_1\,^\nu \polo_2\,^\nu \alpha_4 + \polo_1\,^\mu \polo_4\,^\nu \polo_3\,^\nu \alpha_2 + 
\polo_2\,^\mu  
\polo_3\,^\nu \polo_4\,^\nu \alpha_1 + \polo_4\,^\mu \polo_1\,^\nu \polo_2\,^\nu \alpha_3
\Bigg) 
\end{split} 
\end{equation}
The term in the first line of \eqref{algedem} is manifestly an $S_4$ (and so an $S_3$) singlet. Though it is less manifest, the same is also true of the term in the second line of \eqref{algedem}. This result follows upon
using the identity 
\begin{equation}\label{idemn}
\polo_i\,^\nu \polo_j\,^\nu \polo_k\,^\mu 
= \polo_j\,^\mu \polo_i\,^\nu \polo_k\,^\nu 
\end{equation}
(\eqref{idemn} holds because $\polo_i$ all point along the same direction in $D=4$). 
Using the rules of subsubsection \ref{etp}  the $D=4$ parity odd structures transform in the 
\begin{equation} \label{poddd4}
2 \cdot {\bf 1_S}.
\end{equation}

\subsection*{$D=3$:} In this case, $\polo_i$ does not exist. The S-matrix for photons becomes the same as the S-matrix for scalars. For scalars there is a single parity even structure as well as a single parity odd structure. The parity even and parity odd structures are,
\be
e_{\bf S}=1,\qquad\qquad  o_{\bf A}^{D=3}=\varepsilon_{\alpha\beta\gamma} p_1^\alpha p_2^\beta p_3^\gamma.
\ee
respectively. Remember that they have implicit factors of $\alpha_1\alpha_2\alpha_3\alpha_4$.

\subsection{Gravitons}
\subsection*{$D=6$:} 
The parity even structures for $D \geq 7$ were all constructed and diagrammatically
depicted in Fig. \ref{graviton-structures}. The structures depicted 
in Fig. \ref{graviton-structures} transformed in the representations 
listed in \eqref{gravstruct} which we reproduce for convenience
\begin{eqnarray} \label{gravstructa} 
29 &=& 10 \cdot {\bf 1_S}+ 9 \cdot {\bf 2_M}+1 \cdot {\bf 1_A}.
\end{eqnarray}

The parity even structures in $D=6$ differ from those in $D \geq 7$ in 
only one respect. The equation 
\be \label{sonstds}
(\polo_1\wedge\polo_2\wedge \polo_3\wedge\polo_4)^2=0.
\ee
holds in $D=6$ (though not for $D \geq 7$) because the four transverse polarizations lie in only three dimensions hence their anti-symmetrized product vanishes. The equation \eqref{sonstds} is parity even and clearly transforms 
in the ${\bf 1_S}$ representation. Since the equation involved terms  with 
eight factors of $\polo$, it follows that \eqref{sonstds} implies that a 
particular linear combination of the ${\bf 1_S}$ part of subfigures 1) and 2) in 
Fig. \ref{graviton-plethystic} vanish. It follows that the parity even 
structures in $D=6$ have one less ${\bf 1_S}$ structure than their counterpart
in $D \geq 7$. In summary, the parity even bare structures in 6 dimensions transform 
in the 
\begin{eqnarray} \label{gravstructa1} 
28 &=& 9 \cdot {\bf 1_S}+ 9 \cdot {\bf 2_M}+1 \cdot {\bf 1_A}.
\end{eqnarray}

As the ${\tilde \varepsilon}$ tensor has 3 free indices,  parity odd structures 
potentially appear in sectors with $3, 5$ or $7$ free indices (we need at least 3 free indices to saturate those in ${\tilde \varepsilon}$ and the rest can 
contract in pairs), i.e the subsectors $\tv^{\otimes 3}\ts, \tt \tv^{\otimes 3}, \tt^{\otimes 2}\tv\ts$ and  $\tt^{\otimes 3}\tv.$ Note also that all structures 
that appear in each of these sectors are necessarily 
parity odd (the odd number of indices can only all contract if there is a free ${\tilde \varepsilon}$).

We take up these subsectors 
in turn. We will first count the number of 
$SO(3)\times(\Z_2 \times \Z_2)$ invariants in each sector
separately and then explicitly construct these structures 
in order to deduce their $S_3$ representation counting.

In order to count the structures we proceed along the 
lines of Appendix \ref{projection}. To count the singlets 
in any given sector we evaluate a trace of the sort listed in the first of \eqref{sppff}. In this trace we allow 
the indices $i_1,~ i_2,~ i_3,~ i_4$ to range over all 
all values constrained by the net representation content. 
For instance in the channel $\tt^{\otimes 2}\tv\ts$ two 
of the four indices $i_1,~ i_2,~ i_3,~ i_4$ should range 
over the polarizations of the tensor, one of them should 
range over the polarizations of the vector while the last 
one should be the (unique) polarization of the scalar.
One possibility is that $i_1$ is the scalar, $i_2$ is the 
vector and $i_3$ and $i_4$ range over values for the tensor. Of course the $is$ can be distributed between 
representations in different ways. In this sector the total number of ways in which this  assignment can be made is $4\times 3=12$. In each of the other three sectors namely  $\tv^{\otimes 3}\ts, \tt \tv^{\otimes 3}$ and  $\tt^{\otimes 3}\tv$, it is easy to check that the $i$'s can be distributed between representations in 4 ways. 
Each of the different ways of assignment runs in the trace
in the first line of \eqref{sppff}. 

Now the first line of \eqref{sppff} involves inner products of states sandwiched with a $\Z_2 \times \Z_2$ 
permutation operator. It is easy to verify that {\it no} 
distribution of $i$'s among available representations 
is left invariant by {\it any} $\Z_2 \times \Z_2$ 
permutation; this is true for each choice of sectors. It follows that every matrix element involving a non-trivial permutation element in the first line of 
\eqref{sppff} vanishes. The only matrix elements that do 
not vanish are those involving the identity operator. 
It follows that the RHS of the first line of \eqref{sppff} can be replaced by 
$$ A \sum_{i_1,i_2, i_3, i_4} \langle i_i i_2 i_3 i_4 | \left( \prod_m y_m^{J_m} \right)   | i_1 i_2 i_3 i_4 \rangle $$
where $i_1,~ i_2,~ i_3,~ i_4$ are assigned to the 
representations in any given particular way and the 
number $A$ is the total number of ways of making this assignment divided by 4. In other words $A=3$ in 
the $\tt^{\otimes 2}\tv\ts$ sector but is equal to unity 
in the $\tv^{\otimes 3}\ts, \tt \tv^{\otimes 3}$ and  $\tt^{\otimes 3}\tv$ sectors. 

We thus conclude that the total number of simultaneous $SO(3)$ and $\Z_2 \times \Z_2$ singlets in any given 
sector is simply $A$ times the number of $SO(3)$  singlets in the Clebsch-Gordon decomposition of the 
corresponding representations (the CG decomposition is 
done without imposing any symmetry constraints). 
The procedure described above has a  generalization
to every $D$\footnote{When $D$ is odd, however, the separation into 
	parity even and parity odd structures is more difficult for the bare module as ${\tilde \varepsilon}$
	then has an even number of free indices.}. The case $D=6$ is particularly simple as in this case  $\tt,\tv,\ts$ are simply spin $2$, spin $1$ and spin $0$ representations of $SO(3)$ and their fusion rules are very familiar. 

\begin{figure}
	\centering
	\includegraphics[scale=0.35]{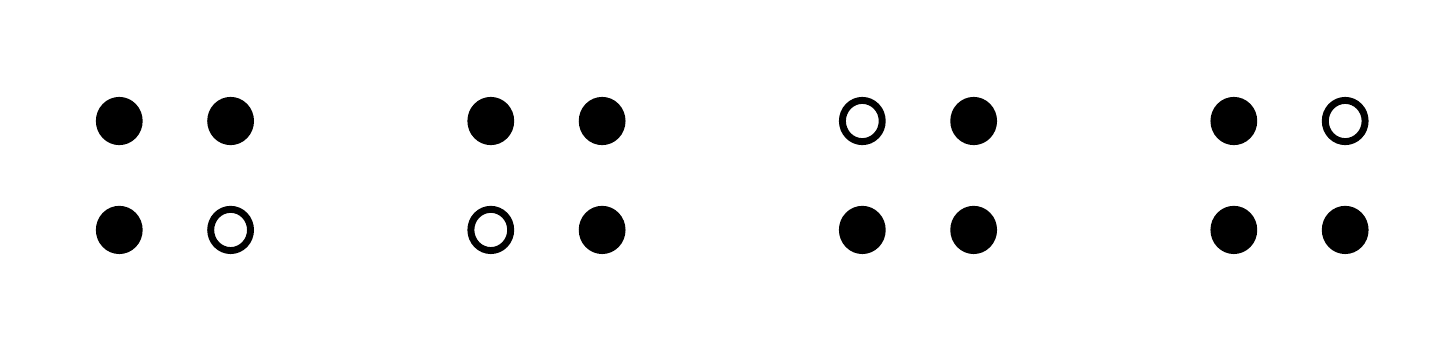}\qquad\qquad
	\includegraphics[scale=0.35]{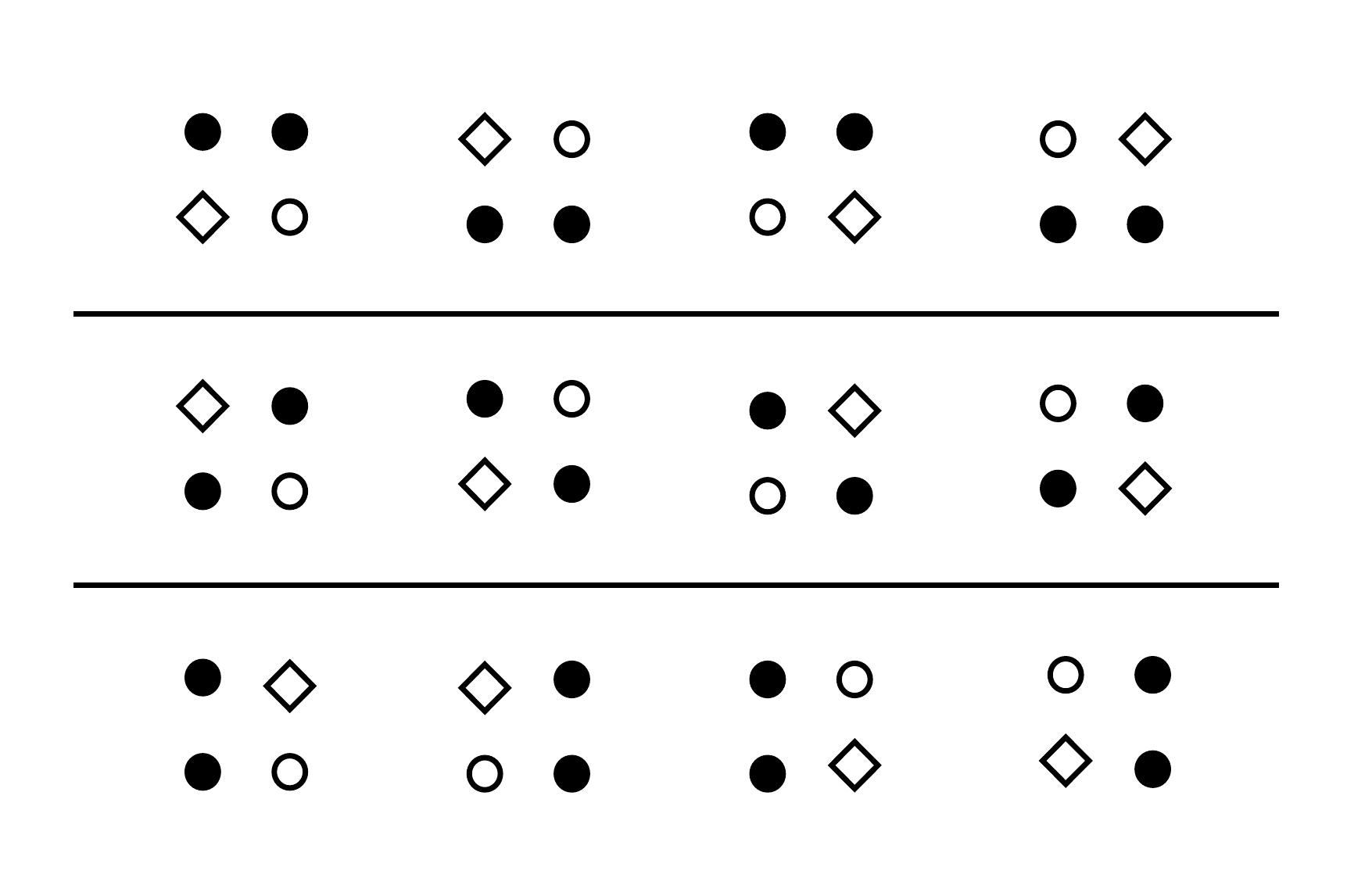}
	\caption{The left figure shows that there is a single $\Z_2\times \Z_2$ orbit in subsectors $\tt^{\otimes 3}\tv, \tt \tv^{\otimes 3}, \tv^{\otimes 3}\ts$. The right figure shows  the three distinct $\Z_2\times \Z_2$ orbits in the subsector $\tt^{\otimes 2}\tv\ts$.}
	\label{t2vs}
\end{figure}

There is another more combinatorial way of understanding the conclusion of the last paragraph. For  each of the subsectors $\tt^{\otimes 3}\tv, \tt \tv^{\otimes 3}, \tv^{\otimes 3}\ts$ all assignments of $is$ to 
representations are related by  $\Z_2\times \Z_2$ transforms. In the subsector  $\tt^{\otimes 2}\tv\ts$,
distinct assignments of $is$ to representations lie in three distinct $\Z_2\times \Z_2$ orbits. This is explained graphically in figure \ref{t2vs}. This is why number of 
simultaneous $SO(3)$ and  $\Z_2\times \Z_2$ singlets in $\tt^{\otimes 2}\tv\ts$ is thrice the number of CG singlets. 

We now consider the sectors one at a time. 
Consider  $\tv^{\otimes 3}\ts$. The number of CG singlets
in the product of 3 vectors is unity. As $A=1$ 
we have a single structure. It is easy to 
construct the corresponding structure; it is given by 
\begin{equation} \label{usds} \begin{split} 
&N \left( {\tilde \varepsilon}^3 \right)_{\mu \nu \rho}\ \Bigg( \polo_1\,^\mu \polo_2\,^\nu \polo_3 \,^\rho  
\alpha_1 \alpha_2 \alpha_3 \alpha_4^2 + \polo_2\,^\mu \polo_1\,^\nu \polo_4 \,^\rho  
\alpha_2 \alpha_1 \alpha_4 \alpha_3^2 + 
\polo_3\,^\mu \polo_4\,^\nu \polo_1 \,^\rho  
\alpha_3 \alpha_4 \alpha_1 \alpha_2^2  \\
&+ 
\polo_4\,^\mu \polo_3\,^\nu \polo_2 \,^\rho  
\alpha_4 \alpha_3 \alpha_2 \alpha_1^2 \Bigg). 
\end{split} 
\end{equation} 
According to the rules of subsubsection \ref{etp} this structure transforms in the ${\bf 1_A}$ representation. 

Next let us turn to the $\tt^{\otimes 3}\tv$. Once again 
in this case $A=1$ and the number of CG singlets is $3$ 
so we have $3$ structures. These structures are easily 
constructed; they are given by
\begin{equation}\label{strconst} \begin{split}
& N \left( {\tilde {\varepsilon}^{3}} \right)_{\mu \nu \rho} \Bigg( \polo_1\,^\mu \polo_2\,^\nu \polo_3 \,^\rho \alpha_4+ \polo_2\,^\mu \polo_1\,^\nu \polo_4 \,^\rho \alpha_3+\polo_3\,^\mu \polo_4\,^\nu \polo_1 \,^\rho \alpha_2+\polo_4\,^\mu \polo_3\,^\nu \polo_2 \,^\rho \alpha_1\Bigg) \left(\polo_1\,.\polo_2\,\polo_3.\polo_4\,\right),\nonumber\\ 
& N \left( {\tilde {\varepsilon}^{3}} \right)_{\mu \nu \rho} 
\Bigg( \polo_1\,^\mu \polo_3\,^\nu \polo_2 \,^\rho \alpha_4+ \polo_3\,^\mu \polo_1\,^\nu \polo_4 \,^\rho \alpha_2+\polo_2\,^\mu \polo_4\,^\nu \polo_1 \,^\rho \alpha_3+\polo_4\,^\mu \polo_2\,^\nu \polo_3 \,^\rho \alpha_1\Bigg) \left(\polo_1\,.\polo_3\,\polo_2\,.\polo_4\,\right),\nonumber\\
&N \left( {\tilde {\varepsilon}^{3}} \right)_{\mu \nu \rho} \Bigg( \polo_1\,^\mu \polo_4\,^\nu \polo_3 \,^\rho \alpha_2+ \polo_4\,^\mu \polo_1\,^\nu \polo_2 \,^\rho \alpha_3+\polo_3\,^\mu \polo_2\,^\nu \polo_1 \,^\rho \alpha_4+\polo_2\,^\mu \polo_3\,^\nu \polo_4 \,^\rho \alpha_1\Bigg) \left(\polo_1\,.\polo_4\,\polo_3\,.\polo_2\,\right)\nonumber\\
\end{split} 
\end{equation} 
Using the rules of subsubsection \ref{etp} the three structures in \eqref{strconst} clearly transform in ${\bf 3_A}$. 

In $\tt^{\otimes 2}\tv\ts$ we have $1$ CG singlet but
$A=3$ and so we have three structures. These structures 
transform in the ${\bf 3_A}$ representation and are given by 
\begin{eqnarray}\label{strconstb} 
& N \left( {\tilde {\varepsilon}^{3}} \right)_{\mu \nu \rho}\left( \left( \polo_1\,^\mu \polo_2\,^\nu \polo_3 \,^\rho \alpha_4^2\alpha_3 + \polo_2\,^\mu \polo_1\,^\nu \polo_4 \,^\rho \alpha_3^2\alpha_4\right)\polo_1\,.\polo_2+ \left( \polo_3\,^\mu \polo_4\,^\nu \polo_1\,^\rho \alpha_2^2\alpha_1\,+ \polo_4\,^\mu \polo_3\,^\nu \polo_2 \,^\rho \alpha_1^2\alpha_2\right)\right.\nonumber\\
&\left.\polo_3\,.\polo_4\right),\nonumber\\ 
&N \left( {\tilde {\varepsilon}^3} \right)_{\mu \nu \rho}\left( \left( \polo_1\,^\mu \polo_3\,^\nu \polo_2 \,^\rho \alpha_4^2\alpha_2 + \polo_3\,^\mu \polo_1\,^\nu \polo_4 \,^\rho \alpha_2^2\alpha_4\right)\polo_1\,.\polo_3+ \left( \polo_2\,^\mu \polo_4\,^\nu \polo_1\,^\rho \alpha_3^2\alpha_1\,+ \polo_4\,^\mu \polo_2\,^\nu \polo_3 \,^\rho \alpha_1^2\alpha_3\right)\right.\nonumber\\
&\left.\polo_2\,.\polo_4\right),\nonumber\\
&N \left( {\tilde {\varepsilon}^3} \right)_{\mu \nu \rho}\left( \left( \polo_1\,^\mu \polo_4\,^\nu \polo_3 \,^\rho \alpha_2^2\alpha_3 + \polo_4\,^\mu \polo_1\,^\nu \polo_2 \,^\rho \alpha_3^2\alpha_2\right)\polo_1\,.\polo_4+ \left( \polo_3\,^\mu \polo_2\,^\nu \polo_1\,^\rho \alpha_4^2\alpha_1\,+ \polo_2\,^\mu \polo_3\,^\nu \polo_4 \,^\rho \alpha_1^2\alpha_4\right)\right.\nonumber\\
&\left.\polo_3\,.\polo_2\right),\nonumber\\  
\end{eqnarray}

 Turning to the $\tt \tv^{\otimes 3}$ sector, $A=1$ and there are $2$ CG singlets so we have 2 structures.  As it is clear that there is no $\tt$ in the totally symmetric or anti-symmetric cube of $\tv$, these $2$ structures necessarily transform in ${\bf 2_M}$ representation. 
 Explicitly they take the form,
 
 \begin{eqnarray}\label{strconstc}
 & N \left( {\tilde {\varepsilon}}^3 \right)_{\mu \nu \rho}\left( \left( \polo_1\,^\mu \polo_4\,^\nu \polo_3 \,^\rho \alpha_4\alpha_3\alpha_2 + \polo_2\,^\mu \polo_3\,^\nu \polo_4 \,^\rho \alpha_3\alpha_4\alpha_1\right)\polo_1\,.\polo_2+\left( \polo_4\,^\mu \polo_1\,^\nu \polo_2 \,^\rho \alpha_3\alpha_2\alpha_1 + \polo_3\,^\mu \polo_2\,^\nu \polo_1 \,^\rho \alpha_2\alpha_1\alpha_4\right)\right.\nonumber\\
 &\left.\polo_3\,.\polo_4\right),\nonumber\\ 
 & N \left( {\tilde {\varepsilon}}^3 \right)_{\mu \nu \rho}\left( \left( \polo_1\,^\mu \polo_4\,^\nu \polo_2 \,^\rho \alpha_4\alpha_2\alpha_3 + \polo_3\,^\mu \polo_2\,^\nu \polo_4 \,^\rho \alpha_2\alpha_4\alpha_1\right)\polo_1\,.\polo_3+\left( \polo_4\,^\mu \polo_1\,^\nu \polo_3 \,^\rho \alpha_2\alpha_3\alpha_1 + \polo_2\,^\mu \polo_3\,^\nu \polo_1 \,^\rho \alpha_3\alpha_1\alpha_4\right)\right.\nonumber\\
&\left.\polo_4\,.\polo_2\right),\nonumber\\  
 & N \left( {\tilde {\varepsilon}}^3 \right)_{\mu \nu \rho}\left( \left( \polo_1\,^\mu \polo_2\,^\nu \polo_3 \,^\rho \alpha_2\alpha_3\alpha_4 + \polo_4\,^\mu \polo_3\,^\nu \polo_2 \,^\rho \alpha_3\alpha_2\alpha_1\right)\polo_1\,.\polo_4+\left( \polo_2\,^\mu \polo_1\,^\nu \polo_4 \,^\rho \alpha_3\alpha_4\alpha_1 + \polo_3\,^\mu \polo_4\,^\nu \polo_1 \,^\rho \alpha_4\alpha_1\alpha_2\right)\right.\nonumber\\
 &\left.\polo_2\,.\polo_3\right),\nonumber\\  
 \end{eqnarray} 
 
In summary, the parity odd part of the $D=6$ bare module has 9 structures which transform under $S_3$ as 
\be \label{repcontentd6}
9=3\cdot {\bf 1_A}\oplus 3\cdot{\bf 2_M}.
\ee

\subsection*{$D=5$:} 

Let us discuss parity even case first. As the transverse symmetry is $SO(2)$ it is best to consider $\tt,\tv$ and $\ts$ as states with charge $\pm2, \pm 1$ and $0$ respectively. In the asymptotic dimension, in the $\tt^{\otimes 4}$ subsector, there are six parity even $\Z_2\times \Z_2$ symmetric singlets as denoted in the first two lines of figure \ref{graviton-structures}; these transformed in the $2\cdot {\bf 3} = 2\cdot{\bf 2_M} + 2\cdot{\bf 1_S}$. In the previous 
subsection the relationship \eqref{sonstds} removed  one of these ${\bf 1_S}$
representations leaving us with $2\cdot{\bf 2_M} + {\bf 1_S}$. The number of states
in the $\tt^{\otimes 4}$ sector is further reduced as we now see. In $D=5$ all states in this sector have charges $\pm 2$ states. The only neutral combination has charge assignments $2, 2, -2,-2$. This set of charges has a 3 dimensional 
orbit which transform in ${\bf 3}$ of $S_3$\footnote{The tensor structures in question can be taken to be either that represented in subfig 1) or subfig 2) of Fig \ref{graviton-structures}
(these two tensor structures, which are distinct in 
$D\geq 7$ are proportional to each other in $D=5$; 
this is what leads to the reduction of $6$ to $3$ 
structures). }.

In the same way the neutral charge assignments the sector  $\tt^{\otimes 2}\tv^{\otimes 2}$ is $2,-2, 1, -1$.  We have 6 neutral combinations of 
these charges that transform in the ${\bf 6_{\rm left}}$ under $S_3$.  
The number $6$ is three less than the 9 parity invariant 
structures in the $\tt^{\otimes 2}\tv^{\otimes 2}$ 
sector in  $D \geq 6$ (see subfigs 3) and 4) ) in 
Fig. \ref{graviton-structures})\footnote{ In this case the independent 
tensor structures can be taken to be that depicted in 
subfig 4) of Fig. \ref{graviton-structures} (the tensor 
structure of subfig 3) is linearly related to that of 
subfig 4) in $D=5$ - this accounts for the reduction 
of structures from $9$ to $6$.}.

Next, the subsector $\tt^{\otimes 3} s$ clearly doesn't contain any singlets in $D=5$ (recall that this sector had one singlet for $D \geq 6$ , see sub fig 5)  of Fig. \ref{graviton-structures}. 

It is easy to do a similar counting in the sectors $\tt^{\otimes 2} s^{\otimes 2}$, $\tv^{\otimes 2} s^{\otimes 2}$, $\tv^{\otimes 4}$ and $s^{\otimes 4}$. 
In these sectors we find $3$, $3$, $3$ and $1$ states respectively. All these 
results are the same as for $D \geq 6$; the corresponding structures are all  
parity even and all transform in the same representations of $S_3$ 
(and in fact are given by the same expressions) as for $D \geq 6$. 

The chief new result is in the sector $\tt \tv^{\otimes 2} s$. In this sector 
we have two possible charge assignments; the $(2, -1, -1, 0)$ and the 
$(-2, 1, 1, 0)$. One of these charge assignments generate a ${\bf 3_S}$ and the other ${\bf 3_A}$. 
giving us a total of six states, twice the number in $D \geq 6$. 3 of these 
states are parity even, transform in the ${\bf 3}$ and are depicted in 
subfig 7) of Fig \ref{graviton-structures}.  The other ${\bf 3_A}$ is parity 
odd and is given by the explicit expression

\begin{eqnarray}\label{hgnb}
& {\tilde {\varepsilon}}^2_{\mu \nu}\left( \left( \polo_4\,^\mu \polo_2\,^\nu \alpha_2\alpha_1\alpha_3^2 + \polo_1\,^\mu \polo_3\,^\nu \alpha_3\alpha_4\alpha_2^2\right)\polo_4\,.\polo_1+\left( \polo_2\,^\mu \polo_4\,^\nu  \alpha_4\alpha_3\alpha_1^2 + \polo_3\,^\mu \polo_1\,^\nu  \alpha_1\alpha_2\alpha_4^2\right)\polo_2\,.\polo_3\right),\nonumber\\
&-({\tilde {\varepsilon}}^2_{\mu \nu}\left( \left( \polo_4\,^\mu \polo_3\,^\nu \alpha_3\alpha_1\alpha_2^2 + \polo_1\,^\mu \polo_2\,^\nu \alpha_2\alpha_4\alpha_3^2\right)\polo_4\,.\polo_1+\left( \polo_3\,^\mu \polo_4\,^\nu  \alpha_4\alpha_2\alpha_1^2 + \polo_2\,^\mu \polo_1\,^\nu  \alpha_1\alpha_3\alpha_4^2\right)\polo_3\,.\polo_2\right)),\nonumber\\
& -({\tilde {\varepsilon}}^2_{\mu \nu}\left( \left( \polo_2\,^\mu \polo_4\,^\nu \alpha_4\alpha_1\alpha_3^2 + \polo_1\,^\mu \polo_3\,^\nu \alpha_3\alpha_2\alpha_4^2\right)\polo_2\,.\polo_1+\left( \polo_4\,^\mu \polo_2\,^\nu  \alpha_2\alpha_3\alpha_1^2 + \polo_3\,^\mu \polo_1\,^\nu  \alpha_1\alpha_4\alpha_2^2\right)\polo_4\,.\polo_3\right)).\nonumber\\
\end{eqnarray} 
In summary, in $D=5$ we have 22 parity even structures which transform in 
\begin{eqnarray} \label{gravstructa2} 
22 &=& 7 \cdot {\bf 1_S}+ 7 \cdot {\bf 2_M}+1 \cdot {\bf 1_A}.
\end{eqnarray}
On the other hand we have 3 parity odd structures which transform in the 
\begin{eqnarray} \label{gravstructa3} 
3 &=& {\bf 1_A}+  {\bf 2_M}
\end{eqnarray}

\subsection*{$D=4$:}

As in the case of photons, the graviton polarizations are essentially numbers. Hence the result for the counting of structures is the same as the one in the case of photons. Physically this is to be expected since the massless particles in $D=4$ have two helicities regardless of their spin. More concretely, due the fact  that $\polo_i\,$s are basically numbers, any appearance of the quantity $\polo_i\,^2$ in the bare module can be replaced by $\alpha_i^2$ due to the tracelessness of the graviton fluctuations (see eqn \eqref{constrem}). Recall that in order to enumerate the possible $SO(D-3)$ singlets in subsection \ref{enumeration}, we had decomposed the effective polarization as $\rho = (\ts \oplus \tv \oplus \tt)$ (see \eqref{cgpg}). In $D=4$, the $\tt$ part can be replaced by $\ts$ using \eqref{constrem}. Hence at the level of group theoretic enumeration undertaken in subsection \ref{enumeration}, the counting for gravitons is the same as that of photons. In terms of explicit bare module structures, consider the following two possible graviton bare module structures in $D=4$
\begin{eqnarray}
(\polo_1\cdot\polo_2) \alpha_1\alpha_2\alpha_3^2\alpha_4^2|_{\Z_2 \otimes \Z_2 }, \qquad (\polo_1\cdot\polo_2)(\polo_2\cdot\polo_3) \alpha_1\alpha_3\alpha_4^2|_{\Z_2 \otimes \Z_2 }
\end{eqnarray} 
Using \eqref{constrem}, we can replace $\polo_2\,^2$ appearing in the second structure by $\a_2^2$ and hence is equal to one of the structures present in the $S_3$ orbit of the first structure. The bare module for the gravitons in $D=4$ therefore is given by that of the photons after the replacement $\polo_i \rightarrow \alpha_i\polo_i$ and $\alpha_i \rightarrow \alpha_i^2$. 
In equations the parity even module is given by, 
\begin{eqnarray}
&&g_{{\bf 3} \to {\bf S}}^{D=4}=(\polo_1\,^\mu \polo_2\,^\mu)(\polo_3\,^\nu\polo_4\,^\nu) \alpha_1\alpha_2\alpha_3\alpha_4, \nonumber\\
\nonumber\\
&&g_{{\bf 3},2}^{(1)}=( \polo_1\,^\mu \polo_2\,^\nu\, \a_1 \a_2\alpha_3^2 \alpha_4^2 + \polo_3\,^\mu\polo_4\,^\mu \a_3\a_4 \alpha_1^2 \alpha_2^2), \qquad g_{{\bf 3},2}^{(2)}=( \polo_1\,^\mu \polo_3\,^\nu\, \a_1 \a_3\alpha_2^2 \alpha_4^2 + \polo_2\,^\mu\polo_4\,^\mu \a_2\a_4 \alpha_1^2 \alpha_3^2) \nonumber\\
&&g_{{\bf 3},2}^{(3)}=( \polo_1\,^\mu \polo_4\,^\nu\, \a_1 \a_4\alpha_3^2 \alpha_2^2 + \polo_3\,^\mu\polo_2\,^\mu \a_3\a_2 \alpha_1^2 \alpha_4^2)\nonumber\\
\nonumber\\
&&g_{{\bf S},1}=(\a_1^2\a_2^2\a_3^2\a_4^2)
\end{eqnarray}
Similarly the parity odd module is generated by,
\be 
h_{{\bf S},1}^{D=4}=\left( N({\tilde \varepsilon})_{\mu} {\polo_{4\mu}} \right) \alpha_1^2\alpha_2^2\alpha_3^2\a_4|_{\Z2\times \Z2} \qquad
{\rm and} \qquad h_{{\bf S},2}^{D=4}=\left( N({\tilde \varepsilon})_{\mu} {\polo_{4\mu}} \right) \polo_{1\nu}\polo_{2\nu} \a_1\a_2\a_4\alpha_3^2||_{\Z2\times \Z2}. 
\ee

\section{Details concerning photon Lagrangians}\label{A}

\subsection{Triviality of $3$-F structures}
In this appendix we prove that any $3$-F structure and its descendants identically vanish when we impose equations of motion, and Bianchi identity.
we write them here for convenience
\begin{equation}
\partial_{\mu} F_{\mu\nu}=0, \ \ \text{and} \ \ \ \partial_aF_{bc}+\partial_bF_{ca}+\partial_cF_{ab}=0
\end{equation} 
Both together imply the following, as shown
\begin{equation}\label{res1}
\begin{split}
&\partial_{\mu} \partial_{\mu} F_{ab}=0\\
\text{Using Bianchi Identity :}\ \text{L.H.S.}\ \ \ &= \ \ \partial_{\mu}(-\partial_a F_{b\mu}-\partial_{b}F_{\mu a})\\
\text{Using E.o.M.  :}\ \ \ & = 0
\end{split}
\end{equation}
Now, let's consider a general descendant term, where some derivatives are contracted among themselves. If they act on the same $F$, \eqref{res1} shows that any such term is just $0$. In case they act on different $F$s, we use momentum conservation to write it as
\begin{equation}
k^{i}_{\mu}k^{j}_{\mu} = \frac{1}{2}\left((k^{s})^2-(k^{i})^2-(k^j)^2\right)
\end{equation}
where $s$ in the superscript is for the particle label apart from $i,j$. Each of the three terms on r.h.s. are $0$ by \eqref{res1}.  Therefore, all the descendants of any $3$-F structure are $0$.
Next we consider the contraction of derivatives with $F$s.  
\subsubsection{$6$ derivatives}
There is only one term that one can write down. Firstly, both the indices of an $F$ can not contract with the same particle momenta, because of anti-symmetry of $F$. Therefore, one index of $F^1$ contracts with $k^2$ and and other with $k^3$ \footnote{the possibility that both indices of $F^1$ contract with $F^2$ or $F^3$ is equivalent to this up to momentum conservation. }. Similarly for $F^2$ and $F^3$. The term is
\begin{equation}
k^1_c k^1_e F^1_{ab}k^2_a k^2_f F^2_{cd}k^3_b k^3_d F^3_{ef}
\end{equation}
Under $1\leftrightarrow 2$ this is antisymmetric, therefore 0.
\subsubsection{4 derivatives}
Next consider a basic structure with 4 derivatives. As there are a total of 4 derivatives, both free indices of at least one field strength must contract with derivatives. Actually both 
free indices of exactly one field strength must contract with 
derivatives\footnote{If the four were to contract with the 4 indices of 2 field strengths then the indices of the remaining field strength would have to contract with each other, and that is not allowed.}. Let this special field strength be $F^3$. 
By use of momentum conservation we can choose the momenta that contract the indices of $F^3$ to be $k^1$ and $k^2$, again no contraction with the same $k^i$. One momentum index must also contract with one of the indices of 
$F^1$ - we can use momentum conservation to set this momentum 
to either $k_1$ or $k_3$. Similarly we can choose to ensure that the momentum that contracts $F^2$ is either $k_2$ or $k_3$. This gives us a total of 4 basic structures 
\begin{equation}\label{bst}
\begin{split} 
& F^{3}_{\mu\nu}  \left( k^1_{\mu}k^2_{\nu} k^1_{a}k^2_{b}F^1_{a \beta}F^2_{b \beta}  \right) \\
& F^{3}_{\mu\nu}  \left( k^1_{\mu}k^2_{\nu} k^3_{a}k^3_{b}F^1_{a \beta}F^2_{b \beta}  \right) \\
& F^{3}_{\mu\nu}  \left( k^1_{\mu}k^2_{\nu} \left(  k^1_{a}k^3_{b} + k^3_{a}k^2_{b} \right) F^1_{a \beta}F^2_{b \beta}  \right) \\
& F^{3}_{\mu\nu}  \left( k^1_{\mu}k^2_{\nu} \left(  k^1_{a}k^3_{b} -k^3_{a}k^2_{b} \right) F^1_{a \beta}F^2_{b \beta}  \right) \\
\end{split}
\end{equation}
The expressions in the first three lines of \eqref{bst} involve
the contraction of $F^3_{\mu\nu}$ with structures that are 
symmetric under $1 \leftrightarrow 2$. These structures all vanish after integrating over $k_1$ and $k_2$.  On the other 
hand the structure in the last line  of \eqref{bst} 
vanishes because of equations of motion  ($k^1_a F^1_{a\beta}=0$ and $k^2_b F^2_{b \beta}=0$).
\subsubsection{2 derivatives}
First, is the case when both the derivatives are contracted with the same $F$, say $F^3$. There is one possible term, up to momentum conservation
\begin{equation}
\begin{split}
&k^1_{\mu}k^2_{\nu}F^1_{ab}F^2_{ba}F^3_{\mu\nu}\\
\end{split}
\end{equation}
Again, indices of $F$ can not contract with the same particle momenta. This term is anti-symmetric under $1\leftrightarrow 2$, and therefore 0.

Next, consider the case when both the derivatives are contracted with different $F$s, say $F^1$ and $F^2$. Similar to the argument for $4$ derivative case, there are $4$ terms
\begin{equation}\label{bst1}
\begin{split} 
& F^{3}_{\mu\nu}  \left( k^1_{a}k^2_{b}F^1_{a \mu}F^2_{b \nu}  \right) \\
& F^{3}_{\mu\nu}  \left(  k^3_{a}k^3_{b}F^1_{a \mu}F^2_{b \nu}  \right) \\
& F^{3}_{\mu\nu}  \left( \left(  k^1_{a}k^3_{b} + k^3_{a}k^2_{b} \right) F^1_{a \mu}F^2_{b \nu}  \right) \\
& F^{3}_{\mu\nu}  \left( \left(  k^1_{a}k^3_{b} -k^3_{a}k^2_{b} \right) F^1_{a \mu}F^2_{b \nu}  \right) \\
\end{split}
\end{equation}
As for the case of $4$ derivatives, the first three of these terms are 0 because of anti-symmetry under $1 \leftrightarrow 2$ and the last one is 0 because of equations of motion.

\subsection{Local photon Lagrangians and  polynomial S-matrices} \label{smatpho}

In order to make contact with the S-matrices described earlier in this paper we now specialize to Lorentz gauge $\partial.A=0$. In this case the Maxwell equation reduces to\footnote{While $A_\mu$ is harmonic only in Lorentz gauge, 
	it follows from the Bianchi identity and the Maxwell 
	equations that 
	$\partial^2F_{\mu\nu}=0$ in any gauge. Consequently if we 
	choose to work in a gauge invariant manner, Lagrangians that differ from each other by terms that include a factor of $\partial^2F_{\mu\nu}=0$ are 
	also in the same equivalence class.} 
\begin{equation} \label{meq}
\partial^2 A_\mu =0, ~~~~~~~\partial.A=0.
\end{equation} 

In this gauge it follows that different $L_4$ belong to the same equivalence class if they differ only by total derivatives once one imposes \eqref{meq} 
In momentum space
\begin{equation}\label{ftms1} \begin{split}
A_\mu(x)&= \int \frac{d ^dk}{(2 \pi)^d} e^{i k.x } ~\epsilon_\mu(k) \\
L_4&= \int \prod_i \frac{d ^dk}{(2 \pi)^d} e^{i \left( \sum_{j} k_j x_j \right) } ~{\tilde L}_4^{\mu_1 \mu_2 \mu_3 \mu_4} (k_1, k_2 , k_3, k_4) ~\epsilon_{\mu_1}(k_1) \epsilon_{\mu_2}(k_2) \epsilon_{\mu_3}(k_3) \epsilon_{\mu_4}(k_4)\\
&k_i. \epsilon_i =0, ~~~~(k_1)_\mu {\tilde L}_4^{\mu_1 \mu_2 \mu_3 \mu_4} (k_1, k_2 , k_3, k_4)=0 ~~~~{\rm similar ~eq ~for ~1\leftrightarrow i}
\end{split} 
\end{equation}
The equations of motion \eqref{meq} and momentum conservation 
merely impose \eqref{condits}. As  ${\tilde L}_4^{\mu_1 \mu_2 \mu_3 \mu_4} (k_1, k_2 , k_3, k_4)$, subject to \eqref{condits}, is the gauge invariant tree level S-matrix obtained from our theory, it follows immediately that equivalence classes of $L_4$ are labelled by their S-matrices as in the scalar case studied in the previous subsubsection. We believe\footnote{\eqref{ftms1} can be re expressed entirely in terms 
	of Field strengths using \eqref{epsilondecomp}, \eqref{perpp}, \eqref{manifest} and \eqref{perpinv}. 
	It should be possible to demonstrate that resulting expression is a local functional of 
	Field strengths (at least in parity invariant situations). 
	We have not undertaken this exercise; we leave its completion to the interested reader.}
that every local gauge invariant S-matrix defines an equivalence class of local 
and {\it manifestly gauge invariant} quartic Lagrangians 
$L_4$. It follows that local 4-photon S-matrices are 
in one to one correspondence with equivalence classes $L_4$ in \eqref{mjpha}. In subsection \ref{pleth} later in this section we turn to the problem of enumerating inequivalent quartic 
Lagrangians $L_4$, and so, effectively, inequivalent 4 photon S-matrices.

\section{S-matrices from Lagrangians no more than cubic in $R_{\alpha \beta \gamma \delta}$}

\subsection{Identities of the Riemann tensor}
\paragraph{ } In this section we list all the identities and symmetry properties of the Riemann tensor that we will be using in the subsequent analysis to determine the diffeomorphism invariant Lagrangian structures.
\begin{enumerate}\label{riemannidentity}
	\item The symmetry properties of the Riemann tensor,
	\begin{equation}
	R_{abcd}=-R_{bacd}=-R_{abdc}, \qquad R_{abcd}=R_{cdab}
	\end{equation}
	\item The algebraic Bianchi identity 
	\begin{equation}
	R_{abcd} + R_{acdb} +R_{adbc} =0 
	\end{equation} 
	\item The differential Bianchi identity
	\begin{equation} \label{dbi}
	\nabla_a R_{bcde} + \nabla_b R_{cade} + \nabla_c R_{abde} =0 
	\end{equation}
	\item The contracted Bianchi identity, which we get from the differential Bianchi identity by contracting the Riemann tensor appropriately.
	\begin{equation}
	\nabla_a R_{ce} + \nabla^b R_{cabe} - \nabla_c R_{ae} =0 
	\end{equation}
	\item Commutator of derivatives 
	\begin{equation}
	[\nabla_f, \nabla_e] R_{abcd} = R_{pbcd}R^p_{afe}+ R_{apcd}R^p_{bfe}+R_{abpd}R^p_{cfe}+R_{abcp}R^p_{dfe}  
	\end{equation}
	
\end{enumerate}
We will be using these identities and symmetry properties of the Riemann tensor along with integration by parts judiciously to fix the independent Lagrangian terms. 

Through this Appendix we focus only on results that are 
true in every dimension; in particular we ignore dimension dependent structures like $\epsilon$ and never make use of special identities for the Riemann tensor that work only in 
special dimensions. 

\subsection{Terms quadratic in $R_{\alpha \beta \gamma \delta}$}

The unique Lagrangian that is linear in $R_{\alpha \beta \gamma \delta}$ is clearly the familiar two derivative Einstein Lagrangian itself.

Let us now turn to terms built out of two 
copies of the Riemann tensor. One Lagrangian of this form 
is the Gauss-Bonnet Lagrangian \eqref{gblag} which we reproduce here for convenience:
\begin{equation} \begin{split}  \label{gblaga}
S_{GB} &= \int \sqrt{-g} \,\, \delta_{[a}^{g}\delta_{b}^{h}\delta_{c}^{i}\delta_{d]}^{j}\,\, R_{ab}^{\phantom {ab}gh} R_{cd}^{\phantom{cd}ij}\\
& \propto
\int \sqrt{-g}\left(R^2 -4R^{\mu \nu}R_{\mu\nu}+R^{\mu\nu\rho\sigma}R_{\mu\nu\rho\sigma}\right).
\end{split}
\end{equation}
Since we are classifying terms up to `equations of motion' 
this Lagrangian is equivalent to 
\begin{equation}\label{tworeim}
S= \int \sqrt{-g} R_{abcd} R^{abcd}
\end{equation}
We will now argue that every other Lagrangian term 
quadratic in two Riemann tensors is equivalent to 
\eqref{tworeim}, plus terms that are cubic or higher 
order in $R_{abcd}$. 

To see how this works, we first note that the symmetries 
of the Riemann tensor and the algebraic Bianchi identity 
ensure that the second potentially independent index 
contraction of 4 derivative two Riemann terms 
\begin{eqnarray} \label{ununu}
S= \int \sqrt{-g} R_{abcd} R^{acbd}
\end{eqnarray}
is in fact 
proportional to \eqref{tworeim}.
Using \eqref{riemannidentity}, we can systematically relate \eqref{ununu}  to \eqref{tworeim} as follows
\begin{eqnarray}
R_{abcd} R^{acbd} &=& R_{abcd} \left(-R^{abdc}-R^{adcb}\right) \nonumber\\ 
&=& R_{abcd} R^{abcd} - R_{abcd} R^{acbd}\nonumber\\
\therefore R_{abcd} R^{acbd} &=& \frac{1}{2} R_{abcd} R^{abcd}
\end{eqnarray} 
where in the first line we have used the algebraic Bianchi identity and in the second line we have used the symmetry properties of the Riemann tensor.

We next turn two Riemann terms involving derivatives. 
In all such terms the derivative indices are either contracted with indices in one of the Riemann tensor, or 
with some other derivative. In the first case an integration 
by parts can always be used to have the derivative whose 
index contracts with (say) the $a$ index of $R_{abcd}$, 
act on (derivatives of)  $R_{abcd}$ itself.
We can then move $\nabla_a$ through all the other derivatives acting on $R_{abcd}$ - along the way generating three Riemann terms (the extra factor of Riemann comes from commuting derivatives) which we are allowed at this stage to ignore - until we obtain an expression that is some derivative of $\nabla^a R_{abcd}$. The contracted  Bianchi 
identity \eqref{dbi}  then sets this term equal to expressions involving 
$R_{\mu\nu}$ which are field redefinition trivial. We present
two examples of such manipulations
\begin{itemize} 
	\item \begin{eqnarray}
	S&=&\int \nabla_a R_{bcfe} \nabla^b R^{acfe} \nonumber\\
	&=& -\int \left(\nabla^b \nabla_a R_{bcfe}\right) R^{acfe} + C_{\partial\cal{M}} \nonumber\\
	&=& \hat{C}_{R_{\mu\nu}} + C_{\mathcal{\partial M}} + \tilde{C}_{R^3} \nonumber
	\end{eqnarray}
\item 
\begin{eqnarray}
S &=& \int \sqrt{g} \nabla_a \nabla_b R_{ecfd} \nabla^e \nabla^f R^{acbd} \nonumber\\
&=& \int \sqrt{g} \nabla^e \nabla^f \nabla_a \nabla_b R_{ecfd} R^{acbd} + C_{\mathcal{\partial M}}\nonumber\\
&=& \hat{C}_{R_{\mu\nu}} + C_{\mathcal{\partial M}} + \tilde{C}_{\nabla  R \nabla R R } \nonumber
\end{eqnarray}
\end{itemize} 
(here ${\hat C}_{R_{\mu\nu}}$ denotes a Lagrangian term 
which has at least one factor of $R_{\mu\nu}$ or its derivatives. $C_{\mathcal{\partial M}}$ denotes a total 
derivative and $\tilde{C}_{R^3}$ denotes a term that is 
cubic or higher order in Riemann tensors). 

Next, terms in which derivative indices are contracted with one 
another can be converted into the first kind of terms 
(those in which derivative indices are contracted with 
an a free index in the Riemann tensor) by use of the differential 
Bianchi identity. For example  
\begin{eqnarray}
S&=& \int \sqrt{g} \nabla_\mu R_{abcd} \nabla^\mu R^{acbd} \nonumber\\
&=& \int \sqrt{g} \left( -\nabla_a R_{b\mu cd}-\nabla_b R_{\mu acd} \right) \nabla^\mu R^{acbd} \nonumber\\
&=& \int \sqrt{g} \nabla^\mu \left( -\nabla_a R_{b\mu cd}-\nabla_b R_{\mu acd} \right) R^{acbd} + C_{\mathcal{\partial M}}\nonumber\\
&=& \int \sqrt{g} \left( -\nabla_a \nabla^\mu R_{b\mu cd}-\nabla_b \nabla^\mu  R_{\mu acd} \right) R^{acbd} + C_{\mathcal{\partial M}} + \tilde{C}_{R^3} \nonumber\\
&=& \hat{C}_{R_{\mu\nu}} + C_{\mathcal{\partial M}} + \tilde{C}_{R^3} \nonumber
\end{eqnarray}
where in the second line we have used the differential Bianchi identity. In the third line we have used integration by parts to get the total derivative term $C_{\mathcal{\partial M}}$. In the fourth line we have used commutator of derivatives to get the $\tilde{C}_{R^3}$ term which is of higher order in Riemann tensor and hence will be dealt with when we classify terms with three Riemann tensors. Finally we use the contracted Bianchi identity to relate this to the Ricci tensor term $\hat{C}_{R_{\mu\nu}}$. 

For another example of such terms consider 
\begin{eqnarray}
S &=& \int \sqrt{g} \nabla_\mu \nabla_\nu R_{abcd} \nabla^\mu \nabla^\nu R^{acbd} \nonumber\\
&=& \int \sqrt{g} \nabla_\nu \nabla^\mu\nabla_\mu  R_{abcd} R^{acbd} + C_{\partial\cal{M}} \nonumber\\
&=& \int \sqrt{g} \nabla_\nu \nabla^\mu\nabla_\mu  \left(-\nabla_a R_{b\nu cd}-\nabla_b R_{\nu a cd}\right) R^{acbd} + C_{\partial\cal{M}} \nonumber\\ 
&=& \hat{C}_{R_{\mu\nu}} + C_{\mathcal{\partial M}} + \tilde{C}_{\nabla  R \nabla R R } \nonumber
\end{eqnarray}

Similar manipulations can be used to show that all 
expressions involving derivatives of two Riemann tensors 
can be turned into terms involving three or more Riemann 
tensor, up to terms that can be removed by field redefinitions. 

\subsection{Terms cubic in the Riemann tensor}

Finally let us consider terms built out of 
three copies of the Riemann tensor. Let us 
first study terms with no derivatives. 
It is easy to convince oneself that there 
are two independent terms of this form (\cite{Fulling:1992vm}).
\begin{eqnarray}
R^{pqrs}R_{pq}^{\phantom{pq}tu}R_{rstu}, \qquad R^{pqrs}R_{p\phantom{t}r}^{\phantom{p}t\phantom{r}u}R_{qtsu}
\end{eqnarray}
All possible non-zero index contractions of three Riemann tensors to yield scalars can be obtained from this two by repeated application of the symmetry properties and the algebraic Bianchi identities. Let us consider an explicit example: $R_{pq}^{\phantom{pq}rs}R_{rtsu}R^{tu}_{\phantom{tu}pq}$ 
\begin{eqnarray}
R_{pq}^{\phantom{pq}rs}R_{rtsu}R^{tu}_{\phantom{tu}pq} &=& R_{pq}^{\phantom{pq}rs}(-R_{rsut}-R_{ruts})R^{tu}_{\phantom{tu}pq}\nonumber\\
&=& R_{pq}^{\phantom{pq}rs}R_{rstu}R^{tu}_{\phantom{tu}pq}-R_{pq}^{\phantom{pq}rs}R_{rtsu}R^{tu}_{\phantom{tu}pq} \nonumber\\
\therefore R_{pq}^{\phantom{pq}rs}R_{rtsu}R^{tu}_{\phantom{tu}pq} &=& \frac{1}{2}R_{pq}^{\phantom{pq}rs}R_{rstu}R^{tu}_{\phantom{tu}pq}
\end{eqnarray}
In the second line we have used the algebraic Bianchi identity and subsequently used the symmetry properties of the Riemann tensor. Similar manipulations can be performed  for any non-zero contraction of three Riemann tensors (and no additional derivatives) to reduce it to one of the two forms or a linear combination of both. 

We note that the combination 
\begin{equation}
R^{pqrs}R_{pq}^{\phantom{pq}tu}R_{rstu}-2R^{pqrs}R_{p\phantom{t}r}^{\phantom{p}t\phantom{r}u}R_{qtsu}, 
\end{equation}
which up to terms proportional to  $R_{\mu\nu}$
is proportional to the second Lovelock term 
\begin{equation}
\begin{split}
\chi_6 &= \frac{1}{8}\epsilon_{abcdef}\epsilon^{ghijkl}R_{ab}^{\phantom {ab}gh} R_{cd}^{\phantom{cd}ij}R_{ef}^{\phantom{ef}kl}\\
&=4 R_{ab}^{\phantom{ab}cd}R_{cd}^{\phantom{cd}ef}R_{ef}^{\phantom{ef}ab}-8R_{a\phantom{c}b}^{\phantom{a}c\phantom{b}d}R_{c\phantom{e}d}^{\phantom{c}e\phantom{d}f}R_{e\phantom{a}f}^{\phantom{e}a\phantom{f}b}-24R_{abcd}R^{abc}_{\phantom{abc}e}R^{de}+3R_{abcd}R^{abcd}R\\
&\quad \quad \quad \quad +24 R_{abcd}R^{ac}R^{bd}+16R_{a}^{\phantom{a}b}R_{b}^{\phantom{b}c}R_{c}^{\phantom{c}a}-12R_{a}^{\phantom{a}b}R_{b}^{\phantom{b}a}R+R^3
\end{split}
\end{equation}  
does not contribute to the graviton 3 point function. 
This may be verified using the fact that with our standard choice of polarization, the on-shell Riemann tensor 
evaluates to 
\begin{eqnarray}
R_{abcd} &=&\frac{1}{2} F_{ab}F_{cd} \nonumber\\
F_{ab}&=&(k_a e_b -k_b e_a), \qquad h_{ab} = e_ae_b
\end{eqnarray} 
an expression that is both gauge invariant and satisfies the necessary symmetry properties of the Riemann tensor (also - on-shell - both the algebraic Bianchi identity and the differential Bianchi identity are satisfied).

We now turn to a discussion of terms with derivatives on three copies of the Riemann tensor. The basis for structures with two derivatives on three Riemann tensors have been discussed in \cite{Fulling:1992vm}.
\begin{eqnarray}
R^{pqrs}R_{p}^{\phantom{p}tuv}\nabla_v \nabla_s R_{qtru}, \qquad R^{pqrs}\nabla_q R_{\phantom{tuv}p}^{tuv}\nabla_s R_{tuvr} \nonumber\\
R^{pqrs}\nabla_r R_{\phantom{tuv}p}^{tuv}\nabla_s R_{tuvq}, \qquad R^{pqrs}\nabla^v R^{t\phantom{p}u}_{\phantom{t}p\phantom{u}r}\nabla_v R_{tqus}
\end{eqnarray}

We will now demonstrate that all such terms are field redefinition equivalent to terms involving four or more 
Riemann tensors.
\begin{enumerate}
	\item[1] \begin{eqnarray}
	\int \sqrt{g} R^{pqrs}R_{p}^{\phantom{p}tuv}\nabla_v \nabla_s R_{qtru} &=& \int \sqrt{g} R^{pqrs}R_{p}^{\phantom{p}tuv}\nabla_v (-\nabla_r R_{qtus}-\nabla_u R_{qtsr})\nonumber\\
	&=&-\int \sqrt{g} R^{pqsr}R_{p}^{\phantom{p}tuv}\nabla_v \nabla_r R_{qtsu} + C'_{R^4} \nonumber\\
	\therefore \int \sqrt{g} R^{pqrs}R_{p}^{\phantom{p}tuv}\nabla_v \nabla_s R_{qtru} && \sim  \frac{1}{2}C'_{R^4}
	\end{eqnarray}
	\item[2] \begin{eqnarray}
	\int \sqrt{g} R^{pqrs}\nabla_r R_{\phantom{tuv}p}^{tuv}\nabla_s R_{tuvq} &=& -\int \sqrt{g} R^{pqrs}\nabla_s \nabla_r R_{\phantom{tuv}p}^{tuv} R_{tuvq}  + C_{\partial\cal{M}}\nonumber\\
	&&\sim C'_{R^4}+ C_{\partial\cal{M}}\nonumber\\
	\end{eqnarray}
	\item[3] \begin{eqnarray}
	\int \sqrt{g} R^{pqrs}\nabla^v R^{t\phantom{p}u}_{\phantom{t}p\phantom{u}r}\nabla_v R_{tqus} &=& -\int \sqrt{g}(\nabla^v R^{pqrs} R^{t\phantom{p}u}_{\phantom{t}p\phantom{u}r}\nabla_v R_{tqus} +  R^{pqrs} R^{t\phantom{p}u}_{\phantom{t}p\phantom{u}r}\nabla^v\nabla_v R_{tqus})+ C_{\partial\cal{M}}\nonumber\\
	\int \sqrt{g} R^{pqrs}\nabla^v R^{t\phantom{p}u}_{\phantom{t}p\phantom{u}r}\nabla_v R_{tqus} &=& \frac{-1}{2}\int \sqrt{g} R^{pqrs} R^{t\phantom{p}u}_{\phantom{t}p\phantom{u}r}\nabla^v\nabla_v R_{tqus} + C_{\partial\cal{M}}\nonumber\\ 
	&&\sim \hat{C}_{R_{\mu\nu}}+  C'_{R^4}+ C_{\partial\cal{M}}\nonumber\\ 
	\end{eqnarray}
	\item[4] \begin{eqnarray}
	\int \sqrt{g} R^{pqrs}\nabla_q R_{tuvp}\nabla_s R^{tuv}_{\phantom{tuv}r} &=& -\int \sqrt{g} R^{pqrs}(\nabla_v R_{tupq}+\nabla_p R_{tuqv})\nabla_s R^{tuv}_{\phantom{tuv}r}\nonumber\\
	\int \sqrt{g} R^{pqrs}\nabla_q R_{tuvp}\nabla_s R^{tuv}_{\phantom{tuv}r} &=& \frac{-1}{2} \int \sqrt{g} R^{pqrs}\nabla_v R_{tupq}\nabla_s R^{tuv}_{\phantom{tuv}r}\nonumber\\ 
	&=&  \frac{-1}{2} \int \sqrt{g} R^{pqrs}\nabla_v R_{tupq}(-\nabla^v R^{tu}_{\phantom{tu}rs}-\nabla_r R^{tu\phantom{s}v}_{\phantom{tu}s})\nonumber\\ 
	\int \sqrt{g} R^{pqrs}\nabla_q R_{tuvp}\nabla_s R^{tuv}_{\phantom{tuv}r} &=& \frac{1}{4} \int \sqrt{g} R^{pqrs}\nabla_v R_{tupq}\nabla^v R^{tu}_{\phantom{tu}rs}\nonumber\\
	&& \sim \hat{C}_{R_{\mu\nu}}+  C'_{R^4}
	\end{eqnarray}
\end{enumerate} 
Although we have only explicitly demonstrated the triviality 
of derivative terms involving two derivatives above, 
the method used to demonstrate this triviality is general. In order to demonstrate this, we have used the differential Bianchi identity and total derivatives judiciously to bring these structures to the schematic form (say) $$R^{pqrs}R_{p}^{\phantom{p}tuv}\nabla_v \nabla_u R_{qtsr}$$
As the reader can see this is a higher point function due to the antisymmetry of the $\nabla_v$  and $\nabla_u$.
This extends to higher derivative terms as well. 

\section{Single letter index for photons and gravitons}\label{single-letter-sum}

In section \ref{pleth} and \ref{pleth-spin}, we have argued that the single letter indices for scalars, photons and gravitons are,
\begin{eqnarray} \label{all-series}
i_{\ts}(x,y)&=& 1+x\,\chi_{\syng{1}}+x^2\, \chi_{\syng{2}}+x^3\, \chi_{\syng{3}}+x^4\, \chi_{\syng{4}}+x^5\, \chi_{\syng{5}}+\ldots \nonumber \\
i_{\tv}(x,y)&=& x\,\chi_{\syng{1,1}}+x^2\, \chi_{\syng{2,1}}+x^3\, \chi_{\syng{3,1}}+x^4\, \chi_{\syng{4,1}}+x^5\, \chi_{\syng{5,1}}+\ldots \nonumber \\
i_{\tt}(x,y)&=& x^2\, \chi_{\syng{2,2}}+x^3\, \chi_{\syng{3,2}}+x^4\, \chi_{\syng{4,2}}+x^5\, \chi_{\syng{5,2}}+\ldots
\end{eqnarray}
respectively. For scalars, it is easy to see that this series is summed to
\be
i_{\ts}(x,y)=(1-x^2)\denom(x,y) \nonumber
\ee
where $\denom(x,y)$ is a function given in \eqref{scalar-single}. In this section, we will sum the series corresponding to photons and gravitons.
\begin{eqnarray}
i_{\ts}(x,y) (x \,\chi_{\syng{1}})&=&(x \,\chi_{\syng{1}}) (1+x\,\chi_{\syng{1}}+x^2\, \chi_{\syng{2}}+x^3\, \chi_{\syng{3}}+x^4\, \chi_{\syng{4}}+\ldots) \nonumber\\
&=& x\,\chi_{\syng{1}}+x^2\, \chi_{\syng{2}}+x^3\, \chi_{\syng{3}}+x^4\, \chi_{\syng{4}}+\ldots \nonumber\\
&& \qquad \quad x^2+x^3\,\chi_{\syng{1}}+x^4\, \chi_{\syng{2}}+x^5\, \chi_{\syng{3}}+x^6\, \chi_{\syng{4}}+\ldots \nonumber\\
&&  \qquad \quad  x^2\,\chi_{\syng{1,1}}+x^3\, \chi_{\syng{2,1}}+x^4\, \chi_{\syng{3,1}}+x^5\, \chi_{\syng{4,1}}+x^6\, \chi_{\syng{5,1}}+\ldots
\end{eqnarray}
The second equality is organized such that first line is the traceless symmetrized product, second line is the trace and the third line is the antisymmetric product. Comparing to \eqref{all-series},
\begin{eqnarray}
i_{\ts}(x,y) (x \,\chi_{\syng{1}})&=& (i_{\ts}(x,y)-1)+x^2(i_{\ts}(x,y))+x \,i_{\tv}(x,y),\nonumber\\
\Rightarrow x\,i_\tv(x,y)&=&((x-x^3)\chi_{\syng{1}}-(1-x^4))\denom(x,y)+1.
\end{eqnarray}

We proceed in the same way to sum the graviton single letter index.
\begin{eqnarray}
i_{\ts}(x,y) (x^2 \,\chi_{\syng{2}})&=&(x^2 \,\chi_{\syng{2}}) (1+x\,\chi_{\syng{1}}+x^2\, \chi_{\syng{2}}+x^3\, \chi_{\syng{3}}+x^4\, \chi_{\syng{4}}+\ldots) \nonumber\\
&=& x^2\, \chi_{\syng{2}}+x^3\, \chi_{\syng{3}}+x^4\, \chi_{\syng{4}}+\ldots \nonumber\\
&& \qquad \quad \,\,\, x^3\,\chi_{\syng{1}}+x^4\, \chi_{\syng{2}}+x^5\, \chi_{\syng{3}}+x^6\, \chi_{\syng{4}}+\ldots \nonumber\\
&& \qquad \quad\,\, \, x^3\, \chi_{\syng{2,1}}+x^4\, \chi_{\syng{3,1}}+x^5\, \chi_{\syng{4,1}}+x^6\, \chi_{\syng{5,1}}+\ldots \nonumber\\
&&  \qquad \qquad \qquad \quad \,\, x^4\,\chi_{\syng{1,1}}+x^5\, \chi_{\syng{2,1}}+x^6\, \chi_{\syng{3,1}}+x^7\, \chi_{\syng{4,1}}+\ldots\nonumber\\
&& \qquad \qquad \qquad \quad \,\, x^4\, \chi_{\syng{2,2}}+x^5\, \chi_{\syng{3,2}}+x^6\, \chi_{\syng{4,2}}+x^7\, \chi_{\syng{5,2}}+\ldots\nonumber\\
&& \qquad \qquad \qquad \quad \,\, x^4\, +x^5\, \chi_{\syng{1}}+x^6\, \chi_{\syng{2}}+x^7\, \chi_{\syng{3}}+\ldots
\end{eqnarray}
In the second equality, the first line is the product that is symmetrized in both indices. The second line is where one index is contracted and other is symmetrized. Third is where one index is symmetrized and other anti-symmetrized. In the fourth line one index is contracted and other is anti-symmetrized. In the fifth line both the indices are anti-symmetrized and finally in the sixth line both indices are contracted. Using \eqref{all-series}, this equality becomes,
\begin{eqnarray}
i_{\ts}(x,y) (x^2 \,\chi_{\syng{2}})&=& (i_{\ts}(x,y)-1-x \,\chi_{\syng{1}})+x^2(i_{\ts}(x,y)-1)+x \,(i_{\tv}(x,y)-x\,\chi_{\syng{1,1}})\nonumber\\
&& +x^3 \,i_{\tv}(x,y) +x^2 \,i_{\tt}(x,y) +x^4 \,i_{\ts}(x,y) ,\nonumber\\
\Rightarrow x^2 i_{\tt}(x,y) &=& ((x^2-x^4)(1+\chi_{\syng{2}})-(x-x^5)\chi_{\syng{1}})\denom(x,y)+x^2\chi_{\syng{1,1}}+x\chi_{\syng{1}}.
\end{eqnarray}

\section{Evaluating plethystic integrals}\label{epi} 
\paragraph{}In this section, we present the details of the Haar integral (see \eqref{singlet-proj}) for scalars, photons and gravitons. The Haar integral that we need to evaluate is the following 

\be
I^D_j(x):=\oint \prod_{i=1}^{\lfloor D/2\rfloor}dy_i \,\Delta(y_i)\,  i_j^{(4)}(x,y)/\denom(x,y).
\ee 
where $i_j^{(4)}(x,y)$ denotes the four particle partition function and $j$ denotes whether we are considering the partition function for scalars, photons or gravitons. The integral over $y_i$ in \eqref{singlet-proj} is a closed anti-clockwise circular contour about $y_i=0$. From \eqref{4-particle}, $i_j^{(4)}(x,y)$ can be expressed in terms of the single letter partition function,
\be
i_j^{(4)}(x,y)=\frac{1}{24}\Big(i^4_j(x,y) +6 i^2_j(x,y) i_j(x^2,y^2)+3i^2_j(x^2,y^2)+8i_j(x,y)i_j(x^3,y^3)+6i_j(x^4,y^4)\Big).
\ee 
The quantity $\denom(x,y)$ is given by   
\begin{eqnarray}
\denom(x,y) &=&\Big(\prod_{i=1}^{D/2}(1-x y_i)(1-x y_i^{-1})\Big)^{-1}\qquad \qquad \qquad \qquad{\rm for \,\, D\,\, even}\nonumber \\
&=&\Big((1-x)\prod_{i=1}^{\lfloor D/2\rfloor}(1-x y_i)(1-x y_i^{-1})\Big)^{-1}\qquad \qquad \,\,{\rm for \,\, D\,\, odd}.
\end{eqnarray}  
where $y_i$ are the charges under the cartan subgroup of $SO(D)$. $\Delta(y_i)$ is the Van der Monde determinant for $SO(D)$. For even dimensions ($D=2N$), the Haar measure is given by, 

\begin{equation}\label{haare}
\Delta_e(y_i) =\frac{2 \left(\prod _{j=1}^{N} \left(\prod _{i=1}^{j-1} \left(y_i+\frac{1}{y_i}-y_j-\frac{1}{y_j}\right)\right)\right)^2}{(2\pi i)^N 2^NN!\prod_{i=1}^{N}y_i}
\end{equation} 
For odd dimensions ($D=2N+1$), the Haar measure is given by,                         
\begin{equation}\label{haaro}
\Delta_o(y_i) =\frac{\left(\prod _{k=1}^N \left(1-y_k-\frac{1}{y_k}\right)\right) \left(\prod _{j=1}^N \left(\prod _{i=1}^{j-1}  \left(y_i+\frac{1}{y_i}-y_j-\frac{1}{y_j}\right)\right)\right)^2}{(2\pi i)^N N!\prod_{i=1}^{N}y_i}
\end{equation}
and the integral over $y_i$ in \eqref{singlet-proj} is a closed circular contour about $y_i=0$.

\subsection{$D\geq 10$}

We expect the result of the plethystic integral to stabilize for $D \geq D^*$ 
where $D^*$ is a critical integer. From numerical experiments (i.e. evaluating 
the plethystic integrals on ${\mathtt {Mathematica}}$ in a power series in $x$ to high 
orders in $x$ - see the next subsection) we have found strong evidence that 
$D^*=10$. Assuming this is the case, the plethystic integral for $D \geq 10$ 
can be evaluated in the large $D$ limit. In this subsection we will proceed to 
perform this evaluation. To be specific we work with the case $D=2N$ (the case 
$D=2N+1$ turns out to give the same answer). 

Setting $y_i= e^{i \theta_i}$ and ignoring overall constants, we  note that the Haar measure exponentiates as follows. 
\begin{eqnarray}\label{largeNmeasure}
&&\prod_{1\leq i<j\leq N} \left( \cos_{\theta_j}-\cos_{\theta_i} \right)^2 \nonumber\\
&&\propto \prod_{1\leq i<j\leq N}  \sin^2 \frac{\theta_i+\theta_j}{2}\sin^2 \frac{\theta_i-\theta_j}{2} \nonumber\\
&&\propto \prod_{1\leq i<j\leq N}  \Big|  e^{\frac{i(\theta_i+\theta_j)}{2}}-e^{\frac{-i(\theta_i+\theta_j)}{2}}\Big|^2\Big|e^{\frac{i(\theta_i-\theta_j)}{2}}-e^{\frac{-i(\theta_i-\theta_j)}{2}}\Big|^2 \nonumber\\
&&\propto \left(\prod_{1\leq i< j\leq N}\Big|  1-e^{-i(\theta_i+\theta_j)}\Big|\Big|  1-e^{i(\theta_i+\theta_j)}\Big|\right)~\left(\prod_{1\leq i< j\leq N}\Big|  1-e^{i(\theta_j-\theta_i)}\Big|\Big|  1-e^{-i(\theta_j-\theta_i)}\Big|\right)
\nonumber\\
&&\propto \frac{\left(\prod_{1\leq i\leq j\leq N}\Big|  1-e^{-i(\theta_i+\theta_j)}\Big|\Big|  1-e^{i(\theta_i+\theta_j)}\Big|\right)~\left(\prod_{1\leq i< j\leq N}\Big|  1-e^{i(\theta_j-\theta_i)}\Big|\Big|  1-e^{-i(\theta_j-\theta_i)}\Big|\right)}{\prod_{i=1}^N \Big|1-e^{2i\theta_i} \Big|\Big|1-e^{-2i\theta_i} \Big|}\nonumber\\
&&\propto e^{-\frac{1}{n}\sum_n\left( \sum_{1\leq i \leq j\leq N} \left(e^{i(\theta_i+\theta_j)}+e^{-i(\theta_i+\theta_j)}\right)+\sum_{1\leq i < j\leq N} \left(e^{i(\theta_j-\theta_i)}+e^{-i(\theta_j-\theta_i)}\right)-\sum_{i =1}^N \left(e^{2i\theta_i}+e^{-2i\theta_i}\right)\right)}\nonumber\\
&&\propto e^{-\frac{1}{2n}\sum_n\left(({\rm Tr}{\cal O}^n)^2-{\rm Tr}{\cal O}^{2n}\right)}
\end{eqnarray}
where $O$ is the orthogonal matrix in diagonal form 
$${\cal O} \sim\left( \begin{matrix}
y_1 & 0 &0 &\cdots &0  \\
0 & \frac{1}{y_1} &0 &\cdots &0\\
0 & 0 & y_2 &\cdots &0 \\
0 & 0 &0 &\frac{1}{y_2}\cdots &0\\
\cdots&\cdots& \cdots& \cdots&\cdots\\
0 & 0 &\cdots&y_N &0\\
0 & 0 &0 &\cdots &\frac{1}{y_{N}}\\
\end{matrix}\right)$$
In going from the fourth line to the fifth line of \eqref{largeNmeasure} 
we have converted the range of the product  in the first term in the numerator
from  $i<j$ to $i \leq j$ (and correspondingly divided out the extra terms); 
this manipulation was needed in order to rewrite the numerator as a trace
after exponentiation.

As an aside it is interesting to compare the measure \eqref{largeNmeasure} with the measure 
for a $U(2N)$ matrix integral, evaluated for a matrix whose eigenvalues 
happen to fall into complex conjugate pairs as is necessarily the case for 
an orthogonal matrix. In other words consider a  $U(2N)$ matrix with eigenvalues 
$e^{i \alpha_m}$ where $\alpha_1= \theta_1$, $\alpha_2= -\theta_1$, $\alpha_3= \theta_2$, $\ldots \alpha_{2N}= -\theta_N$. It is well  known that the Haar measure 
for a Unitary matrix integral is proportional to 
\be \begin{split} \label{unmeas}
&\prod_{m \neq n}\Big|  1-e^{-i(\alpha_m-\alpha_n)}\big| \\
&= e^{-\frac{1}{n}\sum_n {\rm Tr}{U}^n {\rm Tr}{U}^{-n}  }\\
\end{split} 
\ee
We note that at leading order in the large $N$ limit, (i.e. at order $N^2$) the $O(2N)$ measure \eqref{largeNmeasure} is the square root of the $U(2N)$ measure. 
At first sub-leading order (i.e. order $N$), however, the $O(2N)$ measure 
has an extra contribution - the terms proportional to ${\rm Tr} O^{2m}$ in 
the last line of \eqref{largeNmeasure} that have no analogue in the case of 
$U(2N)$ matrix integrals. The additional terms in the $O(2N)$ measure 
come from exponentiating  denominator in the 5th line of \eqref{largeNmeasure}\footnote{To see this completely explicitly we note that once we insert the values of $\alpha_m$ described above into  \eqref{unmeas} it turns into   
\begin{equation} 
\begin{split} 
& \left(\prod_{1\leq i\leq j\leq N}\Big|  1-e^{-i(\theta_i+\theta_j)}\Big|\Big|  1-e^{i(\theta_i+\theta_j)}\Big|\right)^2~\left(\prod_{1\leq i< j\leq N}\Big|  1-e^{i(\theta_j-\theta_i)}\Big|\Big|  1-e^{-i(\theta_j-\theta_i)}\Big|\right)^2 \\
\end{split} 
\end{equation} }.
The physical origin of this this term lies in the fact that for every eigenvalue 
$\theta_i$, the $O(2N)$ matrix automatically has another eigenvalue $-\theta_i$; 
there is no volume factor proportional to $|1-e^{2 i \theta_i}|$ associated 
with this happenstance. From a more formal mathematical point of view, 
in \eqref{largeNintegrand} the trace 
over the adjoint representation (antisymmetric matrices) of $O(2N)$ 
$$ \frac{1}{2}\left(({\rm Tr}{\cal O})^2-{\rm Tr}{\cal O}^{2} \right) $$
replaces the trace over the $U(N)$ adjoint representation  
$$ ({\rm Tr}{U})({\rm Tr}{U})^\dagger $$
in \eqref{unmeas}. The order $N$ contribution to the $O(2N)$ measure 
will play an important role in our evaluation below.

 In the large $N$ limit, $({\rm Tr}{\cal O}^n)^2$ scales as $N^2$ while  ${\rm Tr}{\cal O}^{2n}$ scales as $N$. For this reason we find it convenient to 
 work below with the order unity  variables $\rho_n$  by the relations 
$$({\rm Tr}{\cal O}^n)^2 =N^2\rho_n^2,\qquad {\rm Tr}{\cal O}^{2n} = N\rho_{2n}$$  

We will now explain how the $O(2N)$ matrix integrals that appear in the 
plethystic integrals can be explicitly evaluated in the large $N$ limit.  For simplicity we present all details for how this works only for the plethystic
integral over scalar Lagrangian; the photon and graviton 
integrals work in a similar way. 

The four-particle scalar partition function is given by, 
\begin{eqnarray}\label{multpartN}
i_s^{(4)}(x,y)&=&\frac{1}{24}\Big(i^4_s(x,y) +6 i^2_s(x,y) i_s(x^2,y^2)+3i^2_s(x^2,y^2)+8i_s(x,y)i_s(x^3,y^3)+6i_s(x^4,y^4)\Big) \nonumber\\
i_s(x,y) &=& (1-x^2)\denom(x,y)
\end{eqnarray}
The Haar integrand for the scalar case becomes,
\begin{eqnarray}\label{largeNintegrand}
i_s^{(4)}(x,y)&=&\frac{1}{24} \left(\frac{6 \denom(x^4,y^4)}{\denom\left(x,y\right)}-6 \left(x^2-1\right)^3 \left(x^2+1\right)\denom(x,y) \denom\left(x^2,y^2\right)+\left(x^2-1\right)^4\denom(x,y)^3-\frac{6 x^8 \denom(x^4,y^4)}{\denom\left(x,y\right)}\right.\nonumber\\
&&\left. +\frac{3 \left(x^4-1\right)^2 \denom(x^2,y^2)^2}{\denom\left(x,y\right)} +8 \left(x^2-1\right)^2 \left(x^4+x^2+1\right)\denom\left(x^3,y^3\right)\right)
\end{eqnarray}

The quantity $\denom \left(x^n,y^n\right)^m$ can be exponentiated as follows.
\begin{eqnarray}
\denom\left(x^n,y^n\right)=e^{\sum_{m}\frac{N\rho_{mn}x^{mn}}{m}}=e^{\sum_{m}\frac{nN\a^n_m\rho_{m}x^{m}}{m}} 
\end{eqnarray}
where $\a^n_m$ is non-zero only if $n=0 \mod m$. After putting together \eqref{largeNmeasure}and \eqref{largeNintegrand}, \eqref{singlet-proj} becomes,
\begin{eqnarray}\label{largeNscaint}
I_s^D(x) &=& \frac{\int {\cal D}\rho_n e^{-\frac{1}{2n}\left(N^2\rho_n^2- N\rho_{2n}\right)}i_s^{(4)}(x,y)/\denom(x,y)}{\int {\cal D}\rho_n e^{-\frac{1}{2n}\left(N^2\rho_n^2- N\rho_{2n}\right)}}
\end{eqnarray}
where we have defined the measure ${\cal D}\rho_n \propto \prod_n d\rho_n$ and we have been careful enough to factor out by the group volume. Each of the 
6 terms in \eqref{largeNintegrand} can now be converted into a Gaussian 
integral over the variables $\rho_n$. For example the first term in \eqref{largeNintegrand} can be evaluated as
\begin{eqnarray} \label{sampplet}
\frac{\int {\cal D}\rho_n e^{-\frac{1}{2n}\left(N^2\rho_n^2- N\rho_{2n}\right)}\denom(x^4,y^4)/\denom (x,y)}{\int {\cal D}\rho_n e^{-\frac{1}{2n}\left(N^2\rho_n^2- N\rho_{2n}\right)}}	&=& \frac{\int {\cal D}\rho_n e^{-\frac{1}{2n}\left(N^2\rho_n^2- 2N\b^2_n \rho_{n}-8 N \rho _n x^n \alpha^4_n +2N \rho _n x^n \right)}}{\int {\cal D}\rho_n e^{-\frac{1}{2n}\left(N^2\rho_n^2- 2N\b^2_n \rho_{n}\right)}}  \nonumber\\
&=& e^{\left(\sum _{n=1}^{\infty } \frac{x^{2 n} (4 \alpha^4_n-1)^2}{2 n}+\sum _{n=1}^{\infty } \frac{x^n (4 \alpha^4_n-1) \beta^2_n}{n}\right)}\nonumber\\
&=&\frac{1}{\left(x^4-1\right)^2 \left(x^4+1\right)}
\end{eqnarray}
where in the second line we have introduced the fictitious counting parameter $\b^2_n$ which is non-zero only if $n=0 \mod 2$, and in going from the second 
to the third lines in \eqref{sampplet} we have performed the Gaussian integrals
over $\rho_n$. 

Each of the six terms in \eqref{largeNintegrand} may be integrated in a similar 
manner. Summing all the results  we find that the integral  \eqref{largeNscaint} 
evaluates to 
\begin{eqnarray}\label{largeNscaintres}
I_s^D(x) &=& \frac{1}{(1-x^4)(1-x^6)}=\denom
\end{eqnarray}
In the case of scalars we believe (and numerical experiments indicate) that 
\eqref{largeNscaintres} is correct not just for $D \geq 10$ but in fact for 
$D \geq 4$. 

In order to count photon Lagrangians we need to perform a  similar integral. 
\begin{eqnarray}\label{largeNvecint}
I_v^D(x) &=& \frac{\int {\cal D}\rho_n e^{-\frac{1}{2n}\left(N^2\rho_n^2- N\rho_{2n}\right)}i_v^{(4)}(x,y)/\denom(x,y)}{\int {\cal D}\rho_n e^{-\frac{1}{2n}\left(N^2\rho_n^2- N\rho_{2n}\right)}}
\end{eqnarray}
where $i_v^{(4)}(x,y)$ is given by \eqref{multpartN} and the single letter partition function is given by \eqref{photon-sl},
\be
i_\tv(x,y)=((x-x^3)\chi_{\syng{1}}-(1-x^4))\denom(x,y)+1.
\ee
where $\chi_{\syng{1}}=N\rho_1$. The only difference in this case is the appearance of factors of  $\chi_{\syng{1}}(y^m)$ or equivalently $N\rho_m$ in the Haar integrand due to the multi-particle partition function. This is taken care of in the following manner: We introduce the following fictitious term in the integrand 
$$e^{\sum_n\frac{N\g_n^1 \rho_n}{n}}$$   
where we will set $\g_n^1$ to zero at the end of our computation. The factors of $\chi_{\syng{1}}(y^m)$ appearing in the Haar integrand can then be rewritten as derivatives with respect to $\g_n^1$. Operationally, we do the saddle point analysis first and then  take the derivatives with respect to $\g_n^1$ to account for the $\chi_{\syng{1}}(y^m)$ terms in the Haar integrand\footnote{As an example we work out the number of singlets in the product of two vector representations in arbitrarily high dimensions. 
\begin{eqnarray}
I_{\chi_v}(x) &=& \frac{\int {\cal D}\rho_n e^{-\frac{1}{2n}\left(N^2\rho_n^2- N\rho_{2n}\right)}\chi_{\syng{1}}^2}{\int {\cal D}\rho_n e^{-\frac{1}{2n}\left(N^2\rho_n^2- N\rho_{2n}\right)}}\nonumber\\
&=&\frac{\int {\cal D}\rho_n ~\partial_{\g^1_1}^2\left(e^{-\frac{1}{2n}\left(N^2\rho_n^2- N\rho_{2n}\right)+\frac{N\g_n^1 \rho_n}{n}}\right)}{\int {\cal D}\rho_n e^{-\frac{1}{2n}\left(N^2\rho_n^2- N\rho_{2n}\right)}}\Big|_{\g^i_j=0}\nonumber\\
&=& \partial_{\g^1_1}^2 ~e^{\sum_{n=1}^\infty \left(\frac{2\b_n^2\g_n^1+(\g_n^1)^2}{2n}\right)}\Big|_{\g^i_j=0}\nonumber\\
&=&(2 \beta^2_1 \gamma^1_1+(\beta^2_1)^2+(\gamma^1_1)^2+1)~e^{\sum_{n=1}^\infty \left(\frac{2\b_n^2\g_n^1+(\g_n^1)^2}{2n}\right)}\Big|_{\g^i_j=0}\nonumber\\
&=&1
\end{eqnarray}
This is consistent with our expectation that only one singlet is there in the tensor product of two vector representation of $SO(D)$.}. Following this
algorithm and performing a fair amount of algebra we finally obtain
\begin{eqnarray}\label{largeNvecintres}
I_v^D(x) &=&\frac{x^4 \left(2 x^4+3 x^2+2\right)}{x^{10}-x^6-x^4+1}=x^4 \left(2 x^4+3 x^2+2\right)\denom 
\end{eqnarray}
The generalization of this method to the case of the graviton plethystic integral is a straightforward one and we find 
\begin{eqnarray}\label{largeNtenintres}
I_T^D(x) &=& \frac{\int {\cal D}\rho_n e^{-\frac{1}{2n}\left(N^2\rho_n^2- N\rho_{2n}\right)}i_T^{(4)}(x,y)/\denom(x,y)}{\int {\cal D}\rho_n e^{-\frac{1}{2n}\left(N^2\rho_n^2- N\rho_{2n}\right)}}\nonumber\\
&=&\frac{(x^8 (7 + 10 x^2 + 10 x^4 + 2 x^6 - x^8 + 
	x^{10}))}{(1 - x^4 - x^6 + x^{10})}\nonumber\\
&=&x^8 (7 + 10 x^2 + 10 x^4 + 2 x^6 - x^8 + 
x^{10})\denom \nonumber\\
\end{eqnarray}

\subsection{$D < 10$}

For $D< 10$ the results \eqref{largeNvecintres} and \eqref{largeNtenintres}
are no longer correct. In these dimensions we have not been able to perform
the Haar integrals in for the photon or graviton plethystic formulae 
analytically. We nonetheless have strongly motivated conjectures for the 
exact results for these integrals using numerical integration as we now explain. 

As above we make the change of variables 
$$y_i = e^{i\theta_i}$$
As above, the the contour integral over $y_i$ becomes and angular integral over $\theta_i\sim (0 , 2\pi)$. The Haar measure becomes,
\begin{equation}
\begin{split}
&\Delta_e(\theta_i) =\frac{2^{1-N} \left(\prod _{j=1}^{N} \left(\prod _{i=1}^{j-1} 2 \left(\cos \left(\theta _i\right)-\cos \left(\theta _j\right)\right)\right)\right)^2}{(2\pi)^N N!}\nonumber\\
&\Delta_o(\theta_i) =\frac{(-1)^N \left(\prod _{k=1}^N \left(\cos \left(\theta _k\right)-1\right)\right) \left(\prod _{j=1}^N \left(\prod _{i=1}^{j-1} 2 \left(\cos \left(\theta _i\right)-\cos \left(\theta _j\right)\right)\right)\right)^2}{(2\pi)^N N!}
\end{split}
\end{equation}
The quantity $\denom$ changes to
\begin{equation}\label{dde}
\denom_{e}(x,\theta_i)=\frac{1}{\prod _{i=1}^N \left(1-2 x \cos \left(\theta _i\right)+x^{2}\right)}
\end{equation}
\begin{equation}\label{ddo}
\denom_{o}(x,\theta_i)=\frac{1}{(1-x)\prod _{i=1}^N \left(1-2 x \cos \left(\theta _i\right)+x^{2}\right)}
\end{equation}
The Haar integral now becomes 
\begin{equation}
I^D_j(x) =\int_{0}^{2\pi} \ldots \int_0^{2\pi}  \Delta_{e,o}(\theta_i)i^{(4)}_j(x,\theta_i)/\denom_{e,o}(x,\theta_i)
\end{equation}

In order to perform this integral, we first first note that 
we expect the result of the plethystic integral to be a finite linear sum over the partition functions $Z_{{\bf 1_S}}(x)$, $Z_{{\bf 2_M}}$ and $Z_{{\bf 1_A}}(x)$ (see \eqref{partfnsss}) plus (maybe) a finite polynomial 
accounting for the `errors' in the plethystic procedure 
(i.e. the difference between the plethystic partition function and the partition function over S-matrices). 
Given that each of $Z_{{\bf 1_S}}(x)$, $Z_{{\bf 2_M}}$ and $Z_{{\bf 1_A}}(x)$ equals a polynomial times $\denom$ and 
given that $1/\denom$ is itself a polynomial, it follows, 
in other words, that we expect the plethystic partition function times $1/\denom$  to be a finite polynomial in 
$x$. Motivated by these observations we multiply the 
plethystic integrand by $1/\denom$, Taylor series expand the 
result in $x$ around $0$. The coefficient of every power of $x$ in the result is an integral over $\theta_i$. 
We then evaluate these numerically using the Gauss-Kronrod method. As the numerical integration procedure is very accurate and can be performed very rapidly, we are able to perform this integral up to $x^{26}$. We thus able to verify that the polynomials in $x$ are indeed finite (they terminate) and to evaluate all nonzero coefficients. 
Our final results are summarized in \ref{scalar-plethystic}, \ref{photon-plethystic}  and \ref{graviton-plethystic} for $D\leq 10$.

\section{Most general quartic photon Lagrangian}\label{Lag}

In this section we will demonstrate that the Lagrangians 
\eqref{lagstruct} generate all parity even photon S-matrices
in every dimension. 

All parity even  gauge invariant contributions to the 
Lagrangian that contribute to 4 photon scattering 
consist of products of derivatives (of arbitrary 
number) multiplying 4 $F_{\mu\nu}$ fields in such a way that all indices contract so that the Lagrangian 
is a scalar.

\subsection{Terms with 6 derivatives on 4 $F_{\mu\nu}$ are 
	all descendants} \label{sider}

 It is very easy to see that every term
involving 6 or more derivatives (distributed and contracted in any manner among the 4 $F_{\mu\nu}$ 
operators) is a `descendant' Lagrangian (i.e. the module 
elements `dual' to these Lagrangians are always descendants 
of more elementary generators). In order to see why this is 
the case, suppose it were not true. Then there must
be a scalar expression built out of 4 $F_{\mu\nu}s$ 
and 6 derivatives in which none of the derivatives 
contract with each other. It follows that both indices of at least two $F_{ab}$ operators must contract with 
derivatives. A candidate term of this term might be 
\begin{equation}\label{mkk}
\partial_a F_{\mu \nu} \partial_\mu F_{ab}
\partial_b \partial _\nu \partial^p F_{m n} 
\partial ^mF_{pn}
\end{equation}
in which the indices of the first two field strength operators are both contracted with derivatives. 
In order to see that the term above is trivial  we use the the Bianchi identity
$$ \partial_a F_{\mu\nu}= -\partial_\mu F_{\nu a} - 
\partial_\nu F_{a\mu}$$
to re-express the first field field strength in 
\eqref{mkk} as a sum of two other terms. This gives 
us a sum of terms, each of which is a product of four field strengths. Note, however, that both of these terms have a pair of derivatives with contracted indices, and so both terms are descendants as we wanted to show.

The reader can easily convince herself that exactly 
the same argument can be made whenever two separate 
field strength operators have both their indices contracted with derivatives. Let the two field strengths of this form be the `first' and the second 
$F_{\mu\nu}$ operators in the expression. The two 
derivatives that contract with the second $F_{\mu\nu}$ must act on distinct $F_{\alpha \beta}$ fields 
(else the expression would vanish by the antisymmetry
of $F_{ab}$ ). Moreover neither of these derivatives 
can act on the second field itself (else the expression 
would vanish by the equations of motion). An integration by parts can be used to ensure that none of 
the derivatives act on the `fourth' $F_{\mu\nu}$. 
With this convention it follows that one of the 
two derivatives that contracts with the second $F_{\mu\nu}$ must act on the `third' $F_{\mu\nu}$ while
the second derivative must act on the first 
$F_{\mu\nu}$. We can now replace the expression involving the derivative acting on the 
first $F_{\mu\nu}$ by two different terms via the 
Bianchi identity. A moment's consideration will convince the reader that both these terms are 
descendants.

\subsection{Terms with 4 derivatives on 4 $F_{\mu\nu}$ are also
	all descendants} \label{fder}

Let us now turn to terms involving four derivatives
acting on the four $F_{\mu\nu}$ operators. The reader 
can quickly convince herself that there are five 
terms of this sort that are not obviously trivial. 
These terms are 
\begin{equation} \begin{split} \label{tfour}
T^4_1&= \partial_\delta \partial_\nu F_{\alpha a} F_{\beta a} \partial_\alpha \partial_\beta F_{\nu \beta} F_{\delta b}\\
T^4_2&= \partial_\nu F_{\alpha a}\partial_\delta F_{\beta a} \partial_\alpha F_{\nu b}\partial_\beta F_{\delta b}\\
T^4_3&= \partial_{\delta}\partial_{\gamma}F_{a\alpha} F_{a \beta} \partial_\alpha F_{b\gamma} \partial_\beta F_{b \delta} \\
T^4_4&= \partial_\beta \partial_d F_{ab} F_{bc} \partial_\alpha \partial_a F_{cd} F_{\alpha \beta}\\
T^4_5&= \partial_\gamma \partial_\alpha F_{ab} \partial_\delta\partial_\beta F_{ba} F_{\alpha \beta} F_{\gamma \delta}\\
\end{split}
\end{equation}
(Any other expression that the reader may care to 
write down can be manipulated into one of the five 
forms above up to total derivatives - without, at this 
stage, the use of Bianchi identities).

It is now possible to employ Bianchi identities to find relations between the structures $T^4_i$  ($i = 1 \to 5$). The relations we obtain turn out to be 
strong enough to allow us to deduce that each of 
the terms listed in \eqref{tfour} actually are 
actually trivial. The algebra involved in these demonstrations is lengthy - so we only report one sample computation
\begin{eqnarray}
T^4_1 &=& k^3_\alpha F^1_{\alpha a}k^3_\beta F^2_{\beta a} k^1_\gamma F^3_{\gamma b} k^1_\delta F^4_{\delta b}\nonumber\\
&=& -k^3_\alpha F^1_{\gamma \alpha }k^3_\beta F^2_{\beta a} k^1_a F^3_{\gamma b} k^1_\delta F^4_{\delta b}-k^3_\alpha F^1_{a \gamma }k^3_\beta F^2_{\beta a} k^1_\alpha  F^3_{\gamma b} k^1_\delta F^4_{\delta b}\nonumber\\
&=& -k^3_\alpha F^1_{\gamma \alpha }k^3_\beta F^2_{\beta a} k^1_a F^3_{\gamma b} k^1_\delta F^4_{\delta b} +I_{\textrm{desc}} \nonumber\\
&=& k^3_\gamma F^1_{\gamma \alpha }k^3_\beta F^2_{\beta a} k^1_a F^3_{ b\alpha} k^1_\delta F^4_{\delta b} + k^3_b F^1_{\gamma \alpha }k^3_\beta F^2_{\beta a} k^1_a F^3_{\alpha \gamma} k^1_\delta F^4_{\delta b} +I_{\textrm{desc}} \nonumber\\
&=& k^3_\alpha F^1_{\gamma \alpha }k^3_\beta F^2_{\beta a} k^1_a F^3_{ \gamma b} k^1_\delta F^4_{\delta b} + k^3_b F^1_{\gamma \alpha }k^3_\beta F^2_{\beta a} k^1_a F^3_{\alpha \gamma} k^1_\delta F^4_{\delta b} +I_{\textrm{desc}} \nonumber\\
&=& k^3_\alpha F^1_{\gamma \alpha }k^3_\beta F^2_{\beta a} k^1_a F^3_{ \gamma b} k^1_\delta F^4_{\delta b} - k^2_b F^1_{\gamma \alpha }k^3_\beta F^2_{\beta a} k^1_a F^3_{\alpha \gamma} k^1_\delta F^4_{\delta b} +I_{\textrm{desc}} \nonumber\\
\therefore T^4_1 &\sim & - k^2_b F^1_{\gamma \alpha }k^3_\beta F^2_{\beta a} k^1_a F^3_{\alpha \gamma} k^1_\delta F^4_{\delta b} +I_{\textrm{desc}} \sim \tilde{I}_{\textrm{desc}}\nonumber\\
\end{eqnarray}
In deriving this we have used Bianchi identity between the first $k^1$ and $F^1$ in the second line. In the fourth line we use Bianchi identity between first $k^3$ and $F^3$. The structure in the sixth line is due to momentum conservation and antisymmetry of $F_{ab}$. In the final step, to go from $I_{\textrm{desc}}$ to $\tilde{I}_{\textrm{desc}}$, we have used Bianchi identity between $k^2$ and $F^2$. 

Similar manipulations can be used to prove that all of the $T^4_i$s are descendants of structures with no more than 
two derivatives on $F_{\mu\nu}$. It follows that there is no 
Lagrangian structure built out of four field strengths
and four derivatives that generates a `primary' 
S-matrix. 

\subsection{Primary structures with two derivatives on 
	four field strengths}

The situation is a bit more complicated with terms 
involving two derivatives of the four field strengths. 
By using the equivalence of terms that differ by 
total derivatives, the reader can convince herself that there there are fourteen naively inequivalent structures
at this order. They are 
\begin{equation} \begin{split} \label{ttwo}
T^2_1&= \partial_b F_{\beta a}\partial _a F_{\alpha b}F_{\theta \alpha}F_{\theta \beta}\\
T^2_2&= \partial_b F_{\beta a} \partial_a F_{\beta b} F_{\mu \nu} F_{\mu \nu}\\
T^2_3&=  \partial_b F_{\beta a}\partial_aF_{\alpha b}F_{\theta\beta}F_{\theta\alpha}\\
T^2_4&=  F_{\beta a}\partial_a F_{\alpha b}\partial_b F_{\theta \alpha}F_{\theta \beta}\\
T^2_5&= F_{\beta a} \partial_aF_{\beta b}\partial_b F_{\mu\nu}F_{\mu\nu}\\
T^2_6&= F_{\beta a}\partial_a F_{\alpha b}\partial_bF_{\theta \beta}F_{\theta \alpha}\\
T^2_7&=\partial_{\beta}F_{b a}F_{\theta\beta}\partial_a F_{b \alpha}F_{\theta \alpha}\\
T^2_8&=F_{\alpha a}F_{\beta b}\partial_a F_{\alpha \theta} \partial_b F_{\beta \theta}\\
T^2_9&=F_{\alpha a}F_{\beta b}\partial_b \partial_a F_{\alpha \theta}F_{\beta \theta}\\
T^2_{10}&=F_{b\alpha}F_{a \alpha}\partial_b F_{\mu\nu}\partial_a F_{\mu\nu}\\
T^2_{11}&=F_{b \alpha}F_{a\alpha}\partial_a \partial_b F_{\mu\nu}F_{\mu\nu}\\
T^2_{12}&= F_{\alpha a}F_{\beta b}\partial_a F_{\beta\theta}\partial_b F_{\alpha \theta}\\
T^2_{13}&= F_{\alpha a}F_{\beta b}\partial_a \partial_b F_{\beta\theta} F_{\alpha \theta}\\
T^2_{14}&= F_{ab}  \partial_a F_{\mu\nu} \partial_b F_{\nu\rho} 
F_{\rho \mu}
\end{split}
\end{equation}
$T^2_1$ and $T^2_3$ are equivalent up to re-labelling.

Once again these naively independent structures are not
really all distinct; once again Bianchi identities may 
be used to relate these 14 structures. It turns out that  Bianchi identities generates 13 non-trivial identities between the structures listed in \eqref{ttwo}. These identities  can use used to relate each of these structures to 
a single independent term which we choose to be 
\begin{equation}\label{indterm}
T^2_I= F_{ab}  \partial_a F_{\mu\nu} \partial_b F_{\nu\rho} 
F_{\rho \mu}
\end{equation} 
Once again the algebra to establish these results is to 
lengthy to record in entirety; once again we only present 
some (in this case 2)  sample manipulations.
\begin{eqnarray}
T^2_1 &=& k^2_a F^1_{\beta a} k_b^1 F^2_{\alpha b} F^3_{\theta \alpha} F^4_{\theta \beta} \nonumber\\
&\equiv& k^1_a F^2_{\beta a} k_b^2 F^1_{\alpha b} F^3_{\theta \alpha} F^4_{\theta \beta} \nonumber\\
&=& k^2_b F^1_{\alpha b} k_a^1 F^2_{\beta a} F^3_{\theta \alpha} F^4_{\theta \beta} \nonumber\\ 
&=& T^2_3 \nonumber\\
&=& -k^2_a F^1_{ab} k_\beta^1 F^2_{\alpha b} F^3_{\theta \beta} F^4_{\theta \alpha} -k^2_a F^1_{b \beta} k_a^1 F^2_{\alpha b} F^3_{\theta \beta} F^4_{\theta \alpha} \nonumber\\ 
&=& -k^2_a F^1_{b a } k_\beta^1 F^3_{\theta \beta} F^2_{b \alpha}  F^4_{\theta \alpha} + {\cal O} \left( \textrm{desc}(\textrm{Tr} F^4) \right)  \nonumber\\
&=& -k^2_a F^1_{b a } k_\beta^1 F^3_{\theta \beta} F^2_{b \alpha}  F^4_{\theta \alpha} + {\cal O} \left( \textrm{desc}(\textrm{Tr} F^4) \right) \nonumber\\
&=&  k^2_b F^1_{b a } k_\beta^1 F^3_{\theta \beta} F^2_{ \alpha a}  F^4_{\theta \alpha} +k^2_\alpha F^1_{b a } k_\beta^1 F^3_{\theta \beta} F^2_{ a b}  F^4_{\theta \alpha}+ {\cal O} \left( \textrm{desc}(\textrm{Tr} F^4) \right) \nonumber\\
&=&  k^2_a F^1_{a b } k_\beta^1 F^3_{\theta \beta} F^2_{ \alpha b}  F^4_{\theta \alpha} + k^2_\alpha F^4_{\theta \alpha} k_\beta^1 F^3_{\theta \beta} F^1_{b a }  F^2_{ a b}  + {\cal O} \left( \textrm{desc}(\textrm{Tr} F^4) \right) \nonumber\\
&=& k^2_a F^1_{a b } k_\beta^1 F^3_{\theta \beta} F^2_{ \alpha b}  F^4_{\theta \alpha} + T^2_{10} + {\cal O} \left( \textrm{desc}(\textrm{Tr} F^4) \right) \nonumber\\
\therefore T^2_1 &=& T^2_3 \sim  T^2_{10} + {\cal O} \left( \textrm{desc}(\textrm{Tr} F^4) \right) \nonumber\\ 
\end{eqnarray}
The steps in the manipulation are as follows. In the second line we relabel $(1 \leftrightarrow 2)$ to establish the fact that  $T^2_1 \sim T^2_3$ in momentum space. In the fifth line, we use Bianchi identity corresponding to particle 1. In the eighth line we use Bianchi identity corresponding to particle 2. Equating the seventh and tenth line we obtain the final identity.   Hence we have the Lagrangian term $T^2_1$ is identical to $T^2_{10}$ up to descendants of four photon Lagrangians of derivative order 4. 

Let us now look at a second example
\begin{eqnarray}
T^2_6 &=& k^2_a F^1_{\beta a} k_b^3 F^2_{\alpha b} F^3_{\theta \beta} F^4_{\theta \alpha} \nonumber\\
&=& -k^2_\alpha F^1_{\beta a} k_b^3 F^2_{ b a} F^3_{\theta \beta} F^4_{\theta \alpha}  + {\cal O} \left( \textrm{desc}(\textrm{Tr} F^4) \right) \nonumber\\
&=& -k^2_\alpha F^1_{\theta \alpha} k_b^3 F^2_{ b a} F^4_{\beta a} F^3_{\theta \beta}  + {\cal O} \left( \textrm{desc}(\textrm{Tr} F^4) \right) \nonumber\\
&=& -k^2_\alpha F^1_{\theta \alpha} k_b^3 F^2_{ ab}  F^3_{\beta \theta } F^4_{\beta a} + {\cal O} \left( \textrm{desc}(\textrm{Tr} F^4) \right) \nonumber\\
\therefore T^2_6 &\sim &  {\cal O} \left( \textrm{desc}(\textrm{Tr} F^4) \right)
\end{eqnarray}

where we have used Bianchi identity corresponding to the 
second particle and re-labelling of $(1 \leftrightarrow 4)$. In this way all the Lagrangian structures can be represented in terms of $T^2_{10}$ and descendants of $\textrm{Tr} F^4$ and $(\textrm{Tr} F^2)^2$.

\section{Most general parity even photon 
S-matrix that grows no faster than $s^2$ 
in the Regge limit}\label{phorb}

In this subsection we will classify all local 4 photon S-matrices that grow no faster than $s^2$ in the Regge limit for every choice of polarization vectors. Consider the most general S-matrix from \eqref{expparphdte}.

\bea
S^{E_{{\bf 3},1}^{(1)}}&=&\cf^{{{\bf 3},1}}(t,u)(-8s^2 e_{{\bf 3},1}^{(1)}+8s^2e_{{\bf 3},2}^{(1)}-8s^2e_{{\bf S}}),\nonumber\\
S^{E_{{\bf 3},1}^{(2)}}&=&\cf^{{{\bf 3},1}}(u,s)(-8t^2 e_{{\bf 3},1}^{(2)}+8t^2e_{{\bf 3},2}^{(2)}-8t^2e_{{\bf S}}),\nonumber\\
S^{E_{{\bf 3},1}^{(3)}}&=&\cf^{{{\bf 3},1}}(s,t)(-8u^2 e_{{\bf 3},1}^{(3)}+8u^2e_{{\bf 3},2}^{(3)}-8u^2e_{{\bf S}}),\nonumber\\
\nonumber\\
S^{E_{{\bf 3},2}^{(1)}}&=&\cf^{{{\bf 3},2}}(t,u)(-2(u^2 e_{{\bf 3},1}^{(2)}+t^2 e_{{\bf 3},1}^{(3)})+2(u(s-t)e_{{\bf 3},2}^{(2)}+t(s-u)e_{{\bf 3},2}^{(3)})-2(t^2+u^2)e_{{\bf S}}),\nonumber\\
S^{E_{{\bf 3},2}^{(2)}}&=&\cf^{{{\bf 3},2}}(u,s)(-2(s^2 e_{{\bf 3},1}^{(3)}+u^2 e_{{\bf 3},1}^{(1)})+2(s(t-u)e_{{\bf 3},2}^{(3)}+u(t-s)e_{{\bf 3},2}^{(1)})-2(u^2+s^2)e_{{\bf S}}),\nonumber\\
S^{E_{{\bf 3},2}^{(3)}}&=&\cf^{{{\bf 3},2}}(s,t)(-2(t^2 e_{{\bf 3},1}^{(1)}+s^2 e_{{\bf 3},1}^{(2)})+2(t(u-s)e_{{\bf 3},2}^{(1)}+s(u-t)e_{{\bf 3},2}^{(2)})-2(s^2+t^2)e_{{\bf S}}),\nonumber\\
\nonumber\\
S^{E_{\bf S}}&=&(\cf^{E_{\bf S}}(t,u))(3\,stu\,(e_{{\bf 3},2}^{(1)}+e_{{\bf 3},2}^{(2)}+e_{{\bf 3},2}^{(3)}-2e_{{\bf S}})).
\eea
For the purpose of analyzing the Regge growth, let us rearrange the terms in the S-matrix \eqref{expparphdte} in the following way. We consider terms in the S-matrix which are proportional to the bare generator $e$'s from each of the local generators $E$'s. 
\begin{eqnarray}\label{regphoton1}
S^{E_{{\bf 3},1}}|_{e_{{\bf 3},1}} &=& -8\cf^{{{\bf 3},1}}(t,u)s^2 e_{{\bf 3},1}^{(1)}-8\cf^{{{\bf 3},1}}(u,s)t^2 e_{{\bf 3},1}^{(2)}-8\cf^{{{\bf 3},1}}(s,t)u^2 e_{{\bf 3},1}^{(3)},\nonumber\\
S^{E_{{\bf 3},2}}|_{e_{{\bf 3},1}} &=& -2\bigg((\cf^{{{\bf 3},2}}(u,s)u^2+\cf^{{{\bf 3},2}}(s,t)t^2)e_{{\bf 3},1}^{(1)} + (\cf^{{{\bf 3},2}}(t,u)u^2 +\cf^{{{\bf 3},2}}(s,t)s^2)e_{{\bf 3},1}^{(2)} \nonumber\\
&&+ (\cf^{{{\bf 3},2}}(t,u)t^2+\cf^{{{\bf 3},2}}(u,s)s^2) e_{{\bf 3},1}^{(3)}\bigg),\nonumber\\
S^{E_{{\bf S}}}|_{e_{{\bf 3},1}} &=& 0.\nonumber\\
S|_{e_{{\bf 3},1}}&=&S^{E_{{\bf 3},1}}+S^{E_{{\bf 3},2}}+S^{E_{{\bf S}}}|_{e_{{\bf 3},1}}.
\end{eqnarray}  
For the terms proportional to $e_{{\bf 3},2}^{(1)}$ and its permutations, we obtain,
\begin{eqnarray}\label{regphoton2}
S^{E_{{\bf 3},1}}|_{e_{{\bf 3},2}} &=&8\cf^{E_{{\bf 3},1}}(t,u)s^2 e_{{\bf 3},2}^{(1)}+8\cf^{E_{{\bf 3},1}}(s,u)t^2 e_{{\bf 3},2}^{(2)}+8\cf^{E_{{\bf 3},1}}(t,s)u^2 e_{{\bf 3},2}^{(3)},\nonumber\\
S^{E_{{\bf 3},2}}|_{e_{{\bf 3},2}} &=& 2\bigg((\cf^{E_{{\bf 3},2}}(s,u)u(t-s)+\cf^{E_{{\bf 3},2}}(t,s)t(u-s))e_{{\bf 3},2}^{(1)} \nonumber\\
&+& (\cf^{E_{{\bf 3},2}}(t,u)u(s-t) +\cf^{E_{{\bf 3},2}}(t,s)s(u-t))e_{{\bf 3},2}^{(2)} \nonumber\\
&&+ (\cf^{E_{{\bf 3},2}}(t,u)t(s-u)+\cf^{E_{{\bf 3},2}}(s,u)s(t-u)) e_{{\bf 3},2}^{(3)}\bigg),\nonumber\\
S^{E_{{\bf S}}}|_{e_{{\bf 3},2}} &=& \cf^{E_{{\bf S}}}(t,u)(3\,stu\,(e_{{\bf 3},2}^{(1)}+e_{{\bf 3},2}^{(2)}+e_{{\bf 3},2}^{(3)}))\nonumber\\
S|_{e_{{\bf 3},2}}&=& S^{E_{{\bf 3},1}}+S^{E_{{\bf 3},2}}+S^{E_{{\bf S}}}|_{e_{{\bf 3},2}}.
\end{eqnarray}
And finally, for terms proportional to $e_{\bf S}$,
\begin{eqnarray}\label{regphoton3}
S^{E_{{\bf 3},1}}|_{e_{\bf S}} &=&(-8\cf^{{E_{{\bf 3},1}}}(t,u)s^2 -8\cf^{{E_{{\bf 3},1}}}(s,u)t^2 -8\cf^{{E_{{\bf 3},1}}}(t,s)u^2)e_{\bf S},\nonumber\\
S^{E_{{\bf 3},2}}|_{e_{\bf S}} &=& -2\bigg(\cf^{{E_{{\bf 3},2}}}(s,u)(u^2+s^2)+\cf^{{E_{{\bf 3},2}}}(t,s)(s^2+t^2) + \cf^{{E_{{\bf 3},2}}}(t,u)(t^2+u^2) \bigg)\, e_{\bf S},\nonumber\\
S^{E_{{\bf S}}}|_{e_{\bf S}} &=& -6stu\, \cf^{{E_{{\bf S}}}}(t,u)\,e_{\bf S},\nonumber\\
S|_{e_{\bf S}} &=&S^{E_{{\bf 3},1}}+S^{E_{{\bf 3},2}}+S^{E_{{\bf S}}}|_{e_{\bf S}}.
\end{eqnarray}
Let us first study terms proportional to $e_{{\bf 3},1}^{(3)}$ of the S-matrix (These are given by the last line in \eqref{regphoton1}).
Provided that $D\geq 4$ the condition that the S-matrix 
grow no faster than $s^2$ is only 
met provided the same condition 
holds independently for 
the coefficients of 
$e_{{\bf 3},1}^{(1)}$ and $e_{{\bf 3},1}^{(2)}$.  Our
condition is thus simply that 
\begin{equation} \label{bag}
-\left( 8u^2 \cf^{E_{{\bf 3},1}}(t,s) +2 s^2 \cf^{E_{{\bf 3},2}}(s,u) + 2t^2 \cf^{E_{{\bf 3},2}}(t,u)\right)
\end{equation} 
(together with the two crossing 
related expressions) grow no faster
than $s^2$ in the Regge limit. 

If we assume that $\cf^{E_{{\bf 3},1}}$ and $\cf^{E_{{\bf 3},2}}$ are polynomials then the expression in \eqref{bag} is a 
polynomial of degree 2 or greater. 
Moreover it is symmetric under 
interchange of $s$ and $t$. The only polynomials that meet these 
conditions and still do not grow 
faster than $s^2$ are 
$s^2+t^2$, $st$, $s^2t+t^2s$ and 
$s^2 t^2$. Now if the polynomial 
in \eqref{bag} were to evaluate 
to $s^2t^2$ then the (permutation
related) polynomial that occurs in 
the bracket of the first line of \eqref{regphoton1} (proportional to $e_1^{(1)}$) would evaluate to 
$s^2u^2$. As this expression grows 
faster than $s^2$ and so is disallowed. We conclude that 
the expression in \eqref{bag} 
must be a linear combination of 
$s^2+t^2$, $st$ and $s^2t+t^2s$.

Let us now turn to the result of \eqref{regphoton3} above. Notice that the part 
of this answer that depends on 
$\cf^{E_{{\bf 3},1}}$ and $\cf^{E_{{\bf 3},2}}$ is 
proportional to the term in 
\eqref{bag} completely symmetrized
(i.e. is proportional to the sum 
of the three brackets in the last 
three lines of \eqref{regphoton1}). 
Given the conditions of the last 
paragraph, this term automatically 
grows no faster than $s^2$ in the 
Regge limit. It follows that \eqref{regphoton3} grows no faster than $s^2$ in the 
Regge limit provided the same is 
true of $-s tu \cf^{E_{{\bf S}}}(t,u)$.
This condition immediately 
forces $\cf^{E_{{\bf S}}}(t,u)$ to be a 
constant. 

Finally, let us turn to the expression in \eqref{regphoton2} above. 
The coefficient of $e_{{\bf 3},2}^{(3)}$ in that 
expression is given by 
\be\label{twoo}
8u^2 \cf^{E_{{\bf 3},1}}(t,s) +2 s(t-u) \cf^{E_{{\bf 3},2}}(s,u) + 2t(s-u)\cf^{E_{{\bf 3},1}}(t,u) +3stu\,\cf^{E_{{\bf S}}}(t,u)
\ee
As we now know that $\cf^{E_{{\bf S}}}(t,u)$ is a constant, the 
term proportional to $\cf^{E_{{\bf S}}}(t,u)$ in this expression 
is proportional to $stu$ and so automatically grows no 
faster than $s^2$ in the Regge limit. In order that our 
S-matrix grow no faster than $s^2$ at fixed $t$, it must
be that the same is true of the expression 
\begin{equation}\label{bagt}
8u^2 \cf^{E_{{\bf 3},1}}(t,s) +2 s(t-u) \cf^{E_{{\bf 3},2}}(s,u) + 2t(s-u) \cf^{E_{{\bf 3},2}}(t,u)
\end{equation}
By repeating the reasoning in the paragraph under 
\eqref{bag} it must be that \eqref{bagt}, like 
\eqref{bag}, is a linear combination of the polynomials 
$s^2+t^2$, $st$, and $stu$. 

In summary we require that the expressions in \eqref{bag} and \eqref{bagt} must simultaneously 
be (possibly different) linear combinations of the three polynomials listed above. If $\cf^{E_{{\bf 3},1}}$ and $\cf^{E_{{\bf 3},2}}$ 
are constants, then both \eqref{bag} and \eqref{bagt}
are automatically linear combinations of $s^2+t^2$ and 
$st$. The only other possibility for $\cf^{E_{{\bf 3},1}}$ and 
$\cf^{E_{{\bf 3},2}}$ is that they are proportional to the (unique symmetric degree one) polynomial $s+t$. If we suppose 
that $\cf^{E_{{\bf 3},1}}(s,t)=a(s+t)$ and that $\cf^{E_{{\bf 3},2}}(s,t)=b(s+t)$ 
then \eqref{bag} evaluates to 
$$-4a u^3 -b \left( s^2t  + t^2s  \right) $$
The condition that this expression grow no faster than 
$s^2$ in the Regge limit sets $a=0$. With this condition 
\eqref{bagt} evaluates to  
$$-st(s-u) -st(t-u)=-3stu$$
So $b$ is allowed to be non-zero. In conclusion, the most general photon S-matrix that grows no faster than $s^2$ is given by \eqref{expparphdte} with the momenta polynomials specified by 
\begin{eqnarray}
	\cf^{E_{{\bf 3},1}}(s,t)=a_1, \qquad \cf^{E_{{\bf 3},2}}(s,t)=b_1 + b_2 (s+t), \qquad \cf^{E_{{\bf S}}}(s,t)= c_1
 \end{eqnarray}
where $a_1, b_1, b_2$ and $c_1$ are all arbitrary constants.

\section{Explicit S-matrices for four graviton scattering}

In this appendix we provide a completely explicit listing 
of gravitational S-matrices in every dimension.

\subsection{$D\geq 8$}\label{localgravsmatrix}

When $D \geq 8$ 4 gravity scattering is necessarily parity even. In these dimensions the generators of the local 
S-matrix module are $G_{{\bf S},1}=\chi_6$ (see \eqref{r31} and \eqref{4gsm}) along with  $G_{{\bf 3},1}, \ldots G_{{\bf 3},8}$, $G_{{\bf 3_A}}$ and $G_{{\bf S},2}$ (see \eqref{lablaggravgen}). 

It is relatively straightforward to list the $S$ matrices generated by $G_{{\bf 3},1}, \ldots G_{{\bf 3},8}$, $G_{{\bf 3_A}}$, $G_{{\bf S},2}$ as well as the Lagrangians 
that generate these S-matrices. It is also straightforward 
to list the S-matrices generated by $G_{{\bf S},1}$. As explained 
in subsection \ref{mgef}, however, the fact that the 
Lagrangian $G_{{\bf S},1}$ has only three factors of the Riemann 
tensor complicates the listing of Lagrangians dual to these 
S-matrices. As explained in subsection \ref{mgef}, one can 
find a solution to this problem by adopting an alternate 
view of the S-matrix module; a view in which we include generator $G_{{\bf 3},9}$ instead of $G_{{\bf S},1}$ but the module 
is now not freely generated but is instead subject to 
the relations \eqref{chigen}. 

In this subsection we first perform the simple part of our listing. We list the S-matrices generated by $G_{{\bf 3},1}, \ldots G_{{\bf 3},8}$, $G_{{\bf 3_A}}$, $G_{{\bf S},2}$. 
After this is done we return to question of listing the 
Lagrangians that generate the `descendant' S-matrices of 
$G_{{\bf S},1}$.  

\begin{itemize}

	\item The S-matrix corresponding to $G_{{\bf 3},1}$ in \eqref{elo} is specified by the polynomial $\cf^{G_{{\bf 3},1}} (t,u)$ which exhibits a $\mathbb{Z}_2$ symmetry ($t \leftrightarrow u$). In equations,
	\begin{equation}\label{b5smat} 
	\begin{split} 
	&S^{G_{{\bf 3},1}}=\frac14\big(\cf^{G_{{\bf 3},1}}(t,u)\left[\left( p^1_p \epsilon^1_q - p^1_q \epsilon^1_p  
	\right) \left( p^2_p \epsilon^2_q - p^2_q \epsilon^2_p  \right)
	\left( p^3_r \epsilon^3_s - p^3_s \epsilon^3_r  
	\right) \left( p^4_r \epsilon^4_s - p^4_s \epsilon^4_r  \right)\right.\\
	&\left.\left( p^1_a \epsilon^1_b - p^1_b \epsilon^1_a  
	\right) \left( p^2_a \epsilon^2_b - p^2_b \epsilon^2_a  \right)
	\left( p^3_c \epsilon^3_d - p^3_d \epsilon^3_c  
	\right) \left( p^4_c \epsilon^4_d - p^4_d \epsilon^4_c  \right)\right] \\
	&+\cf^{G_{{\bf 3},1}}(s,u)\left[3\leftrightarrow 2\right]+\cf^{G_{{\bf 3},1}}(s,t)\left[2\leftrightarrow 4\right]\big).
	\end{split}
	\end{equation}
	The most general descendant which gives rise to S-matrix in \eqref{b5smat} is given by,
	\begin{equation}\label{lagb5smat}
	\begin{split}
	L^{G_{{\bf 3},1}}&=\sum_{m, n} \left(\cf^{G_{{\bf 3},1}}\right)_{m,n} 2^{m+n}\left(\prod_{i=1}^m\prod_{j=1}^n\left(\partial_{\mu_i}\partial_{\nu_j} R_{abpq}\right)R_{baqp}\left(\partial^{\mu_i}R_{cdrs}\right)\left(\partial^{\nu_j}R_{dcsr}\right)\right).
	\end{split}
	\end{equation}
	We have defined the momenta polynomials as,
	\begin{eqnarray}
	\cf^{G_{{\bf 3},1}}(t,u)&=& \sum_{m,n}\left(\cf^{G_{{\bf 3},1}}\right)_{m,n} t^m u^n.
	\end{eqnarray}
		
	In order to see the fact that Lagrangian \eqref{lagb5smat} results in the S-matrix \eqref{b5smat}, we note that 
	\begin{equation}
	R_{abpq}R_{baqp}R_{cdrs}R_{dcsr}
	\end{equation}
	 linearizes to give $\text{Tr}(F^{1}F^{2})\text{Tr}(F^{3}F^{4})\text{Tr}(F^{1}F^{2})\text{Tr}(F^{3}F^{4})$ plus permutations. Once linearized, it is clear that the structure has extra $\mathbb{Z}_2$ symmetry of $1$ to $2$ exchange. The descendant Lagrangian \eqref{lagb5smat} therefore linearizes to, 
\begin{equation}
\begin{split}
L^{G_{{\bf 3},1}}&=	\frac{1}{16}\sum_{m, n} \left(\cf^{G_{{\bf 3},1}}\right)_{m,n} 2^{m+n}\left(\prod_{i=1}^m\prod_{j=1}^n\partial_{\mu_i}\partial_{\nu_j}\left( F^1_{ab}F^1_{pq}\right)F^2_{ba}F^2_{qp}\partial^{\mu_i}\left(F^3_{cd}F^3_{rs}\right)\partial^{\nu_j}\left(F^4_{dc}F^4_{sr}\right)\right).
\end{split}
\end{equation}
plus permutations.
	
\item The S-matrix corresponding to $G_{{\bf 3},2}$ in \eqref{elo} is specified by the momenta polynomial $\cf^{G_{{\bf 3},2}}(s,u)$ which has the $\mathbb{Z}_2$ symmetry ($s \leftrightarrow u$). Explicitly it is given by
		\begin{equation}\label{b6smat} 
	\begin{split} 
	&S^{G_{{\bf 3},2}}=\frac14\big(\cf^{G_{{\bf 3},2}} (s,u)\left[\left( p^1_p \epsilon^1_q - p^1_q \epsilon^1_p  
	\right) \left( p^2_p \epsilon^2_q - p^2_q \epsilon^2_p  \right)
	\left( p^3_v \epsilon^3_w - p^3_w \epsilon^3_v  
	\right) \left( p^4_v \epsilon^4_w - p^4_w \epsilon^4_v  \right)\right.\\
	&\left.\left( p^1_r \epsilon^1_s - p^1_s \epsilon^1_r  
	\right) \left( p^4_r \epsilon^4_s - p^4_s \epsilon^4_r  \right)
	\left( p^2_t \epsilon^2_u - p^2_u \epsilon^2_t  
	\right) \left( p^3_t \epsilon^3_u - p^3_u \epsilon^3_t  \right)\right] \\
	&+\cf^{G_{{\bf 3},2}}(t,u)\left[3\leftrightarrow 2\right]+\cf^{G_{{\bf 3},2}}(s,t)\left[3\leftrightarrow 4\right]\big).
	\end{split}
	\end{equation}
	 The most general descendant is
	\begin{equation}\label{lagb6smat}
	\begin{split}
	L^{G_{{\bf 3},2}}&=\sum_{m, n} \left(\cf^{G_{{\bf 3},2}}\right)_{m,n} 2^{m+n}\left(\prod_{i=1}^m\prod_{j=1}^n\left(\partial_{\mu_i}\partial_{\nu_j} R_{pqrs}\right)\left(\partial^{\mu_i}R_{pqtu}\right)R_{tuvw}\left(\partial^{\nu_j}R_{rsvw}\right)\right).	
	\end{split}
	\end{equation}

	That the descendant Lagrangian \eqref{lagb6smat} generates the S-matrix \eqref{b6smat} is easy to see; the Lagrangian
	\begin{equation}
	R_{pqrs}R_{pqtu}R_{tuvw}R_{rsvw}
	\end{equation} 
	linearizes to give $\text{Tr}(F^{1}F^{2})\text{Tr}(F^{3}F^{4})\text{Tr}(F^{1}F^{4})\text{Tr}(F^{2}F^{3})$ plus permutations. This structure has an obvious extra $\mathbb{Z}_2$ symmetry of $1$ to $3$ exchange. The descendant Lagrangian \eqref{lagb6smat} then linearizes to give, \begin{equation}
	\begin{split}
	\frac{1}{16}\sum_{m, n} \left(\cf^{G_{{\bf 3},2}}\right)_{m,n} 2^{m+n}\left(\prod_{i=1}^m\prod_{j=1}^n\partial_{\mu_i}\partial_{\nu_j}\left( F^1_{pq}F^1_{rs}\right)\partial^{\mu_i}\left(F^2_{pq}F^2_{tu}\right)F^3_{tu}F^3_{vw}\partial^{\nu_j}\left(F^4_{rs}F^4_{vw}\right)\right).
		\end{split}
	\end{equation}
plus permutations.

\item The most general S-matrix corresponding $G_{{\bf 3},3}$ in \eqref{elo} is specified by the momenta polynomials $\cf^{G_{{\bf 3},3}}(s,u)$ which is symmetric under ($s \leftrightarrow u$).
	\begin{equation}\label{b3smat} 
	\begin{split} 
	&S^{G_{{\bf 3},3}}=\frac14\big(\cf^{G_{{\bf 3},3}} (s,u)\left[\left( p^1_p \epsilon^1_q - p^1_q \epsilon^1_p  
	\right) \left( p^2_p \epsilon^2_t - p^2_t \epsilon^2_p  \right)
	\left( p^3_t \epsilon^3_v - p^3_v \epsilon^3_t  
	\right) \left( p^4_q \epsilon^4_v - p^4_v \epsilon^4_q  \right)\right.\\
	&\left.\left( p^1_r \epsilon^1_s - p^1_s \epsilon^1_r  
	\right) \left( p^2_r \epsilon^2_u - p^2_u \epsilon^2_r  \right)
	\left( p^3_u \epsilon^3_w - p^3_w \epsilon^3_u  
	\right) \left( p^4_s \epsilon^4_w - p^4_w \epsilon^4_s  \right)\right] \\
	&+\cf^{G_{{\bf 3},3}}(t,u)\left[3\leftrightarrow 2\right]+\cf^{G_{{\bf 3},3}}(s,t)\left[3\leftrightarrow 4\right]\big).
	\end{split}
	\end{equation}
	The most general descendant Lagrangian that gives rise to the S-matrix \eqref{b3smat} is as follows
	\begin{equation}\label{lagb3smat}
	\begin{split}
	L^{G_{{\bf 3},3}}&=\sum_{m, n} \left(\cf^{G_{{\bf 3},3}}\right)_{m,n} 2^{m+n}\left(\prod_{i=1}^m \prod_{j=1}^n 
	\left(\partial_{\nu_j} \partial_{\mu_i} R_{pqrs}\right) 
	\left(\partial^{\mu_i} R_{ptru}\right)    R_{tvuw}\left(\partial^{\nu_j}R_{qvsw}\right)\right).
	\end{split}
	\end{equation} 
		
	In order to see the fact that Lagrangian \eqref{lagb3smat} results in the S-matrix \eqref{b3smat}, we note that 
	\begin{equation}
	R_{pqrs}R_{ptru}R_{tvuw}R_{qvsw}
	\end{equation}
	linearizes to $\text{Tr}(F^{1}F^{2}F^{3}F^{4})\text{Tr}(F^1F^2F^3F^4)$ plus permutations. This structure, like $\text{Tr}(F^4)$ again has $\mathbb{Z}_2$ symmetry of $1\leftrightarrow 3$, which manifests in the $ u \leftrightarrow s$ symmetry of the momenta functions $\cf^{G_{{\bf 3},3}}(s,u)$. It follows therefore the Lagrangian \eqref{lagb3smat} linearizes to 
\begin{equation}
\begin{split}
	\frac{1}{16}\sum_{m, n} \left(\cf^{G_{{\bf 3},3}}\right)_{m.n} 2^{m+n}\left(\prod_{i=1}^m \prod_{j=1}^n 
\partial_{\nu_j} \partial_{\mu_i} (F^{1}_{pq}F^{1}_{rs}) 
\partial^{\mu_i} (F^{2}_{pt}F^{2}_{ru})    F^{3}_{tv}F^{3}_{uw}\partial^{\nu_j}(F^{4}_{qv}F^{4}_{sw})\right)
\end{split}
\end{equation}
plus permutations.	
	
\item 
The S-matrix corresponding to $G_{{\bf 3},4}$ in \eqref{elo} is specified by the momenta polynomials $\cf^{G_{{\bf 3},4}}(s,t)$ with the $\mathbb{Z}_2$ symmetry $(s \leftrightarrow t)$. The explicit S-matrix is as follows 
		\begin{equation}\label{b4smat} 
	\begin{split} 
	&S^{G_{{\bf 3},4}}=\frac14\big(\cf^{G_{{\bf 3},4}} (s,t)\left[\left( p^1_p \epsilon^1_q - p^1_q \epsilon^1_p  
	\right) \left( p^2_p \epsilon^2_t - p^2_t \epsilon^2_p  \right)
	\left( p^3_t \epsilon^3_v - p^3_v \epsilon^3_t  
	\right) \left( p^4_q \epsilon^4_v - p^4_v \epsilon^4_q  \right)\right.\\
	&\left.\left( p^1_r \epsilon^1_s - p^1_s \epsilon^1_r  
	\right) \left( p^2_u \epsilon^2_w - p^2_w \epsilon^2_u  \right)
	\left( p^3_w \epsilon^3_s - p^3_s \epsilon^3_w  
	\right) \left( p^4_r \epsilon^4_u - p^4_u \epsilon^4_r  \right)\right] \\
	&+\cf^{G_{{\bf 3},4}}(s,u)\left[3\leftrightarrow 4\right]+\cf^{G_{{\bf 3},4}}(u,t)\left[2\leftrightarrow 4\right]\big).
	\end{split}
	\end{equation}
	 The most general descendant which gives rise to this S-matrix is given by
	\begin{equation}
	\begin{split}\label{lagb4smat}
	L^{G_{{\bf 3},4}}=\sum_{m, n} \left(\cf^{G_{{\bf 3},4}}\right)_{m,n} 2^{m+n}\left(\prod_{i=1}^m \prod_{j=1}^n 
	\left(\partial_{\nu_j} \partial_{\mu_i} R_{pqrs}\right) 
	\left(\partial^{\mu_i} R_{ptuw}\right)\left(\partial^{\nu_j}R_{tvws}\right)R_{qvru}\right).
	\end{split}
	\end{equation} 
		
	In order to see the fact that Lagrangian \eqref{lagb4smat} results in the S-matrix \eqref{b4smat}, we note that 
	\begin{equation}
	R_{pqrs}R_{ptuw}R_{tvws}R_{qvru}
	\end{equation}
	linearizes to $\text{Tr}(F^{1}F^{2}F^{3}F^{4})\text{Tr}(F^1F^3F^2F^4)$ plus permutations. This structure has $\mathbb{Z}_2$ symmetry of $2\leftrightarrow 3$ (and hence the $\mathbb{Z}_2$ symmetry of the momenta polynomials $\cf^{G_{{\bf 3},4}}(s,t)$). When linearized, the most general descendant Lagrangian \eqref{lagb4smat} becomes,
	\begin{equation} 
	\begin{split}
		\frac{1}{16}\sum_{m, n} \left(\cf^{G_{{\bf 3},4}}\right)_{m,n} 2^{m+n}\left(\prod_{i=1}^m \prod_{j=1}^n 
		\partial_{\nu_j} \partial_{\mu_i} (F^{1}_{pq}F^{1}_{rs}) 
		\partial^{\mu_i} (F^{2}_{pt}F^{2}_{uw})    \partial^{\nu_j}(F^{3}_{tv}F^{3}_{ws})F^{4}_{qv}F^{4}_{ru}\right).
	\end{split}
\end{equation} 
plus permutations.

	\item The S-matrix corresponding to $G_{{\bf 3},5}$ in \eqref{elo} is specified by momenta polynomials $\cf^{G_{{\bf 3},5}}(t,u)$ which has a $\mathbb{Z}_2$ symmetry in $(t \leftrightarrow u)$.
	\begin{equation}\label{b2smat} 
	\begin{split} 
	&S^{G_{{\bf 3},5}}=\frac14\bigg(\cf^{G_{{\bf 3},5}}(t,u)\left[\left( p^1_p \epsilon^1_q - p^1_q \epsilon^1_p  
	\right) \left( p^2_p \epsilon^2_q - p^2_q \epsilon^2_p  \right)
	\left( p^3_v \epsilon^3_w - p^3_w \epsilon^3_v  
	\right) \left( p^4_v \epsilon^4_w - p^4_w \epsilon^4_v  \right)\right.\\
	&\left.\left( p^1_r \epsilon^1_s - p^1_s \epsilon^1_r  
	\right) \left( p^2_t \epsilon^2_u - p^2_u \epsilon^2_t  \right)
	\left( p^3_r \epsilon^3_t - p^3_t \epsilon^3_r  
	\right) \left( p^4_s \epsilon^4_u - p^4_u \epsilon^4_s  \right)\right] \\
	&+\cf^{G_{{\bf 3},5}}(u,s)\left[2\leftrightarrow 3\right]+\cf^{G_{{\bf 3},5}}(s,t)\left[2\leftrightarrow 4\right]\bigg)
	\end{split}
	\end{equation}
	The S-matrix \eqref{b2smat} is produced by the general descendant Lagrangian 
		\begin{equation}\label{lagb2smat}
		\begin{split}
		L^{G_{{\bf 3},5}}&=\sum_{m, n} \left(\cf^{G_{{\bf 3},5}}\right)_{m,n} 2^{m+n}\left(\prod_{i=1}^{m}\prod_{j=1}^{n} \left(\partial_{\nu_j}\partial_{\mu_i}R_{pqrs}\right)R_{pqtu}\left(\partial^{\mu_i}R_{rtvw}\right)\left(\partial^{\nu_i}R_{suvw}\right)\right).
		\end{split}
		\end{equation}

	In order to see the fact that Lagrangian \eqref{lagb2smat} results in the S-matrix \eqref{b2smat}, we note that 
	\begin{equation} 
	R_{pqrs}R_{pqtu}R_{rtvw}R_{suvw}
	\end{equation}
	linearizes to  $\text{Tr}(F^{1}F^{2})\text{Tr}(F^{3}F^{4})\text{Tr}(F^1F^3F^2F^4)$ plus permutation. The general descendant Lagrangian \eqref{lagb2smat} therefore linearizes to 
		\be
\frac{1}{16}\sum_{m, n} \left(\cf^{G_{{\bf 3},5}}\right)_{m,n} 2^{m+n}\left(\prod_{i=1}^{m}\prod_{j=1}^{n} \partial_{\nu_j}\partial_{\mu_i}\left(F^{1}_{pq}F^{1}_{rs}\right)F^{2}_{pq}F^{2}_{tu}\partial^{\mu_i}\left(F^{3}_{rt}F^{3}_{vw}\right)\partial^{\nu_i}\left(F^{4}_{su} F^{4}_{vw}\right)\right)\\
\ee
plus permutations.

	\item The most general S-matrix generated by $G_{\bf 6}=G_{{\bf 3},6}\oplus G_{\bf 3_A}$ 
	is specified by an arbitrary polynomial $\cf^{G_{\bf 6}}(t,u)$  with 
	no symmetry restrictions. The corresponding S-matrix is  
	 \begin{equation} \label{nonttt}
	 \begin{split}  
	 &S^{G_{\bf 6}}=\frac14 \cf^{G_{\bf 6}}(s,t)\left[\left( p^1_p \epsilon^1_q - p^1_q \epsilon^1_p  
	 \right) \left( p^2_p \epsilon^2_q - p^2_q \epsilon^2_p  \right)
	 \left( p^3_u \epsilon^3_v - p^3_v \epsilon^3_u  
	 \right) \left( p^4_u \epsilon^4_v - p^4_v \epsilon^4_u  \right)\right.\\
	 &\left.\left( p^1_r \epsilon^1_s - p^1_s \epsilon^1_r  
	 \right) \left( p^2_r \epsilon^2_t - p^2_t \epsilon^2_r  \right)
	 \left( p^3_w \epsilon^3_t - p^3_t \epsilon^3_w  
	 \right) \left( p^4_w \epsilon^4_s - p^4_s \epsilon^4_w  \right)\right] \\
	 &+ \text{$S_3$ permutations (also act on $s,t,u$).}
	 \end{split}
	 \end{equation}
	 This S-matrix \eqref{nonttt} is produced (up to proportionality) by the
	 Lagrangian
	 \begin{equation} \label{lagnon}
	\begin{split}
	 L^{G_{\bf 6}}=\sum_{m, n} \left(\cf^{G_{\bf 6}}\right)_{m.n} 2^{m+n}\left( \prod_{i=1}^{m}\prod_{j=1}^{n} \left(\partial_{\mu_i}\partial_{\nu_j}R_{pqrs}\right)\left(\partial^{\mu_i}R_{pqrt}\right)\left(\partial^{\nu_i}R_{uvwt}\right)R_{uvws}\right)
	\end{split}
	 \end{equation} 
	The fact that \eqref{lagnon} yields the S-matrix \eqref{nonttt} follows from 
	the fact that 
	    \begin{eqnarray}
	 	R_{pqrs}R_{pqrt}R_{uvwt}R_{uvws}
	 	\end{eqnarray} 
	 	linearizes to $\text{Tr}(F^{1}F^{2})\text{Tr}(F^{3}F^{4})\text{Tr}(F^1F^2F^3F^4)$ 
	 	plus permutations (where the superscript, as usual,  labels particles). 
	 	It follows that the Lagrangian \eqref{lagnon} linearizes to 
	 \begin{equation}\label{b1smat}
	\frac{1}{16}\sum_{m, n} \left(\cf^{G_{\bf 6}}\right)_{m.n} 2^{m+n}\left( \prod_{i=1}^{m}\prod_{j=1}^{n} \partial_{\mu_i}\partial_{\nu_j}\left(F^{1}_{pq}F^{1}_{rs}\right)\partial^{\mu_i}\left(F^{2}_{pq}F^{2}_{rt}\right)\partial^{\nu_i}\left(F^{3}_{uv}F^{3}_{wt}\right)F^{4}_{uv} F^{4}_{ws}\right)
	 \end{equation} 
	 plus permutations. The replacement rule $ \partial_\mu \rightarrow i k_\mu$
	 then turns \eqref{b1smat} into \eqref{nonttt}.

		\item The S-matrix corresponding to $G_{{\bf 3},7}$ in \eqref{elo} is specified by $\cf^{G_{{\bf 3},7}}(t,u)$ which has $\mathbb{Z}_2$ symmetry of ($t \leftrightarrow u$). Explicitly, 
	
	\begin{equation}\label{b8smat} 
	\begin{split} 
	&S^{G_{{\bf 3},7}}=\frac{1}{16}\left(\cf^{G_{{\bf 3},7}} (t,u)\left( p^1_p \epsilon^1_q - p^1_q \epsilon^1_p  
	\right) \left( p^2_p \epsilon^2_q - p^2_q \epsilon^2_p  \right)
	\left( p^3_r \epsilon^3_s - p^3_s \epsilon^3_r  
	\right) \left( p^4_r \epsilon^4_s - p^4_s \epsilon^4_r  \right)\right. \\
	+&\left.\cf^{G_{{\bf 3},7}} (s,u)\left( p^1_p \epsilon^1_q - p^1_q \epsilon^1_p  
	\right) \left( p^3_p \epsilon^3_q - p^3_q \epsilon^3_p  \right)
	\left( p^2_r \epsilon^2_s - p^2_s \epsilon^2_r  
	\right) \left( p^4_r \epsilon^4_s - p^4_s \epsilon^4_r  \right) \right.\\
	+&\left.\cf^{G_{{\bf 3},7}} (t,s)\left( p^1_p \epsilon^1_q - p^1_q \epsilon^1_p  
	\right) \left( p^4_p \epsilon^4_q - p^4_q \epsilon^4_p  \right)
	\left( p^3_r \epsilon^3_s - p^3_s \epsilon^3_r  
	\right) \left( p^2_r \epsilon^2_s - p^2_s \epsilon^2_r  \right) \right)\\
	&\left( \left( p_a^1 \epsilon_b^1-p_b^1 \epsilon_a^1 \right)
	p^2_a\left( p_\mu^2 \epsilon_\nu^2-p_\nu^2 \epsilon_\mu^2 \right) p^3_b\left( p_\nu^3 \epsilon_\alpha^3-p_\alpha^3 \epsilon_\nu^3 \right) \left( p_\alpha^4 \epsilon_\mu^4- p_\mu^4 \epsilon_\alpha^4 \right) \right.\\
	&\left.+\left( p_a^2 \epsilon_b^2-p_b^2 \epsilon_a^2 \right)
	p^1_a\left( p_\mu^1 \epsilon_\nu^1-p_\nu^1 \epsilon_\mu^1 \right) p^4_b\left( p_\nu^4 \epsilon_\alpha^4-p_\alpha^4 \epsilon_\nu^4 \right) \left( p_\alpha^3 \epsilon_\mu^3- p_\mu^3 \epsilon_\alpha^3 \right)\right.\\
	&\left.+\left( p_a^3 \epsilon_b^3-p_b^3 \epsilon_a^3 \right)
	p^4_a\left( p_\mu^4 \epsilon_\nu^4-p_\nu^4 \epsilon_\mu^4 \right) p^1_b\left( p_\nu^1 \epsilon_\alpha^1-p_\alpha^1 \epsilon_\nu^1 \right) \left( p_\alpha^2 \epsilon_\mu^2- p_\mu^2 \epsilon_\alpha^2 \right)\right.\\
	&\left.+\left( p_a^4 \epsilon_b^4-p_b^4 \epsilon_a^4 \right)
	p^3_a\left( p_\mu^3 \epsilon_\nu^3-p_\nu^3 \epsilon_\mu^3 \right) p^2_b\left( p_\nu^2 \epsilon_\alpha^2-p_\alpha^2 \epsilon_\nu^2 \right) \left( p_\alpha^1 \epsilon_\mu^1- p_\mu^1 \epsilon_\alpha^1 \right)\right)
	\end{split} 
	\end{equation}
	
	The S-matrix \eqref{b8smat}	is generated by the descendant Lagrangian 
	\begin{equation}\label{lagb8smat}
	\begin{split}
	L^{G_{{\bf 3},7}}	&=-\sum_{m, n} \left(\cf^{G_{{\bf 3},7}}\right)_{m,n} 2^{m+n}\left(\prod_{i=1}^{m}\prod_{j=1}^{n}\left(\partial_{\mu_i}\partial_{\nu_j}R_{pqab}\right)\left(\partial_{a}R_{qp\mu \nu}\right)\left(\partial_b\partial^{\mu_i}R_{rs\nu \alpha}\right) \left(\partial^{\nu_j}R_{sr\alpha\mu}\right)\right)
	\end{split}
	\end{equation}

	In order to see that S-matrix \eqref{b8smat} is generated by \eqref{lagb8smat}, we note that
	\begin{equation}
	R_{pqab}\partial_a R_{qp \mu\nu} \partial_b R_{rs\nu \alpha} R_{sr \alpha \mu}
	\end{equation} 
	linearizes to give $\text{Tr}(F^{1}F^{2})\text{Tr}(F^{3}F^{4})F^{1}_{ab}\text{Tr}(p^2_a F^{2}p^3_b F^{3}F^{4})$ plus permutations. This structure again has only $\mathbb{Z}_2$ symmetry of $3\leftrightarrow 4$. The descendant Lagrangian \eqref{lagb8smat} then linearizes to give	
		\begin{equation}
	\begin{split}
	-\frac{1}{16}\sum_{m, n} \left(\cf^{G_{{\bf 3},7}}\right)_{m,n} 2^{m+n}\left(\prod_{i=1}^{m}\prod_{j=1}^{n}\partial_{\mu_i}\partial_{\nu_j}\left(F^1_{pq}F^1_{ab}\right)\partial_{a}F^2_{qp}F^2_{\mu\nu}\partial_b\partial^{\mu_i}\left(F^3_{rs}F^3_{\nu\alpha}\right) \partial^{\nu_j}\left(F^4_{sr}F^4_{\alpha\mu}\right)\right)
	\end{split}
	\end{equation}
plus permutations. 

	\item The S-matrix corresponding to $G_{{\bf 3},8}$ in \eqref{elo} is specified by the momenta functions $\cf^{G_{{\bf 3},8}}(s,u)$ which has the $\mathbb{Z}_2$ symmetry ($s \leftrightarrow u$). In equations, 
	\begin{equation}\label{b7smat} 
	\begin{split} 
	&S^{G_{{\bf 3},8}}=\frac{1}{16}\left(\cf^{G_{{\bf 3},8}}(s,u)\left( p^1_p \epsilon^1_q - p^1_q \epsilon^1_p  
	\right) \left( p^2_q \epsilon^2_r - p^2_r \epsilon^2_q  \right)
	\left( p^3_r \epsilon^3_s - p^3_s \epsilon^3_r  
	\right) \left( p^4_s \epsilon^4_p - p^4_p \epsilon^4_s  \right) \right. \\	
	&\left.\cf^{G_{{\bf 3},8}} (t,u)\left( p^1_p \epsilon^1_q - p^1_q \epsilon^1_p  
	\right) \left( p^3_q \epsilon^3_r - p^3_r \epsilon^3_q  \right)
	\left( p^2_r \epsilon^2_s - p^2_s \epsilon^2_r  
	\right) \left( p^4_s \epsilon^4_p - p^4_p \epsilon^4_s  \right) \right.\\	
	+&\left.\cf^{G_{{\bf 3},8}} (t,s)\left( p^1_p \epsilon^1_q - p^1_q \epsilon^1_p  
	\right) \left( p^3_q \epsilon^3_r - p^3_r \epsilon^3_q  \right)
	\left( p^4_r \epsilon^4_s - p^4_s \epsilon^4_r  
	\right) \left( p^2_s \epsilon^2_p - p^2_p \epsilon^2_s  \right)\right) \\
	&\left( \left( p_a^1 \epsilon_b^1-p_b^1 \epsilon_a^1 \right)
	p^2_a\left( p_\mu^2 \epsilon_\nu^2-p_\nu^2 \epsilon_\mu^2 \right) p^3_b\left( p_\nu^3 \epsilon_\alpha^3-p_\alpha^3 \epsilon_\nu^3 \right) \left( p_\alpha^4 \epsilon_\mu^4- p_\mu^4 \epsilon_\alpha^4 \right) \right.\\
	&\left.+\left( p_a^2 \epsilon_b^2-p_b^2 \epsilon_a^2 \right)
	p^1_a\left( p_\mu^1 \epsilon_\nu^1-p_\nu^1 \epsilon_\mu^1 \right) p^4_b\left( p_\nu^4 \epsilon_\alpha^4-p_\alpha^4 \epsilon_\nu^4 \right) \left( p_\alpha^3 \epsilon_\mu^3- p_\mu^3 \epsilon_\alpha^3 \right)\right.\\
	&\left.+\left( p_a^3 \epsilon_b^3-p_b^3 \epsilon_a^3 \right)
	p^4_a\left( p_\mu^4 \epsilon_\nu^4-p_\nu^4 \epsilon_\mu^4 \right) p^1_b\left( p_\nu^1 \epsilon_\alpha^1-p_\alpha^1 \epsilon_\nu^1 \right) \left( p_\alpha^2 \epsilon_\mu^2- p_\mu^2 \epsilon_\alpha^2 \right)\right.\\
	&\left.+\left( p_a^4 \epsilon_b^4-p_b^4 \epsilon_a^4 \right)
	p^3_a\left( p_\mu^3 \epsilon_\nu^3-p_\nu^3 \epsilon_\mu^3 \right) p^2_b\left( p_\nu^2 \epsilon_\alpha^2-p_\alpha^2 \epsilon_\nu^2 \right) \left( p_\alpha^1 \epsilon_\mu^1- p_\mu^1 \epsilon_\alpha^1 \right)\right)
	\end{split} 
	\end{equation}
	The descendant is of the general form
	\begin{equation}\label{lagb7smat} 
	\begin{split}
	L^{G_{{\bf 3},8}}	&=-\sum_{m, n} \left(\cf^{G_{{\bf 3},8}}\right)_{m,n} 2^{m+n}\left(\prod_{i=1}^{m}\prod_{j=1}^{n}\left(\partial_{\mu_i}\partial_{\nu_j}R_{pqab}\right)\left(\partial_{a}\partial^{\mu_i}R_{qr\mu \nu}\right)\left(\partial_b R_{rs\nu\alpha}\right) \left(\partial^{\nu_j}R_{sp\alpha\mu}\right)\right)
	\end{split}
	\end{equation}
		Reader can convince himself/herself that the descendant Lagrangian \eqref{lagb7smat} gives rise to the S-matrix \eqref{b7smat} by noting that 
	\begin{equation}
	R_{pqab}\partial_a R_{qr \mu\nu} \partial_b R_{rs\nu \alpha}R_{sp\alpha\mu}
	\end{equation}
	linearizes to give $\text{Tr}(F^{1}F^{2}F^{3}F^{4})F^{1}_{ab}\text{Tr}(p^2_a F^{2}p^3_b F^{3}F^{4})$ plus permutations. This structure has neither $\mathbb{Z}_2\times \mathbb{Z}_2$, which although is preserved by the first trace but broken by the $F\text{Tr}(...)$ part, nor it has $S^3$ which is preserved by the $F\text{Tr}(...)$ part but broken by the $\text{Tr}(F^4)$ part. Only $\mathbb{Z}_2$ is preserved, that is just $2\leftrightarrow 4$ flip symmetry. Consequently the $\mathbb{Z}_2\times \mathbb{Z}_2$ symmetrization had to be done explicitly in \eqref{b7smat}. The descendant Lagrangian \eqref{lagb7smat} then linearizes to  
	\begin{equation} 
	\begin{split}
		-\frac{1}{16}\sum_{m, n} \left(\cf^{G_{{\bf 3},8}}\right)_{m,n} 2^{m+n}\left(\prod_{i=1}^{m}\prod_{j=1}^{n}\partial_{\mu_i}\partial_{\nu_j}\left(F^1_{pq}F^1_{ab}\right)\partial_{a}\partial^{\mu_i}\left(F^2_{qr}F^2_{\mu\nu} \right)\partial_b\left(F^3_{rs}F^3_{\nu\alpha}\right) \partial^{\nu_j}\left(F^4_{sp}F^4_{\alpha\mu}\right)\right)|_{\mathbb{Z}_2\times \mathbb{Z}_2}\\
	\end{split}
	\end{equation}

	\item The S-matrix corresponding to  $G_{{\bf S},2}$ in \eqref{elo} is given by the momenta polynomial $\cf^{G_{{\bf S},2}}(s,t)$ which is fully symmetric in $s,t$ and $u$. The explicit expression for the S-matrix is given by,
	\begin{equation}\label{b9smat}
	\begin{split}
	S^{G_{{\bf S},2}}=\frac{1}{16}\left(\cf^{G_{{\bf S},2}}(s,t)\right) \times \\
	\left[ \left( p_a^1 \epsilon_b^1-p_b^1 \epsilon_a^1 \right)
	p^2_a\left( p_\mu^2 \epsilon_\nu^2-p_\nu^2 \epsilon_\mu^2 \right) p^3_b\left( p_\nu^3 \epsilon_\alpha^3-p_\alpha^3 \epsilon_\nu^3 \right) \left( p_\alpha^4 \epsilon_\mu^4- p_\mu^4 \epsilon_\alpha^4 \right) \right.\\
	\left.\left( p_p^1 \epsilon_q^1-p_q^1 \epsilon_p^1 \right)
	p^2_p\left( p_\beta^2 \epsilon_\gamma^2-p_\gamma^2 \epsilon_\beta^2 \right) p^3_q\left( p_\gamma^3 \epsilon_\delta^3-p_\delta^3 \epsilon_\gamma^3 \right) \left( p_\delta^4 \epsilon_\beta^4- p_\beta^4 \epsilon_\delta^4 \right) \right.\\
	+\left.(1\leftrightarrow 2)+(1\leftrightarrow 3)+ (1\leftrightarrow4)\right]
	\end{split}
	\end{equation}
	 The most general descendant Lagrangian giving rise to \eqref{b9smat} is
	\begin{equation}\label{lagb9smat}
	\begin{split}
	L^{G_{{\bf S},2}}	&=\sum_{m, n} \left(\cf^{G_{{\bf S},2}}\right)_{m,n} 2^{m+n}\left(
	\prod_{i=1}^{m}\prod_{j=1}^{n}\left(\partial_{\mu_i}\partial_{\nu_j}R_{abpq}\right) \left(\partial^{\mu_i}\partial_p \partial_a R_{\mu\nu\beta\gamma}\right)\left(\partial^{\nu_j}\partial_q \partial_b R_{\nu\alpha\gamma\delta}\right)R_{\alpha \mu\delta\beta}\right)
	\end{split}
	\end{equation}

	It is easy to see that the descendant Lagrangian \eqref{lagb9smat} generates the S-matrix \eqref{b9smat}. Consider the Lagrangian 
	\begin{equation}
	R_{abpq} \partial_p \partial_a R_{\mu\nu\beta\gamma}\partial_q\partial_b R_{\nu \alpha\gamma \delta}R_{\alpha\mu\delta\beta}
	\end{equation}
	which linearizes to give  $F^1_{pq}\text{Tr}(p^2_{p}F^2p^3_{q}F^3F^4)F^1_{ab}\text{Tr}(p^2_{a}F^2p^3_{b}F^3F^4)$ plus permutations. This structure has $S^3$ symmetry, because $2,3,4$ can be permuted and the structure remains invariant. The descendant Lagrangian \eqref{lagb9smat} linearizes to give,
	 \begin{equation}
	 \begin{split}
	 \frac{1}{16}\sum_{m, n} \left(\cf^{G_{{\bf S},2}}\right)_{m,n} 2^{m+n}\left(
	 \prod_{i=1}^{m}\prod_{j=1}^{n}\partial_{\mu_i}\partial_{\nu_j}(F^1_{ab}F^1_{pq}) \partial^{\mu_i}\partial_p \partial_a (F^2_{\mu\nu}F^2_{\beta\gamma})\partial^{\nu_j}\partial_q \partial_b(F^3_{\nu\alpha}F^3_{\gamma \delta})F^4_{\alpha \mu}F^4_{\delta\beta}\right)
	 \end{split}
	 \end{equation} plus permutations.
	
\item Finally we turn to the specification of the S-matrices 
descended from $G_{{\bf S},1}$. If we are interested in specifying 
only the S-matrix - and not the Lagrangian that gives 
rise to this S-matrix - this job is easily done. In addition 
to the S-matrices already listed in this appendix we have 
one additional contribution specified by $\cf^{G_{{\bf S},1}}$, a  fully symmetric polynomial of $s,t,u$ that is otherwise  
unconstrained. The S-matrix is given by 
\begin{equation}
\label{b0smat}
S^{G_{{\bf S},1}}_{D\geq 7} =24(3 \cf^{G_{{\bf S},1}}(t,u) \epsilon^{ijklmnp}\epsilon^{asdfghj}\epsilon^1_{i}\epsilon^2_{j}\epsilon^3_{k}\epsilon^4_{l}p^1_m p^2_n p^3_p \epsilon^1_{a}\epsilon^2_{s}\epsilon^3_{d}\epsilon^4_{f}p^1_g p^2_h p^3_j)
\end{equation}
At the level of S-matrices we have now completed our listings. The most general sum of \eqref{b0smat}, \eqref{nonttt}, \eqref{b2smat}, \eqref{b3smat}, \eqref{b4smat}, \eqref{b5smat}, \eqref{b6smat}, \eqref{b7smat}, \eqref{b8smat}, \eqref{b9smat} gives the most general 
local S-matrix for gravitational scattering in $D \geq 8$.

\item If, on the other hand, we are also interested in listing 
the Lagrangians that give rise to the S-matrices that descend 
from $G_{{\bf S},1}$ we are forced to proceed differently. In this 
case we do {\it not} include \eqref{b0smat} in our listing of 
S-matrices but instead replace \eqref{b0smat} by the most 
general S-matrix corresponding to  $G_{{\bf 3},9}$ \eqref{sevecha}. 
Such an S-matrix  is given by the momenta polynomial $\cf^{G_{{\bf 3},9}}(t,u)$ which is  symmetric in $s \leftrightarrow u$. As the ${\bf 1}_S$ part of $G_{{\bf 3},9}$ can be eliminated using \eqref{chiei}, only the ${\bf 2}_M$ part suffices for us. This means we further impose,
\be\label{1s-remove}
\cf^{G_{{\bf 3},9}}(t,u)+\cf^{G_{{\bf 3},9}}(u,s)+\cf^{G_{{\bf 3},9}}(s,t)=0.
\ee

The explicit expression for the S-matrix is given by,
\begin{equation}\label{b10smat}
\begin{split}
&S^{G_{{\bf 3},9}}=\frac14\big(\cf^{G_{{\bf 3},9}}(t,u)\left[\left( p^1_p \epsilon^1_q - p^1_q \epsilon^1_p  
\right) \left( p^2_p \epsilon^2_t - p^2_t \epsilon^2_p  \right)
\left( p^3_t \epsilon^3_v - p^3_v \epsilon^3_t  
\right) \left( p^4_u \epsilon^4_v - p^4_v \epsilon^4_u  \right)\right.\\
&\left.\left( p^1_r \epsilon^1_s - p^1_s \epsilon^1_r  
\right) \left( p^4_s \epsilon^4_w - p^4_w \epsilon^4_s  \right)
\left( p^2_r \epsilon^2_u - p^2_u \epsilon^2_r  
\right) \left( p^3_q \epsilon^3_w - p^3_w \epsilon^3_q  \right)\right] \\
&+\cf^{G_{{\bf 3},9}}(u,s)\left[2\leftrightarrow 3\right]+\cf^{G_{{\bf 3},9}}(s,t)\left[2\leftrightarrow 4\right]\big).
\end{split}
\end{equation}
The most general descendant Lagrangian giving rise to \eqref{b10smat} is
\begin{equation}\label{lagb10smat}
\begin{split}
L^{G_{{\bf 3},9}}	&=\sum_{m, n} \left(\cf^{G_{{\bf 3},9}}\right)_{m,n} 2^{m+n}\left(
\prod_{i=1}^{m}\prod_{j=1}^{n}\left(\partial_{\mu_i}\partial_{\nu_j}R_{pqrs}\right) R_{ptru}(\partial^{\mu_i} R_{tvqw})(\partial^{\nu_j}R_{uvsw})\right)
\end{split}
\end{equation}

It is easy to see that the descendant Lagrangian \eqref{lagb10smat} generates the S-matrix \eqref{b10smat}. Consider the Lagrangian 
\begin{equation}
R_{pqrs} R_{ptru} R_{tvqw}R_{uvsw}
\end{equation}
which linearizes to give  $F^1_{pq}F^2_{pt}F^3_{tv}F^4_{uv}F^2_{ru}F^1_{rs}F^4_{sw}$ plus permutations. The descendant Lagrangian \eqref{lagb9smat} linearizes to give,
\begin{equation}
\begin{split}
\frac{1}{16}\sum_{m, n} \left(\cf^{G_{{\bf 3},9}}\right)_{m.n} 2^{m+n}\left(
\prod_{i=1}^{m}\prod_{j=1}^{n}\left(\partial_{\mu_i}\partial_{\nu_j}F^1_{pq}F^1_{rs}\right) F^2_{pt}F^2_{ru}(\partial^{\mu_i} F^3_{tv}F^3_{qw})(\partial^{\nu_j}F^4_{uv}F^4_{sw})\right)
\end{split}
\end{equation} plus permutations.

With this new point of view, the most general gravitational 
S-matrix is now given by a sum of \eqref{nonttt}, \eqref{b2smat}, \eqref{b3smat}, \eqref{b4smat}, \eqref{b5smat}, \eqref{b6smat}, \eqref{b7smat}, \eqref{b8smat}, \eqref{b9smat} and \eqref{b10smat}. However as we have 
explained in subsection \ref{mgef}, this way of listing the 
most general S-matrix - and most general Lagrangian - is 
`over complete'. The fact that 
\be\label{nynya}
r^{(i)}\, G_{{\bf S},1}=4\Big( -G_{{\bf 3},1}^{(i)}-2G_{{\bf 3},2}^{(i)}-16 G_{{\bf 3},3}^{(i)}+16 G_{{\bf 3},4}^{(i)}-2G_{{\bf 3},5}^{(i)}+10 G_{{\bf 3},6}^{(i)}+16 G_{{\bf 3},9}^{(i)}+(4G_{\bf 3_A}^{(i+1)}-4G_{\bf 3_A}^{(i+2)})\Big).
\ee
tells us (see \eqref{coa}) that $\cf$'s  obeys the following equivalence relation.
\begin{eqnarray}\label{redundancyuu}
\cf^{G_{{\bf 3},1}} (t,u)&\sim& \cf^{G_{{\bf 3},1}} (t,u)  -(u-t) h(t,u),\nonumber\\
\cf^{G_{{\bf 3},2}} (t,u)&\sim& \cf^{G_{{\bf 3},2}} (t,u)  -2(u-t) h(t,u),\nonumber\\
\cf^{G_{{\bf 3},3}} (t,u)&\sim& \cf^{G_{{\bf 3},3}} (t,u)  -16(u-t) h(t,u),\nonumber\\
\cf^{G_{{\bf 3},4}} (t,u)&\sim& \cf^{G_{{\bf 3},4}} (t,u)  +16(u-t) h(t,u),\nonumber\\
\cf^{G_{{\bf 3},5}} (t,u)&\sim& \cf^{G_{{\bf 3},5}} (t,u)  -2(u-t) h(t,u),\nonumber\\
\cf^{G_{{\bf 3},6}} (t,u)&\sim& \cf^{G_{{\bf 3},6}} (t,u)  +10(u-t) h(t,u),\nonumber\\
\cf^{G_{{\bf 3},9}} (t,u)&\sim& \cf^{G_{{\bf 3},9}} (t,u)  +16(u-t) h(t,u),\nonumber\\
\cf^{G_{{\bf 3_A}}} (t,u)&\sim& \cf^{G_{{\bf 3_A}}} (t,u)  -4s\, h(t,u).
\end{eqnarray}
where $h(u,t)$ is a totally antisymmetric function under $S_3$. And also $\cf^{G_{{\bf 3},9}} (t,u)$ satisfies \eqref{1s-remove}.
\end{itemize}

We can summarize the above discussion by stating that the general most general parity invariant s-matrix in $D \geq 7$ is given by 
\begin{equation}\label{Dgeq7expsmatrix}
S = \big(\sum_{i=1}^{8} S^{G_{{\bf 3},i}}\big)+S^{G_{{\bf 3_A}}}+S^{G_{{\bf S},1}}+S^{G_{{\bf S},2}}.
\end{equation}
which is obtained from the general Lagrangian

\begin{equation}\label{Dgeq7explag}
L = \big(\sum_{i=1}^{9} L^{G_{{\bf 3},i}}\big)+L^{G_{{\bf 3_A}}}+L^{G_{{\bf S},2}}.
\end{equation}

\subsection{$D=7$}
 
\subsection*{Even}
               In $D=7$ the parity even S-matrices continue to be given by \eqref{Dgeq7expsmatrix}.                    
                                    
\subsubsection*{Odd}     
 The most general S-matrix is specified by the momenta functions $\cf^{H_{{\bf 3},1}^{D=7}}(t,u)$ and $\cf^{H_{{\bf 3},2}^{D=7}}(t,u)$ that has $\Z_2$ symmetry under the two arguments and $\cf^{H_{{\bf S}}^{D=7}}(t,u)$ that is completely symmetric under $S_3$.

 \begin{eqnarray}\label{parityodd7smatrix}
 S^{H_{{\bf 3},1}^{D=7}}&=& -4i(\cf^{H_{{\bf 3},1}^{D=7}}(t,u) \left( p^1_a \epsilon^1_b - p^1_b \epsilon^1_a  
 \right)\left( p^2_b \epsilon^2_a - p^2_a \epsilon^2_b    
 \right)\left( p^3_c \epsilon^3_d - p^3_d \epsilon^3_c    
 \right)\left(  p^4_d \epsilon^4_c - p^4_c \epsilon^4_d  
 \right)\nonumber\\
 &&+ \cf^{H_{{\bf 3},1}^{D=7}}(s,u) \left( p^1_a \epsilon^1_b - p^1_b \epsilon^1_a  
 \right)\left( p^3_b \epsilon^3_a - p^3_a \epsilon^3_b    
 \right)\left( p^2_c \epsilon^2_d - p^2_d \epsilon^2_c    
 \right)\left(  p^4_d \epsilon^4_c - p^4_c \epsilon^4_d  
 \right) \nonumber\\
 && +\cf^{H_{{\bf 3},1}^{D=7}}(t,s) \left( p^1_a \epsilon^1_b - p^1_b \epsilon^1_a  
 \right)\left( p^4_b \epsilon^4_a - p^4_a \epsilon^4_b    
 \right)\left( p^3_c \epsilon^3_d - p^3_d \epsilon^3_c    
 \right)\left(  p^2_d \epsilon^2_c - p^2_c \epsilon^2_d  
 \right))\nonumber\\
 &&(*(8\epsilon^1\wedge  p^2 \wedge\epsilon^2 \wedge p^3 \wedge\epsilon^3\wedge p^4 \wedge\epsilon^4))\nonumber\\
 S^{H_{{\bf 3},2}^{D=7}}&=& -4i(\cf^{H_{{\bf 3},2}^{D=7}}(s,u) \left( p^1_a \epsilon^1_b - p^1_b \epsilon^1_a  
 \right)\left( p^2_b \epsilon^2_c - p^2_c \epsilon^2_b    
 \right)\left( p^3_c \epsilon^3_d - p^3_d \epsilon^3_c    
 \right)\left(  p^4_d \epsilon^4_a - p^4_a \epsilon^4_d  
 \right) \nonumber\\
 &&+ \cf^{H_{{\bf 3},2}^{D=7}}(t,s) \left( p^1_a \epsilon^1_b - p^1_b \epsilon^1_a  
 \right)\left( p^3_b \epsilon^3_c - p^3_c \epsilon^3_b    
 \right)\left( p^4_c \epsilon^4_d - p^4_d \epsilon^4_c    
 \right)\left(  p^2_d \epsilon^2_a - p^2_a \epsilon^2_d  
 \right) \nonumber\\
 && +\cf^{H_{{\bf 3},2}^{D=7}}(u,t) \left( p^1_a \epsilon^1_b - p^1_b \epsilon^1_a  
 \right)\left( p^4_b \epsilon^4_c - p^4_c \epsilon^4_b    
 \right)\left( p^2_c \epsilon^2_d - p^2_d \epsilon^2_c    
 \right)\left(  p^3_d \epsilon^3_a - p^3_a \epsilon^3_d  
 \right))\nonumber\\
 &&(8*(\epsilon^1\wedge  p^2 \wedge\epsilon^2 \wedge p^3 \wedge\epsilon^3\wedge p^4 \wedge\epsilon^4))\nonumber\\
S^{H_{{\bf S}}^{D=7}}&=&4i(\cf^{H_{{\bf S}}^{D=7}}(t,u))\nonumber\\
&&(\left(p^1_\a \epsilon^1_\b - p^1_\b \epsilon^1_\a\right)p^2_\a\left(p^2_\g \epsilon^2_\d - p^2_\d \epsilon^2_\g\right)p^3_\b\left(p^3_\d \epsilon^3_\mu - p^3_\mu \epsilon^3_\d\right)\left(p^4_\d \epsilon^4_\g - p^4_\g \epsilon^4_\d\right)\nonumber\\
&& +\left(p^2_\a \epsilon^2_\b - p^2_\b \epsilon^2_\a\right)p^1_\a\left(p^1_\g \epsilon^1_\d - p^1_\d \epsilon^1_\g\right)p^3_\b\left(p^3_\d \epsilon^3_\mu - p^3_\mu \epsilon^3_\d\right)\left(p^4_\d \epsilon^4_\g - p^4_\g \epsilon^4_\d\right)\nonumber\\
&& + \left(p^3_\a \epsilon^3_\b - p^3_\b \epsilon^3_\a\right)p^2_\a\left(p^2_\g \epsilon^2_\d - p^2_\d \epsilon^2_\g\right)p^1_\b\left(p^1_\d \epsilon^1_\mu - p^1_\mu \epsilon^1_\d\right)\left(p^4_\d \epsilon^4_\g - p^4_\g \epsilon^4_\d\right)\nonumber\\
&& +\left(p^4_\a \epsilon^4_\b - p^4_\b \epsilon^4_\a\right)p^2_\a\left(p^2_\g \epsilon^2_\d - p^2_\d \epsilon^2_\g\right)p^3_\b\left(p^3_\d \epsilon^3_\mu - p^3_\mu \epsilon^3_\d\right)\left(p^1_\d \epsilon^1_\g - p^1_\g \epsilon^1_\d\right))\nonumber\\
  &&(8*(\epsilon^1\wedge  p^2 \wedge\epsilon^2 \wedge p^3 \wedge\epsilon^3\wedge p^4 \wedge\epsilon^4))\nonumber\\
 \end{eqnarray}
  The Lagrangians generating S-matrix \eqref{parityodd7smatrix} are listed in \eqref{lagno}

  \subsection{$D=6$}

\subsubsection*{Even}

As mentioned in the main text, the most general S-matrix and Lagrangian continue to be given by  \eqref{Dgeq7expsmatrix} and \eqref{Dgeq7explag} but with the function $\cf^{G_{{\bf S},1}}(t,u)$ set equal to zero. 

\subsubsection*{Odd}
The most general parity odd S-matrix in $D=6$ is specified by the momenta functions $\cf^{H_{{\bf 3_A},1}^{D=6}}(t,u), \cf^{H_{{\bf 3_A},2}^{D=6}}(t,u)$ and $\cf^{H_{{\bf 3_A},3}^{D=6}}(t,u)$ all of which are antisymmetric in first two arguments. All three structures transform in ${\bf 3_A}={\bf 2_M}\oplus {\bf 1_A}$.

 The Explicit S-matrix is as follows, dropping the $D=6$ superscript.
\begin{eqnarray}\label{oddsmatd6grav}
S^{H_{{\bf 3_A},1}}&=& \frac{1}{16} \left(\cf^{H_{{\bf 3_A},1}}(t,u) \left( p^1_a \epsilon^1_b - p^1_b \epsilon^1_a  
\right)\left( p^2_b \epsilon^2_a - p^2_a \epsilon^2_b    
\right)\left( p^3_c \epsilon^3_d - p^3_d \epsilon^3_c    
\right)\left(  p^4_d \epsilon^4_c - p^4_c \epsilon^4_d  
\right)\right. \nonumber\\
&&\left.+ \cf^{H_{{\bf 3_A},1}}(s,u) \left( p^1_a \epsilon^1_b - p^1_b \epsilon^1_a  
\right)\left( p^3_b \epsilon^3_a - p^3_a \epsilon^3_b    
\right)\left( p^2_c \epsilon^2_d - p^2_d \epsilon^2_c    
\right)\left(  p^4_d \epsilon^4_c - p^4_c \epsilon^4_d  
\right)\right. \nonumber\\
&&\left. \cf^{H_{{\bf 3_A},1}}(t,s) \left( p^1_a \epsilon^1_b - p^1_b \epsilon^1_a  
\right)\left( p^4_b \epsilon^4_a - p^4_a \epsilon^4_b    
\right)\left( p^3_c \epsilon^3_d - p^3_d \epsilon^3_c    
\right)\left(  p^2_d \epsilon^2_c - p^2_c \epsilon^2_d  
\right)\right)\nonumber\\
&&\left( 8\left( p^1_\a \epsilon^1_\b - p^1_\b \epsilon^1_\a\right)p^2_\a p^3_\b ~*(\epsilon^2\wedge k^2\wedge\epsilon^3\wedge k^3\wedge\epsilon^4\wedge k^4) \right.\nonumber\\
&&\left.+8\left( p^2_\a \epsilon^2_\b - p^2_\b \epsilon^2_\a\right)p^1_\a p^3_\b~*( \epsilon^1\wedge k^1\wedge\epsilon^3\wedge k^3\wedge\epsilon^4\wedge k^4) \right.\nonumber\\
&&\left. +8\left(p^3_\a \epsilon^3_\b - p^3_\b \epsilon^3_\a\right)p^2_\a p^1_\b ~*(\epsilon^2\wedge k^2\wedge\epsilon^1\wedge k^1\wedge\epsilon^4\wedge k^4) \right.\nonumber\\
&&\left. 8\left(p^4_\a \epsilon^4_\b - p^4_\b \epsilon^4_\a\right)p^2_\a p^3_\b ~*(\epsilon^2\wedge k^2\wedge\epsilon^3\wedge k^3\wedge\epsilon^1\wedge k^1) \right)\nonumber\\
S^{H_{{\bf 3_A},2}}&=& \frac{1}{16} \left(\cf^{H_{{\bf 3_A},2}}(s,u) \left( p^1_a \epsilon^1_b - p^1_b \epsilon^1_a  
\right)\left( p^2_b \epsilon^2_c - p^2_c \epsilon^2_b    
\right)\left( p^3_c \epsilon^3_d - p^3_d \epsilon^3_c    
\right)\left(  p^4_d \epsilon^4_a - p^4_a \epsilon^4_d  
\right)\right. \nonumber\\
&&\left.+ \cf^{H_{{\bf 3_A},2}}(t,s) \left( p^1_a \epsilon^1_b - p^1_b \epsilon^1_a  
\right)\left( p^3_b \epsilon^3_c - p^3_c \epsilon^3_b    
\right)\left( p^4_c \epsilon^4_d - p^4_d \epsilon^4_c    
\right)\left(  p^2_d \epsilon^2_a - p^2_a \epsilon^2_d  
\right)\right. \nonumber\\
&&\left. \cf^{H_{{\bf 3_A},2}}(u,t) \left( p^1_a \epsilon^1_b - p^1_b \epsilon^1_a  
\right)\left( p^4_b \epsilon^4_c - p^4_c \epsilon^4_b    
\right)\left( p^2_c \epsilon^2_d - p^2_d \epsilon^2_c    
\right)\left(  p^3_d \epsilon^3_a - p^3_a \epsilon^3_d  
\right)\right)\nonumber\\
&&\left( 8\left( p^1_\a \epsilon^1_\b - p^1_\b \epsilon^1_\a\right)p^2_\a p^3_\b~ *(\epsilon^2\wedge k^2\wedge\epsilon^3\wedge k^3\wedge\epsilon^4\wedge k^4) \right.\nonumber\\
&&\left.+8\left( p^2_\a \epsilon^2_\b - p^2_\b \epsilon^2_\a\right)p^1_\a p^3_\b ~*(\epsilon^1\wedge k^1\wedge\epsilon^3\wedge k^3\wedge\epsilon^4\wedge k^4) \right.\nonumber\\
&&\left.+8\left(p^3_\a \epsilon^3_\b - p^3_\b \epsilon^3_\a\right)p^2_\a p^1_\b ~*(\epsilon^2\wedge k^2\wedge\epsilon^1\wedge k^1\wedge\epsilon^4\wedge k^4) \right.\nonumber\\
&&\left. 8\left(p^4_\a \epsilon^4_\b - p^4_\b \epsilon^4_\a\right)p^2_\a p^3_\b ~*(\epsilon^2\wedge k^2\wedge\epsilon^3\wedge k^3\wedge\epsilon^1\wedge k^1) \right)\nonumber\\
S^{H_{{\bf 3_A},3}}&=& \frac{1}{16}\left(\cf^{H_{{\bf 3_A},3}}(s,t)\bigg(\epsilon^{abcdef}\left( p^1_\g \epsilon^1_\d - p^1_\d \epsilon^1_\g  \right)\left( p^2_e \epsilon^2_\rho - p^2_\rho \epsilon^2_e  \right)\left( p^3_f \epsilon^3_\rho - p^3_\rho \epsilon^3_f\right)\left( p^4_\d \epsilon^4_\g - p^4_\g \epsilon^4_\d  \right)\right.\nonumber\\
&&\left.\left(\left(p^1_a \epsilon^1_b - p^1_b \epsilon^1_a\right)k^2_c\left(p^2_\mu \epsilon^2_\nu - p^2_\nu \epsilon^2_\mu\right)k^3_d\left(p^3_\nu \epsilon^3_\a - p^3_\a \epsilon^3_\nu\right)\left(p^4_\a \epsilon^4_\mu - p^4_\mu \epsilon^4_\a\right)-\left(p^4_a \epsilon^4_b - p^4_b \epsilon^4_a\right)\right.\right.\nonumber\\
&&\left.\left.k^3_c\left(p^3_\mu \epsilon^3_\nu - p^3_\nu \epsilon^3_\mu\right)k^2_d\left(p^2_\nu \epsilon^2_\a - p^2_\a \epsilon^2_\nu\right)\left(p^1_\a \epsilon^1_\mu - p^1_\mu \epsilon^1_\a\right)\right)\right.\nonumber\\
&&\left.+\left( p^2_\g \epsilon^2_\d - p^2_\d \epsilon^2_\g  \right)\left( p^1_e \epsilon^1_\rho - p^1_\rho \epsilon^1_e  \right)\left( p^4_f \epsilon^4_\rho - p^4_\rho \epsilon^4_f\right)\left( p^3_\d \epsilon^3_\g - p^3_\g \epsilon^3_\d  \right)\right.\nonumber\\
&&\left.\left(\left(p^2_a \epsilon^2_b - p^2_b \epsilon^2_a\right)k^1_c\left(p^1_\mu \epsilon^1_\nu - p^1_\nu \epsilon^1_\mu\right)k^4_d\left(p^4_\nu \epsilon^4_\a - p^4_\a \epsilon^4_\nu\right)\left(p^3_\a \epsilon^3_\mu - p^3_\mu \epsilon^3_\a\right)-\left(p^3_a \epsilon^3_b - p^3_b \epsilon^3_a\right)\right.\right.\nonumber\\
&&\left.\left.k^4_c\left(p^4_\mu \epsilon^4_\nu - p^4_\nu \epsilon^4_\mu\right)k^1_d\left(p^1_\nu \epsilon^1_\a - p^1_\a \epsilon^1_\nu\right)\left(p^2_\a \epsilon^2_\mu - p^2_\mu \epsilon^2_\a\right)\right)\bigg)+\cf^{H_{{\bf 3_A},3}}(u,t)(2 \leftrightarrow 4)+\cf^{H_{{\bf 3_A},3}}(s,u)(3 \leftrightarrow 4)\right)\nonumber\\
\end{eqnarray}

The most general descendant Lagrangian which gives rise to \eqref{oddsmatd6grav} is given by

\begin{eqnarray}\label{oddlagd6grav}
L^{H_{{\bf 3_A},1}}&=& -\sum_{m, n} \left(\cf^{H_{{\bf 3_A},1}}\right)_{m,n} 2^{m+n}\left(\prod_{i=1}^m \prod_{j=1}^n 
\epsilon^{abcdef}(\partial_{\mu_i} \partial_{\nu_j}R_{\mu\nu\a\b }) \partial_\a  R_{\nu\mu ab} \partial^{\mu_i}\partial_\b R_{\g\d cd}  \partial^{\nu_j}R_{\d\g ef}\right),\nonumber\\
L^{H_{{\bf 3_A},2}}&=& -\sum_{m, n} \left(\cf^{H_{{\bf 3_A},2}}\right)_{m.n} 2^{m+n}\left(\prod_{i=1}^m \prod_{j=1}^n 
\epsilon^{abcdef}(\partial_{\mu_i} \partial_{\nu_j}R_{\mu\nu\a\b }) \partial^{\mu_i} \partial_\a  R_{\nu\g ab}\partial_\b R_{\g\d cd}  \partial^{\nu_j}R_{\d\mu ef}\right),\nonumber\\
L^{H_{{\bf 3_A},3}}&=& -\sum_{m, n} \left(\cf^{H_{{\bf 3_A},3}}\right)_{m.n} 2^{m+n}\left(\prod_{i=1}^m \prod_{j=1}^n\epsilon^{abcdef}(\partial_{\mu_i} \partial_{\nu_j}R_{ab\g\d})(\partial^{\mu_i} \partial_cR_{\mu\nu e\rho})(\partial^{\nu_j}\partial_d R_{\nu\alpha f\rho})R_{\a \mu \d\g}\right),\nonumber\\
\end{eqnarray}  
Consider the following Lagrangian structures, 
\begin{eqnarray}
&&\epsilon^{abcdef}R_{\mu\nu\a\b } \partial_\a  R_{\nu\mu ab} \partial_\b R_{\g\d cd} R_{\d\g ef}, \qquad \epsilon^{abcdef}R_{\mu\nu\a\b } \partial_\a  R_{\nu\g ab} \partial_\b R_{\g\d cd} R_{\d\mu ef}\nonumber\\
&& \epsilon^{abcdef}R_{ab\g\d}\partial_cR_{\mu\nu e\rho}\partial_d R_{\nu\alpha f\rho}R_{\a \mu \d\g}.
\end{eqnarray}
The linearized structure in momentum space is proportional to  
\begin{eqnarray}
&&\frac{1}{16}(\text{Tr}(F^1F^2)\text{Tr}(F^3F^4))(\epsilon^{abcdef}F^1_{\a\b}\partial^\a F^2_{ab}\partial^\b F^3_{cd}F^4_{ef}),\qquad \frac{1}{16}(\text{Tr}(F^1F^2F^3F^4))(\epsilon^{abcdef}F^1_{\a\b}\partial^\a F^2_{ab}\partial^\b F^3_{cd}F^4_{ef})\nonumber\\
&&\frac{1}{16} \epsilon^{abcdef}F^1_{ab}(\partial_cF^2_{\mu\nu}\partial_d F^3_{\nu\alpha}F^{4}_{\a \mu})(F^1_{\g\d}F^2_{e\rho}F^3_{f\rho}F^4_{\d\g}).
\end{eqnarray} 
plus permutations. It is therefore easy to see that the descendant Lagrangian \eqref{oddlagd6grav} gives rise to S-matrix \eqref{oddsmatd6grav}.

    \subsection{$D=5$}
\subsubsection*{Even}
There is further reduction of 2 $\Z_2$ symmetric structures in this dimension. To be precise there all the Lagrangians giving rise to the labelled Lagrangians $G_{{\bf 3},1}-G_{{\bf 3},6}$ are not linearly independent. We find the following relation between the s-matrices.
\begin{eqnarray}\label{5dgravrelation1}
\sum_{i=2}^6 S^{G_{{\bf 3},i}} =0 
\end{eqnarray}
with 
\begin{eqnarray}\label{5dgravrelation2}
&\cf^{G_{{\bf 3},6}}(t,u)= \frac{5\a}{2}-2\b,\qquad \cf^{G_{{\bf 3},5}}(t,u)= -\frac{\a}{2}, \qquad \cf^{G_{{\bf 3},3}}(t,u)= 2\b \nonumber\\
&\cf^{G_{{\bf 3},4}}(t,u)=2\a,\qquad \cf^{G_{{\bf 3},1}}(t,u)=\frac{-6\a+5\b}{8}.
\end{eqnarray} 
The structures we choose to reduce: $L^{G_{{\bf 3},3}}$ and $L^{G_{{\bf 3},4}}$.

\subsubsection*{Odd}

 The most general parity odd S-matrix in $D=5$ is specified by the momentum function $\cf^{H_{\bf 3_A}^{D=5}}(t,u)$  which transforms in ${\bf 2_M}+ {\bf 1_A}={\bf 3_A}$ of $S_3$. As a result this function is antisymmetric in the two arguments.
 
 The explicit S-matrix is as follows,
    \begin{eqnarray}\label{oddsmatd5grav}
 S^{H_{\bf 3_A}^{D=5}}&=& -i\left(\cf^{H_{\bf 3_A}^{D=5}}(s,t)\left(4*(p^4\wedge\e^4\wedge p^1\wedge\e^1\wedge p^2)\left(p^2_f\e^2_g-p^2_g\e^2_f\right)\left(p^3_g\e^3_h-p^3_h\e^3_g\right)\left(p^4_h\e^4_f-p^4_f\e^4_h\right)\right.\right.\nonumber\\
 &&\left.\left.  \left(p^1_\a \e^1_\b-p^1_\b\e^1_\a\right)p_\a^2p_\b^3\left(p^2_\mu \e^2_\nu-p^2_\nu\e^2_\mu\right)\left(p^3_\nu\e^3_\mu-p^3_\mu\e^3_\nu\right) + (1\leftrightarrow 2, 3\leftrightarrow 4)+ (1\leftrightarrow 3, 3\leftrightarrow 4)\right.\right.\nonumber\\
 &&\left.\left.+(1\leftrightarrow 4, 2\leftrightarrow 3)\right)\right.\nonumber\\
 &&\left. + \cf^{H_{\bf 3_A}^{D=5}}(t,u)\left(4*(p^2\wedge\e^2\wedge p^1\wedge\e^1\wedge p^3)\left(p^3_f\e^3_g-p^3_g\e^3_f\right)\left(p^4_g\e^4_h-p^4_h\e^4_g\right)\left(p^2_h\e^2_f-p^2_f\e^2_h\right)\right.\right.\nonumber\\
 &&\left.\left.  \left(p^1_\a \e^1_\b-p^1_\b\e^1_\a\right)p_\a^3p_\b^4\left(p^3_\mu \e^3_\nu-p^3_\nu\e^3_\mu\right)\left(p^4_\nu\e^4_\mu-p^4_\mu\e^4_\nu\right) + (1\leftrightarrow 2, 3\leftrightarrow 4)+ (1\leftrightarrow 3, 3\leftrightarrow 4)\right.\right.\nonumber\\
 &&\left.\left.+(1\leftrightarrow 4, 2\leftrightarrow 3)\right)\right.\nonumber\\
 &&\left.+ \cf^{H_{\bf 3_A}^{D=5}}(u,s)\left(4*(p^3\wedge\e^3\wedge p^1\wedge\e^1\wedge p^4)\left(p^4_f\e^4_g-p^4_g\e^4_f\right)\left(p^2_g\e^2_h-p^2_h\e^2_g\right)\left(p^3_h\e^3_f-p^3_f\e^3_h\right)\right.\right.\nonumber\\
 &&\left.\left.  \left(p^1_\a \e^1_\b-p^1_\b\e^1_\a\right)p_\a^4p_\b^2\left(p^4_\mu \e^4_\nu-p^4_\nu\e^4_\mu\right)\left(p^2_\nu\e^2_\mu-p^2_\mu\e^2_\nu\right) + (1\leftrightarrow 2, 3\leftrightarrow 4)+ (1\leftrightarrow 3, 3\leftrightarrow 4)\right.\right.\nonumber\\
 &&\left.\left.+(1\leftrightarrow 4, 2\leftrightarrow 3)\right) \right) 
 \end{eqnarray}
The S-matrix \eqref{oddsmatd5grav} is obtained from the following descendant Lagrangian
The most general descendant is given by 
\begin{eqnarray}\label{oddlagd5grav}
L^{H_{\bf 3_A}^{D=5}}&=& \sum_{m, n} \left(\cf^{H_{\bf 3_A}^{D=5}}\right)_{m,n} 2^{m+n}\left(\prod_{i=1}^m \prod_{j=1}^n 
\epsilon^{abcde}((\partial_{\mu_i} \partial_{\nu_j}R_{\a\b cd})\partial^{\mu_i}\partial_\a\partial_e R_{\mu\nu fg} \partial^{\nu_j}\partial_\b R_{\nu\mu gh} R_{abhf} \right),\nonumber\\
\end{eqnarray}
      The $\Z_2$ antisymmetry of the momentum function $\cf^{H_{\bf 3_A}^{D=5}}(t,u)$ is explicit in momentum space and can be seen by re-labelling the second and the third Riemann tensor in \eqref{oddlagd5grav} and using conservation equation. 
      
 That the S-matrix \eqref{oddsmatd5grav} is obtained from the Lagrangian \eqref{oddlagd5grav} is evident when one considers the Lagrangian
         $$\epsilon^{abcde}R_{\a\b cd}\partial_\a\partial_e R_{\mu\nu fg} \partial_\b R_{\nu\mu gh} R_{abhf} $$ 
  The linearized structure in momentum space is given by,
  \begin{eqnarray}
  	\epsilon^{abcde}\left( F^1_{\a\b} \partial_\a F^2_{\mu\nu}\partial_\b F^3_{\nu\mu} F^4_{ab}\right)\left( F^1_{cd}\partial_e F^2_{fg}F^3_{gh}F^4_{hf}\right)
  \end{eqnarray}    
  plus permutations.

\subsection{$D=4$}
\subsubsection*{Even}
The independent parity even structures are just three.
\begin{eqnarray}
&&(R_{abcd}R_{abcd})^2, \qquad R_{pqrs}R_{pqtu}R_{rtvw}R_{suvw}\nonumber\\
&& R_{abcd}\nabla^a\nabla^cR_{efe'f'}\nabla^b\nabla^d R_{fgf'g'}R_{geg'e'}
\end{eqnarray}
We show the reduction of the local module as follows.
From \cite{Fulling:1992vm}, we know that the S-matrix corresponding to $G_{\bf 6}=G_{{\bf 3},6}\oplus G_{{\bf 3_A}}$ vanishes. The relations that among the rest 8 derivative Lagrangians are as follows. 
\begin{eqnarray}\label{parityevenreld4}
&&S^{G_{{\bf 3},5}}+S^{G_{{\bf 3},1}}+S^{G_{{\bf 3},2}}=0 \nonumber\\
&&{\rm with},~~  \cf^{G_{{\bf 3},5}}=1,~~ \cf^{G_{{\bf 3},1}}=-\frac{1}{4},~~\cf^{G_{{\bf 3},2}}=\frac{1}{2}.
\end{eqnarray}
We use this to eliminate $G_{{\bf 3},2}$ from our basis. We also find linear relations between the s-matrix at 10 derivative level,
\begin{eqnarray}\label{parityevenreld42}
&&S^{G_{{\bf 3},8}}+S^{G_{{\bf 3},7}}=0\nonumber\\
&&{\rm with},~~  \cf^{G_{{\bf 3},8}}=-4,~~ \cf^{G_{{\bf 3},7}}=1.
\end{eqnarray}
Thus at 10 derivative level we only have $G_{{\bf 3},7}$ as the basis structure. Finally we find the following relation between the descendants of $G_{{\bf 3},1}$, $G_{{\bf 3},5}$ and $G_{{\bf 3},7}$.
\begin{eqnarray}\label{parityevenreld43}
	&&S^{G_{{\bf 3},7}}+S^{G_{{\bf 3},1}}+S^{G_{{\bf 3},5}}=0\nonumber\\
	&&{\rm with}~~\cf^{G_{{\bf 3},7}}=18,~~\cf^{G_{{\bf 3},5}}(t,u)=-4(t+u),~~\cf^{G_{{\bf 3},1}}(t,u)=(t+u). 
\end{eqnarray}
This can be used to eliminate $G_{{\bf 3},7}$. Finally we are left with $G_{{\bf 3},1},G_{{\bf 3},5}$ and $G_{{\bf 3},9}$ as our local module generator in $D=4$. 
The most general parity even s-matrix is given by,
\begin{equation}
S^{D=4}_{\rm even} =  S^{G_{{\bf 3},1}}+S^{G_{{\bf 3},5}}+S^{G_{{\bf 3},9}}.
\end{equation}

\subsubsection*{Odd}
 The most general parity odd S-matrix in $D=4$ is parameterized by the momenta functions $\cf^{O_{\bf 3}^{D=4}}(t,u)$ and $\cf^{O_{\bf S}^{D=4}}(t,u)$, where the first function is symmetric in the two arguments while the second one is completely symmetric under $S_3$.
 The explicit S-matrix is as follows,   
 \begin{eqnarray}\label{oddlagd4grav}
 S^{O_{\bf 3}^{D=4}}&=&  \frac{1}{8}\cf^{O_{\bf 3}^{D=4}}(t,u) \left(4*(p_1\wedge \epsilon_1\wedge p_2 \wedge \epsilon_2) \left( p^3_\mu \epsilon^3_\nu - p^3_\nu \epsilon^3_\mu  
 \right)\left( p^4_\mu \epsilon^4_\nu - p^4_\nu \epsilon^4_\mu  
 \right)\left( p^1_a \epsilon^1_b - p^1_b \epsilon^1_a  
 \right)\right.\nonumber\\
 &&\left.\left( p^2_b \epsilon^2_a - p^2_a \epsilon^2_b    
 \right)\left( p^3_c \epsilon^3_d - p^3_d \epsilon^3_c    
 \right)\left(  p^4_d \epsilon^4_c - p^4_c \epsilon^4_d  
 \right)+\left(1\rightarrow 3, 2\rightarrow 4\right)\right)\nonumber\\
 &&+\frac{1}{8}\cf^{O_{\bf 3}^{D=4}}(u,s) \left(4*(p_1\wedge \epsilon_1\wedge p_3 \wedge \epsilon_3) \left( p^2_\mu \epsilon^2_\nu - p^2_\nu \epsilon^2_\mu  
 \right)\left( p^4_\mu \epsilon^4_\nu - p^4_\nu \epsilon^4_\mu  
 \right)\left( p^1_a \epsilon^1_b - p^1_b \epsilon^1_a  
 \right)\right.\nonumber\\
 &&\left.\left( p^3_b \epsilon^3_a - p^3_a \epsilon^2_b    
 \right)\left( p^2_c \epsilon^2_d - p^2_d \epsilon^2_c    
 \right)\left(  p^4_d \epsilon^4_c - p^4_c \epsilon^4_d  
 \right)+\left(1\rightarrow 2, 3\rightarrow 4\right)\right)\nonumber\\
 &&+\frac{1}{8}\cf^{O_{\bf 3}^{D=4}}(s,t) \left(4*(p_1\wedge \epsilon_1\wedge p_4 \wedge \epsilon_4) \left( p^3_\mu \epsilon^3_\nu - p^3_\nu \epsilon^3_\mu  
 \right)\left( p^2_\mu \epsilon^2_\nu - p^2_\nu \epsilon^2_\mu  
 \right)\left( p^1_a \epsilon^1_b - p^1_b \epsilon^1_a  
 \right)\right.\nonumber\\
 &&\left.\left( p^4_b \epsilon^4_a - p^4_a \epsilon^4_b    
 \right)\left( p^3_c \epsilon^3_d - p^3_d \epsilon^3_c    
 \right)\left(  p^2_d \epsilon^2_c - p^2_c \epsilon^2_d  
 \right)+\left(1\rightarrow 2, 3\rightarrow 4\right)\right)\nonumber\\
 S^{O_{\bf S}^{D=4}}&=& \frac{1}{16}\left(\cf^{O_{\bf S}^{D=4}}(t,u) \right) \nonumber\\
 &&\left( 2*(p^2\wedge p^3\wedge p^1\wedge\epsilon^1) 
 \left(p^2_c \epsilon^2_d - p^2_d \epsilon^2_c\right)\left(p^3_d \epsilon^3_e - p^3_e \epsilon^3_d\right)\left(p^4_e \epsilon^4_c - p^4_c \epsilon^4_e\right)\left(p^1_\a \epsilon^1_\b - p^1_\b \epsilon^1_\a\right)p^2_\a\left(p^2_\g \epsilon^2_\d - p^2_\d \epsilon^2_\g\right)\right.\nonumber\\
 &&\left.p^3_\b\left(p^3_\d \epsilon^3_\mu - p^3_\mu \epsilon^3_\d\right)\left(p^4_\d \epsilon^4_\g - p^4_\g \epsilon^4_\d\right)\right.\nonumber\\
 &&\left. +2*(p^1\wedge p^3\wedge p^2\wedge\epsilon^2) 
 \left(p^1_c \epsilon^1_d - p^1_d \epsilon^1_c\right)\left(p^3_d \epsilon^3_e - p^3_e \epsilon^3_d\right)\left(p^4_e \epsilon^4_c - p^4_c \epsilon^4_e\right)\left(p^2_\a \epsilon^2_\b - p^2_\b \epsilon^2_\a\right)p^1_\a\left(p^1_\g \epsilon^1_\d - p^1_\d \epsilon^1_\g\right)\right.\nonumber\\
 &&\left.p^3_\b\left(p^3_\d \epsilon^3_\mu - p^3_\mu \epsilon^3_\d\right)\left(p^4_\d \epsilon^4_\g - p^4_\g \epsilon^4_\d\right)\right.\nonumber\\
 &&\left. + 2*(p^2\wedge p^1\wedge p^3\wedge\epsilon^3) 
 \left(p^2_c \epsilon^2_d - p^2_d \epsilon^2_c\right)\left(p^1_d \epsilon^1_e - p^1_e \epsilon^1_d\right)\left(p^4_e \epsilon^4_c - p^4_c \epsilon^4_e\right)\left(p^3_\a \epsilon^3_\b - p^3_\b \epsilon^3_\a\right)p^2_\a\left(p^2_\g \epsilon^2_\d - p^2_\d \epsilon^2_\g\right)\right.\nonumber\\
 &&\left.p^1_\b\left(p^1_\d \epsilon^1_\mu - p^1_\mu \epsilon^1_\d\right)\left(p^4_\d \epsilon^4_\g - p^4_\g \epsilon^4_\d\right)\right.\nonumber\\
 &&\left. +2*(p^2\wedge p^3\wedge p^4\wedge\epsilon^4) 
 \left(p^2_c \epsilon^2_d - p^2_d \epsilon^2_c\right)\left(p^3_d \epsilon^3_e - p^3_e \epsilon^3_d\right)\left(p^1_e \epsilon^1_c - p^1_c \epsilon^1_e\right)\left(p^4_\a \epsilon^4_\b - p^4_\b \epsilon^4_\a\right)p^2_\a\left(p^2_\g \epsilon^2_\d - p^2_\d \epsilon^2_\g\right)\right.\nonumber\\
 &&\left.p^3_\b\left(p^3_\d \epsilon^3_\mu - p^3_\mu \epsilon^3_\d\right)\left(p^1_\d \epsilon^1_\g - p^1_\g \epsilon^1_\d\right) \right) 
 \end{eqnarray}
 The most general descendant corresponding to the S-matrix \eqref{oddsmatd4grav} is given by
 
 \begin{eqnarray}\label{oddsmatd4grav}
 L^{O_{\bf 3}^{D=4}}&=& \sum_{m, n} \left(\cf^{O_{\bf 3}^{D=4}}\right)_{m.n} 2^{m+n}\left(\prod_{i=1}^m \prod_{j=1}^n 
 \epsilon^{abij} (\partial_{\mu_i} \partial_{\nu_j}R_{abef})R_{ijfe}\partial^{\mu_i}R_{cdgh}\partial^{\nu_j}R_{dchg}\right),\\
 L^{O_{\bf S}^{D=4}}&=& \sum_{m, n} \left(\cf^{O_{\bf S}^{D=4}}\right)_{m.n} 2^{m+n}\left(\prod_{i=1}^m \prod_{j=1}^n 
 \epsilon^{abmn}(\partial_{\mu_i} \partial_{\nu_j}R_{mngh})\partial^{\mu_i} \partial_a \partial_g R_{cdij}\partial^{\nu_j}\partial_b \partial_h R_{dejk} R_{ecki}\right),\nonumber
 \end{eqnarray}
  
It can be easily seen that the descendant Lagrangian \eqref{oddlagd4grav} generates the S-matrix \eqref{oddsmatd4grav}. Consider the following parity odd Lagrangians
$$\epsilon^{abij} R_{abef}R_{ijfe}R_{cdgh}R_{dchg}, \qquad \epsilon^{abmn}R_{mngh}\partial^a \partial^g R_{cdij}\partial^b \partial^h R_{dejk} R_{ecki}.$$               
Roughly these structures, when linearized, can be represented as the tensor product of the electromagnetism structures. The symmetries also become manifest.
$$\text{Tr}(F^1\wedge F^2)\text{Tr}(F^3F^4)(\text{Tr}(F^1F^2)\text{Tr}(F^3F^4)),\qquad (*F^1_{ab})\text{Tr}(\partial_a F^2\partial_a F^3F^4))((F^1_{ab})\text{Tr}(\partial_a F^2\partial_a F^3F^4)).$$
plus permutations.
Therefore \eqref{oddlagd4grav} generates the S-matrix \eqref{oddsmatd4grav}.

\section{Sample Exchange contributions to four point scattering}\label{sampexch}

\subsection{Most general exchange contribution to four 
	scalar scattering} 

Consider a massive particle in a representation ${\cal P}$ of 
its Lorentz little group $SO(D-1)$. Such a particle can have a nonzero three point functions with either two scalars, two photons or two gravitons, only if 
${\cal P}$ is vectorial rather than spinorial in nature. It follows that 
we can have exchange contributions to four scalar, four photon or four 
graviton scattering only when the exchanged particle transforms in 
a representation ${\cal P}$ that can be built 
by symmetrizing and anti-symmetrizing vectors (and then removing traces from the symmetric parts). Any such representation 
is labelled by a Young Tableaux. 

Now consider a particle with momentum $p_3$ in such a representation of the little group. On-shell states of this particle are labelled by a polarization tensor, whose indices 
are all orthogonal to $p_3$ and are otherwise appropriately 
symmetrized/ anti-symmetrized and trace removed. Now consider the on-shell 3 point function between such a particle and two massless scalars  with momentum $p_1$ and $p_2$. As the amplitude is Lorentz invariant, all indices of the polarization tensor associated with the particle ${\cal P}$ have to be contracted with $p_1-p_2$, the only vector available (recall that the  polarization tensor is orthogonal to $p_1+p_2=-p_3$). As all indices of the polarization tensor are contracted with the same vector, it follows that the three point function vanishes unless 
${\cal P}$ is labelled by a completely symmetric  Young Tableaux, i.e. a Tableaux  with a single row, 
i.e. is a traceless symmetric tensor, $e_{\mu_1 \ldots \mu_l}$.  
In this case the three point function of two massless scalars 
with this representation is necessarily proportional to 
$$  (p_1 -p_2)^\mu_1 (p_1-p_2)^\mu_2 \ldots (p_1-p_2)^\mu_l e_{\mu_1 \ldots \mu_l}.$$
This three point function has the necessary Bose symmetry under 
interchange of 1 and 2 (and so is nonzero) only if $l$ is even. 

The  propagator for a massive 
spin $l$ particle is given by 
$$ \frac{{\cal P}_{\alpha_1 \alpha_2 ...\alpha_l}^{\beta_1 \beta_2 ...\beta_l}}{p^2+m^2} $$
where ${\cal P}$ is the projection operator that first projects 
all vector indices orthogonal to the momentum $p$ that runs 
through the propagator,  and then projects
the resulting  $(D-1)^l$ dimensional Hilbert space (generated by 
the direct product of the $l$ vectors orthogonal to $p$) onto the subspace of spin $l$ traceless symmetric tensors. 
It follows that the scalar four point function that 
results from the exchange of a spin $l$ particle is 
proportional to 
\begin{equation}\label{spinsfp}
\frac{ (p_1-p_2)_{\alpha_1}
	\ldots (p_1 -p_2)_{\alpha_l}	{\cal P}_{\alpha_1 \alpha_2 ...\alpha_l}^{\beta_1 \beta_2 ...\beta_l} (p_3-p_4)_{\beta_1} \ldots (p_3 -p_4)_{\beta_l} }{(p_1+p_2)^2+ m^2}
\end{equation}
The explicit formula \cite{PhysRevD.9.898, Ingraham:1974un} for the propagator  ${\cal P}_{\alpha_1 \alpha_2 ...\alpha_l}^{\beta_1 \beta_2 ...\beta_l}$ is complicated by the requirement of tracelessness. This requirement determines the projector to be \footnote{S.M. would like to  thank P. Nayak, R. Sinha R. Soni and  R. Poojari for detailed discussions - and initial collaboration on a related project \cite{Nayak:2017qru} - on contributions to the S matrix of tree level exchanges to higher spin exchanges, that proved useful while writing this section.}  
\begin{eqnarray} \label{proj}
{\cal P}^{(s)}_{\mu_1\dots\mu_l,\nu_1\dots\nu_l}&=&\Bigg \{ \underset{p=0}{\overset{[l/2]}{\sum}}\frac{(-1)^p l!(2l + D -2p-5)!!}{2^pp!(l-2p)!(2l + D -5)!!} 
\Theta_{\mu_1\mu_2}\Theta_{\nu_1\nu_2}\dots\Theta_{\mu_{2p-1}\mu_{2p}}\Theta_{\nu_{2p-1}\nu_{2p}}\cr
&&\hspace{1cm} \times \Theta_{\mu_{2p+1}\nu_{2p+1}}\dots \Theta_{\mu_s,\nu_s} \Bigg \}_{\text{sym}(\mu),\text{sym}(\nu)},
\end{eqnarray}
where 
\be\label{kkk}
\Theta_{\mu\nu}=\eta_{\mu\nu}-p_\mu p_\nu/p^2
\ee
is the projector onto the spatial slice orthogonal to $p_\mu$ and $[l/2]$ is  largest integer smaller than or equal to $l/2$ and ${\text{sym}(\mu)}$ denotes the 
operation of symmetrizing over the $l$ $\mu$ indices, i.e the operation 
$ \frac{1}{l!} \sum_{P}$ where the sum is over all $l!$ permutations $P$ 
of the indices $\mu_i$ and  ${\text{sym}(\nu)}$ means 
something similar. 
Let us define 
\begin{equation}\label{abcdef}
\begin{split}  
&a= (p_1-p_2)^\mu \Theta_{\mu\nu}  (p_1-p_2)^\nu\\
&b=(p_3-p_4)^\mu \Theta_{\mu\nu} (p_3-p_4)^\nu\\
&c=(p_1-p_2)^\mu \Theta_{\mu\nu} (p_3-p_4)^\nu\\
\end{split}
\end{equation}
It follows from \eqref{spinsfp} and \eqref{proj} that 
the spin $l$ exchange is given, up to a proportionality 
constant, by \cite{PhysRevD.9.898, Ingraham:1974un}
A simple computation yields 
\begin{equation}\label{abc} 
a=s, ~~~b=s, ~~~~c=u-t, ~~~~
\end{equation}
\begin{equation}\label{spsex}
S =\frac{(ab)^\frac{l}{2} }{(p_1+p_2)^2+ m^2} A_{l}^{D}\bigg(\frac{c}{\sqrt{ab}}\bigg) ~~+{\rm crossing} 
\end{equation} 
where we have defined the polynomial
\be
A_{l}^{D}(x)\equiv \underset{p=0}{\overset{[l/2]}{\sum}}\frac{(-1)^p l!(2l + D -2p-5)!!}{2^pp!(l-2p)!(2l + D -5)!!}	\left( x \right)^{l-2p}= \frac{1}{2^l (\frac{D-3}{2})_l}C_{l}^{\frac{D-3}{2}}(x),\qquad (a)_b\equiv \frac{\Gamma(a+b)}{\Gamma(a)}.
\ee
which is proportional to the Gegenbauer polynomial $C_{l}^{\frac{D-3}{2}}(x)$.
Substituting \eqref{abc} into \eqref{spsex} we obtain 
\be\label{spsex1} 
S \propto \frac{ s^l }{m^2-s}A_l^D\bigg(\frac{u-t}{s}\bigg)+\frac{ t^l }{m^2-t}A_l^D\bigg(\frac{s-u}{t}\bigg)+\frac{ u^l }{m^2-u}A_l^D\bigg(\frac{t-s}{u}\bigg).
\ee

\subsubsection*{Regge growth} \label{rgs} 

In the Regge limit the first and third lines in \eqref{spsex1} - that correspond to $s$ channel and $u$ channel exchange -  are qualitatively different from the
second line ($t$ channel exchange).

The leading behavior of the $t$ channel graph is obtained 
from the term $p=0$ in the second line of \eqref{spsex1}. 
and is given by $(s^l)/( m^2-t)$. Note that this 
expression is not a polynomial in $t$. As `impact parameter 
space' is roughly Fourier conjugate to $t$ space, it follows that $t$ channel exchange leads to non-trivial  scattering 
at finite impact parameter in the Regge limit. Note the 
contrast with contact interactions (studied in detail 
earlier in this paper). As the S-matrices that follow from 
contact interactions are polynomial in $s$ as well as 
in $t$, such interactions contribute to scattering in the 
Regge limit only at zero impact parameter. 

Note that, on the other hand, that at leading order in the Regge limit $s$ and $u$ channel scattering amplitudes are proportional to $s^{l-1}$ and $u^{l-1}$. Recall that since $l$ is even, the leading terms from the s and u channel cancel with each other. However the first sub-leading terms 
yield an S-matrix  proportional to 
$$2m^2 s^{l-2}+ 3(l-2p)s^{l-2}t$$
This behavior is analytic in $t$ and so contributes 
only to scattering at zero impact parameter. The 
leading large $s$ behavior is of the sort that can be 
reproduced (and so also cancelled) by a local `counter-term' whenever $l \geq 2$. 

At any rate  the second term of \eqref{spsex1} gives the dominant growth 
of the S-matrix in the Regge limit, and this 
growth is proportional to  $s^{l}$. Note in particular 
that the exchange of a massive spin zero scalar leads to 
scattering with Regge growth $s^0$, while the 
exchange of a massive spin 2 particle leads to an amplitude that grows in the Regge limit like $s^2$. 
All higher spin exchanges yield scattering amplitudes that 
grow faster than $s^2$ in the Regge limit. As the corresponding term is non analytic in $t$ it cannot be 
mimicked or cancelled by a local counter-term. 

It is often asserted that scattering due to exchange of a spin $l$ particle leads to a scattering amplitude that 
grows like $s^l$ in the Regge limit. As we have already 
seen in the simple example of scalar scattering, this 
claim is really correct only in the $t$ channel - the channel that contributes to scattering at nonzero impact parameter. Amplitudes in the $s$ and $u$ 
channels need not grow like $s^l$. In the example we have 
seen so far the growth in these channels is slower than 
$s^l$; below we will also encounter examples in which the 
growth is faster.

\subsubsection{Angular dependence and spherical harmonics}

In this subsubsection we will rewrite the first of the three terms \eqref{spsex1} in the center of mass frame, i.e. the 
frame in which $p_1+p_2$ is a vector that points in the time direction.  In this frame the $\Theta$ projects
onto space. $p_1-p_2$ and $p_3-p_4$ are purely spatial vectors.
In this frame $|{\vec p}_1|=|{\vec p}_2|=p$ and
\begin{equation}\label{varval}
s=4p^2, ~~~t=2p^2(1-\cos \theta), ~~~ u=
2p^2(1+\cos \theta),~~~{\rm ~so~} \frac{u-t}{s}= 
\cos \theta
\end{equation} 
where $\theta$ is the scattering angle.

For the purposes of evaluating \eqref{spinsfp}, 
$\Theta_{\mu\nu} \rightarrow \delta_{ij}$ ($i$ and $j$ are spatial indices). In this frame the projector listed in \eqref{proj} is a traceless symmetric tensor in the indices 
$\mu_i$ for any fixed values of the indices $\nu_j$. It follows that the projector is a spin $l$ scalar spherical harmonic of $SO(D-1)$. The precise spherical harmonic obtained depends on the value of the indices $\nu_i$.
In particular, if we dot all $\nu$ indices of the projector with a fixed vector ${\vec \beta}$ we obtain 
the  unique 
spherical harmonic invariant under all the $SO(D-2)$ rotations that keep ${\vec \beta}$ fixed\footnote{If we choose the fixed vector to be the `z' axis we find the generalization of the $m=0$ spherical harmonic of $SO(3)$.}. If we now dot the the $\mu_i$ 
indices with a second vector ${\vec \alpha}$ such that the angle between ${\vec \beta}$ and ${\vec \alpha}$ is $\theta$, we obtain $|\alpha|^l |\beta|^l$ times the value of this special spherical harmonic at angle $\theta$. Plugging in the explicit formula 
for the projector yields the following explicit formula for the `rotationally symmetric spherical harmonic' as a function of $\cos \theta$
$$A_l^D(\cos\theta) = \underset{p=0}{\overset{[l/2]}{\sum}}\frac{(-1)^p l!(2l + D -2p-5)!!}{2^pp!(l-2p)!(2l + D -5)!!}\left( \cos(\theta)  \right)^{l-2p}$$
Using the last of  \eqref{varval} we see that the the first term in \eqref{spsex1} is proportional to 
$$s^l \frac{A_l(\cos\theta)}{s-m^2} .$$
where $\theta$ is the scattering angle \cite{Bekaert:2009ud}.

\subsection{Four photon scattering from massive 
	spin $s$ exchange}

The three particle photon-photon-${\cal P}$ S-matrix  is, 
in general,  nonzero when ${\cal P}$ belongs to a larger class of representations than the traceless symmetric tensors. Although it is not difficult to enumerate the 
representations ${\cal P}$ which can consistently couple 
to two photons, we will not pause to record the results 
of this exercise, postponing it (and several other aspects of 
a systematic enumeration of exchange contributions 
to four photon scattering) to future work. In this subsection we simply focus on the special case for which ${\cal P}$ is traceless symmetric\footnote{As the $D=4$ little group is $SO(3)$, traceless
	symmetric tensors are the only vectorial representations 
	that exist in this case, so  the limited analysis presented in this subsection is actually exhaustive 
	for the case $D=4$.}. 
Without loss of information we can choose the polarization 
of the exchanged particle to take the form 
$$e_{\mu_1 \ldots \mu_l} =(\epsilon_3)_{\mu_1}
(\epsilon_3)_{\mu_2} \ldots (\epsilon_3)_{\mu_l},~~~~~~
(\epsilon_3)_\mu k_3^\mu=0, ~~~~~\epsilon_3 . \epsilon_3=0$$
We will now construct the most general 3 point coupling 
between two photons and a spin $l$ particle.  Let the two photons have 
polarizations and momenta $\epsilon_1,~ k_1$ and 
$\epsilon_2, ~k_2$. Let the spin $l$ particle have 
mass $m$ and momentum $k_3$. The photon photon spin $l $
on-shell 3 point functions are necessarily polynomials 
of the following Lorentz invariant building blocks
\begin{align}
&A_1= \epsilon_1.(k_2-k_3), ~A_2= \epsilon_2.(k_1-k_3),~A_3=\epsilon_3.(k_1-k_2),\nonumber\\
&b_{12}=\epsilon_1.\epsilon_2,~~~~~b_{23}=\epsilon_2.\epsilon_3,~~~~~b_{13}=\epsilon_1.\epsilon_3,\nonumber \\
\end{align}
The most general polynomial of degree 1 in 
$\epsilon_1$ and $\epsilon_2$ and of degree $l$ in 
$\epsilon_3$ is a linear combination of the five structures 
\begin{equation}\label{mjtf}
A_1 A_2 A_3^l, ~~~b_{12}  A_3^l~~~
b_{13} A_2 A_3^{l-1}~~~b_{23} A_1 A_3^{l-1}~~~b_{13} b_{23} A_3^{l-2}
\end{equation} 
Under the interchange 
\begin{equation}\label{interchange}
(\epsilon_1, k_1) \leftrightarrow (\epsilon_2, k_2)
\end{equation}
\begin{equation}\label{inteff} A_1 \leftrightarrow A_2,~b_{12} \leftrightarrow
b_{12}, ~b_{13} \leftrightarrow b_{23}, ~A_3 \leftrightarrow -A_3
\end{equation}
When $l$ is odd, the only Bose symmetric linear combination
of the five structures listed in \eqref{mjtf} is 
proportional to  
\begin{equation}\label{odds}
(A_2 b_{13}+A_1 b_{23}) A_3^{l-1}
\end{equation} 
When $l$ is even, on the other hand, any 
linear combination of the following four structures 
\begin{equation}\label{mjtfse}
A_1 A_2 A_3^l, ~~~b_{12}  A_3^l~~~
b_{13} A_2 A_3^{l-1} -b_{23} A_1 A_3^{l-1}~~~b_{13} b_{23} A_3^{l-2}
\end{equation} 
is Bose symmetric. 

We now turn to the constraints imposed by the photonic gauge invariance. Under gauge transformation $\epsilon_i \rightarrow \epsilon_{i}+ c_{i} k_i $ the building blocks transform as   
\begin{equation}\label{gtpar} \begin{split}
& \delta A_1 = -c_1 m^2, ~ \delta A_2 = -c_2 m^2, ~ \delta A_3 = 0, \\
& \delta  b_{12} = \frac{1}{2} c_1 A_2 + \frac{1}{2} c_2 A_1, ~ \delta b_{23} = -\frac{1}{2} c_2 A_3, ~ 
\delta b_{13} = \frac{1}{2} c_1 A_3
\end{split}
\end{equation}
It is easily verified that the odd structure \eqref{odds} 
is not gauge invariant. As a consequence the three point 
functions between two photons and a traceless symmetric 
$l$ rank tensor vanishes when $l$ is odd.  

When $l$ is even, on the other hand, it is easily verified
that the most general Bose symmetric gauge invariant 3 point function 
is a linear combination of the two structures 
$C_1$ and $C_2$ where 
\begin{equation}\label{combinations}
\begin{split}
&C_1= A_3^l \left(A_1 A_2 +2 m^2 b_{12} \right) \\
&C_2= A_3^{l-2} \left( -b_{12} A_3^2 + 
\left( b_{13} A_2  -b_{23} A_1 \right) A_3 + 2 m^2 
b_{13} b_{23}  \right) \\
\end{split} 
\end{equation} 
$C_1$ is proportional to the S-matrix that follows from the Lagrangian 
$$ \left( \partial_{\mu_1} \ldots \partial_{\mu_l} F_{\alpha \beta }
\right) F^{\alpha \beta} S_{\mu_{1} \ldots \mu_{l}}$$
while $C_2$ is proportional to the S-matrix that follows from the Lagrangian  
$$ \left(  \partial_{\mu_3} \ldots \partial_{\mu_l} F_{\mu_1 \alpha} \right)
F_{\mu_2 \alpha} S_{\mu_1 \ldots \mu_l}$$ 
where $S_{\mu_1 \ldots \mu_{l}}$ represents the linearized 
spin $l$ field. 

The four point function that follows by stitching two 
$C_1$ exchanges through the spin $l$ propagator is particularly simple\footnote{The most general exchange 
	contribution of a massive spin $l$ particle is given 
	by stitching together two three point functions of the 
	form $a C_1 + b C_2$. We leave the analysis of this 
	general case to future work. }. 
It is proportional to $E_{{\bf 3},1}$. The S-matrix is 
\eqref{expparphdte} with the function $\cf^{E_{{\bf 3},2}}(t,u)=0$, $\cf^{E_{{\bf S}}}(t,u)=0$ but 
$$\cf^{E_{{\bf 3},1}}(t,u) =
\frac{ c \,s^l }{m^2-s}  
A_{l}^D \bigg(\frac{u-t}{s}\bigg).$$
where $c$ is a real number.

\subsubsection*{Regge growth} 
Unlike the scalar case the analysis here is a bit different because of the momenta dependence of the tensor structure $E_{{\bf 3},1}$ multiplying $\cf^{E_{{\bf 3},1}}$. The Regge limit i.e. large $s$ at fixed $t$ limit of the S-matrix yields the following behavior for each channel.  
\begin{eqnarray}
\CS_s &\to &s^{l-1}\Big(s^2e_{{\bf S}} -s^2 e_{{\bf 3},2}^{(1)}+s^2 e_{{\bf 3},1}^{(1)}\Big)  \nonumber\\ 
\CS_t &\to &\frac{s^{l}}{m^2-t}\Big(t^2e_{{\bf S}} -t^2 e_{{\bf 3},2}^{(2)}+t^2 e_{{\bf 3},1}^{(2)}\Big) \nonumber\\ 
\CS_u &\to &- s^{l-1}\Big(s^2e_{{\bf S}} -s^2 e_{{\bf 3},2}^{(3)}+s^2 e_{{\bf 3},1}^{(3)}\Big)  \\
\CS_s + \CS_u &\to & s^{l+1} \left( \left(e_{{\bf 3},2}^{(3)}-e_{{\bf 3},2}^{(1)}
\right) + \left( e_{{\bf 3},1}^{(1)} - e_{{\bf 3},1}^{(3)} \right)  \right). 
\end{eqnarray}

As in the case of 4 scalar scattering, the $t$ channel 
exchange gives rise to a term that scales like $s^l$; 
the coefficient of this term is non analytic in $t$. It follows that the exchange contribution for $l > 2$ 
grows faster than $s^2$ (in a way that cannot be cancelled by a local counter-term). Of course the $t$ channel exchange contributions from $l=2$ and $l=0$ grow no 
faster than $s^2$.

Let us now study $s$ and $u$ channel exchanges. Unlike the case of scalars, these channels yield the dominant 
contribution - $\propto s^{l+1}$ - to photon scattering. As in the case of scalar scattering subsection \ref{rgs}, the contributions from $s$ and $u$ exchange appear with opposite signs. In this case, however, the two contributions appear with different polarization dependences, and hence do not cancel for generic choice of the polarization vectors. As for the 
scalars, however, this leading behavior can be cancelled 
by a contact term proportional to 
$$ {\rm Tr} 
\left( \partial_{\mu_1} \ldots \partial_{\mu_{l-1}} F 
\partial_{\mu_1} \ldots \partial_{\mu_{l-1}} F \right) 
{\rm Tr} 
\left( F^2 \right) $$
provided that $l \geq 2$. It follows that the combined 
contribution of spin $l$ exchange plus the canceling counter-term grows no faster than $s^2$ for $l=2$ 
(the same is true for $l=0$; in this case there is no 
canceling counter-term).

\subsection{Massive spin $l$ exchange contribution to 4 graviton scattering}

The scalar Lagrangian \eqref{massive scalar Lagrangian} and 
the Spin 2 Lagrangian \eqref{Lagrangian B} are the first 
two members of a one parameter Lagrangians that describe 
the three point coupling between massive spin $l$ particles 
- i.e. massive particles transforming in the traceless symmetric $l$ index representation of $SO(D-1)$ with even $l$- 
with two gravitons. The Lagrangians in 
question are 
\begin{equation}\label{hspin}
\nabla_{\mu_1}\nabla_{\mu_2}......\nabla_{\mu_l}R_{abcd}R^{abcd}S^{\mu_1 \mu_2 . . . . . . .\mu_l}
\end{equation} 
In this subsubsection we compute the contribution of exchange 
amplitudes sewing two of these three point functions to 
4 graviton scattering\footnote{Two gravitons cannot couple to spin $l$ particles when $l$ is odd. When $l$ is even there are 3 independent 
	coupling structures. In this subsubsection we focus on 
	only one of these three structures as an example. We hope 
	to return to the general case in future work.}.
The resultant S-matrix clearly has the same index structure 
as the Lagrangian $(R_{abcd}R^{abcd})^2$. As a consequence, 
when the S-matrix is written in the form \ref{b5smat}, the 
only nonzero function is $\cf^{G_{{\bf 3},1}}(t,u)$. Explicitly we find 

\be
\cf^{G_{{\bf 3},1}}(t,u)=\frac{ s^l }{m^2-s} A_l^D\bigg(\frac{u-t}{s}\bigg).
\ee

\subsubsection*{Regge growth}
Here we analyze the Regge behavior from sewing the three point functions arising from \eqref{hspin}. We list the behavior channel by channel :
\begin{align}
\CS_s\to  &s^{l-1}\Big(s^4(-1+\epsilon_1^\perp\cdot \epsilon_2^\perp)^2(-1+\epsilon_3^\perp\cdot \epsilon_4^\perp)^2\Big)  \\ 
\CS_t\to  &s^l\Big(t^4(-1+\epsilon_1^\perp\cdot \epsilon_3^\perp)^2(-1+\epsilon_2^\perp\cdot \epsilon_4^\perp)^2\Big)  \\
\CS_u\to  & - s^{l-1}\Big(s^4(-1+\epsilon_1^\perp\cdot \epsilon_4^\perp)^2(-1+\epsilon_2^\perp\cdot \epsilon_3^\perp)^2\Big)  
\end{align}
As expected, the contribution from the $t$ channel scales 
like $s^l$. For $l\geq 4$ this manifestly non-local  contribution grows faster than $s^2$. The special 
cases $l=0$ and $l=2$ have already been analyzed in 
detail above; recall that we found that in these cases
the combined contribution of $s$ and $u$ channel exchange
grows faster than $s^2$ even after any possible local 
counter-term subtraction.

\end{document}